\documentclass[12pt, a4paper, twoside, english, final]{ETHthesis}

\usepackage[12pt]{ETHthesis}
\usepackage{amsmath, amssymb, amsthm}
\usepackage{hyperref}
\usepackage{hyperxmp}
\hypersetup{
    pdfauthor={Wenqi Jiang},
    pdftitle={Vector-Centric Machine Learning Systems: A Cross-Stack Approach},
    pdfsubject={A thesis submitted to attain the degree of Doctor of Sciences to ETH Zurich},
    pdfinfo={
        Copyright={Copyright (C) 2025, Wenqi Jiang. Permission to print for personal and academic use, as well as permission for electronic reproduction and dissemination in unaltered and complete form are granted. All other rights reserved.}
    },
    pdflang={en},
    colorlinks=false,
    hidelinks
}

\usepackage{graphicx}
\usepackage{booktabs}
\usepackage{xcolor}    
\usepackage{array}     

\usepackage{cleveref}
\usepackage{pifont}
\usepackage{multirow}
\usepackage{enumitem}
\usepackage{algorithm}
\usepackage{algpseudocode} 

\usepackage{float}
\usepackage{subcaption}
\usepackage{soul}  
\usepackage[NoDate]{currvita} 
\usepackage[numbers]{natbib}  
\usepackage[T1]{fontenc}

\usepackage{color}
\usepackage{lmodern}
\usepackage{titlesec}
\definecolor{light-gray}{gray}{0.65}

\titleclass{\subsubsubsection}{straight}[\subsubsection]
\newcounter{subsubsubsection}[subsubsection]
\renewcommand\thesubsubsubsection{\thesubsubsection.\arabic{subsubsubsection}}
\titleformat{\subsubsubsection}
  {\normalfont\normalsize\bfseries}{\thesubsubsubsection}{1em}{}
\titlespacing*{\subsubsubsection}{0pt}{3.25ex plus 1ex minus .2ex}{1.5ex plus .2ex}

\usepackage{tabularx}
\newcolumntype{M}[1]{>{\centering\arraybackslash}m{#1}} 
\newcolumntype{R}[1]{>{\raggedleft\arraybackslash}m{#1}} 
\newcolumntype{L}[1]{>{\raggedright\arraybackslash}m{#1}} 

\titleformat{\chapter}
   [display]
   {\bfseries\Large}
   {\filleft{\fontsize{80}{100}\selectfont\color{light-gray}\thechapter}}
   {-25pt}
   {\filleft\fontsize{26}{35}\selectfont}
   [\vspace{5pt}\titlerule]

\iftwoside
\else
\fi
\newcommand{\cleardoublepageempty}{\clearpage\thispagestyle{empty}\cleardoublepage}

\floatstyle{boxed}
\newfloat{algo}{t!}{alg}
\floatname{algo}{Algorithm}

\usepackage{pdflscape} 
\usepackage{tikz}
\usepackage{epsfig,color,endnotes,alltt}
\usepackage{listings}
\usepackage[toc,page]{appendix}
\usepackage{bytefield}
\usepackage[most]{tcolorbox}

\newtcolorbox[auto counter]{summary}[1][]{title={\bfseries Summary~\thetcbcounter},enhanced,drop shadow={black!50!white},
  coltitle=black,
  top=0.3in,
  attach boxed title to top left=
  {xshift=1.5em,yshift=-\tcboxedtitleheight/2},
  boxed title style={size=small,colback=pink},#1}

\definecolor{dkgreen}{rgb}{0,0.6,0}
\definecolor{gray}{rgb}{0.5,0.5,0.5}
\definecolor{mauve}{rgb}{0.58,0,0.82}

\newcommand{\ballnumber}[1]{\tikz[baseline=(myanchor.base)] \node[circle,fill=.,inner sep=1pt] (myanchor) {\color{-.}\bfseries\footnotesize #1};}
\newcommand{\blueballnumber}[1]{\tikz[baseline=(myanchor.base)] \node[circle,fill=blue!80,inner sep=1pt] (myanchor) {\color{-.}\bfseries\footnotesize #1};}

\newcommand{\whiteballnumber}[1]{\tikz[baseline=(myanchor.base)] \node[circle,fill=black!0,inner sep=1pt, draw=black!100, thick] (myanchor) {\color{black}\bfseries\footnotesize #1};}

\usepackage{xspace}
\newcommand{\ours}{\texttt{RAGO}\xspace}
\newcommand{\ragdesc}{\texttt{RAGSchema}\xspace}
\newcommand{\oursemph}{\textbf{\texttt{RAGO}}\xspace}
\newcommand{\ragdescemph}{\textbf{\texttt{RAGSchema}}\xspace}
\newcommand{\niparagraph}[1]{\vspace{2pt}\noindent\textbf{#1}}
\newcommand{\egc}{e.g.,\xspace}
\newcommand{\iec}{i.e.,\xspace}
\newcommand{\bench}[1]{\textsf{#1}\xspace}
\newcommand{\xx}[1]{\textsf{#1}\xspace}

\makeatletter
\newcommand{\myLabeledItem}[1][]{
        \protected@edef\@currentlabel{#1}%
\item[#1]
}
\makeatother

\initETHthesis

\begin{document}


\dissnum{31293}
\title{Vector-Centric Machine Learning Systems: A Cross-Stack Approach}
\author{Wenqi Jiang}
\dateofbirth{23.11.1996}
\citizen{China}
\examiners{
    Prof.\,Dr. Gustavo Alonso
    \linebreak
    Prof.\,Dr. Torsten Hoefler
    \linebreak
    Prof.\,Dr. Ana Klimovic
    \linebreak
    Prof.\,Dr. Christos Kozyrakis
}

\pagestyle{empty}
\maketitle

\pagestyle{plain}
\pagenumbering{roman}
\sloppy
\chapter*{Abstract}



The advancement of computing infrastructure has been a key driver of recent  machine learning (ML) breakthroughs.
With hundreds of billions of dollars invested into ML software and hardware infrastructure every year, the efficiency of machine learning systems is more important than ever.

Today, two major trends are shaping the evolution of ML systems. 

First, \textit{modern AI systems are becoming increasingly complex, often integrating components beyond the model itself --- vector data systems play a critical role.} 
A notable example is \textit{Retrieval-Augmented Generation} (RAG), which incorporates not only multiple model components but also retrieval systems based on \textit{vector search on vector databases.}
The heterogeneity of both system components (models and vector data systems) and the underlying hardware (ML accelerators for models versus alternative hardware for retrievals) sets these workloads apart from conventional model serving and training.  

Second, with the end of Moore's Law, \textit{computer systems with specialized or heterogeneous hardware are becoming increasingly prevalent.}  
The presence of diverse hardware components—with varying compute capabilities, memory hierarchies, and interconnect architectures—introduces new challenges: achieving high system efficiency is no longer feasible without accounting for the rapid evolution of the hardware landscape.

Building on the two observations above, this thesis addresses \textit{three key research questions} around machine learning system efficiency. 
First, how can we design \textit{efficient systems for emerging, complex ML workloads such as RAG serving}? 
Second, how can we develop more \textit{efficient vector data systems and hardware}? 
Finally, how can we achieve \textit{synergistic optimizations across algorithms, systems, and hardware} in the post-Moore era?

To address these questions, \textit{this thesis adopts a cross-stack approach and makes three major contributions to improving ML system efficiency, presenting solutions that span algorithms, systems, and hardware.}  
First, it introduces several pioneering works about \textit{RAG serving efficiency} across the computing stack. \textit{PipeRAG} focuses on algorithm-level improvements, \textit{RAGO} introduces system-level optimizations, and \textit{Chameleon} explores heterogeneous accelerator systems for RAG.  
Second, this thesis investigates \textit{algorithm-hardware co-design for vector search}, which is essential not only in RAG systems but in search engines, recommender systems, etc. Specifically, \textit{FANNS} and \textit{Falcon} optimize quantization-based and graph-based vector search, the two most popular paradigms of retrieval algorithms. 
Third, this thesis addresses the serving efficiency of \textit{recommender systems}, another example of vector-centric ML systems, where the memory-intensive lookup operations on embedding vector tables often represent a major performance bottleneck. \textit{MicroRec} and \textit{FleetRec} propose solutions at the hardware and system levels, respectively, optimizing both data movement and computation to enhance the efficiency of large-scale recommender models.

The work presented in this dissertation is not a one-time effort but rather a foundational step toward the ongoing evolution of ML infrastructure.
It highlights the importance of (a) optimizing vector data systems within modern ML pipelines and (b) pursuing performance optimization across the computing stack.  
The ideas introduced in this thesis will serve as a solid foundation for the design and implementation of next-generation ML systems.

\chapter*{Zusammenfassung}

Der Fortschritt in der Recheninfrastruktur war ein entscheidender Treiber für die jüngsten Durchbrüche im Bereich des maschinellen Lernens (ML).  
Mit jährlichen Investitionen in Höhe von mehreren hundert Milliarden US-Dollar in ML-Software- und Hardwareinfrastrukturen steht außer Zweifel, dass die Effizienz von Systemen für maschinelles Lernen von großer Bedeutung ist.

Heute prägen zwei zentrale Trends die Weiterentwicklung von ML-Systemen.  
Erstens: \textit{Moderne KI-Systeme werden zunehmend komplexer und integrieren häufig mehrere Systemkomponenten.} Ein bemerkenswertes Beispiel ist \textit{Retrieval-Augmented Generation} (RAG), das nicht nur verschiedene Modellkomponenten, sondern auch Retrieval-Systeme umfasst.  
Die Heterogenität sowohl der Systemkomponenten als auch der zugrunde liegenden Hardware unterscheidet diese Arbeitslasten deutlich von herkömmlichem Model Serving und Training.  
Zweitens: Mit dem Ende von Moore’s Law wird \textit{spezialisierte und heterogene Hardware zunehmend verbreitet.}  
Die Vielfalt an Hardwarekomponenten — mit unterschiedlichen Rechenkapazitäten, Speicherhierarchien und Interconnect-Architekturen — stellt neue Herausforderungen dar: Eine hohe Systemeffizienz lässt sich nicht mehr erreichen, ohne die rasante Entwicklung der Hardwarelandschaft zu berücksichtigen.

Aufbauend auf den oben genannten Beobachtungen behandelt diese Dissertation \textit{drei zentrale Forschungsfragen} zur Effizienz von Systemen für maschinelles Lernen.  
Erstens: Wie lassen sich \textit{effiziente Systeme für neuartige und komplexe ML-Arbeitslasten wie RAG-Serving} entwerfen?  
Zweitens: Wie können wir \textit{effizientere Vektordatensysteme und Hardware} entwickeln?  
Und drittens: Wie lassen sich \textit{synergetische Optimierungen über Algorithmen, Systeme und Hardware hinweg} im Zeitalter nach Moore’s Law realisieren?

Zur Beantwortung dieser Fragen \textit{leistet diese Dissertation Beiträge in den folgenden drei Bereichen und präsentiert Lösungen, die Algorithmen, Systeme und Hardware umfassen. Sie zeigt damit die Notwendigkeit und Wirksamkeit eines bereichsübergreifenden Ansatzes zur Optimierung der Effizienz von ML-Systemen auf.}  
Erstens werden drei wegweisende Arbeiten zur \textit{Optimierung der RAG-Serving-Performance über die gesamte Rechenarchitektur hinweg} vorgestellt. \textit{PipeRAG} konzentriert sich auf Verbesserungen auf der Algorithmusebene, \textit{RAGO} führt Optimierungen auf Systemebene ein, und \textit{Chameleon} untersucht Hardwarebeschleunigung für großskalige Retrievals.  
Zweitens untersucht die Dissertation das Thema \textit{Algorithmus-Hardware-Co-Design für Vektorsuche}, eine zentrale Komponente moderner KI-Systeme. Konkret optimieren \textit{FANNS} und \textit{Falcon} quantisierungsbasierte und graphbasierte Vektorsuchverfahren, die zwei populärsten Paradigmen in der Retrieval-Algorithmenforschung.  
Drittens behandelt diese Arbeit die \textit{effiziente Verwaltung von Embedding-Tabellen in Empfehlungssystemen}, bei denen speicherintensive Lookups oft einen bedeutenden Leistungsengpass darstellen. \textit{MicroRec} und \textit{FleetRec} schlagen hierfür Lösungen auf Hardware- bzw. Systemebene vor, die sowohl die Datenbewegung als auch die Rechenleistung optimieren, um die Effizienz großskaliger Empfehlungssysteme zu verbessern.

Die in dieser Dissertation vorgestellte Arbeit ist kein einmaliger Beitrag, sondern ein grundlegender Schritt in Richtung der fortlaufenden Weiterentwicklung von ML-Systemen.
Sie unterstreicht die Bedeutung von (a) der Optimierung von Vektordatensystemen in modernen ML-Pipelines und (b) der Leistungsoptimierung über die gesamte Rechenarchitektur hinweg.  
Die in dieser Arbeit vorgestellten Konzepte und Methoden bilden eine solide Grundlage für die Entwicklung und Implementierung der nächsten Generation von ML-Systemen.

\chapter*{Acknowledgments}

The past five years I spent at the Systems Group have been an incredible journey. I am deeply grateful to have met exceptional advisors, mentors, colleagues, and friends who have supported me both intellectually and personally. Their encouragement and companionship made this path both rewarding and meaningful.

First and foremost, I would like to express my sincere gratitude to my advisor, Gustavo, who has profoundly shaped my research style. His ability to bring together ideas from diverse areas --- such as databases and computer architecture --- has inspired me to explore problems that lie at the intersections of traditionally separate research communities.
Learning and adopting this cross-stack approach has been intellectually demanding, as it requires a broad base of knowledge before producing anything meaningful. But I found it deeply rewarding, as I was able to identify and tackle a lot of exciting research problems across these conventional boundaries thanks to Gustavo’s guidance.

Gustavo’s mentorship has been invaluable, both in research and beyond.
He provided me a rare combination of broad research freedom and deep technical guidance. 
Gustavo is exceptional at assessing the potential of a research idea, even when I explored topics that were not widely pursued within the group.
Whenever I presented an idea, he would grasp it immediately and provide sharp, on-point feedback, which was highly valuable in shaping the research direction at an early stage. 
I’m glad that many of these explorations eventually led to positive outcomes. Even for those that didn’t take off --- as Gustavo often warned me before --- I still gained valuable lessons that helped me grow into a more independent researcher.
Beyond research, Gustavo has provided me with tremendous support in career planning and in helping me navigate the broader academic landscape. I am deeply grateful for his unwavering support and mentorship throughout my PhD.

I am also deeply grateful to my co-advisor, Torsten, for his mentorship and guidance. Torsten’s breadth and depth of knowledge is truly inspiring. It often feels as though he knows everything — from machine learning to parallel computing, distributed systems, computer architecture, and beyond.  It has been a privilege to learn from such a brilliant and versatile researcher. 
I especially appreciate his thoughtful advice on research aesthetics and academic writing, which has not only improved how I communicate my work but also shaped how I think about research itself.

I would like to thank Ana and Christos for serving on my thesis committee. It has been an honor to receive thoughtful feedback and engage in constructive discussions with world-leading experts in systems, architecture, and machine learning.

I would also like to express my gratitude to my mentors and collaborators in industry, whose complementary expertise has greatly enriched my research journey. In particular, I would like to thank Shuai Zhang and Boran Han at AWS for their guidance and deep expertise in machine learning and natural language processing. I am also grateful to Vidushi Dadu, Amir Yazdanbakhsh, and Suvinay Subramanian at Google for their valuable insights into computer architecture and ML systems.

My sincere thanks go to all my friends and colleagues at ETH who have supported me throughout this journey, both in and outside of research. In particular, I would like to thank:
Maximilian Böther, Monica Chiosa, Jonas Dann, Johannes de Fine Licht, Timon Fercho, Ghislain Fourny, Michal Friedman, Dan Graur, Yongjun He, Zhenhao He, Maximilian Heer, Hang Hu, Marko Kabic, Yevhen Khavrona, Dario Korolija, Dimitrios Koutsoukos, Tom Kuchler, Shigang Li, Antonio Lopardo, Vasileios Mageirakos, Fabio Maschi, Martin Parvanov, Javier Moya Paya, Thomas Preusser, Abishek Ramdas, Benjamin Ramhorst, Theo Rekatsinas, Cédric Renggli, Runbin Shi, Foteini Strati, Michal Wawrzoniak, Bowen Wu, Xiaozhe Yao, Marco Zeller, Yazhuo Zhang, Ce Zhang, and Yu Zhu.

I would also like to thank our supportive administration, especially Natasha Vaidya, Macy Savic-Bederka, Nadia Mouci Menard, and Simonetta Zysset, for their generous support throughout my time in Zurich.

Lastly, but certainly not least, I would like to thank my family and my friends. Your love and support have been the foundation that made these last five years truly worthwhile.

\cleardoublepageempty

\pagestyle{headings}
\pdfbookmark[0]{\contentsname}{contents}
\tableofcontents
\cleardoublepageempty

\pagenumbering{arabic}
\sloppy
\chapter{Introduction}

\section{Motivation and Problem Statement}

The advancement of computing infrastructure has been a key driver of recent machine learning breakthroughs, alongside algorithmic innovations and the increasing availability of data.
When the landmark AlexNet model was introduced in 2012, it was trained on a relatively modest setup --- just two commercial NVIDIA GPUs designed for gaming~\cite{krizhevsky2012imagenet}.  
A decade later, OpenAI’s GPT-4 model was trained on a system consisting of 25,000 GPUs~\cite{gpt4gpu}, each offering roughly 1,000× more floating-point operations per second (FLOPS) than the GPUs used for AlexNet~\cite{nvidia_ampere}.  
%
In fact, hardware vendors are not the only ones interested in building large-scale, high-performance AI computing infrastructure: all major hyperscalers --- Google, Meta, Amazon, and Microsoft --- are making substantial investments in their own AI hardware and software systems to meet the growing computational demands of machine learning~\cite{jouppi2017datacenter, microsoft_maia, meta_mita}.

The scale of investment on AI infrastructure is staggering.
For example, Microsoft has committed \$80 billion in 2025 alone to develop its AI infrastructure~\cite{microsoft_80b}.
In the server industry, AI servers now account for 70\% of the total market value~\cite{aiserver_70perc}.
This marks a fundamental shift in computing, as investments in AI servers have now surpassed those in traditional CPU-based servers, which have been the backbone of computer science research for decades.
Given such massive financial commitments, optimizing the efficiency of these expensive machine learning systems is more crucial than ever.

\textbf{This thesis targets ML system efficiency optimization, a topic that spans machine learning, data management, computer systems, and computer architecture.}
Rather than focusing on a single layer of the computing stack, this thesis highlights the necessity of cross-stack optimizations due to the increasing complexity of ML workloads.
Specifically, the research projects presented in this thesis are driven by two key observations that I will elaborate in the following sections:
\begin{itemize}
    \item Future machine learning systems extend beyond training and serving a single model --- workloads involving multiple model components and vector data systems are becoming increasingly pervasive.
    \item Modern computer systems increasingly rely on specialized and heterogeneous hardware, driven by the end of Moore's Law and the need for better performance.
\end{itemize}

\subsection{Beyond Models: The Rise of Vector Data Systems in Modern ML Infrastructure}

%
Modern machine learning applications are becoming increasingly complex, often integrating multiple model and non-model components within a single system --- a paradigm known as \textit{compound AI systems}~\cite{compoundai, compoundai_microsoft}.
Multiple models can be responsible for different tasks, such as LLMs for language processing and diffusion models for image generation, while non-model components provide additional functionality. 
For instance, a model can invoke a search engine to retrieve real-time information from the internet, interface with a programming environment to compile and execute generated code, or utilize a calculator for mathematical computations.
%
%

\begin{figure}[t]
  \centering
  \includegraphics[width=0.8\linewidth]{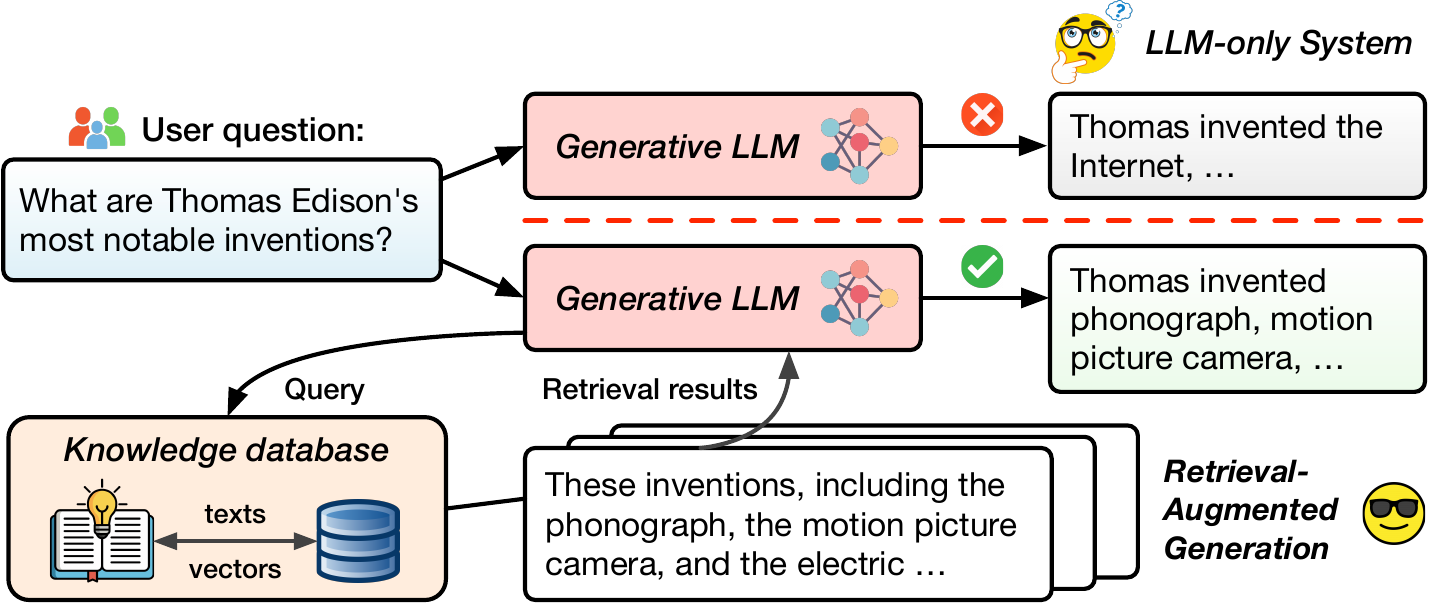}
  \caption{LLM-only system (top) versus retrieval-augmented generation (bottom).}
  \label{fig:rag-example}
\end{figure}

A notable example of a compound AI system is \textit{Retrieval-Augmented Generation (RAG)}, which \textbf{combines LLMs and vector data systems for knowledge retrieval}.
Figure~\ref{fig:rag-example} illustrates an example of RAG-based question answering, where a user asks a question about Thomas Edison’s inventions. 
Rather than generating an answer directly, the system first retrieves relevant knowledge from a database. This process is conceptually similar to invoking a search engine like Google or Bing, where a query returns related web pages.  
More specifically, the retrieval is performed through \textit{vector search} on a \textit{vector database}. Each document in the database is processed by a neural network and encoded into a {vector representation} that captures its semantic meaning. When a query is received, it is also converted into a query vector and compared against the stored document vectors. The database then returns the most relevant documents based on {vector similarity}.  
For instance, in response to a query about Thomas Edison, the retrieval process may return documents highlighting his contributions to the phonograph, cameras, and other inventions.  
The retrieved documents are subsequently incorporated into the prompt, allowing the model to generate a more well-informed and contextually relevant response.

By having the retrieval system, RAG offers several key advantages over standard LLMs.  
First, the model is less likely to hallucinate~\cite{lewis2020retrieval, li2023dark}, as it can leverage retrieved knowledge to generate more accurate responses.  
Second, RAG can be integrated with a private database, enabling the generation of personalized answers based on domain-specific or proprietary data~\cite{xiong2024benchmarking, woodbridge2016improving}.  
Third, it provides a flexible mechanism for knowledge updates~\cite{OpenAI_assistant_2, OpenAI_assistant_1}. Instead of retraining the model to incorporate new information~\cite{borgeaud2022improving, lewis2020retrieval}, updates can be made directly to the underlying database, significantly reducing the computational cost of model maintenance.  
Finally, RAG can lower {generation costs} by allowing for smaller model sizes~\cite{lewis2020retrieval, izacard2020leveraging, khandelwal2019generalization, lewis2020pre}. Since knowledge is retrieved dynamically at inference time, the model no longer needs to store all factual information within its parameters, reducing the necessity for excessively large models.


Designing efficient ML systems for RAG serving is interesting yet challenging due to \textbf{component and hardware heterogeneity within a single system}.
Vector databases for retrievals play a crucial role in RAG pipelines, meaning that the system is not only about on model serving.  
Beyond the simplified example presented here, additional models may be involved in the system, such as a database encoder, a query rewriter, and a retrieval result reranker.  
These diverse model and database components often run on \textit{heterogeneous hardware}, with machine learning accelerators serving various model components, while CPUs or specialized retrieval accelerators responsible for vector search.

\subsection{Hardware Specialization Drives the Necessity for Full-Stack Performance Optimizations}

Moore’s Law, which historically predicted the doubling of transistor density approximately every two years, has reached its limits due to fundamental physical constraints in semiconductor scaling~\cite{hennessy2019new, esmaeilzadeh2011dark}. 
As a result, the traditional approach of relying on general-purpose CPUs for continuous performance improvements is no longer sustainable. 
Specifically, single-core performance has plateaued, as Dennard Scaling --- which allowed for higher clock speeds without increasing power consumption --- broke down in the mid-2000s~\cite{hennessy2019new, esmaeilzadeh2011dark}. 
While multi-core architectures have provided a temporary solution for sustaining performance growth, the performance scalability is fundamentally constrained by Amdahl’s Law, which states that the maximum speedup of a parallel system is limited by the fraction of the workload that remains inherently serial.
These limitations have led to diminishing returns from CPU-based architectures, especially as modern workloads, such as machine learning, demand exponentially higher compute efficiency.

To meet the computational demands of emerging applications such as machine learning (ML), the era of \textit{one-size-fits-all computing} has given way to \textit{heterogeneous architectures}, where specialized accelerators are deployed to optimize task-specific computations. 
Examples of such accelerators include \textit{GPUs} (Graphics Processing Units) for massively parallel workloads~\cite{nvidia_volta, nvidia_ampere}, \textit{TPUs} (Tensor Processing Units) for deep learning~\cite{jouppi2017datacenter, tpuv5e}, \textit{FPGAs} (Field-Programmable Gate Arrays) for reconfigurable computing~\cite{nurvitadhi2017can, qasaimeh2019comparing}, and various \textit{domain-specific ASICs} (Application-Specific Integrated Circuits) designed for ML inference and training~\cite{microsoft_maia, meta_mita}.

While this specialization significantly improves computational efficiency, it also introduces new system design challenges. The presence of heterogeneous hardware components with varying compute capabilities, memory hierarchies, and interconnect architectures necessitates a fundamental shift in system and algorithm design. Thus, achieving high system efficiency is no longer feasible without accounting for the rapid evolution of the underlying hardware landscape; instead, \textbf{holistic cross-stack co-design is more important than ever for maximizing the computational efficiency of modern AI workloads.}  


\section{Contributions and Thesis Outline}

\begin{figure}[t]
  \centering
  \includegraphics[width=1.0\linewidth]{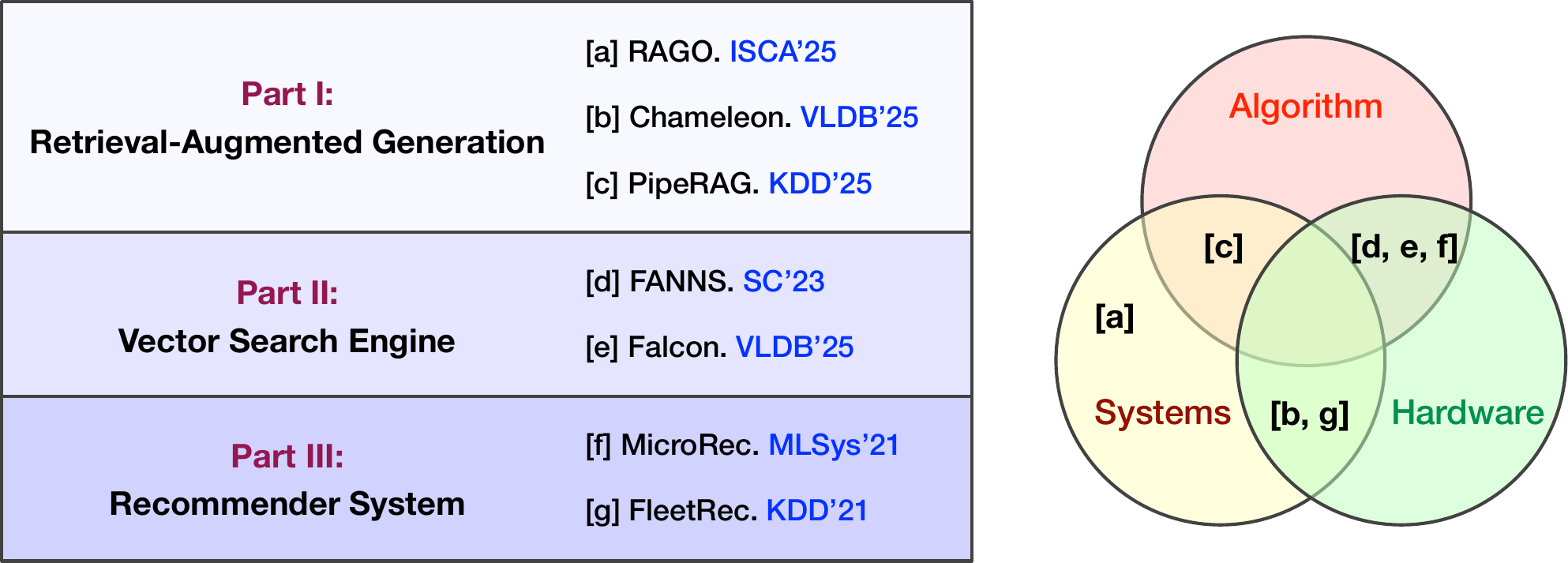}
  \caption{An overview of thesis contributions. It addresses the efficiency of vector-centric machine learning systems via a cross-stack approach, exploring the interplay across the algorithm, system, and hardware layers.}

  \label{fig:thesis-overview}
\end{figure}

Building on the above observations regarding the complexity of machine learning workloads and the heterogeneity of underlying hardware, this thesis focuses on addressing the following research questions:

\vspace{-0.5em}
\begin{itemize}[leftmargin=*, itemsep=0.em]
    \item How can we build efficient systems for emerging, complex ML workloads such as RAG, which integrate multiple system components running on heterogeneous hardware?
    \item How can we design more efficient vector data systems and specialized hardware?
    \item How can we achieve synergistic optimizations across algorithms, systems, and hardware in the post-Moore era?
\end{itemize}

\textbf{In this thesis, I explore the intricate interplay between algorithms, systems, and hardware, demonstrating how cross-layer optimizations can yield significant performance improvements in compound machine learning systems.}  
Specifically, this thesis makes the following contributions, organized into three main parts as shown in Figure~\ref{fig:thesis-overview}:

\begin{itemize}[leftmargin=*, itemsep=0.em]
    \item \textit{Part I} focuses on performance optimization for RAG serving across the entire computing stack. 
    Chapter~\ref{chap:rago} presents \textit{RAGO}~\cite{jiang2025rago}, the first systematic RAG performance optimization framework at the systems level. 
    Chapter~\ref{chap:chameleon} addresses large-scale retrieval bottlenecks in RAG at the hardware level by introducing \textit{Chameleon}~\cite{jiang2023chameleon}, the first heterogeneous and disaggregated accelerator system for RAG serving. 
    Chapter~\ref{chap:piperag} proposes \textit{PipeRAG}, the first algorithm-system co-design solution that improves the serving efficiency of RAG with iterative retrievals.

    \item \textit{Part II} introduces performance optimization for vector search, a fundamental component not only in RAG systems but also in modern search engines and recommender systems. 
    I explore vector search through an algorithm–hardware co-design approach, focusing on the two most widely adopted paradigms. Chapter~\ref{chap:fanns} and Chapter~\ref{chap:falcon} present \textit{FANNS}~\cite{jiang2023co} and \textit{Falcon}~\cite{jiang2024accelerating}, which optimize quantization-based and graph-based retrieval algorithms, respectively.

    \item \textit{Part III} investigates recommender system serving, another important vector-centric machine learning workload involving embedding vector table management. 
    Chapter~\ref{chap:microrec} introduces \textit{MicroRec}~\cite{jiang2020microrec}, which optimizes embedding table lookup performance through both hardware and data structure solutions.  
    Chapter~\ref{chap:fleetrec} presents \textit{FleetRec}~\cite{jiang2021fleetrec}, which extends \textit{MicroRec} by integrating heterogeneous hardware to handle both embedding table management and model inference efficiently.

\end{itemize}

The following sections introduce each major contribution of the thesis in details.



\subsection{Full-stack Optimization for RAG Serving}

\textit{Part I} of this thesis focuses on performance optimization for RAG serving across the computing stack, encompassing algorithm-, system-, and hardware-level solutions.

I first introduce \textit{RAGO}~\cite{jiang2025rago}, \textbf{the first work to systematically study performance optimization for RAG serving}.  
Specifically, RAGO addresses three fundamental research questions: \textit{How should the RAG serving problem be formulated given the diversity of emerging RAG algorithms? What does the performance taxonomy look like across different RAG algorithms? How should systems be designed to efficiently serve diverse RAG workloads?}  
To navigate the complex algorithm landscape, RAGO introduces \textit{RAGSchema}, an abstraction designed to systematically encapsulate performance-related RAG attributes.  
Using \textit{RAGSchema}, case studies on several RAG paradigms are conducted, demonstrating how variations in workload characteristics across different RAG algorithms significantly impact system design choices.  
Finally, \textit{RAGO} is introduced as a RAG system scheduler that optimizes task placement, resource allocation, and batching policies to meet performance objectives across various RAG configurations.  
Some observations made in this chapter, e.g., retrieval can be a significant bottleneck in RAG pipelines, lay a foundation for heterogeneous accelerator systems for RAG (Chapter~\ref{chap:chameleon}) and algorithm-system co-design for RAG with iterative retrievals (Chapter~\ref{chap:piperag}).

I then present \textit{Chameleon}~\cite{jiang2023chameleon}, \textbf{the first heterogeneous accelerator system for RAG serving}.  
The motivation for hardware heterogeneity arises from two key observations: (1) the workload characteristics of LLM inference and vector search differ significantly, and (2) large-scale retrieval can become a performance bottleneck in RAG pipelines, as demonstrated in \textit{RAGO}~\cite{jiang2025rago}, particularly given the rapid advancements in model accelerators.  
To address these challenges, I designed and implemented a heterogeneous accelerator system, including a distributed accelerator system for vector search and a multi-GPU LLM inference engine.  
Beyond leveraging hardware heterogeneity, the system is designed with disaggregated accelerators across the network, enabling flexible handling of both inference-bound and retrieval-bound workloads.

I further explore algorithm-level optimizations for RAG serving through \textit{PipeRAG}, \textbf{the first algorithm-system co-design approach for iterative retrieval-augmented generation.} 
Periodic retrievals from large databases can cause significant stalls during generation, leaving ML accelerators idle while waiting for retrieval results. To mitigate this inefficiency, \textit{PipeRAG} introduces several key algorithmic improvements. 
First, it employs approximate data prefetching, enabling retrieval and generation processes to be executed concurrently, thereby reducing latency and improving overall system throughput. Second, it utilizes flexible retrieval intervals to maximize pipeline execution efficiency, adjusting retrieval frequency based on workload characteristics. Finally, it integrates a performance model that automatically balances retrieval quality and latency by adapting to the current generation state and underlying hardware constraints.

\subsection{Algorithm-Hardware Co-Design for Vector Search}

As software-level optimizations for vector search become increasingly challenging due to the convergence of retrieval algorithms, \textit{Part II} of this thesis explores algorithm-hardware co-design for two major classes of vector search algorithms: quantization-based and graph-based vector search.

I first introduce \textit{FANNS}~\cite{jiang2023co}, a hardware-algorithm co-design solution for product quantization (PQ), a widely used large-scale vector search algorithm. 
The product quantization (PQ) algorithm is commonly combined with an inverted file index (IVF), forming the widely used IVF-PQ algorithm.
Given the numerous design possibilities for an IVF-PQ accelerator due to varying algorithm parameters, \textit{FANNS} leverages the reconfigurability of FPGAs to explore different design points. Given a dataset, a target recall requirement, and an FPGA device, \textit{FANNS} automatically (a) identifies the optimal combination of parameter settings and hardware design and (b) generates a ready-to-deploy accelerator. Specifically, \textit{FANNS} first evaluates the relationship between IVF-PQ parameters and recall for the given dataset. It then enumerates all valid accelerator designs under the FPGA hardware resource constraints. Next, the \textit{FANNS} performance model predicts the queries-per-second (QPS) throughput across different combinations of algorithm parameters and hardware configurations. Finally, using the best combination determined by the performance model, the \textit{FANNS} code generator produces FPGA code, which is compiled into an executable FPGA bitstream for deployment.

I then present \textit{Falcon}~\cite{jiang2024accelerating}, which focuses on low-latency graph-based vector search. Due to the fine processing granularity of each graph traversal iteration, achieving low search latency requires both algorithmic improvements and hardware specialization. In addition to developing the \textit{Falcon} accelerator, I also introduce \textit{delayed-synchronization traversal} (DST), a hardware-efficient traversal algorithm designed to enhance search performance and quality by relaxing the strict order of graph traversal, thereby improving accelerator utilization. The design of \textit{DST} is based on two key observations. First, the synchronous and greedy nature of software-oriented best-first search (BFS) limits the amount of parallelism the accelerator can exploit, leading to significant under-utilization. Second, relaxing the order of candidate evaluations does not compromise recall. Inspired by label-correcting algorithms used in parallel shortest path computation, \textit{DST} relaxes synchronization constraints that enforce greedy traversal order, increasing the volume of parallel workloads the accelerator can process. As a result, \textit{DST} not only reduces search latency through improved accelerator utilization but also enhances recall by enabling the exploration of search paths that a greedy BFS might otherwise overlook.

\subsection{Vector Table Management for Recommender Systems} 

\textit{Part III} of this thesis addresses performance optimization of deep learning recommender models (DLRMs), another important machine learning workload that involves vector data management.
In \textit{Part I}, I show that RAG is a heterogeneous system, both in terms of its components and underlying hardware.  
Similarly, in recommender systems, heterogeneity can exist even within a single model: in addition to the neural network, the many vector embedding tables form a critical system component.  
During each inference, these embedding tables must be accessed, and the performance of DLRM serving is often limited by the high cost of random memory accesses required for embedding lookups.

To address this challenge, I first introduce \textit{MicroRec}~\cite{jiang2021microrec}, a system designed to accelerate recommendation inference by optimizing embedding data structures and leveraging a heterogeneous memory hierarchy. \textit{MicroRec} reduces the number of lookup operations by restructuring the data layout and efficiently mapping embedding tables across different memory tiers, including DDR memory, HBM, and SRAM. This approach reflects the skewed distribution of embedding table sizes and access frequencies, ensuring that frequently accessed embeddings reside in faster memory while less critical embeddings are stored in lower-cost memory. The proposed design is implemented on an FPGA platform, integrating both the embedding lookup step and the complete inference process.

Building upon these optimizations, I further introduce \textit{FleetRec}~\cite{jiang2021fleetrec}, a high-performance and configurable heterogeneous computing cluster for recommendation inference. While \textit{MicroRec} significantly accelerates embedding lookups, the deep neural network (DNN) inference stage on FPGA emerges as the new bottleneck. To overcome this limitation, \textit{FleetRec} adopts a hybrid accelerator strategy, leveraging the strengths of both FPGA and GPU architectures. Specifically, FPGA-based solutions, such as \textit{MicroRec}, are utilized for embedding table lookups, while GPUs are dedicated exclusively to DNN computation.
\textit{FleetRec} treats GPUs, FPGAs, and CPU servers as end devices interconnected through a high-speed network, allowing for flexible configuration of node types and quantities to accommodate various workload scales, embedding table sizes, and computational demands.

\section{Related Publications} 
This dissertation is primarily based on work that has also been presented in the following publications (* - equal contribution):

\begin{itemize}
    \item \textbf{RAGO: Systematic Performance Optimization for Retrieval-Augmented Generation Serving} \textit{by \textbf{Wenqi Jiang}, Suvinay Subramanian, Cat Graves, Gustavo Alonso, Amir Yazdanbakhsh, and Vidushi Dadu, in ACM/IEEE 52nd Annual International Symposium on Computer Architecture (ISCA'25)~\cite{jiang2025rago}.}

    \item \textbf{Chameleon: a Heterogeneous and Disaggregated Accelerator System for Retrieval-Augmented Language Models} \textit{by \textbf{Wenqi Jiang}, Marco Zeller, Roger Waleffe, Torsten Hoefler, and Gustavo Alonso, in Proceedings of the VLDB Endowment (VLDB'25)~\cite{jiang2023chameleon}.}

    \item \textbf{PipeRAG: Fast Retrieval-Augmented Generation via Algorithm-System Co-Design} \textit{by \textbf{Wenqi Jiang}, Shuai Zhang, Boran Han, Jie Wang, Bernie Wang, and Tim Kraska, in Proceedings of the 31st ACM SIGKDD International Conference on Knowledge Discovery and Data Mining (KDD'25)~\cite{jiang2024piperag}.}
    
    \item \textbf{Accelerating Graph-based Vector Search by Hardware Acceleration and Delayed-Synchronization Traversal} \textit{by \textbf{Wenqi Jiang}, Hang Hu, Torsten Hoefler, and Gustavo Alonso, in Proceedings of the VLDB Endowment (VLDB'25)~\cite{jiang2024accelerating}.}

    \item \textbf{Co-Design Hardware and Algorithm for Vector Search} \textit{by \textbf{Wenqi Jiang}, Shigang Li, Yu Zhu, Johannes de Fine Licht, Zhenhao He, Runbin Shi, Cedric Renggli, Shuai Zhang, Theodoros Rekatsinas, Torsten Hoefler, and Gustavo Alonso, in The International Conference for High Performance Computing, Networking, Storage and Analysis (SC'23)~\cite{jiang2023co}.}

    \item \textbf{FleetRec: Large-Scale Recommendation Inference on Hybrid GPU-FPGA Clusters} \textit{by \textbf{Wenqi Jiang}*, Zhenhao He*, Shuai Zhang, Kai Zeng, Liang Feng, Jiansong Zhang, Tongxuan Liu, Yong Li, Jingren Zhou, Ce Zhang, and Gustavo Alonso, in Proceedings of the 27th ACM SIGKDD International Conference on Knowledge Discovery and Data Mining (KDD'21)~\cite{jiang2021fleetrec}.}
    
    \item \textbf{MicroRec: Efficient Recommendation Inference by Hardware and Data Structure Solutions} \textit{by \textbf{Wenqi Jiang}, Zhenhao He, Shuai Zhang, Thomas B. Preußer, Kai Zeng, Liang Feng, Jiansong Zhang, Tongxuan Liu, Yong Li, Jingren Zhou, Ce Zhang, and Gustavo Alonso, in 4th Conference on Machine Learning and Systems (MLSys'21)~\cite{jiang2021microrec}.}
\end{itemize}

Apart from the above-mentioned publications, participation in other research projects has resulted in the following publications:

\begin{itemize}

    \item \textbf{SwiftSpatial: Spatial Joins on Modern Hardware} \textit{by \textbf{Wenqi Jiang}, Martin Parvanov, and Gustavo Alonso, in International Conference on Management of Data (SIGMOD'25)~\cite{jiang2025swiftspatial}.}

    \item \textbf{MS MARCO Web Search: A Large-scale Information-rich Web Dataset with Millions of Real Click Labels} \textit{by Qi Chen, Xiubo Geng, Corby Rosset, Carolyn Buractaon, Jingwen Lu, Tao Shen, Kun Zhou, Chenyan Xiong, Yeyun Gong, Paul Bennett, Nick Craswell, Xing Xie, Fan Yang, Bryan Tower, Nikhil Rao, Anlei Dong, \textbf{Wenqi Jiang}, Zheng Liu, Mingqin Li, Chuanjie Liu, Zengzhong Li, Rangan Majumder, Jennifer Neville, Andy Oakley, Knut Magne Risvik, Harsha Vardhan Simhadri, Manik Varma, Yujing Wang, Linjun Yang, Mao Yang, and Ce Zhang, in International World Wide Web Conference (WWW'24)~\cite{chen2024ms}.}

    \item \textbf{Data-Informed Geometric Space Selection} \textit{by Shuai Zhang and \textbf{Wenqi Jiang}, in Thirty-seventh Conference on Neural Information Processing Systems (NeurIPS'23)~\cite{zhang2023data}.}

    \item \textbf{Data Processing with FPGAs on Modern Architectures} \textit{by \textbf{Wenqi Jiang}, Dario Korolija, and Gustavo Alonso, in Companion of the 2023 International Conference on Management of Data (SIGMOD'23 Tutorial)~\cite{jiang2023data}.}

    \item \textbf{Distributed Recommendation Inference on FPGA Clusters} \textit{by Yu Zhu, Zhenhao He, \textbf{Wenqi Jiang}, Kai Zeng, Jingren Zhou, and Gustavo Alonso, in 31st International Conference on Field-Programmable Logic and Applications (FPL'21)~\cite{fpl_recommendation_cluster}.}

\end{itemize}

\chapter{Preliminary}
\label{chap:prelim}

This chapter provides preliminary knowledge that is relevant across the upcoming chapters.  
I will first discuss the importance of vector data in the current machine learning era by outlining three major use cases: information retrieval via vector search, retrieval-augmented generation, and recommender systems.  
I will then shift focus to the background of system designs in the post-Moore era, highlighting the necessity of hardware heterogeneity.

\section{Fundamentals of Vector Data Management}

\textit{Vector data is fundamental in the deep learning era}, as various unstructured and semi-structured data sources, including images, videos, and text, can be effectively represented in the vector data format.  
In computer vision, deep convolutional neural networks (CNNs) extract high-dimensional feature vectors from images~\cite{liu2007clustering, babenko2016efficient}.  
Similarly, in video analysis, deep models encode spatiotemporal features into vector representations for applications such as action recognition and video retrieval~\cite{garcia2008fast, shao2008batch}.  
For natural language processing, transformer-based models like BERT~\cite{devlin2018bert, reimers2019sentence} can generate contextualized embeddings to represent text, facilitating applications such as search and retrieval.  
In recommender systems, deep learning encodes user profiles, behaviors, and preferences into dense embeddings to improve personalized recommendations~\cite{he2017neural, din_alibaba_attention_fc, facebook_benchmark}.  

In this section, I will introduce three important use cases of vector data, namely vector search for information retrieval (Section~\ref{chap:prelim:retrieval}), retrieval-augmented generation (Section~\ref{chap:prelim:rag}), and recommender systems (Section~\ref{chap:prelim:dlrm}).

\subsection{Vector Search for Information Retrieval}
\label{chap:prelim:retrieval}

One essential use case of vector data is \textit{information retrieval}, which retrieves information from knowledge databases given user queries of various formats (text, images, videos, etc.).
A common approach to performing this retrieval is \emph{vector search}, which has become the cornerstone of recent information retrieval systems~\cite{PQ, malkov2018efficient}.
In the case of documents retrievals, for example, vector search enables the system to assess semantic relevance by encoding both documents and queries as high-dimensional vectors (\egc hundreds to thousands dimensions), where proximity in this vector space reflects semantic similarity between queries and database documents.

In this section, I will present the definition of vector search (Section~\ref{chap:prelim:retrieval:def}) and introduce two classes of most popular vector search algorithms, namely quantization-based vector search for large-scale datasets (Section~\ref{chap:prelim:retrieval:pq}) and graph-based vector search for smaller datasets (Section~\ref{sec:background_gvs}).

\subsubsection{Vector Search Problem Definition}
\label{chap:prelim:retrieval:def}

A $k$ nearest neighbor (\textit{kNN}) search takes a $D$-dimensional query vector $q$ as input and retrieves the $k$ most similar vectors from a database $Y$ containing $D$-dimensional vectors, based on metrics such as L2 distances and cosine similarity. 

Real-world vector search systems typically adopt \textit{approximate nearest neighbor (ANN) search} instead of exact kNN search to boost search performance (latency and throughput) by avoiding exhaustive scans of all database vectors. In this thesis, I will use the terms \textit{vector search} and \textit{ANN search} interchangeably.

The quality of an ANN search is measured by the recall at $k$ ($R@k$). 
Let \( \mathit{NN}_k(q) \) be the set of true \( k \) nearest neighbors to a query \( q \) and \( \mathit{ANN}_k(q) \) be the set of \( k \) results returned by the ANN search, recall at $k$ measures the proportion of the true \( k \) nearest neighbors that are successfully retrieved by the ANN search: \( R@k = \frac{| ANN_k(q) \cap NN_k(q) |}{| NN_k(q) |} \).

\subsubsection{Quantization-Based Vector Search}
\label{chap:prelim:retrieval:pq}


The \emph{IVF-PQ} algorithm, which combines an inverted file (IVF) index with product quantization (PQ)~\cite{PQ}, is one of the most widely used approaches for \textbf{large-scale} vector search~\cite{borgeaud2022improving, izacard2020leveraging, khandelwal2019generalization}.

\begin{table}[t]

  \begin{center}
    \caption{Definitions of vector search and IVF-PQ symbols.}
    \label{tab:table1}
    \scalebox{0.9}{
    \begin{tabular}{L{4.5em} L{25em}} 
      \toprule
      Symbol & \multicolumn{1}{c}{Definition} \\
      \midrule
      $x$ & A query vector.\\
      $y$ & A database vector. \\
      $k$ & The number of most similar vectors to return. \\
      $m$ & The sub-space number of product quantization.  \\
      $nlist$ & The totol Voronoi cell number. \\
      $nprobe$ & The number of cells to be scanned per query. \\
      \bottomrule
    \end{tabular}
    } 
  \end{center}
	\vspace*{-1em} 
\end{table}

\begin{figure}[t]
	\centering
  \includegraphics[width=0.8\linewidth]{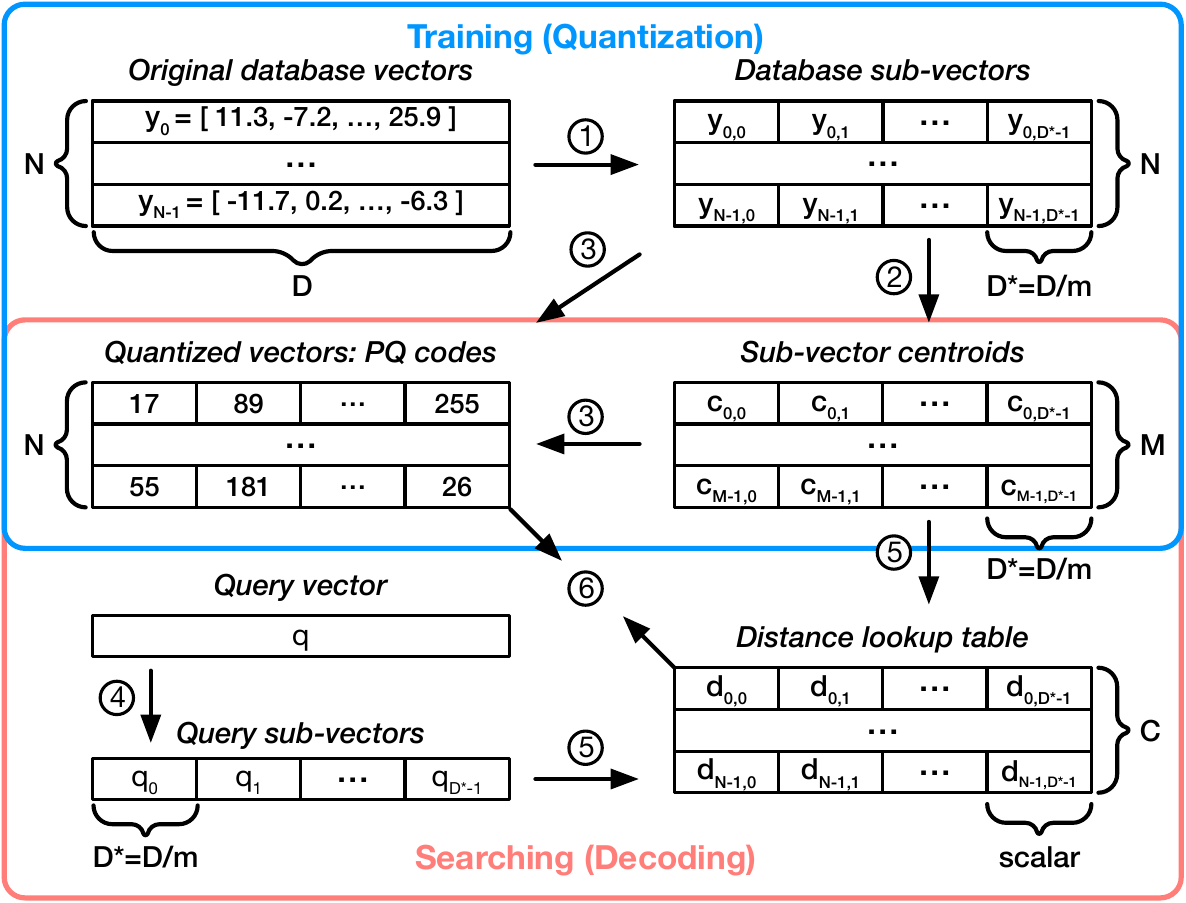}
  \caption{Product quantization (PQ) for vector search.}
  \label{fig:PQ}
\end{figure}


\textbf{Inverted-File (IVF) Index.} An IVF index divides a vector dataset $Y$ into many (\textit{nlist}) disjoint subsets, typically using clustering algorithms like K-means. Each of these subsets is termed an IVF list. At query time, the IVF index is scanned, and only a select few (\textit{nprobe}) IVF lists whose cluster centroids are close to the query vector are scanned, such that the search space is effectively pruned. 

\textbf{Product Quantization (PQ).}
PQ reduces memory usage and computations of vector search by compressing each database vector into $m$-byte PQ codes. Figure~\ref{fig:PQ} overviews the workflow of PQ.

\textit{Training (quantization).} All database vectors are partitioned evenly into $m$ sub-vectors~\whiteballnumber{1}, which possess a dimensionality of $D^{*}=\frac{D}{m}$, typically ranging from 4 to 16 in practice. 
A clustering algorithm is performed in each sub-space~\whiteballnumber{2} to obtain a list of centroids $c$, allowing each database sub-vector to be approximated by its nearest centroid. 
Typically, the number of clusters per sub-space is set as $M=256$, such that a cluster ID can be represented with one byte. Thus, once the cluster centroids are stored, each database vector can be represented by $m$-byte PQ codes.


\textit{Searching (decoding).} A query vector is compared against the quantized database vectors.
The distance computation can be formulated as $\hat{d}(x,y)=d(x,c(y)) =\sum_{i=1}^{m}d(x_i,c_i(y_i))$, where $\hat{d}(x,y)$ is the approximate distance between a query vector $x$ and a quantized database vector $y$, and $c(y)$ is the reconstructed database vector using the PQ codes and the cluster centroid vectors per sub-space. 
To calculate $\hat{d}(x,y)$, the query vector is divided into $m$ sub-vectors ($x_i$)~\whiteballnumber{4} and compared against the reconstructed quantized sub-database-vectors $c_i(y_i)$.
To speed up distance computations given many database vectors, a distance lookup table~\whiteballnumber{5} can be constructed and reused within a query, encompassing all combinations between a sub-query-vector and a cluster centroid within the same sub-space. 
With this table, the value of $d(x_i,c_i(y_i))$ can be swiftly retrieved by looking up the table with the PQ code as the address~\whiteballnumber{6}, leading to improved computational efficiency.



Optionally, one can use optimized product quantization (OPQ) to further improve quantization quality~\cite{OPQ}. The key idea is to rotate the vector space such that the sub-spaces are independent and the variances of each sub-space are normalized. 
At query time, OPQ simply introduces a vector-matrix multiplication between the query and the transformation matrix, while the rest search procedures are identical to PQ. 







\subsubsection{Graph-based Vector Search}
\label{sec:background_gvs}

Graph-based vector search (GVS) is a widely used ANN search algorithm, known for its high search performance and accuracy, particularly for \textit{small-scale} datasets~\cite{malkov2014approximate, malkov2018efficient, fu2017fast, zhao2023towards, zuo2023arkgraph, lu2021hvs, gao2023high}.  
In contrast, PQ-based search is more suitable for large-scale datasets, as GVS requires additional storage for graph edges along with raw vectors, whereas PQ compresses database vectors, reducing storage demands.

GVS involves constructing a proximity graph \( G(V, E) \), where \( V \) represents the set of nodes, each is a database vector, and \( E \) represents the set of edges between nodes, with each edge indicating high similarity between the two connected nodes. 
Once the graph is constructed, query vectors can traverse the graph to find their nearest neighbors.


\begin{figure}[t]
	\centering
  \includegraphics[width=0.75\linewidth]{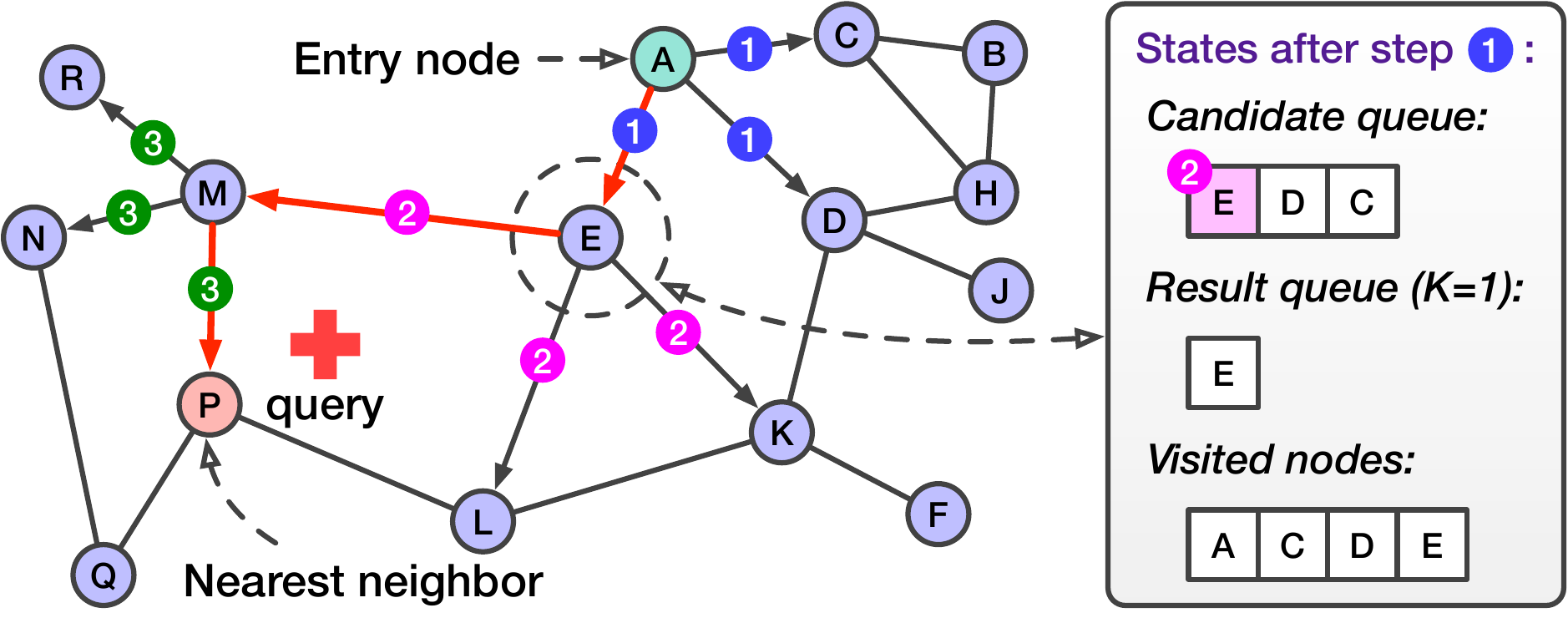}
  \caption{An example of best-first search (BFS) on graphs.}
  \label{fig:bfs}
\end{figure}

Figure~\ref{fig:bfs} illustrates an example of the classic best-first search (BFS) algorithm returning the \(K=1\) nearest neighbor. In this graph, point A serves as the fixed entry point. In the first iteration, the neighbors of A (C, D, and E) are evaluated and inserted into the candidate queue, with E, the closest node to the query, inserted into the result queue. Since E is also the closest candidate in the queue, it is evaluated next in the second iteration. In the third iteration, P, the nearest neighbor to the query, is found and inserted into the result queue. The search may continue beyond three iterations, as long as there are additional qualified candidates that have not yet been evaluated.

\subsection{Retrieval-Augmented Generation}
\label{chap:prelim:rag}

Another essential use case of vector data is \textit{Retrieval-Augmented Generation (RAG)}, an increasingly popular approach to serve generative large language models (LLMs), enabled by advancements in vector search.

\begin{figure}[t]
	\centering
  \includegraphics[width=0.8\linewidth]{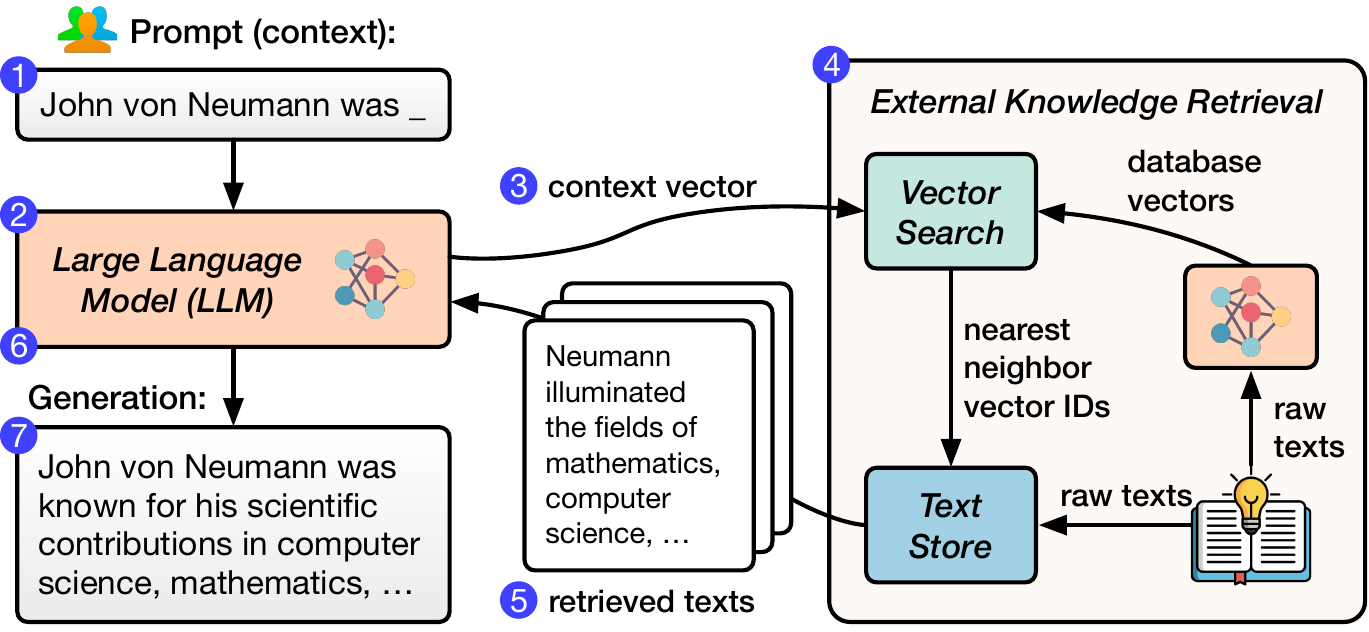}
  \caption{An example of retrieval-augmented generation.}
  \label{fig:ralm}
\end{figure}

The architecture of RAG, as shown in Figure~\ref{fig:ralm}, allows the LLM to focus on learning linguistic structures, while incorporating context-specific knowledge during inference. 
Specifically, the external textual knowledge is encoded as vectors using LLMs and stored in a vector database. 
Given an inference context (e.g., a prompt), the knowledge retriever identifies relevant knowledge in the database via vector search, which assesses relevance by computing the similarity between the context vector and the database vectors. The retrieved texts are then included into the LLM's prompt to facilitate high-quality generation. 

RAG show three major advantages over conventional LLMs.
\textit{First of all}, RAG, even using smaller LLMs with one to two orders of magnitude fewer parameters, can match or surpass the generation quality of conventional LLMs on various tasks~\cite{lewis2020retrieval, izacard2020leveraging, komeili2021internet, guu2020retrieval, khandelwal2019generalization, khandelwal2020nearest, lewis2020pre}, thus significantly lowering the inference cost. 
This is because conventional LLMs rely on a vast number of parameters trained on massive datasets to capture and retain textual knowledge~\cite{brown2020language, chowdhery2022palm, smith2022using, rae2021scaling}, while RAG can integrate retrieved knowledge during inference, not burdening the LLM's parameters.
\textit{Moreover}, knowledge editing in RAG is as straightforward as updating the database, enabling efficient integration of new or private knowledge~\cite{OpenAI_assistant_2, OpenAI_assistant_1}. In contrast, updating knowledge in conventional LLMs is inflexible, requiring additional training~\cite{borgeaud2022improving, lewis2020retrieval}. 
\textit{Finally}, RAG enhance the reliability and interpretability of generated content by sourcing knowledge externally, while conventional LLMs are prone to producing non-factual content, known as hallucination~\cite{lewis2020retrieval, li2023dark}.

\subsection{Recommender Systems}
\label{chap:prelim:dlrm}

In addition to vector search and RAG, recommender systems represent another important use case of vector data, where various features such as user preferences, location, and demographic attributes are encoded as vectors and retrieved during inference.

Deep-learning recommendation model (DLRM) inference comprises a huge portion of the workload in data centers. Thus, it is crucial to optimize its performance to serve these models efficiently.
Such optimizations can lead to instant economic benefits through (a) higher recommendation quality since more candidate items can be scored in the same time frame; and (b) reduced energy consumption as a result of the improved inference efficiency.

\begin{figure}[t]
  \centering
  \includegraphics[width=0.65\linewidth]{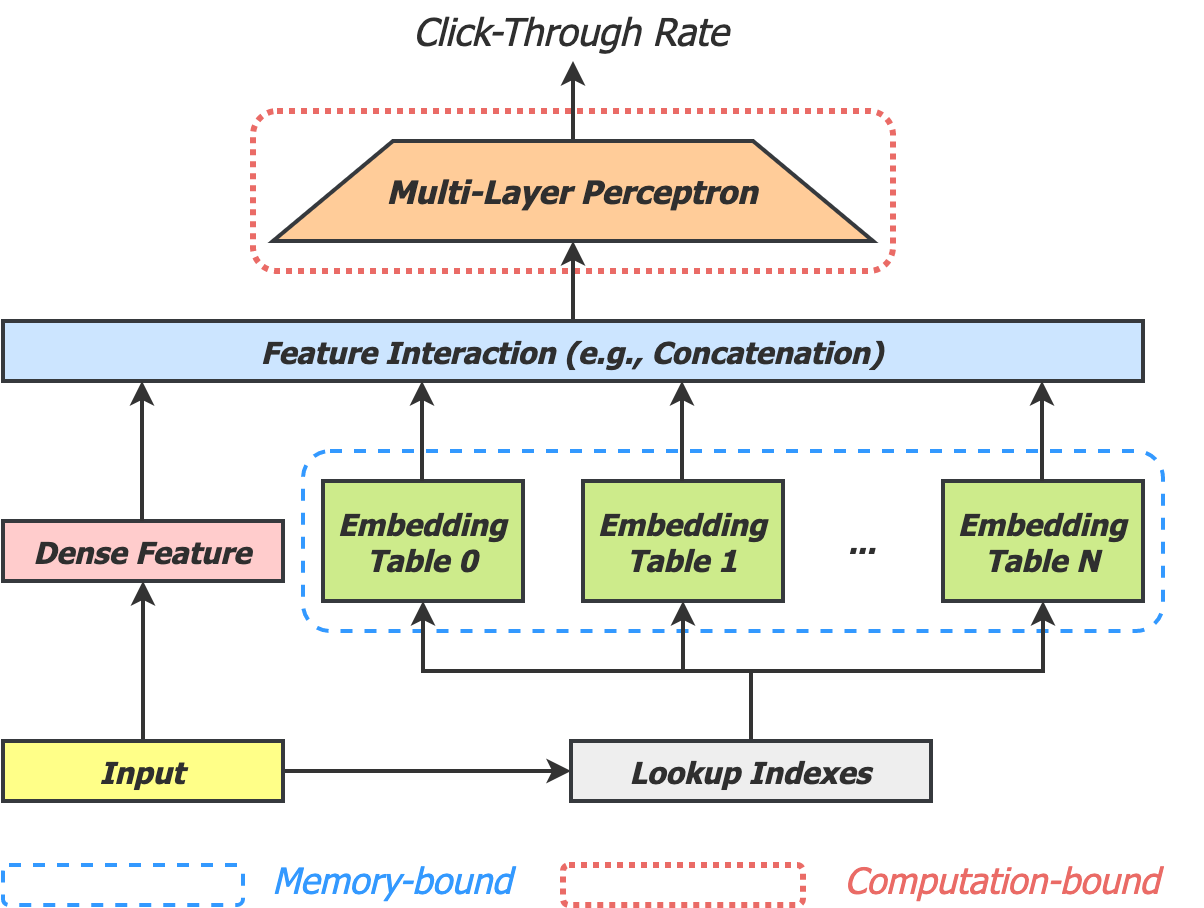}
  \caption{A representative deep recommendation model.}
  \label{fig:embedding}
\end{figure}

Figure~\ref{fig:embedding} shows the architecture of a classical deep recommendation model for \textit{Click-Through Rate} (CTR) prediction, i.e., how likely it is that the user will click on the product.  
While regular DNNs take dense features as input, \citet{he2017neural} proposed to encode user information by embedding tables in recommendation models.
The input feature vector consists of dense features (e.g., age) and sparse features (e.g., advertisement category). The model translates each sparse feature into a dense embedding vector by looking it up in an embedding table. These vectors are then combined with dense features and fed to several fully-connected (FC) layers before the model outputs the predicted CTR. Although there are alternative architecture designs \cite{wide_and_deep_app_store, wide_and_deep_MT_youtube, facebook_benchmark, din_alibaba_attention_fc, he2017neural}, most recommendation systems are built around two major building blocks, i.e., the embedding tables and the DNN classifier.

The specific model design varies from scenario to scenario. Some adjustable parameters include: number of fully-connected layers, number of hidden neurons in each layer, numbers and sizes of embedding tables, feature interaction operations (e.g., concatenation, weighted sum, and element-wise multiplication), whether to include bottom fully-connected (FC) layers. The models this thesis target do not contain bottom FCs, and each table is looked up only once.

{\sloppy
\section{System and Hardware in Post-Moore Era}

As Moore’s Law and Dennard Scaling reached their ends, continuous performance and efficiency gains from general-purpose CPUs is no longer sustainable~\cite{hennessy2019new, esmaeilzadeh2011dark}.
To address the growing computational demands of applications like machine learning (ML), the era of general-purpose computing has shifted towards \textit{heterogeneous architectures}, where specialized accelerators are optimized for specific tasks.
These include GPUs for massively parallel processing, TPUs for deep learning~\cite{jouppi2017datacenter}, FPGAs for reconfigurable computing~\cite{singh2017xilinx, intel_opencl_fpga}, and various domain-specific ASICs designed for ML inference and training~\cite{han2016eie, chen2016eyeriss, chen2014dadiannao}.

In this section, I will introduce GPUs (Section~\ref{chap:prelim:gpu}) and FPGAs (Section~\ref{chap:prelim:fpga}), the two primary accelerators used throughout this thesis.

\subsection{GPU}
\label{chap:prelim:gpu}



Graphics Processing Units (GPUs) have evolved from fixed-function graphics accelerators into highly parallel, programmable processors that are widely used for general-purpose computation. Their architecture is designed to handle massively parallel workloads, making them well-suited for data-intensive applications such as machine learning.

\textit{Architecture.} A modern GPU consists of many streaming multiprocessors (SMs). Each SM contains multiple CUDA cores (for NVIDIA GPUs), registers, and shared memory. 
Unlike CPUs, which are optimized for low-latency sequential execution, GPUs focus on high-throughput execution by scheduling parallel execution across thousands of cores~\cite{nvidia_volta, amd_rdna}. 
In addition to standard CUDA cores, modern GPUs feature specialized tensor cores, first introduced in NVIDIA’s Volta architecture~\cite{nvidia_volta}.
Tensor cores are designed to accelerate matrix multiplications, a fundamental operator in deep learning. 
By performing fused multiply-accumulate (FMA) operations on small matrices in low precision, tensor cores significantly improve the performance of deep learning training and inference workloads~\cite{nvidia_ampere}.

\textit{Programming Model.}
GPUs support several parallel programming models, with CUDA and OpenCL being the most prominent. CUDA, developed by NVIDIA, provides a flexible programming interface with direct access to GPU resources, allowing developers to write highly optimized parallel code~\cite{cuda_guide}. 
OpenCL, an open standard, enables GPU programming across multiple hardware vendors, including AMD and Intel~\cite{stone2010opencl}.

\subsection{FPGA}
\label{chap:prelim:fpga}





Field programmable gate array (FPGA) lies between general-purpose processors (e.g., CPUs) and application-speciﬁc integrated circuits (ASICs): it behaves like ASICs, yet its circuit can be reconfigured virtually infinite number of times, thus leveraging performance and design flexibility.
Developers can design arbitrary micro-architectures given the traits of specific applications, compile the design to a bitstream file representing the circuit configuration, and load the bitstream on FPGAs to start accelerating the applications. 
This is achieved by the hardware compiler which can map a logical design to the physical hardware units on FPGAs. 

\textit{Architecture.} 
FPGAs are consists of several types of hardware building blocks (a) Block-RAM (BRAM) as small yet fast on-chip memory, (b) Flip-Flops (FF) as registers, (c) Digital Signal Processors (DSP) as computing units, and (d) lookup-tables (LUT) as either memory or computing units. 
The customized hardware design described by the program is then mapped to these hardware building blocks on the chip.

\textit{Programming Model.} Developing FPGA accelerators typically requires much more efforts compared with software designs. Traditionally, FPGAs are programmed by hardware description languages (HDL), such as Verilog and VHDL. Developers need to define the behavior of the circuit at the granularity of a single clock cycle. Recently, High-Level Synthesis (HLS) has become popular in the FPGA community. It allows programmers to develop the circuit at a higher level using C/C++ or OpenCL~\cite{vivado_hls, intel_opencl_fpga}. However, developing accelerators by HLS requires extra efforts to learn the specific hardware-friendly coding style and to fine-tune the performance by exploring a wide range of \textit{pragmas}.

}

\part{Full-Stack Retrieval-Augmented Generation Optimization}
\label{part:rag}

\chapter{RAGO: Systematic Performance Optimization for RAG Serving} 
\label{chap:rago}

This chapter explores RAG serving performance optimization from the systems perspective. By evaluating various paradigms of RAG algorithms, I show the complexity of the RAG serving landscape due to the many algorithm variants. 
Some observations made in this chapter, e.g., retrieval can be a significant bottleneck in RAG pipelines, lay a foundation for heterogeneous accelerator systems for RAG (Chapter~\ref{chap:chameleon}) and algorithm-system co-design for RAG with iterative retrievals (Chapter~\ref{chap:piperag}).

\section{Introduction}
\label{sec_rago:intro}


In contrast to conventional LLM-only serving systems, which center predominantly on optimizing the prefix (prompt decoding) and decoding (token generation) stages, RAG presents three challenges:
\textbf{(C1)} RAG systems are intrinsically heterogeneous, comprising a diverse array of system components, including vector search-based retrieval~\cite{borgeaud2022improving, shao2024scaling, lewis2020retrieval}, generative LLMs~\cite{llamaindex, team2024gemini, brown2020language}, and multiple optional models such as database encoders~\cite{reimers2019sentence, lee2024gecko}, query rewriters~\cite{chan2024rq, ma2023query}, and retrieval result rerankers~\cite{glass2022re2g, allahverdiyev2024chunkrag}.
These components often run on heterogeneous hardware platforms.
For example, retrievals are typically performed on CPU servers, whereas ML accelerators (e.g., TPUs or GPUs) are used for model serving.
This interplay of diverse components and hardware platforms amplifies the search space, far surpassing that of LLM-only systems;
\textbf{(C2)} Various RAG configurations defined by factors such as database size, retrieval frequency, model selection, and serving hardware, exhibit substantial performance variability.
This variability can veer the bottleneck between inference and retrieval or among different models within the serving pipeline; and
\textbf{(C3)} A natural consequence of the heterogeneity in components and the variability in performance is the emergence of a new challenge: \emph{how can we design efficient RAG serving systems?}
Addressing this challenge demands meticulously navigating key decisions in scheduling policies across diverse RAG configurations and hardware platforms.

To address these challenges in optimizing RAG serving performance, my ground the proposed approach in three key design principles: 
(1) \textbf{Workload abstraction}: 
Tackling the heterogeneity of RAG systems necessitates an abstraction to encapsulate the diverse RAG workloads.
Without such abstraction, the inherent complexity of RAG configurations become intractable;
(2) \textbf{Critical system design decisions}: 
To unveil the critical system design decisions and illuminate the performance trade-offs inherent in RAG serving, a careful performance characterization of representative RAG workloads is warranted.
Without understanding how these different workloads behave, the optimization process risks becoming guesswork; and
(3) \textbf{Systematic optimization framework}: 
To navigate the large optimization space arising from the Cartesian product of RAG workload and system design dimensions, an optimization framework is essential to uncover and exploit efficiency opportunities in RAG serving systems.

%
To systematically describe RAG workloads, my introduce \ragdescemph (\S\ref{sec_rago:rag-schema}), a RAG serving abstraction that encapsulates a set of essential performance-relevant workload attributes.
\ragdesc includes two key components: 
(a) specification of the RAG pipeline
\,---document encoder, query rewriter, result reranker, and generative LLM---\, and (b) model and retrieval configurations, including model size, database size, the number of query vectors per retrieval, and iterative retrieval frequency.
This abstraction simplifies the representation of complex RAG workloads while providing sufficient information for performance characterization and optimization.

Building on \ragdesc, my perform a detailed 
workload characterization (\S\ref{sec_rago:case-studies}) to identify bottlenecks and key system design decisions.
my analyze four representative RAG paradigms, each with distinct RAG pipeline:
(a) RAG with hyperscale retrieval~\cite{borgeaud2022improving, shao2024scaling, wang2023instructretro};
(b) RAG for long-context sequence processing~\cite{lee2024can, li2024retrieval, yue2024inference};
(c) RAG with iterative retrieval~\cite{borgeaud2022improving, trivedi2022interleaving, jiang2023active}; and
(d) RAG with query rewriter and retrieval reranker models~\cite{chan2024rq, ma2023query, glass2022re2g, allahverdiyev2024chunkrag}.
My analysis reveal \textit{significant performance variability both across and within paradigms}, with a subset of findings summarized as follows.
First, bottlenecks shift between retrieval and inference across RAG paradigms.
For instance, hyperscale retrieval can spend over 80\% in retrieval (\S\ref{subsec_rago:case1}) while in long-context scenarios, retrieval accounts for less than 1\% of the total latency (\S\ref{subsec_rago:case2}).
%
Second, even smaller models within the pipeline can significantly influence system performance.
For example, in long-context processing, a database encoder that is 100$\times$ smaller than the main generative LLM can become the bottleneck due to the large number of tokens it must process (\S\ref{subsec_rago:case2}).
Third, iterative retrievals during decoding can stall the pipeline, as the decoding process waits for retrieval results (\S\ref{subsec_rago:case3}).
The insights from these studies underscore not only the importance of making appropriate system design decisions but also the indispensable need for a tailored optimization framework for RAG serving systems, given their far less predictable performance landscape compared to LLM-only systems.
\begin{figure}[t]
  \centering
  \begin{subfigure}[b]{0.85\linewidth}
    \includegraphics[width=\linewidth]{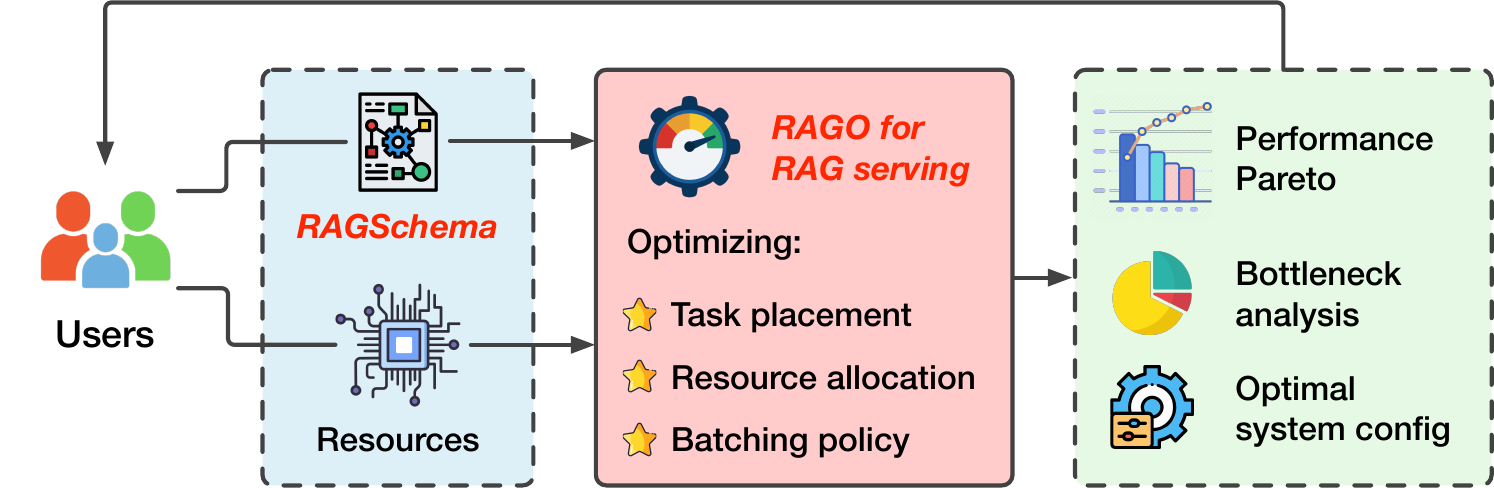}
  \end{subfigure}
  \caption{\oursemph for systematic RAG serving optimization.}
  \label{fig_rago:paper_overview}
\end{figure}

To this end, my introduce \oursemph~(\underline{R}etrieval-\underline{A}ugmented \underline{G}eneration \underline{O}ptimizer), a system performance optimization framework for efficient RAG serving (\Cref{fig_rago:paper_overview}).
%
Given a RAG workload represented by \ragdesc and system resource constraints, this framework explores the scheduling policy space to determine optimal schedules aligned with user-defined performance objectives. 
Key scheduling decisions of \ours include deciding whether inference components are collocated or disaggregated across ML accelerators (\emph{task placement}), assigning the type and quantity of resources to each component (\emph{resource allocation}), and tuning batch sizes for retrieval and inference tasks to balance throughput and latency (\emph{batching policies}). 
\ours uses an analytical cost model, inspired by~\cite{parashar2019timeloop,zhang2022full,huang2024mind}, to identify the performance Pareto frontier and generate corresponding system schedules. 
This cost model is based on XPU, a generic systolic-array ML accelerator~\cite{jouppi2017datacenter,inferentia,zhang2022full}, and serve as the core engine of \ours for evaluating various RAG paradigms and configurations.
Below, I summarize the key contributions of this work:

\begin{itemize}
\item I propose \ragdesc, a RAG workload abstraction that simplifies RAG workload representation and enables systematic performance characterization and optimization.
\item Using \ragdesc, I identify key system design decisions and performance trade-offs from characterizing four representative RAG paradigms and their instantiations. 
%
\item I develop \ours, a systematic optimization framework that optimizes scheduling policies for efficient RAG serving. Results show that \ours delivers up to 2$\times$ improvement in QPS per chip and a 55\% reduction in time-to-first-token latency compared to RAG serving systems built on LLM-only systems.
\end{itemize}
\section{Background}
\label{sec_rago:background}

Since the background of vector search and RAG is already covered in Chapter~\ref{chap:prelim}, this section mainly focuses on the introduction of LLM serving systems. 

%
Serving LLM-only systems typically involves two distinct stages: prefix (prompt computation) and decode (token generation)~\cite{patel2023splitwise, zhong2024distserve}.
The prefix stage processes the input prompt to generate the first output token and populate the associated key-value (KV) cache~\cite{vaswani2017attention}, which holds the encoded representation of the input context.
The decode stage, on the other hand, generates subsequent tokens one at a time in an auto-regressive manner, relying on the KV cache from the prefix stage.

Modern LLM serving systems~\cite{patel2023splitwise, zhong2024distserve} often disaggregate these stages, operating on separate accelerator to accommodate their distinct workload characteristics.
This disaggregated design is essential for performance due to the distinct workload characteristics of the two stages~\cite{patel2023splitwise, zhong2024distserve}.
The prefix stage processes the entire input sequence at once, making it highly compute-intensive.
%
Even with small batches, the prefix stage benefits from accelerators with high computational throughput to handle the full sequence length efficiently~\cite{patel2023splitwise}.
In contrast, the decode stage is memory-bound, as each inference step requires accessing the KV cache of previous tokens, while the amount of computation is small~\cite{patel2023splitwise}.
In addition to workload differences, these two phases affect different performance metrics with different SLAs: time-to-first-token (TTFT) for the prefix phase and time-per-output-token (TPOT) for the decode phase.
Ultimately, optimizing the performance of LLM-only serving often depends on efficient resource allocation between the prefix and decode stages~\cite{patel2023splitwise}.


\section{Structuring the Complex Terrain of RAG Serving}
\label{sec_rago:rag-schema}
In this section, I first describe four representative RAG paradigms with increasingly diverse and complex RAG pipelines. 
I then describe \ragdesc~(\S\ref{sec_rago:rag-schema:schema}), a structured abstraction to capture this workload diversity, serving as a foundation for serving performance characterization~(\S\ref{sec_rago:case-studies}) and optimization~(\S\ref{sec_rago:ragflow}).

\begin{table*}[t]
    \centering
    \begin{footnotesize}
    \setlength{\tabcolsep}{4pt} 
    \renewcommand{\arraystretch}{0.95} 
    \caption{\ragdesc component names, descriptions, and example design parameters.}
    \label{tbl:ragschema}
    \scalebox{0.85}{ 
        \begin{tabular}{p{5cm}!{\vrule width 0.7pt}p{11cm}!{\vrule width 0.7pt}p{2cm}!}
            \hline
            \textbf{\ragdesc Components} & \textbf{Description} & \textbf{Example} \\ \hline
            Document Encoder & Model size (parameters) of the encoder used to convert database documents and queries into vector representations. & 120M \\ \hline
            Vector Dimensionality & The number of dimensions for each database vector. & 768-dim \\ \hline
            Database Vector Number & Number of the database vectors, depends on the corpus size and passage chunk lengths. & 1,000 \\ \hline
            Retrieval Frequency & Whether iterative retrievals are permitted during decoding and number of retrievals per sequence. & Four per sequence \\ \hline
            Queries Per Retrieval & Number of query vectors used per retrieval (one or multiple). & Two per retrieval \\ \hline
            Query Rewriter & Model size of the generative query rewriter, if applied. & 8B \\ \hline
            Query Reranker & Model size of the retrieval results reranker (usually an encoder-only model), if applied. & 120M \\ \hline
            Generative LLM & Represents the model size of the main generative LLM used for answer generation. & 70B \\ \hline
        \end{tabular}
    } 
    \end{footnotesize}
\end{table*}

\subsection{Representative RAG Paradigms}
\label{sec_rago:rag-schema:cases}

I now show the workload diversity by describing the following representative RAG paradigms:

\niparagraph{Paradigm I: Hyperscale Retrieval.}
Hyperscale retrieval plus smaller LLMs can be used as an alternative to larger LLMs~\cite{borgeaud2022improving, shao2024scaling, wang2023instructretro}.
Prior work has shown that RAG systems can match or even surpass the quality of LLM-only systems when database sizes are sufficiently large~\cite{borgeaud2022improving, shao2024scaling}.
This is achieved while using sufficiently smaller models\,---approximately one-tenth the parameters of their LLM-only counterparts~\cite{borgeaud2022improving, wang2023instructretro}.
This quality parity is achieved because LLM-only models rely on their vast parameter sets to encode comprehensive knowledge during training~\cite{brown2020language, chowdhery2022palm, smith2022using, rae2021scaling}, whereas RAG systems dynamically integrate external knowledge at inference time, reducing the need for extensive parameterization within the model itself.

\niparagraph{Paradigm II: Long-Context Sequence Processing.}
Another common paradigm is to use RAGs to facilitate long-context processing~\cite{lee2024can, li2024retrieval, yue2024inference}.
For example, when answering questions based on a lengthy document (e.g., with more than 100K tokens) that a user has uploaded, a straightforward solution is to use the entire context\,---similar to use cases in Gemini 1.5~\cite{team2024gemini}, NotebookLM~\cite{notebooklm}, and ChatGPT~\cite{chatgpt} ---\,as a prompt.
However, this approach is often prohibitively expensive due to the large number of tokens to process.
Instead, an efficient alternative is to treat the user-provided long document as a knowledge database, retrieving only the relevant information needed to answer the questions.
This method substantially reduces the prompt size by avoiding the need to load the full text into the model's context window.
Recent studies~\cite{lee2024can,yue2024inference} demonstrate that this retrieval-based approach achieves similar response quality to using the full document as a prompt, providing a practical balance between cost and quality in handling long contexts.
In contrast to the paradigm I, RAG for long-context processing introduces two key modifications.
First, this setup includes a database encoder, which is necessary for constructing the database when the long context is initially provided.
Second, the database is orders of magnitude smaller.
For example, given a context length of 100K tokens and a passage chunk size of 100 tokens, the database only consists of 1K vectors, compared to tens to hundreds of billions of vectors in paradigm I~\cite{borgeaud2022improving, shao2024scaling}.

\niparagraph{Paradigm III: Iterative Retrievals.}
While a single retrieval at the beginning may suffice in some scenarios, recent studies~\cite{borgeaud2022improving, trivedi2022interleaving, jiang2023active} indicate that iterative retrievals\,---periodically updating retrieved content during generation---\,can significantly enhance model quality.
Such update of the retrieved content is particularly valuable in scenarios requiring multi-hop reasoning, where each retrieval provides additional context to guide the subsequent token generation process~\cite{trivedi2022interleaving, yue2024inference}.
In this configuration, the decoder initiates retrievals at flexible intervals during generation.
Upon issuing a retrieval, the generation of this sequence temporarily pauses the token generation, to process newly retrieved content through the prefix phase.
Only after integrating this additional context does the decoder continue generating the rest of sequence.

\niparagraph{Paradigm IV: Query Rewriter and Reranker}
Users often pose vague or complex queries, making it challenging to retrieve relevant information directly.
To address this, the retrieval process can be significantly improved by incorporating pre-processing and post-processing steps~\cite{chan2024rq, ma2023query, glass2022re2g, allahverdiyev2024chunkrag}.
For pre-processing, recent studies~\cite{chan2024rq, ma2023query} demonstrate that leveraging an LLM to rewrite the user's query can improve retrieval quality.
This LLM may either rephrase the query for clarity or decompose complex questions into multiple simpler queries that cover different aspects of the user's original intent~\cite{chan2024rq, ma2023query}.
Once the initial results are retrieved through vector search, a reranking model can be applied as a post-processing step~\cite{glass2022re2g, allahverdiyev2024chunkrag, rag_github}.
The reranker improves content retrieval quality by scoring each document's relevance beyond simple vector similarity and choosing documents that more closely align with the user's intended question.

\subsection{\ragdescemph for Workload Abstraction}
\label{sec_rago:rag-schema:schema}
\begin{figure}[t]
  \centering
  \includegraphics[width=1.0\linewidth]{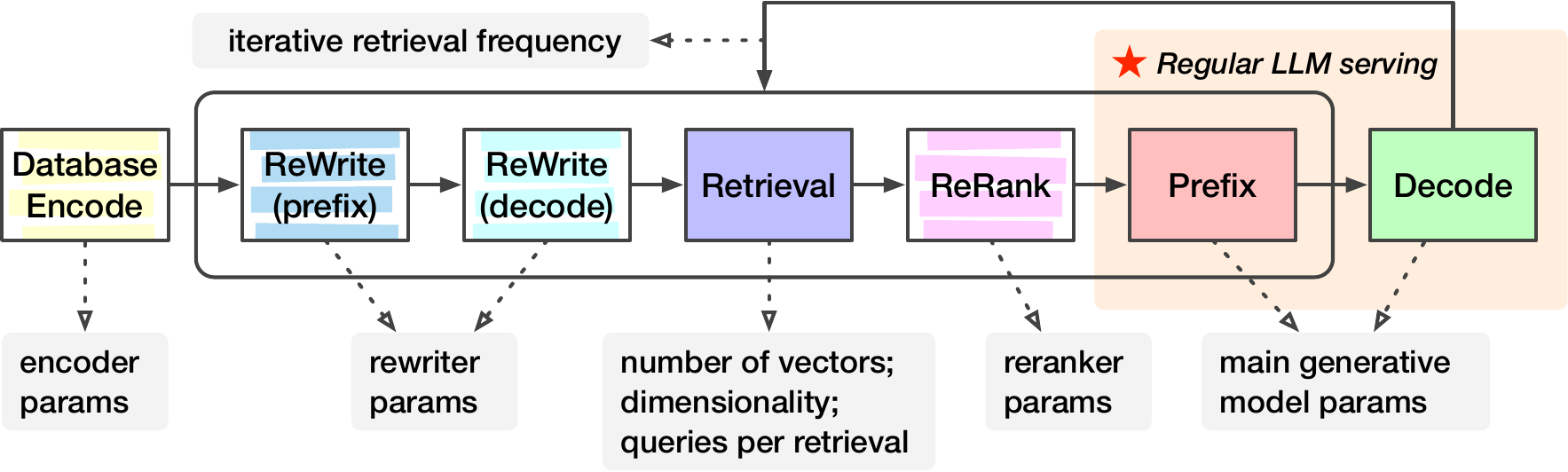}
  \vspace{-1em}
    \caption{Describing general RAG pipelines with \ragdesc.}
  \label{fig_rago:rag-schema}
\end{figure}

Given these diverse paradigms, RAG workloads exhibit significant variability across algorithm configurations in the following ways.
First, retrieval configurations can vary dramatically. 
Database sizes may span several orders of magnitude~\cite{borgeaud2022improving, shao2024scaling, lee2024can, li2024retrieval}; a retrieval may involve not a single query vector~\cite{lewis2020pre, guu2020realm} but multiple ones~\cite{wang2024richrag, besta2024multi, chan2024rq}; and some models support iterative retrievals during the generation process~\cite{borgeaud2022improving, trivedi2022interleaving, jiang2023active}.
Second, a RAG system may include several models in addition to the main generative LLM. 
These auxiliary models include a database encoder for processing real-time uploaded documents~\cite{reimers2019sentence, lee2024gecko}; a query rewriter model~\cite{chan2024rq, ma2023query} to rephrase user queries; and a result reranker model~\cite{glass2022re2g, allahverdiyev2024chunkrag, rag_github} to score retrieved information.

%
To navigate the complex RAG configuration space, I introduce \ragdescemph: \emph{a structured and modular abstraction that captures the key performance-relevant attributes of various RAG serving workloads}.
As visualized in \Cref{fig_rago:rag-schema} and detailed in \Cref{tbl:ragschema}, \ragdesc defines both (1) the execution flow of the RAG pipeline and (2) the configuration of its components.
For the RAG pipeline definition, optional stages\footnote{A "stage" refers to the execution of a RAG pipeline component.}---such as the database encoder, query rewriter, reranker, and iterative retrieval---\,can be included or omitted.
For each included component, \ragdesc specifies relevant configurations, including model parameter counts, vector dimensionality, number of database vectors, queries per vector, and iterative retrieval frequency if applicable.
While \ragdesc abstracts RAG serving workloads, it is not an abstraction for quality, as different models and databases of the same size can lead to varying quality.


%

\subsection{Empirical RAG Performance Trade-off Analysis}
%
Even though precise bottlenecks and tradeoffs depend on the exact RAGSchema, high-level performance bottlenecks in RAG systems are driven by RAG workload pipelines and the Amdahl's law.
In this section, I make general observations about RAG workloads, and quantify in ~\S\ref{sec_rago:case-studies} using detailed performance models.

I represent inference throughput as a function of FLOPs, and retrieval throughput as a function of the bytes of database vectors accessed.
Note the precise throughput depends on CPU server efficiency, accelerator capability, scheduling policies, etc.

\niparagraph{Inference components.}  
For a model with size \( M \) and a sequence length \( L \), the FLOPs required for processing the entire sequence are approximately: \( \text{FLOPs}_{\text{inference}} \approx 2 \cdot M \cdot L \) for short sequences (e.g., \( L \leq 10^3 \)) where the quadratic complexity of the attention mechanism still has negligible impact.
%

%
%

\niparagraph{Retrieval component.}  
The retrieval workload can be approximately described by the number of bytes of database vectors processed per query. 
Unlike model inference, decoding quantized database vectors represents a fundamentally different workload where FLOPs is not an appropriate metric~\cite{jiang2023chameleon, andre2016cache}. 
Given a database with \( N_{\text{dbvec}} \) vectors, where each vector consists of \( B_{\text{vec}} \) bytes, and each query scans a subset of \( P_{\text{scan}} \) percent of database vectors, the total bytes to scan per query is approximately:  \(
\text{B}_{\text{retrieval}} \approx N_{\text{dbvec}} \cdot B_{\text{vec}} \cdot \frac{P_{\text{scan}}}{100}
\).
%
%
Here, \( P_\text{scan} \) is determined by evaluating a set of sample queries and analyzing the relationship between \( P_\text{scan} \) and retrieval quality measured by recall, as a common practice for retrieval configuration tuning~\cite{annbench}. The minimum value of \( P_\text{scan} \) that satisfies the required retrieval quality is then selected.

\niparagraph{End-to-end RAG performance.}  
While the latency of RAG serving is the sum of the latencies of each stage in the RAG pipeline, the throughput of the pipeline is determined by its slowest stage (excluding iterative retrieval paradigm for now, as it follows a different pattern discussed in~\S\ref{subsec_rago:case3}).  
For a RAG pipeline with \( m \) stages, where each stage has a throughput denoted by \( \text{QPS}_i \) (\( i = 1, 2, \ldots, m \)), the end-to-end RAG serving throughput is:  
\(
\text{QPS}_{\text{RAG}} = \max(\text{QPS}_1, \text{QPS}_2, \ldots, \text{QPS}_m)
\).

From this high-level model, we can draw several key insights.
First, retrieval can become a bottleneck when its workload (\( N_{\text{dbvec}} \cdot B_{\text{vec}} \cdot \frac{P_{\text{scan}}}{100} \)) is high while the inference workload (\(  2 \cdot M \cdot L \)) is relatively low. 
%
Second, in paradigms with multiple inference components, any model can become critical depending on its size \( M \) and processed sequence length \( L \), which may vary based on the model's role.
%
Finally, the cumulative effect of multiple inference stages and retrievals can significantly impact overall serving performance.
%
I discuss detailed evaluation methodology and quantitative characterization in the subsequent sections.
%
\section{Methodology}
\label{sec_rago:method}
This section outlines the methodology used to characterize RAG performance (\S\ref{sec_rago:case-studies}) and evaluate \ours  across various configurations (\S\ref{sec_rago:eval}).

\niparagraph{Models and database.}
I evaluate four LLMs---\bench{Llama-3 1B}, \bench{8B}, \bench{70B}, and \bench{405B}~\cite{dubey2024llama}---covering size scales comparable to those used in~\cite{shao2024scaling, li2020improving, wang2023instructretro}. 
As RAG quality continues to benefit from larger knowledge corpora~\cite{borgeaud2022improving, shao2024scaling}, I adopt a hyperscale database~\cite{borgeaud2022improving}.
This database contains 64 billion passages, each encoded as a 768-dimensional vector~\cite{borgeaud2022improving}, making it approximately 400$\times$ larger than the largest academic vector search datasets~\cite{SIFT, babenko2016efficient, simhadri2022results}.
%
I apply product quantization (PQ) as in~\cite{borgeaud2022improving} to compress each vector to 96 bytes (1 byte per 8 dimensions), resulting in a 5.6\,TiB quantized vector database. 
Following the index recommendations of the ScaNN library~\cite{scann_github}, I use a balanced fanout of 4K vectors per node across the three-level tree index~\cite{sun2023automating} ($(64 \times 10^9)^{1/3} = 4 \times 10^3$).
To balance retrieval quality and performance, each query is compared against 0.1$\%$ of the database vectors by default, as this setup has shown high recall (over 90\%) in billion-scale datasets~\cite{jiang2023chameleon}.

\niparagraph{LLM sequence lengths.}
In line with common RAG use cases such as question-answering~\cite{lewis2020pre, trivedi2022interleaving, chan2024rq}, I evaluate sequence lengths derived from QA datasets~\cite{bajaj2016ms, joshi2017triviaqa, rajpurkar2018know}, where the question lengths range from six to 42 tokens.
To simplify the search space, I use 32 tokens as the typical question length.
The input prompt length includes both the question and relevant retrieved content.
The typical nearest neighbor retrieved ranges from two to 10~\cite{asai2023self, jiang2023active, trivedi2022interleaving, borgeaud2022improving}, each with an average length of 100 tokens.
I pick five as a common value for the evaluations. 
Given this, I approximate the average length of input prompt (question + relevant retrieved contents) to 512 tokens.
For generation lengths (decode stage), I rely on data from long-form QA~\cite{fan2019eli5} and chatbot datasets~\cite{sharegpt, vllm}, selecting 256 tokens as a representative decode length. 

\begin{table}[!t]
\caption{Performance specifications of three versions of XPUs. We report performance on XPU-C (\textcolor{blue}{\(\star\)}) by default.}
\label{tbl:xpu:gen}
\begin{footnotesize}
\begin{center}
\setlength{\tabcolsep}{6pt} 
\renewcommand{\arraystretch}{1.2} 
\scalebox{0.9}{
\begin{tabular}{@{\extracolsep{\fill}}l||c|c|c@{\extracolsep{\fill}}}
\hline
& \textbf{XPU-A} & \textbf{XPU-B} & {\textcolor{blue}{\(\star\)}}\,\textbf{XPU-C}\\
\hline\hline
\textbf{TFLOPS}                & 197 & 275 & 459 \\
\textbf{HBM (GB)}              & 16 & 32 & 96 \\
\textbf{Mem. BW (GB/s)}        &  819     &  1200     & 2765 \\
\textbf{Inter-Chip Link BW (GB/s)} & 200   &  300     & 600 \\ \hline
\textbf{Resembles} & TPU v5e~{\cite{tpuv5e}}& TPU v4~{\cite{tpuv4}} & TPU v5p~{\cite{tpuv5p}}\\\hline
\end{tabular}
} 
\end{center}
\end{footnotesize}
\vspace{-0.22cm}
\end{table}
%
%

\niparagraph{System setup.}
The evaluation assumes a data center model-serving environment with abundant resources to support various system configurations.
Across the RAG serving stages (\egc prefix, decode), I allocate a total of 16 to 32 servers hosting 64 to 128 XPUs (4 XPUs per server), as a minimum of 16 servers is required to ensure sufficient host memory capacity for the dataset (5.6\,TiB after quantization).
An XPU refers to a generic systolic-array-based ML accelerator~\cite{jouppi2017datacenter, inferentia}.
The number of XPUs allocated to each model component is configured in powers-of-two scaling factors (e.g., 1, 2, 4, etc.).
Each XPU, inspired by the setup of TPU v5p accelerators~\cite{tpuv5p}, is equipped with 96\,GB of high-bandwidth memory (2.7\,TB/s) and 459\,TFLOPS of compute capacity.
The XPUs are interconnected via a high-bandwidth 3D torus topology, offering 600\,GB/s of inter-chip bandwidth (six 100\,GB/s links per chip).
I also evaluate two other versions of XPUs, as shown in Table~\ref{tbl:xpu:gen}, for ablation studies.
The host CPUs are modeled after AMD EPYC Milan processors, featuring 96 cores, 384\,GB of memory, and 460\,GB/s of memory bandwidth.
I assume that XPU host servers support distributed retrieval across large databases.

\niparagraph{Simulation setup.}
RAG performance is reported by assembling the costs of all model inference and retrieval stages, based on a search across various system configurations (details described in~\S\ref{sec_rago:ragflow}).
I now describe the production-grade simulators used to measure inference and retrieval performance.

\emph{(a) Inference performance modeling.}
I adopt an in-house calibrated XPU simulator for inference simulation. 
The simulator is well-correlated with the production-grade XPU accelerators across a set of real-world ML models.
%
%
The simulator abstracts inference as a sequence of operators, where total latency is computed as the sum of each operator’s execution time and the associated data movement costs, similar to other established ML simulators~\cite{zhang2022full,parashar2019timeloop}.
The cost of each operator is calculated using a roofline model that accounts for compute, memory, and network costs.
For multi-XPU inference, the simulator explores variousevaluates a range of model sharding strategies, including tensor parallelism~\cite{shoeybi2019megatron, rajbhandari2020zero}, pipeline parallelism~\cite{huang2019gpipe, narayanan2019pipedream}, or the hybrid approach. 
Each accelerator is assigned a subset of inference components, with inter-machine communication costs explicitly modeled to ensure realistic latency estimations.

\emph{(b) Retrieval performance modeling.}
The retrieval simulation is based on ScaNN~\cite{guo2020accelerating, scann_github}, a product quantization library~(\S~\ref{chap:prelim:retrieval}) that demonstrates state-of-the-art performance across dozens of algorithms in the ANN benchmark~\cite{annbench}.
I implement the ScaNN performance model described in~\cite{sun2023automating}, which models the search process as a sequence of vector scan operations at each level of a multi-level tree~\cite{sun2023automating, scann_github}. The total retrieval latency is calculated as the sum of the latencies for these scan operations.
ScaNN dedicates one thread per query and parallelizes batches of queries across multiple threads.
The cost of each operator is calculated by a roofline model that factors in batch sizes, the number of CPU cores, per-core CPU processing throughput, and memory bandwidth.
For large databases requiring distributed search across multiple servers, I assume each server holds a shard of the dataset with independent indexes.
Queries are routed to all servers, and results are aggregated. 
The workload is balanced across servers, with negligible overhead for broadcast and gather operations.
To populate simulator parameters, I benchmark the maximum achievable per-core throughput and memory bandwidth by running open-source ScaNN~\cite{scann_github} on smaller datasets configured with the same tree node sizes (4K vectors per node) as the 64-billion vector database.
On AMD EPYC 7R13 CPUs with 24 cores, ScaNN achieved a PQ code scanning throughput of 18\,GB/s per CPU core, with approximately 80\% memory bandwidth utilization.
%
%
I then calibrate the retrieval performance model using internal production datasets comparable in scale to the 64-billion vector dataset used in this study~(\S\ref{sec_rago:method}).

\begin{table}[t]
    \centering
    \begin{footnotesize}
    \setlength{\tabcolsep}{4pt} 
    \renewcommand{\arraystretch}{0.95} 
    \caption{\ragdesc of the workloads used in case studies.}
    \label{tbl:ragschema-example}
    \scalebox{0.92}{ 
        \begin{tabular}{p{5cm}!{\vrule width 0.7pt}p{2.5cm}!{\vrule width 0.7pt}p{2.5cm}!{\vrule width 0.7pt}p{2cm}!{\vrule width 0.7pt}p{2cm}}
            \hline
            \textbf{\ragdesc Components} & \textbf{Case 1} & \textbf{Case 2} & \textbf{Case 3} & \textbf{Case 4} \\ \hline
            Document Encoder & N/A & 120M (768-d)  & N/A & N/A \\ \hline
            Database Vector Number & 64B & 1/10/100K & 64B & 64B  \\ \hline
            Retrieval Frequency & 1 & 1 & 2/4/8 & 1 \\ \hline
            Queries Per Retrieval & 1/2/4/8 & 1 & 1 & 1 \\ \hline
            Query Rewriter & N/A & N/A & N/A & 8B \\ \hline
            Query Reranker & N/A & N/A & N/A & 120M  \\ \hline
            Generative LLM & 1/8/70/405B & 8/70B & 8/70B & 8/70B \\ \hline
        \end{tabular}
    } 
    \end{footnotesize}
\end{table}


\niparagraph{Performance metrics.}
I report common metrics used in the evaluation of LLM systems~\cite{patel2023splitwise,zhong2024distserve}:
\begin{itemize}
\item {[\textbf{TTFT}]\,\textbf{Time-to-First-Token}} $\mapsto$ average latency from request reception to the generation of the first output token.
\item {[\textbf{TPOT}]\,\textbf{Time-Per-Output-Token}} $\mapsto$ average latency between generating each output token in a sequence. 
\item {[\textbf{QPS}]\,\textbf{Queries-Per-Second}} $\mapsto$ maximum throughput, or the number of requests the system can process per second. \footnote{Note that the term ``queries'' here does not refer to retrieval queries, and I use the QPS metric exclusively for end-to-end RAG serving performance in this chapter.}
\item {[\textbf{QPS/Chip}]\,\textbf{Queries-Per-Second/Chip}} $\mapsto$ QPS normalized by chip count, reflecting system cost efficiency. 
\end{itemize}

Since continuous batching~\cite{yu2022orca, vllm} is enabled in the decode stage, I report the worst-case TPOT latency.
This is because sequences in the batch can be at different stages of generation --- some generating the first token and others generating the last ones --- and performance is determined by the latter.
In contrast, prefix operates deterministically, allowing us to report the precise TTFT latency.

\section{RAG Serving Performance Characterization}
\label{sec_rago:case-studies}

In this section, I characterize workloads using four case studies, each representing a \ragdesc instantiation of a distinct RAG paradigm described in~\S\ref{sec_rago:rag-schema:cases}.
These case studies highlight the performance variability across RAG workloads, quantify the impact of paradigm choices on system performance, and motivate the need for an RAG optimization framework~(\S\ref{sec_rago:ragflow}).
While arbitrary RAG configurations can be constructed beyond these studies, they can often be seen as interpolations of the provided cases. 
For instance, a RAG system with a small private database could be viewed as a hybrid of Case I and II, where the large-scale retrieval in Case I is replaced with the small-scale retrieval of Case II.

\niparagraph{Characterization methodology.}
I evaluate performance using the methodology outlined in~\S\ref{sec_rago:method}.
Unless otherwise specified, the end-to-end performance plots depict the performance Pareto across all system scheduling options.
The time breakdown plots are normalized by the resource usage of each component, reflecting time $\times$ resource consumption. 
These plots assume (a) four XPUs per host server and (b) each component operating at its maximum QPS/Chip.
Thus, if a plot shows that retrieval time exceeds the combined time of all inference components, the host servers for retrievals are the bottleneck, leaving XPUs occasionally idle; conversely, if inference dominates, retrieval resources may be underutilized.
The evaluated workloads are summarized in~\Cref{tbl:ragschema-example}. 
While all configurations in the table are analyzed, the plots highlight a representative subset to avoid redundancy when similar trends are observed across multiple configurations (e.g., model sizes).
%


\subsection{{\textup{\textbf{Case I:}}} Hyperscale Retrieval}
\label{subsec_rago:case1}
I adopt configurations similar to those in RETRO~\cite{borgeaud2022improving}, replicating its database setup and similar sized LLMs along with more recent, larger models.
The RAG system performs only \emph{one} retrieval at the beginning, which may involve one or multiple query vectors, as suggested by recent studies~\cite{wang2024richrag, besta2024multi, chan2024rq}.

\begin{figure}[t]
  \centering
  \begin{subfigure}[b]{0.6\linewidth}
    \includegraphics[width=\linewidth]
    {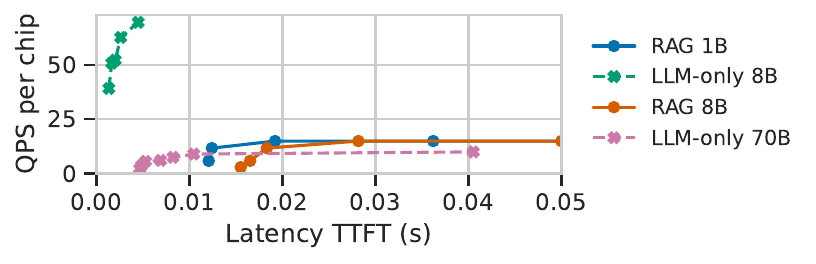}
  \end{subfigure}
  \caption{Larger LLM versus RAG with smaller models.}
    \label{fig_rago:case1:rag_vs_llm}
\end{figure}

\niparagraph{Takeaways:} 
Hyperscale retrieval can pose a significant bottleneck in RAG pipelines. 
%
This bottleneck becomes increasingly dominant with (1) smaller LLM, (2) multi-query retrievals, (3) better inference accelerators, (4) shorter prefix and decode sequence lengths, and (5) higher retrieval quality.

\niparagraph{System performance comparison (RAG vs. LLM-only). }
\Cref{fig_rago:case1:rag_vs_llm} compares RAG and LLM-only systems across different model sizes, with TTFT latency on the x-axis and QPS/Chip on the y-axis.
As shown in the RETRO paper~\cite{borgeaud2022improving}, RAG can achieve similar or superior generation quality to LLM-only systems with an order of magnitude fewer parameters.
Here, I extend this comparison to system performance.
The results indicate that RAG 8B outperforms LLM-only 70B in QPS/Chip by a factor of 1.5$\times$.
Although the model size is reduced by approximately 10$\times$, the benefits of using smaller models in RAG are moderated by the retrieval overhead and the need for longer prompts to integrate retrieved information (512 tokens in RAG versus 32-token questions in LLM-only systems), resulting in only a 3.2$\times$ reduction in inference FLOPs.
Consequently, the QPS/Chip gain is not directly proportional to the reduction in parameter size.
Interestingly, the results suggest that RAG model sizes can be increased up to a certain limit without compromising QPS/Chip, as retrieval performance is the limiting factor.
For example, RAG 1B and RAG 8B exhibit similar QPS, highlighting the importance of system performance analysis in determining how much larger RAG models can scale.
While RAG models offer significant advantages at certain scales, their benefits may diminish at lower parameter counts as retrieval latency becomes a bottleneck. 
For example, despite RAG 1B having only one-eighth the parameters of LLM-only 8B, its QPS/Chip does not scale proportionally, because the retrieval overhead in RAG outweigh the benefits of reduced model size.

\begin{figure}[t]
  \centering
  %
  \begin{subfigure}[b]{0.4\linewidth}
    \includegraphics[width=\linewidth]
    {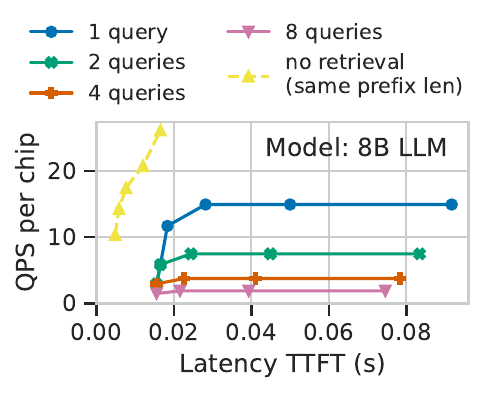}
    \caption{QPS/Chip 8B}
    \label{fig_rago:case1:latency:8b}
  \end{subfigure}
  \begin{subfigure}[b]{0.4\linewidth}
    \includegraphics[width=\linewidth]
    {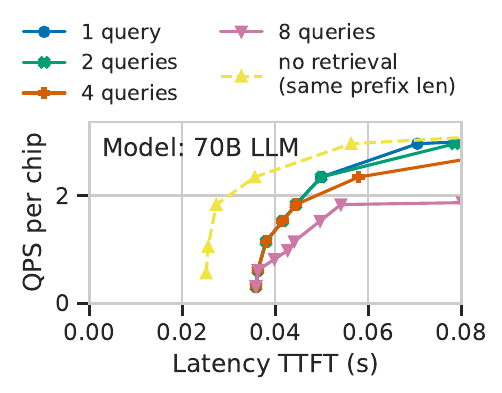}
    \caption{QPS/Chip 70B}
    \label{fig_rago:case1:latency:70b}
  \end{subfigure}
  \begin{subfigure}[b]{0.4\linewidth}
    \includegraphics[width=\linewidth]
    {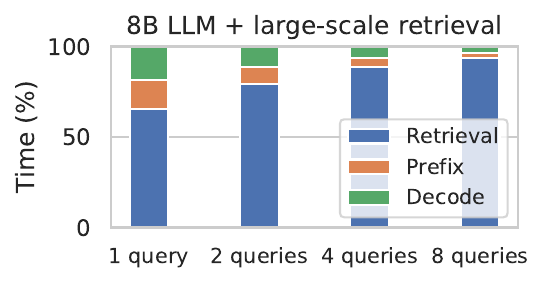}
    \caption{Breakdown 8B}
    \label{fig_rago:case1:break:8b}
  \end{subfigure}
  \begin{subfigure}[b]{0.4\linewidth}
    \includegraphics[width=\linewidth]
    {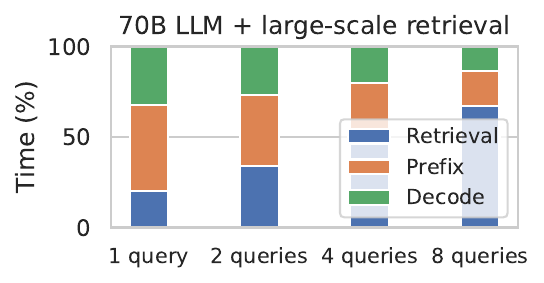}
    \caption{Breakdown 70B}
    \label{fig_rago:case1:break:70b}
  \end{subfigure}

  \caption{RAG performance given various model size and query numbers for hyperscale retrieval.}
  \label{fig_rago:case_1_latency_throughput}
\end{figure}

\niparagraph{Sensitivity to model size.}
Figure~\ref{fig_rago:case1:latency:8b} and Figure~\ref{fig_rago:case1:latency:70b} present the QPS/Chip for the \bench{8B} (left) and \bench{70B} (right) models, alongside time breakdowns for retrieval, prefix, and decode stages in Figure~\ref{fig_rago:case1:break:8b} and Figure~\ref{fig_rago:case1:break:70b}.
The yellow line represents a ``no retrieval'' configuration, where retrieval is omitted while the prefix remains the same length.
For the 8B model, retrieval is the primary bottleneck; as query counts double, QPS nearly halves due to increased retrieval demands.
Conversely, for the 70B model, inference initially limits performance until four queries per retrieval.
At higher query vector counts per retrieval (\egc 8 queries), the bottleneck shifts, and retrieval starts to dominate, as seen in the time breakdown in Figure~\ref{fig_rago:case1:break:70b}.

\begin{figure}[t]
  \centering
  \begin{subfigure}[b]{0.3\linewidth}
    \includegraphics[width=\linewidth]
    {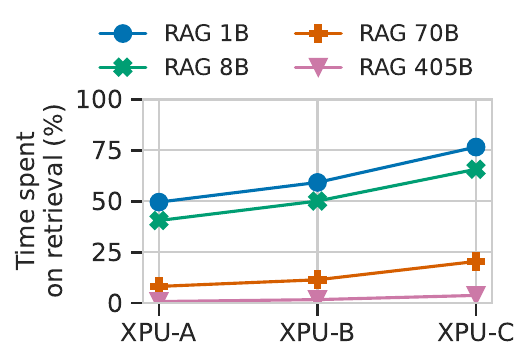}
    \caption{XPU Gen}
    \label{fig_rago:case1:xpu}
  \end{subfigure}
  \begin{subfigure}[b]{0.27\linewidth}
    \includegraphics[width=\linewidth]
    {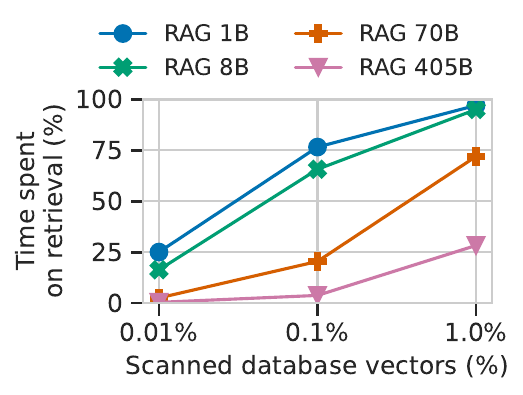}
    \caption{Retrieval Config}
    \label{fig_rago:case1:db_scan}
  \end{subfigure}
  \begin{subfigure}[b]{0.38\linewidth}
    \includegraphics[width=\linewidth]
    {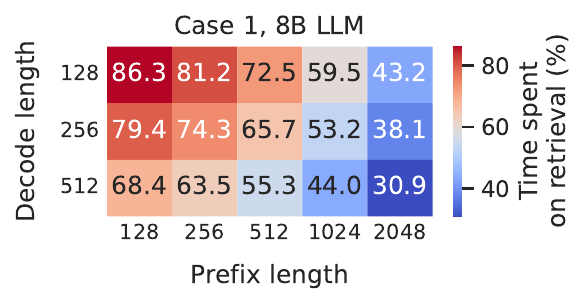}
    \caption{Seq. Length}
    \label{fig_rago:case1:seqlen}
  \end{subfigure}
  \caption{The percentage of retrieval time across hardware, retrieval configurations, and sequence lengths in Case I. }
  \label{fig_rago:case_1_sweep_hardware_models}
\end{figure}

\niparagraph{Sensitivity to XPU versions.}
\Cref{fig_rago:case1:xpu} shows the impact of XPU capability on the percentage of time spent on retrieval for LLMs ranging from 1B to 405B parameters.
As the XPU capabilities advance (from version A to C), the proportion of time spent on retrieval increases by up to 25\%.
While for larger models (e.g. 405B), LLM remains the dominant bottleneck in RAG serving, retrieval is dominant factor for RAG with small models (50\% - 75\% across XPUs versions).
Overall, with more advanced ML accelerators, system efficiency increasingly depends on optimizing retrieval processes.

\niparagraph{Sensitivity to sequence lengths.}
\Cref{fig_rago:case1:seqlen} illustrates the sensitivity of retrieval overhead to changes in decode length and prefix length given for \bench{8B} model.
The retrieval overhead varies significantly with both decode and prefix lengths --- retrieval bottlenecks diminish as sequence lengths increase, shifting retrieval from a primary bottleneck to a secondary factor.
For example, 86.3\% of the time is spent on retrieval at  shorter sequence lengths (\egc 128 or 256), while the retrieval overhead drops to just 30.9\% with longer prefix and decode lengths (2048 and 512).
Adjusting the prefix and decode lengths results in unequal changes in the percentage of retrieval time.
For example, in a setting of 128 tokens for both prefix and decode, increasing the prefix length to 256 tokens reduces retrieval time from 86.3\% to 81.2\%, while increasing the decode length to 256 tokens lowers it to 79.4\%.
This difference occurs because prefix inference is inherently faster than decoding the same number of tokens due to the autoregressive nature of decoding.

\niparagraph{Sensitivity to retrieval configurations.}
Retrieval performance in RAG workflow is highly sensitive to the percentage of database vectors scanned per search.
Regardless of the ANN algorithm used, ANN search does not conform to a fixed workload --- there is an fundamental trade-off between retrieval performance and quality: scanning more vectors improves quality but reduces performance~\cite{PQ, malkov2018efficient}.
This trade-off is further influenced by data distribution; for instance, with the same algorithm, hardware, and QPS, one dataset may achieve over 90\% recall, while another may fall below 50\%~\cite{simhadri2022results}.
While prior evidence suggests that higher recall can enhance generation quality~\cite{jiang2024piperag, 2024ragannquality}, there has been no consensus on the optimal recall threshold.
\Cref{fig_rago:case1:db_scan} illustrates the impact of varying the percentage of scanned database vectors, ranging from 0.01\% to 1\% (with 0.1\% as the default), on the proportion of time spent on retrieval across different model sizes.
For all models, increasing the scanned database vectors significantly amplifies the proportion of time spent on retrieval, highlighting the substantial variability in retrieval performance across RAG configurations.

%

\subsection{{\textup{\textbf{Case II:}}} Long-Context Sequence Processing}
\label{subsec_rago:case2}
\begin{figure}[t]
  \centering
  %
  \begin{subfigure}[b]{0.4\linewidth}
    \includegraphics[width=\linewidth]
    {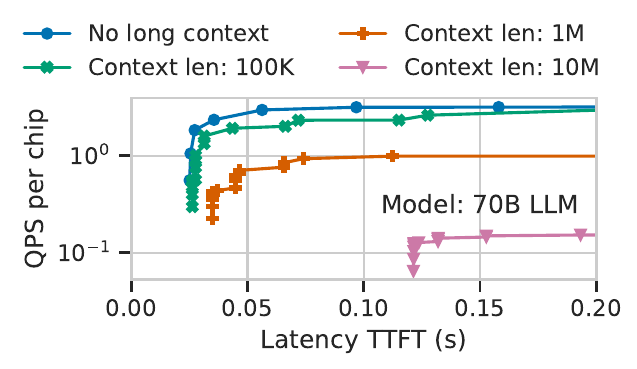}
    \caption{QPS/Chip 70B}
    \label{fig_rago:case2:latency:70b}
  \end{subfigure}
  %
  \begin{subfigure}[b]{0.44\linewidth}
    \includegraphics[width=\linewidth]
    {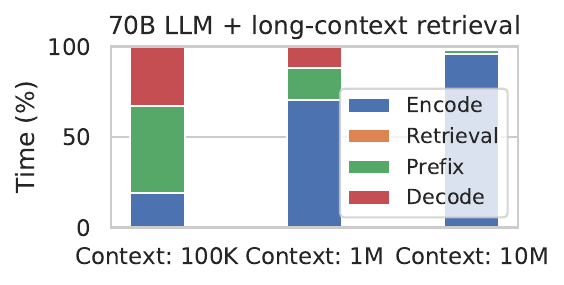}
    \caption{Breakdown 70B}
    \label{fig_rago:case2:break:70b}
  \end{subfigure}
  \caption{RAG performance for long-context processing.}
  \label{fig_rago:case_2_latency_throughput}
\end{figure}
%

%
As shown in \Cref{tbl:ragschema-example}, I evaluate context lengths ranging from 100K to 10M tokens, resulting in a database of 1K to 100K vectors, with each chunk sized at 128 tokens and small overlaps between chunks.
I use a sentence transformer model with 120\,M parameters~\cite{reimers2019sentence} to encode the passages, generating 768-dimensional embeddings, as relatively compact models are sufficient to achieve high retrieval quality~\cite{reimers2019sentence}.
Instead of ANN search, I use brute-force kNN search due to the high indexing costs associated with newly generated embeddings.

\niparagraph{Takeaways:}
In contrast to the Case I, retrieval performance plays a minimal role here.
Instead, the database vector encoding process emerges as the bottleneck, even with a small encoder model, due to the significantly longer context the encoder must process compared to the generative LLM.
%


\niparagraph{RAG vs. long-context LLMs.}
Despite the high encoding cost, RAG is significantly more efficient than processing the entire long context as a prompt (long-context LLM).
For instance, with a 1M-token context and a 70B LLM, RAG reduces the required prefix length to 512 tokens, achieving a speedup of 2852.6$\times$ in TTFT and 6633.9$\times$ in QPS/Chip.
This is even considering an efficient long-context LLM applying global attention to all tokens in only one out of every four layers, while the rest layers only apply local attention to the last 128 tokens.
This cost efficiency arises for two primary reasons: ($\mathrm{I}$) In long-context RAG, the database encoder, typically a small model (\egc 120M parameters), performs the encoding.
This is much less computationally intensive compared to the LLM-only system with billions of parameters, which would require significantly more FLOPs if fed the entire context.
($\mathrm{II}$) 
Long-context LLMs require key-value caches for every token, consuming substantial XPU memory (i.e. cost).
In contrast, RAG significantly reduces prompt lengths, saving XPU memory.
This distinction enables RAG to handle larger batch sizes during generation, increasing QPS/Chip.

\niparagraph{Sensitivity to context length.}
Figure~\ref{fig_rago:case_2_latency_throughput} presents performance trends when the input context length scales from 100K to 10M tokens for the 70B model.
"No long context" line represents the standard prompt length of a 512-token prefix.
As the context length increases, RAG performance gradually degrades due to the increasing cost of context encoding, even though retrieval enables prompt truncation for the generative LLM.
This happens due to database encoding becoming the bottleneck (Figure~\ref{fig_rago:case_2_latency_throughput}), especially at longer context lengths ($>$1M).
Notably, encoding time scales with context length, despite the relatively small encoder applied (120M parameters), due to the sheer volume of data processed.
Therefore, caching the generated embedding for potential reuse can significantly reduce computation with minimal cost.
For instance, caching 10K 768-d database vectors in FP16 format (for 1M tokens) requires only 15 MB of CPU memory or storage.
The retrieval time is minimal even when using brute-force search due to small database (1K-100K vectors vs 64B in other cases).

\subsection{{\textup{\textbf{Case III:}}} Iterative Retrievals + Prefix}
\label{subsec_rago:case3}
\begin{figure}[t]
  \centering
  \begin{subfigure}[b]{0.4\linewidth}
    \includegraphics[width=\linewidth]
    {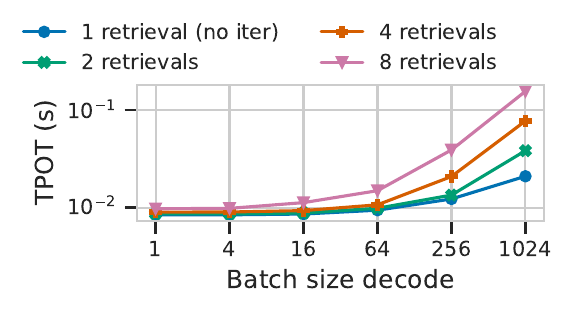}
    \caption{Retrieval frequency}
    \label{fig_rago:case_3_latency_throughput_retrievals}
  \end{subfigure}
  \begin{subfigure}[b]{0.4\linewidth}
    \includegraphics[width=\linewidth]
    {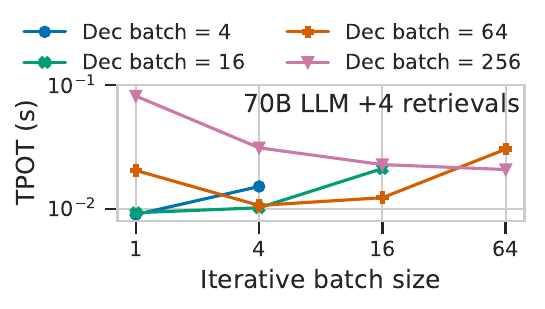}
    \caption{Prefix-retrieval batch size}
    \label{fig_rago:case_3_latency_prefix_decode_batch}
  \end{subfigure}
  \caption{RAG performance with iterative retrievals.}
  \label{fig_rago:case_3_latency_throughput}
\end{figure}
%

%
The iterative retrieval setup allows for 2, 4, or 8 retrievals per sequence generation process.
Each retrieval is triggered at random intervals during the 256-token decoding process, with retrievals uniformly distributed across token positions.

\niparagraph{Takeaways:}
Batch sizes for iterative retrievals must be carefully selected, as they significantly impact TPOT latency.
Larger batch sizes improve retrieval and prefix throughput but may stall decoding, negating the gains.
%

\niparagraph{Sensitivity to retrieval frequency.}
\Cref{fig_rago:case_3_latency_throughput_retrievals} examines the impact of different retrieval frequencies (1-8 per sequence) on TPOT latency as the decode batch size increases from 1 to 1024, as QPS/Chip shows similar trends as multi-query retrieval in Case I.
The results indicate that TPOT latency increases with both retrieval frequency and the decode batch size.
At smaller decode batch sizes (one, four, and 16), the TPOT latency differences between retrieval frequencies are relatively minor.
This is because, at these lower batch sizes, the decode step remains the dominant factor in TPOT latency, contributing approximately 60\%-80\% of the latency, while the effect of additional retrievals remains limited.
At higher batch sizes, however, the decode process achieves higher QPS/Chip, reducing its share of the overall TPOT latency. 
This shift in bottleneck exposes the impact of retrieval frequency, as retrievals become the primary contributor to latency.
Consequently, at larger batch sizes, the latency gap across different retrieval frequencies widens, making the increased time required for multiple retrievals more pronounced.

\niparagraph{Sensitivity to iterative retrieval batch size.}
In \Cref{fig_rago:case_3_latency_prefix_decode_batch}, I observe the nuanced interplay between decode batch size, iterative retrieval-prefix batch size and TPOT latency for a 70B model processing four retrievals per sequence.
At smaller decoding batch sizes (4 and 16), increasing the iterative retrieval batch size results in a noticeable increase in latency.
This is due to the inherent challenge in finding enough retrieval requests to batch within an active set of decoding sequences over a given time interval, introducing stalls.
For decode batch sizes of 256, the relationship reverses.
As the iterative retrieval batch size increases, latency decreases.
Here, the abundance of active decode sequences allow the system to batch retrieval requests more rapidly, enabling improved performance.
The decode batch size of 64 presents a particularly intriguing case: it reaches its lowest TPOT at retrieval batch size of four.
This optimal point represents a balance where idle time is minimized and the batching of retrieval requests is most efficient.
However, beyond this threshold, latency begins to climb again as it becomes progressively harder to amass a sufficient number of retrieval requests for efficient batching.
This behavior illustrates the delicate balance in RAG system performance when trying to balance retrieval performance and decoding efficiency.

\begin{figure}[t]
  \centering
  \begin{subfigure}[b]{0.48\linewidth}
    \includegraphics[width=\linewidth]
    {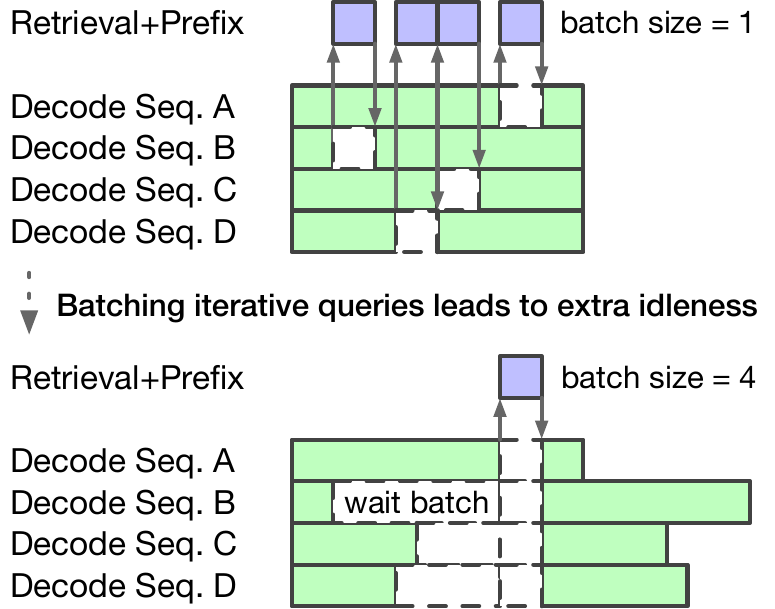}
    \caption{Wait for query batching}
    \label{fig_rago:case_3_batching_idleness}
  \end{subfigure}
  \begin{subfigure}[b]{0.48\linewidth}
    \includegraphics[width=\linewidth]
    {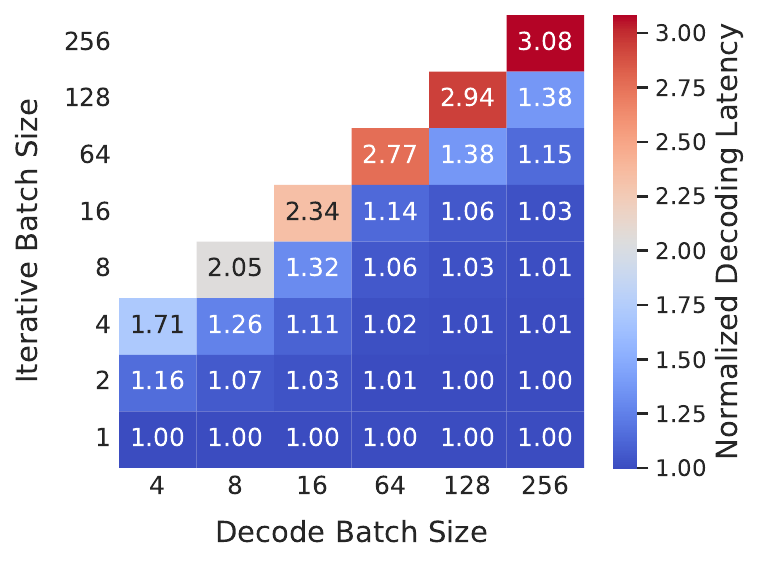}
    \caption{Performance degrade}
    \label{fig_rago:case_3_batching_idleness_latency}
  \end{subfigure}
  \caption{Decode idleness due to batched iterative queries.}
  \label{fig_rago:case_3_batching}
\end{figure}

\Cref{fig_rago:case_3_batching} further illustrates the phenomenon of decoding slowdown caused by idleness.
\Cref{fig_rago:case_3_batching_idleness} visualizes the batching process, while the heatmap (\Cref{fig_rago:case_3_batching_idleness_latency}) shows normalized decoding latency (compared to no retrieval) as a function of the decode batch size (x-axis) and iterative retrieval batch size (y-axis).
In this evaluation, the retrieval and prefix stages are assumed to have zero latency, isolating the slowdown to the batching-induced waiting time.
The results show that the effective latency is highly sensitive to the ratio of decode batch size to iterative retrieval batch size.
When these batch sizes are similar (\egc both set to 64), the normalized decoding latency reaches up to \xx{2.78}.
This increase occurs because one of the requests may generate a larger number of tokens before the next retrieval, resulting in idleness becomes a dominant factor.
For smaller ratios (e.g., decode batch size 64 and retrieval batch size up to 16), latency increases more gradually, indicating a more balanced workload with minimal idleness.
This observation aligns with \Cref{fig_rago:case_3_latency_prefix_decode_batch}, where, for a decode batch size of 64, increasing the iterative retrieval batch size from 16 (\xx{1.14} normalized latency due to idleness) to 64 (\xx{2.78} normalized latency due to idleness) causes a significant increase in TPOT latency.
In summary, the results suggest that (a) when there is a large pool of XPUs that allows for large decoding batches, one can choose the iterative batch size that saturates database throughput, however, (b) with a smaller pool of XPUs and smaller decoding batch sizes, the optimal decoding batch size may actually be lower than the one that fully saturates the database.

\subsection{{\textup{\textbf{Case IV:}}} Query Rewriter and reranker}
\label{subsec_rago:case4}
\begin{figure}[t]
  \centering
  \begin{subfigure}[b]{0.4\linewidth}
    \includegraphics[width=\linewidth]
    {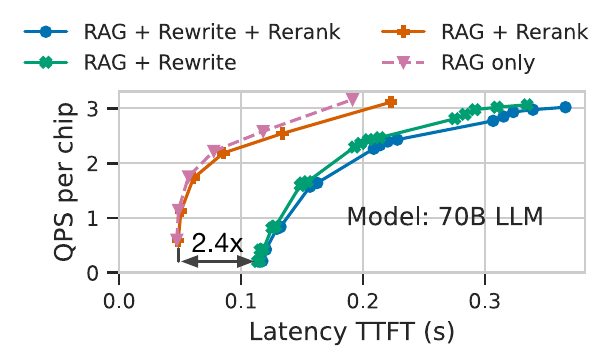}
  \end{subfigure}
  \begin{subfigure}[b]{0.48\linewidth}
    \includegraphics[width=\linewidth]
    {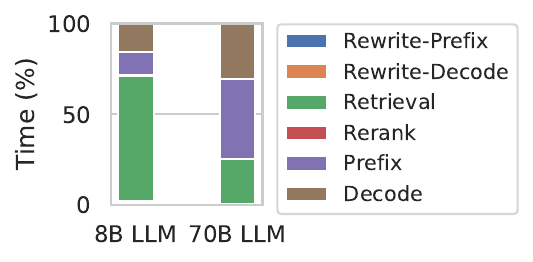}
  \end{subfigure}
  \caption{RAG performance with rewriter and reranker. }
  \label{fig_rago:case_4_latency_throughput}
\end{figure}
In this setup, I extend Case I by integrating an 8B query rewriter model~\cite{dubey2024llama} and a 120M reranker~\cite{reimers2019sentence}.
The rewriter processes a 32-token question and generates a rephrased question of the same length, while the reranker evaluates 16 nearest passages, each containing 100 tokens, and returns the top five nearest neighbors.

\niparagraph{Takeaways:} 
While the reranker has negligible impact on overall RAG performance, the query rewriter can significantly increase TTFT latency due to its autoregressive nature.

\niparagraph{System performance comparison (RAG vs RAG with rewrite and rerank).}
\Cref{fig_rago:case_4_latency_throughput} (left) presents the performance of various RAG configurations with or without rewriter and reranker.
The results indicate that QPS/Chip remains largely unaffected by the addition of the rewriting and reranking modules.
This is further validated from  \Cref{fig_rago:case_4_latency_throughput} which shows that negligible time is spent in rewriter and reranker stages.
However, the TTFT latency increases significantly (2.4$\times$) when the rewriter is included, due to its autoregressive generation nature, while reranking has minimal impact on TTFT.
This highlights the importance of considering an application’s latency tolerance when integrating the rewriter component, as it can substantially affect the user experience in latency-sensitive scenarios.
%

\section{\oursemph: Systematic RAG Serving Optimization}
\label{sec_rago:ragflow}
\begin{figure}[t]
  \centering
  \begin{subfigure}[b]{0.8\linewidth}
    \includegraphics[width=\linewidth]{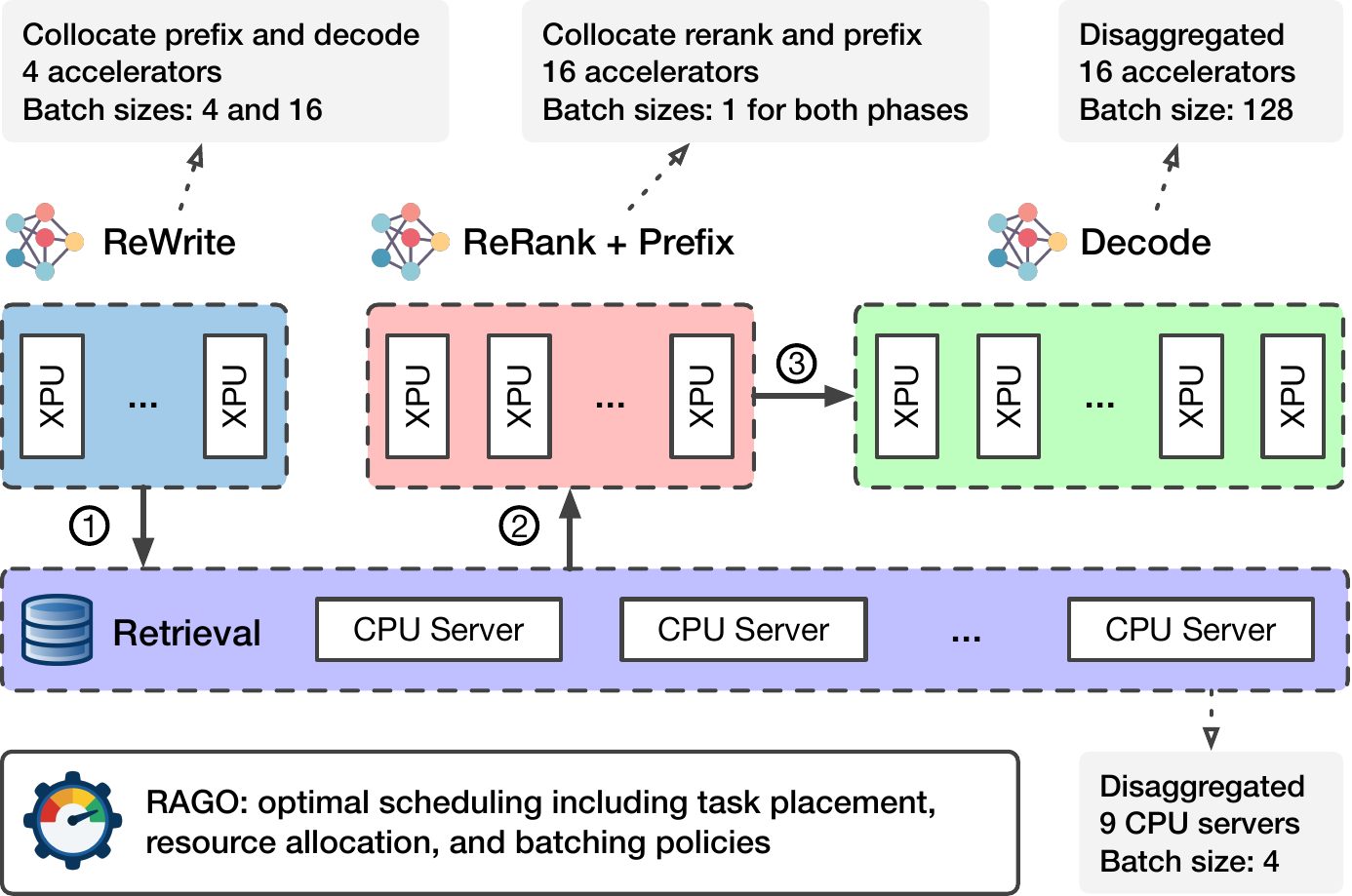}
  \end{subfigure}
  \caption{An example of \ours optimizing placement, allocation, and batching policies for efficient RAG serving.}
  \label{fig_rago:flexrag}
\end{figure}
Given the heterogeneous components and high workload variance across RAG (\S\ref{sec_rago:case-studies}), one-size-fits-all systems are inherently inadequate for achieving optimal serving efficiency.
To overcome this challenge, I introduce \oursemph, a systematic framework to design and optimize RAG serving systems across diverse configurations.
\ours determines an optimized scheduling policy tailored to a specific \ragdesc and a defined performance target. 
The following sections expound on the components of \ours and its overall design.

\subsection{\ours Scheduling Decisions}
Each scheduling solution comprises three pivotal system decisions: \textit{task placement, resource allocation, and batching policy.}
\Cref{fig_rago:flexrag} illustrates an example how these decisions come together to optimize a RAG serving pipeline under the constraint of 36 XPUs.
In this example, \ours adopts a hybrid collocation-disaggregation task placement strategy.
Specifically, the pipeline is organized into two collocated subsystems: (1) the rewrite-prefix and rewrite-decode phases; and (2) the rerank and prefix phases of response generation.
This organization ensures that tightly coupled tasks are efficiently grouped.
Resource allocation is tailored to the computational demands of each subsystem.
For instance, the query rewriter is assigned four XPUs, while the decoding phase, requiring significantly higher computational power, is allocated 16 XPUs.
To further enhance efficiency, \ours assigns batching policies customized to the characteristics of each phase.
For example, the rerank and prefix phases prioritize low-latency processing with a batch size of one, whereas the decoding phase operates with a much larger batch size of 128 to maximize throughput.
By orchestrating these decisions, \ours acts as the engine driving optimized serving performance across diverse RAG configurations.
Below, I formally describe each system scheduling decision, deferring how to search for optimal schedules to~\S\ref{sec_rago:rag_scheduler}.

\niparagraph{[I] Task placement.}
Recent LLM serving systems~\cite{patel2023splitwise,zhong2024distserve} advocate for disaggregating the prefix and decode phases (\S\ref{sec_rago:background}), as these phases exhibit distinct workload characteristics\,---compute-bound vs. memory-bound---\,and impact TTFT versus TPOT.
However, given the multiple components in a RAG pipeline (\Cref{fig_rago:rag-schema}), a natural question arises: \emph{should RAG systems adhere to the convention of fully disaggregated designs common in LLM-only systems?}
While prefix-decode disaggregation often proves beneficial (\S~\ref{sec_rago:background}), RAG pipelines may benefit more from adopting a collocation or hybrid collocation-disaggregation strategy, particularly for components leading up to the prefix phase.
%
%
First, several components in the pipeline --- such as the database encoder, reranker, and the prefix phases of both the query rewriter and the main LLM --- share a similar profile of high computational intensity, and thus time-multiplexing these components on the same set of XPUs can inherently mitigate workload imbalances among them.
%
Second, components up to the prefix phase directly influence TTFT latency: while a fully disaggregated design, constrained by limited accelerator resources per stage, can prolong TTFT, collocation mitigates this by allowing all components to share the available resources, thereby reducing latency.
%

%
\begin{figure}[t]
  \centering
  \begin{subfigure}[b]{0.65\linewidth}
    \includegraphics[width=\linewidth]{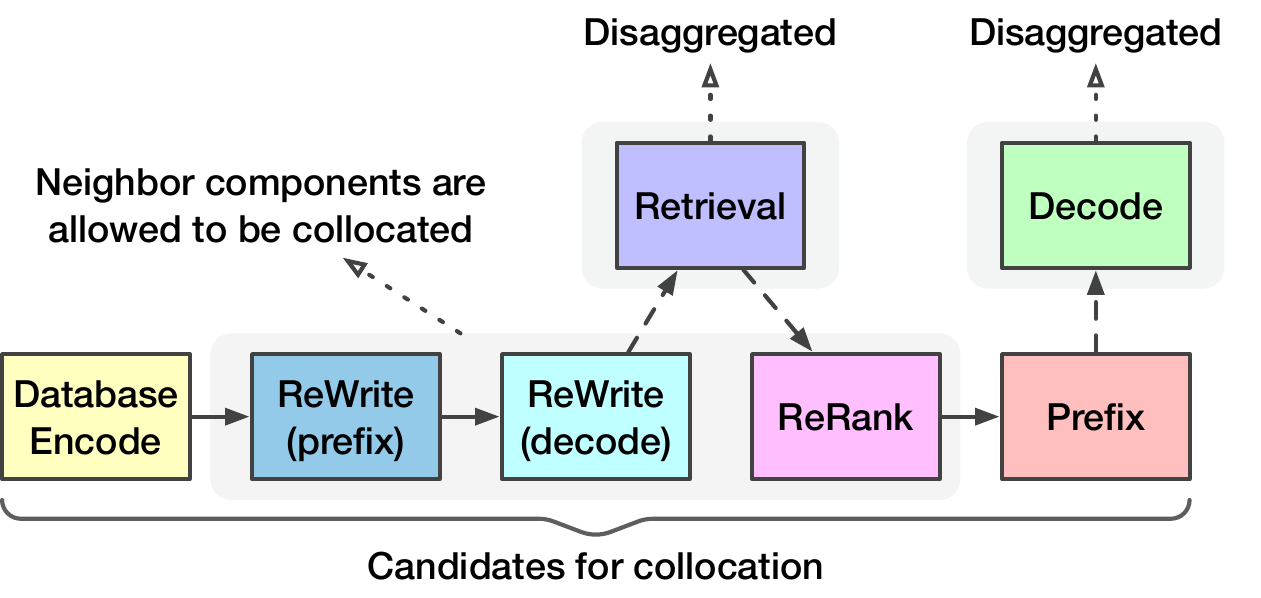}
  \end{subfigure}
  \caption{\ours allows collocation of neighbor models.}
  \label{fig_rago:collocation_example}
\end{figure}

That said, the decision between collocation and disaggregation depends on the specific characteristics of the RAG pipeline. 
For instance, the decoding phase of the query rewriter is autoregressive, and scales pooly with small batch sizes even with additional XPUs~\cite{vllm, yu2022orca}. 
Thus, collocating it with the prefix phase across many chips risks underutilizing hardware resources, as analyzed in~\S\ref{sec_rago:eval}.
%
%
To address these challenges, \ours supports hybrid collocation-disaggregation task placement policies. 
This approach balances flexibility and performance, as outlined as follow.
Firstly, the main LLM’s prefix and decode phases remain disaggregated, consistent with the strategies in~\cite{patel2023splitwise,zhong2024distserve}; 
Secondly, retrieval is always treated as a disaggregated task, as it operates on CPUs rather than XPUs.
Finally, neighboring phases up to the prefix can be collocated (\Cref{fig_rago:collocation_example}). Collocation is restricted to consecutive neighbors to avoid excessively complicating the search space.

\niparagraph{[II] Resource allocation.}
After determining task placement, \ours assigns resources to each pipeline phase based on its computational and memory requirements.
For collocated inference phases, this involves selecting the appropriate number of accelerators to ensure efficient execution. 
Similarly, for retrieval operations, \ours determines the number of CPU servers required to meet workload demands.
The framework balances throughput requirements and latency constraints to ensure optimal performance.
Additionally, \ours ensures that each component has sufficient accelerator or CPU memory capacity to store the required models or database segments while meeting the specified performance targets.

\niparagraph{[III] Batching policy.}
Given a batch of incoming requests, \ours enables each stage of the pipeline to adopt varying batch sizes, offering flexibility to balance latency and throughput at each stage.
%
For the decode stage, \ours leverage continuous batching~\cite{yu2022orca, vllm} to use larger batch sizes than individual requests, thereby improving throughput, as I evaluate in \S~\ref{sec_rago:eval}.
%
%
Moreover, in the case of iterative retrievals (\S~\ref{subsec_rago:case3}), \ours allows distinct batch sizes for the \emph{initial} retrieval/prefix pair and the \emph{subsequent} decoder-initiated retrieval/prefix iterations.
This differentiation is crucial because the initial retrieval and prefix phases directly affect TTFT, while the iterative ones primarily impact TPOT during token generation (see \S~\ref{subsec_rago:case3}).
%

\begin{figure}[t]
  \centering
  \begin{subfigure}[b]{0.65\linewidth}
    \includegraphics[width=\linewidth]{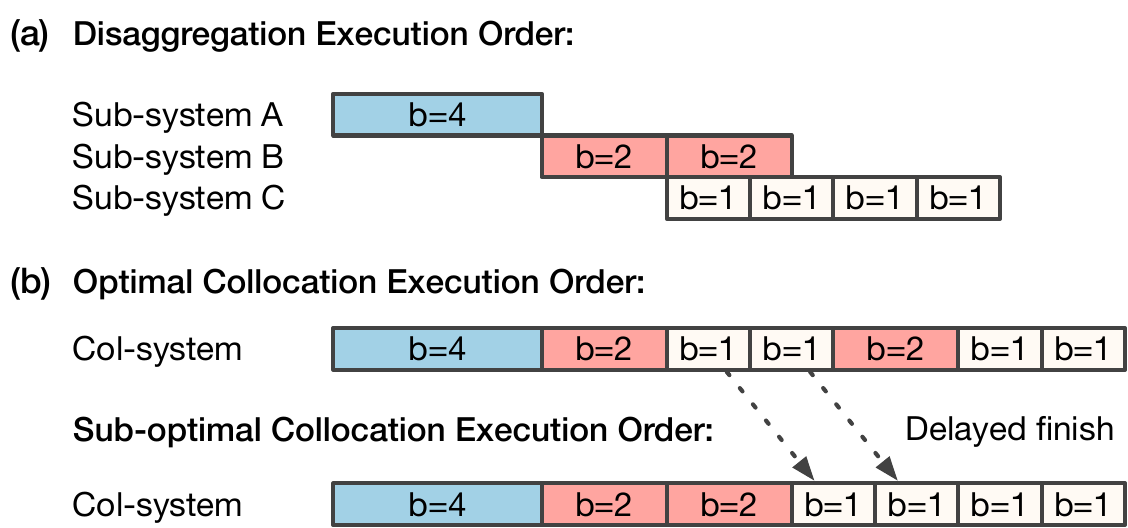}
  \end{subfigure}
  \caption{Execution order of batched requests until prefix.}
  \label{fig_rago:execution_order}
\end{figure}
Once batch sizes are determined, \ours organizes their execution order to maximize efficiency based on the task placement strategy.
Here, I discuss the order of stages up to prefix, as the generative LLM decode always apply continuous batching~\cite{yu2022orca, vllm}.
In a fully disaggregated design (\Cref{fig_rago:execution_order}(a)), execution is straightforward. As soon as (1) sufficient inputs arrive for a subsystem and (2) the subsystem completes its previous batch, it processes the new batch and forwards the output to the next subsystem.
On the other hand, \Cref{fig_rago:execution_order}(b) shows the execution order of the collocated design. 
For simplicity, I use time-multiplexing strategy and leave more complex strategies such as simultaneous execution as future work.
In time-multiplexed designs, the throughput of the collocated system is fixed once batch sizes are set for each stage.
In such cases, a stage begins execution as soon as it accumulates enough inputs. 
As illustrated in \Cref{fig_rago:execution_order}, the optimal execution order prioritizes completing the final stage (b=1) early over processing another round of the second stage (b=2), thereby minimizing the average completion time of the final stage.
If a retrieval operation is required between collocated stages (e.g., between the rewrite and prefix stages), the system pauses until the retrieval phase is complete before resuming the next collocated model inference phase.
By optimizing batching policies and execution order, \ours ensures that the trade-offs between latency and throughput are carefully managed for each stage, enabling efficient end-to-end RAG serving.
\subsection{Searching for Optimal Scheduling Policies}
\label{sec_rago:rag_scheduler}
Given a \ragdesc and hardware resource constraints, \ours performs an exhaustive search across potentially millions of schedules to identify Pareto frontier for key performance metrics. 
%
Given \( m \) model inference stages, the size of the search space can be expressed as \( C \cdot R^m \cdot T^m \), where \( C \) represents the number of collocation options, \( R \) denotes the number of resource allocation options per stage, and \( T \) is the batching options per stage.
%
\ours operates in three main steps.
First, it performs performance profiling by evaluating each RAG component individually (\egc retrieval, prefix, etc.) under varying resource allocations and batch sizes.
This evaluation relies on the calibrated analytical model introduced in~\S\ref{sec_rago:method}.
Next, it proceeds to schedule generation, where it explores \emph{all} possible RAG schedules by considering (a) collocation strategies, (b) resource allocation strategies within the overall resource budget, and (c) batching strategies for each component.
Finally, \ours calculates the end-to-end performance for each schedule and identifying the Pareto frontier along with the corresponding schedule configurations.

\ours relies on several simplification assumptions to calculate the end-to-end performance.
Data movement is assumed to be negligible, as transferring retrieved tokens to accelerators requires minimal bandwidth.
For instance, a 1K-token prompt involves only a few kilobytes of data movement, which incurs a latency in the range of microseconds\,---negligible compared to the milliseconds required to decode a single token in the RAG pipeline.
Additionally, key-value (KV) cache movement between prefix and decode accelerators can be performed layer-by-layer~\cite{patel2023splitwise}, allowing it to overlap with computation, thanks to high inter-chip interconnect bandwidth (\iec 600\,GB/s, see \Cref{tbl:xpu:gen}).
%
I follow the same methodology described in \S~\ref{sec_rago:method} to compute decode latency per step (\iec constant latency of the final token~\cite{yu2022orca}).
%
Finally, I simplify granularity by searching hardware resource quantities and batch sizes in powers of two.

\ours allows users to constrain the search space to reduce search time by specifying batch size ranges and setting overall and per-stage resource constraints.
For example, in the RAG paradigm I, which involves three stages and up to 128 chips, the search process takes approximately one minute.
Adding an encode stage in the RAG paradigm II increases the complexity, resulting in a search time of about ten minutes, which still remains computationally tractable.
For the most complex case (RAG paradigm IV), which includes six stages, the search time is kept under an hour by limiting the resource allocation options.
For instance, I first fix the number of XPUs for the main generative LLM's prefix and decode stages, as they are the primary bottlenecks~(\Cref{subsec_rago:case4}), and only explore resource allocation options for the remaining stages (rewrite-prefix, rewrite-decode, and rerank).
I leave search space pruning optimizations as future work.
\section{Evaluation}
\label{sec_rago:eval}
I evaluate the effectiveness of \ours by revisiting the four RAG case studies in \S\ref{sec_rago:case-studies}.
I begin with an analysis of the performance overview across all scheduling policies, followed by a detailed examination of each scheduling decision: placement, allocation, and batching.

\niparagraph{Evaluation methodology.}
For evaluating placement and resource allocation decisions, I focus on case study II (C-II)\,---long-context sequence---\,and case study IV (C-IV)\,---RAG with rewriter and reranker.
I select these case studies because of their additional components, which visibly distinguish them from LLM-only systems and introduce unique optimization challenges.
For micro-batching policy evaluations under bursts of user requests, I include case study I (C-I) with hyperscale retrieval, alongside C-II and C-IV. 
I exclude case study III (C-III), which focuses on iterative retrievals during decoding, as it was evaluated in details in~\S\ref{subsec_rago:case3}.

\subsection{Overall Performance}
\label{sec_rago:eval:e2e}

\niparagraph{Baseline system.}
The baseline is an extension of LLM-only systems, where additional RAG components are collocated with the prefix system of the generative LLM.
Rather than arbitrarily assigning chips to prefix and decode, I carefully tune the ratio based on their time consumption. 
In this tuned baseline, the prefix and decode stages are allocated in a 1:1 chip ratio, reflecting their similar time requirements in the pipeline (1.2$\sim$1.4:1 across the 8B and 70B models). 
\begin{table*}[t]
    \centering
    \begin{footnotesize}
    \setlength{\tabcolsep}{4pt} 
    \renewcommand{\arraystretch}{0.85} 
    \caption{Comparison of \ours and baseline system schedules (placement, allocation, and batching strategies) in Case II.}
    \label{tbl:rag-configurations}
    \scalebox{0.75}{ 
        \begin{tabular}{p{4.5cm}!{\vrule width 0.7pt}cc!{\vrule width 0.7pt}cccc!{\vrule width 0.7pt}cccc}
            \hline
            \multirow{2}{*}{\textbf{Schedules}} & \multicolumn{2}{c!{\vrule width 0.7pt}}{\textbf{Performance}} & \multicolumn{4}{c!{\vrule width 0.7pt}}{\textbf{Batch Sizes}} & \multicolumn{4}{c}{\textbf{Num XPUs}} \\ \cline{2-11}
             & \textbf{TTFT (s)} & \textbf{QPS/Chip} & \textbf{Encode} & \textbf{Retrieval} & \textbf{Prefix} & \textbf{Decode} & \textbf{Encode} & \textbf{Prefix} & \textbf{Decode} & \textbf{Total} \\ \hline
            RAGO (Max QPS/Chip)       & 2.47 & 1.08 & 2   & 2   & 128  & 1024 & 64 & 16 & 16 & 96  \\ \hline
            RAGO (Min TTFT)           & 0.03 & 0.22 & 1   & 1   & 1    & 128  & 64 (col) & 64 (col) & 64 & 128 \\ \hline
            Baseline (Max QPS/Chip)   & 1.54 & 0.65 & 2   & 2   & 128  & 256  & 64 (col) & 64 (col) & 64 & 128 \\ \hline
            Baseline (Min TTFT)       & 0.03 & 0.22 & 1   & 1   & 1    & 128  & 64 (col) & 64 (col) & 64 & 128 \\ \hline
        \end{tabular}
    } 
    \end{footnotesize}
\end{table*}

\begin{figure}[t]
  \centering
  \begin{subfigure}[b]{0.35\linewidth}
    \includegraphics[width=\linewidth]
    {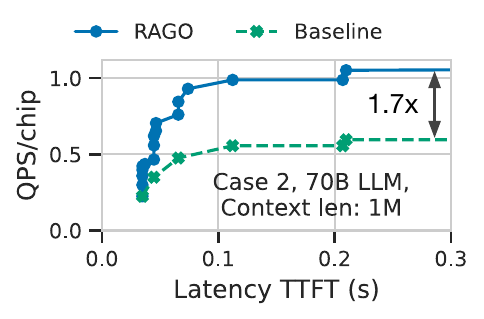}
    \caption{Case II}
    \label{fig_rago:eval:e2e_case_2}
  \end{subfigure}
  \begin{subfigure}[b]{0.32\linewidth}
    \includegraphics[width=\linewidth]
    {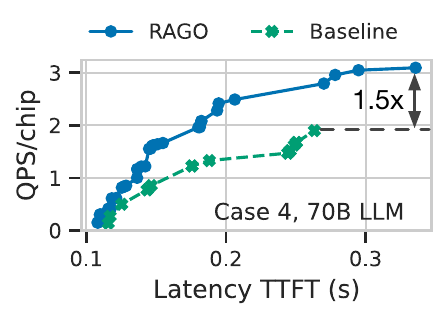}
    \caption{Case IV}
    \label{fig_rago:eval:e2e_case_4}
  \end{subfigure}
  \caption{\ours versus LLM-only system extension.}
  \label{fig_rago:eval:e2e}
\end{figure}

\niparagraph{Impact of scheduling policies on QPS/Chip.}
\Cref{fig_rago:eval:e2e_case_2} illustrates the Pareto performance comparison between \ours and the baseline in terms of QPS/Chip across two distinct RAG case studies.
In C-II, \ours achieves a \xx{1.7$\times$} improvement in maximum QPS/Chip over the baseline. 
This speedup underscores the inefficiencies of the baseline approach, particularly in handling the encoding stage for long-context sequences.
The encoder, while smaller than the generative LLM, becomes a critical bottleneck as context lengths grow. 
Specifically, in the baseline configuration, encoding is collocated with the prefix stage, leading to resource imbalance: decoding XPUs (\xx{50\%} of all XPUs) remain idle, while encode-prefix XPUs are overloaded. 
This imbalance can theoretically reduce QPS/Chip by up to \xx{2.0$\times$} in the baseline, which aligns with the observed reduction of \xx{1.94$\times$} for a large 10M-token context, though this specific data point is not plotted.
On the other hand, \ours achieves high QPS/Chip by allocating 64 out of the 96 XPUs to encoding (\Cref{tbl:rag-configurations}), reflecting the high time consumption of this stage. 

A similar inefficiency of the baseline is observed in C-IV (Figure~\ref{fig_rago:eval:e2e_case_4}), where the rewriter and reranker models, despite their relatively small size (8B and 120M), significantly impact throughput in the baseline system. 
This QPS drop can be attributed to two primary factors.
First, collocating rewriter-decode stage and the prefix stage of the main generative LLM leads to XPU under-utilization due to the low computational intensity of the autoregressive decoding stage, particularly when handling small batch sizes.
%
Second, retrieval operations are introduced between the rewriting and prefix stages add wait times for retrieval results
(e.g., 10 ms with a batch size of one given 32 host servers), further reducing throughput.
In contrast, \ours demonstrates its ability to mitigate these bottlenecks through optimized task placement, resource allocation, and batching strategies. 
These results highlight the importance of disaggregating smaller pipeline stages and balancing resource distribution to unlock the full throughput potential of RAG systems.
%

%
%

\begin{figure}[t]
  \centering
  \begin{subfigure}[b]{0.4\linewidth}
    \includegraphics[width=\linewidth]
    {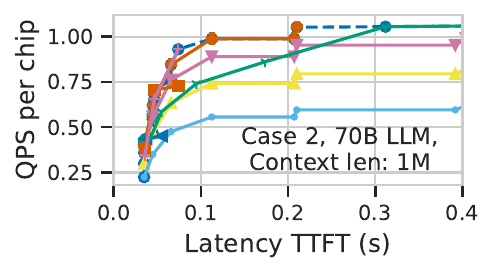}
    \caption{Case II}
    \label{fig_rago:eval:pareto_breakdown_case_2}
  \end{subfigure}
  \begin{subfigure}[b]{0.37\linewidth}
    \includegraphics[width=\linewidth]
    {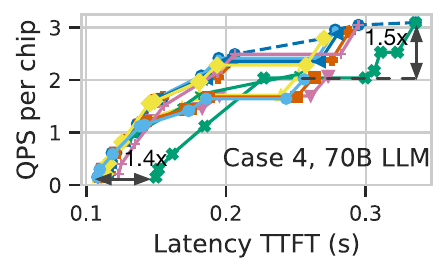}
    \caption{Case IV}
    \label{fig_rago:eval:pareto_breakdown_case_4}
  \end{subfigure}
  \caption{Performance Pareto across multiple placement and allocation plans in case 2 and 4.}
  \label{fig_rago:eval:pareto_breakdown}
\end{figure}

\niparagraph{Pareto composition analysis.}
%
\Cref{fig_rago:eval:pareto_breakdown_case_2} and \Cref{fig_rago:eval:pareto_breakdown_case_4} reveal how diverse placement and allocation plans contribute to the overall Pareto frontier.
The dashed lines represent the global Pareto frontier, while each solid line corresponds to the Pareto frontier of a specific combination of placement and allocation strategies, with each point on a line representing a batching policy. 
%
%
The overall Pareto frontier is constructed from multiple distinct plans, each embodying a unique trade-off between TTFT and QPS/Chip.
%
This diversity underscores the importance of tailoring placement and allocation strategies to the specific performance priorities of a deployment.
For instance, as shown in Figure \ref{fig_rago:eval:pareto_breakdown_case_4}, selecting the most throughput-optimized plan results in a trade-off, with TTFT approximately \xx{40\%} higher compared to the most latency-optimized plan, while achieving 1.5$\times$ QPS/Chip.
This is because the throughput-optimized plan allocates only one chip to the query rewriter, given its minimal contribution to the end-to-end generation latency, as analyzed in \S\ref{subsec_rago:case4}.
In contrast, the latency-optimized plan allocates 32 chips to the query rewriter, resulting in low resource utilization since a significant number of chips are assigned to this non-bottleneck stage.
These findings emphasize that there is no one-size-fits-all strategy. 
Instead, the optimal placement and allocation plans must be aligned with the operational objectives, whether minimizing latency, maximizing throughput, or achieving a balance between the two.

\subsection{Scheduling Policy Sensitivity Analysis}
\label{sec_rago:eval:schedule}

I now delve into a detailed analysis of the performance implication of each scheduling decision.

\niparagraph{Task placement sensitivity.}
\begin{figure}[t]
  \centering
  \begin{subfigure}[b]{0.4\linewidth}
    \includegraphics[width=\linewidth]
    {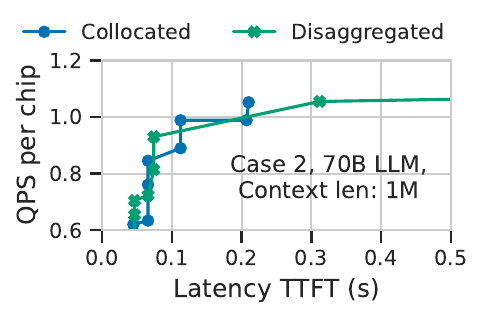}
    \caption{Case II}
    \label{fig_rago:eval:placement_case_2}
  \end{subfigure}
  \begin{subfigure}[b]{0.48\linewidth}
    \includegraphics[width=\linewidth]
    {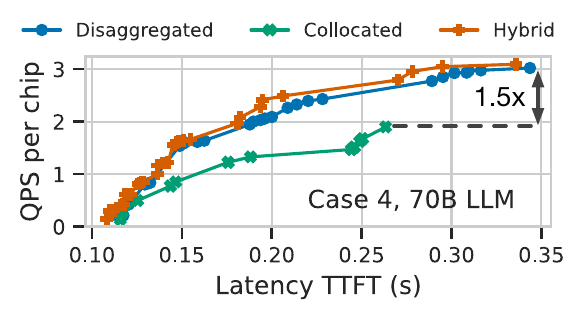}
    \caption{Case IV}
    \label{fig_rago:eval:placement_case_4}
  \end{subfigure}
  \centering
  \caption{Comparison of task placement plans.}
  \label{fig_rago:eval:placement}
\end{figure}
\Cref{fig_rago:eval:placement} compares the impact of different task placement policies on system performance across C-II and C-IV.
Each line in \Cref{fig_rago:eval:placement_case_2} and \Cref{fig_rago:eval:placement_case_4} represents the Pareto frontier for a specific placement strategy, illustrating the relationship between QPS/Chip and TTFT latency under these policies.

In C-II (\Cref{fig_rago:eval:placement_case_2}), task placement sensitivity is minimal.
Both collocated and disaggregated strategies yield comparable performance, as the encoder and prefix stages are computationally intensive. 
Whether these stages are time-multiplexed (collocated) or spatially multiplexed (disaggregated), performance remains consistent (only 2\% difference in max QPS/Chip) as long as the accelerator ratio between stages is appropriately chosen.
This demonstrates that task placement decisions in this case have little effect on system efficiency, provided resources are balanced effectively.

In contrast, C-IV (\Cref{fig_rago:eval:placement_case_4}) shows a more pronounced sensitivity to placement policies. 
Here, a hybrid placement strategy\,---combining elements of disaggregation and collocation---\,slightly outperforms the fully disaggregated approach and significantly surpasses the collocated plan, achieving up to a \xx{1.5$\times$} improvement in QPS/Chip. 
The key advantage of the hybrid and disaggregated strategies lies in their ability to mitigate the underutilization of prefix chips, which occurs when the rewriter model is collocated with the prefix stage. 
By separating the rewriter model from the prefix system, these strategies prevent resource bottlenecks and enable optimal throughput.

\niparagraph{Resource allocation sensitivity.}
\begin{figure}[t]
  \centering
  \begin{subfigure}[b]{0.49\linewidth}
    \includegraphics[width=\linewidth]{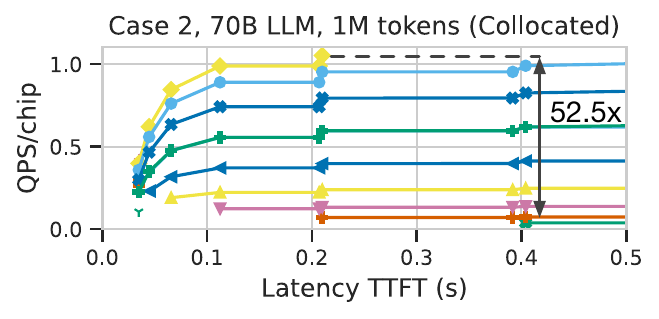}
    \caption{Collocated}
    \label{fig_rago:eval:resource_case_2_collocated}
  \end{subfigure}
  \begin{subfigure}[b]{0.49\linewidth}
    \includegraphics[width=\linewidth]{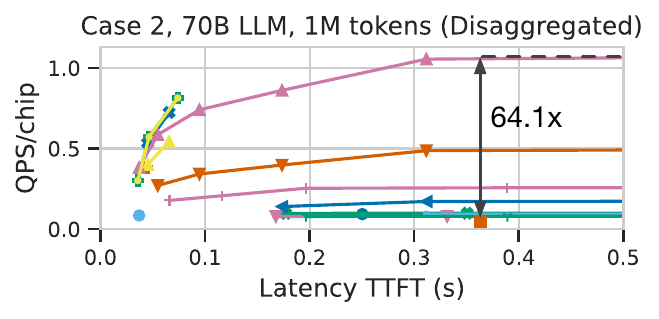}
    \caption{Disaggregated}
    \label{fig_rago:eval:resource_case_2_disaggregated}
  \end{subfigure}
  \centering
  \caption{Comparison of resource allocation plans (case II).}
  \label{fig_rago:eval:resource_case_2}
\end{figure}
\Cref{fig_rago:eval:resource_case_2} shows the Pareto frontier for different resource allocation policies in C-II, including both collocated and disaggregated placement plans.
For collocated plans, the maximum QPS/chip can vary by up to 52.5$\times$ if insufficient resources are allocated to high-workload stages, when other stages have surplus capacity. 
For example, in the collocated plan (\Cref{fig_rago:eval:resource_case_2_collocated}), imbalanced resource distribution across the encoder and prefix stages leads to underutilization of available accelerators, limiting throughput.
This effect may amplify to 64.1$\times$ QPS/chip difference for disaggregated plans, as disaggregated stages rely heavily on precise balancing to maximize performance. 
Tailoring resource distribution to the specific demands of each stage is essential for optimizing both latency and throughput in RAG systems.

%
\begin{figure}[t]
  \centering
  \begin{subfigure}[b]{0.32\linewidth}
    \includegraphics[width=\linewidth]
    {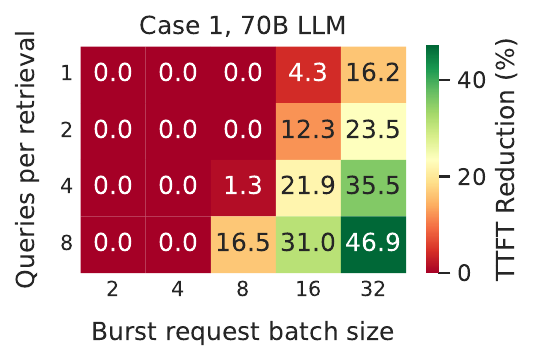}
    \caption{Case I}
    \label{fig_rago:eval:microbatch_case_1}
  \end{subfigure}
  \begin{subfigure}[b]{0.35\linewidth}
    \includegraphics[width=\linewidth]
    {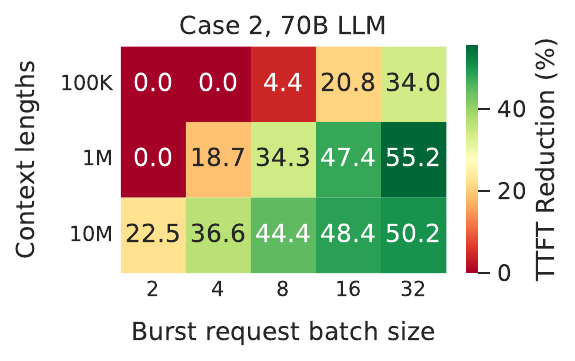}
    \caption{Case II}
    \label{fig_rago:eval:microbatch_case_2}
  \end{subfigure}
  \begin{subfigure}[b]{0.3\linewidth}
    \includegraphics[width=\linewidth]
    {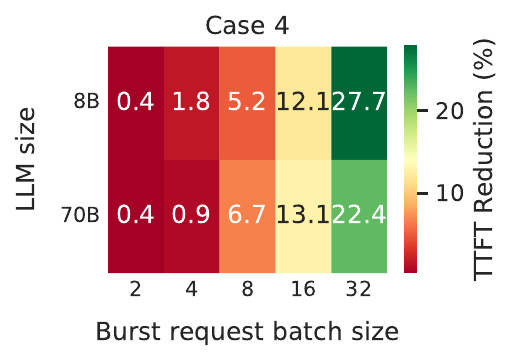}
    \caption{Case IV}
    \label{fig_rago:eval:microbatch_case_4}
  \end{subfigure}
  \caption{TTFT latency reduction by micro-batching.}
  \label{fig_rago:eval:microbatch}
\end{figure}

\niparagraph{Impact of micro-batching on TTFT latency.} 
The effectiveness of micro-batching is determined by the throughput sensitivity to batch sizes at each pipeline stage.
%
%
\Cref{fig_rago:eval:microbatch} compares the impact of micro-batching on TTFT reduction across the case studies. 
For C-II (\Cref{fig_rago:eval:microbatch_case_2}), micro-batching is effective even with a small batch size of two, reducing TTFT by 22\%. 
This is because both the encoding and prefix stages are computationally intensive, achieving reasonable throughput even with smaller batch sizes. 
With larger batches of 32, the TTFT reduction increases further to 55\% for 1M tokens.
In C-I (\Cref{fig_rago:eval:microbatch_case_1}), micro-batching only becomes effective with larger batch sizes, such as eight or 16.
This inefficiency at smaller batch sizes arises from the vector search system, where reducing the query batch size below 16 fails to improve latency.
However, with batch sizes increasing to 32, micro-batching still achieves a significant latency reduction of 46\% for eight queries per vector.
For C-IV (\Cref{fig_rago:eval:microbatch_case_4}), TTFT reduction is moderate, with a maximum improvement of approximately \xx{25\%} at a batch size of 32. 
This modest improvement is primarily due to the query rewriter decoding stage, which exhibits little latency reduction with smaller sequence batch sizes.
\section{Related Work}
\label{sec_rago:related_work}

\niparagraph{RAG performance optimization.} 
As an emerging research area, RAG performance optimization remains underexplored, with existing studies targeting specific configurations.
For instance, retrieval acceleration~\cite{jiang2023chameleon} is effective when retrieval costs dominate, document prefetching~\cite{jiang2024piperag, zhang2024accelerating} benefits iterative retrievals, and prefix state caching~\cite{yao2024cacheblend, jin2024ragcache} is efficient when the length ratio of prefix-to-decode is high. 
If these techniques are adopted, the workload distribution within a RAG system evaluated by \ours is expected to shift.
For example, retrieval acceleration~\cite{jiang2023chameleon} will shift the workload toward being more inference-bound. 
By contrast, caching KV states of retrieved documents~\cite{yao2024cacheblend, jin2024ragcache} will increase the importance of retrieval and decoding performance. 
Additionally, supporting iterative retrievals through data prefetching~\cite{jiang2024piperag, zhang2024accelerating} will reduce decoding engine idleness during retrieval operations.

\niparagraph{LLM and retrieval optimization.}  
Extensive research has been devoted to optimizing LLM systems and their underlying hardware~\cite{patel2023splitwise, zhong2024distserve, vllm, yu2022orca, qin2024mecla, zhang2024llmcompass, zhao2024alisa, lee2024tender, li2024large, bang2023vtrain, yun2024duplex}. 
Similarly, significant efforts have focused on enhancing retrieval efficiency on modern hardware, spanning product-quantization-based ANN algorithms~\cite{jiang2023co, johnson2019billion, lee2022anna, liu2023juno} and graph-based approaches~\cite{jiang2024accelerating, zeng2023df, groh2022ggnn, zhao2020song}.
However, the complexity and heterogeneity of RAG pipelines far surpass those of LLM-only or retrieval-only systems, rendering direct extensions of these systems inadequate for efficient RAG serving.

\section{Conclusion}
\label{sec_rago:conclusion}

This work represents an early exploration of RAG through a systems lens, establishing a foundation for this rapidly evolving field.
\ragdesc provides a structured abstraction for RAG serving, facilitating systematic workload characterization and bridging the gap between algorithms and system design.
Leveraging \ragdesc, I proposed \ours, a system optimization framework that delivers up to a 2$\times$ improvement in QPS per chip and a 55\% reduction in TTFT compared to a strong baseline.

As the field advances, the characterization results in this chapter can guide the design of future RAG systems and hardware. 
For instance, this chapter finds that retrieval can become a bottleneck in certain RAG paradigms, particularly at hyperscale, underscoring the need for further retrieval optimizations.
Moreover, with multiple model components in RAG pipelines, efficient support for collocated models on accelerators will be increasingly critical.
Finally, scaling to extremely complicated RAG and agentic AI systems introduces challenges related to a broader optimization search space and additional efficiency metrics, such as energy and cost efficiency. 
I leave these investigations to future work.
\chapter{Chameleon: Heterogeneous Accelerator System for RAG Serving} 
\label{chap:chameleon}

As identified in Chapter~\ref{chap:rago}, both retrieval and model inference can become significant bottlenecks in RAG serving.  
This chapter explores efficient RAG serving from both hardware and system perspectives by introducing Chameleon, a heterogeneous and disaggregated accelerator system including not only LLM accelerators but also retrieval accelerators.

\section{Introduction}
\label{sec_chameleon:intro}

As shown in Chapeter~\ref{chap:rago}, efficient RAG serving presents \textbf{two challenges}. \textit{First, the workload characteristics of the LLM and the retriever are distinct.} While the LLM inference primarily relies on rapid tensor operations, the vector search system --- often utilizing fast and memory-efficient search algorithms like Product Quantization (PQ)~\cite{PQ} --- demands both substantial memory capacity to hold the vectors and fast processing of quantized database vectors during query time.
\textit{Second, the diverse range of RAG configurations leads to shifting system requirements and bottlenecks.}
{Regarding retrieval frequency, some models retrieve once per generated token~\cite{khandelwal2019generalization, meng2021fast, alon2022neuro}, while others retrieve only once per entire sequence~\cite{lewis2020retrieval, izacard2020leveraging}. 
In terms of scale, database sizes vary from millions~\cite{lewis2020retrieval, guu2020realm} to tens of billions of vectors (92 TB)~\cite{borgeaud2022improving, wang2023instructretro}, and model sizes range from hundreds of millions~\cite{guu2020realm, lewis2020pre} to tens of billions of parameters~\cite{wang2023instructretro}.}


I envision a high-performance and efficient RAG system to adhere to \textbf{two key design principles} to address the two aforementioned challenges. 
\textit{Firstly, RAG should incorporate \textbf{heterogeneous accelerators}, employing not only inference accelerators such as GPUs but also vector search accelerators}, such that both RAG components are fast and efficient.
\textit{Secondly, the heterogeneous accelerators should be \textbf{disaggregated} to support diverse RAG demands efficiently}, in contrast to a monolithic approach where a fixed number of LLM and retrieval accelerators reside on the same server. 
The rationale is twofold: (a) performance bottlenecks shift between various RAG configurations of different retrieval frequencies, database sizes, and model sizes, thus requiring a case-specific optimal balance between the two types of accelerators; and (b) {a huge database (e.g., with tens of TBs of vectors~\cite{borgeaud2022improving, wang2023instructretro}) may necessitate more retrieval accelerators than a single server can accommodate.}

To materialize this vision, I propose \textit{Chameleon}, a heterogeneous and disaggregated accelerator system for efficient, flexible, and high-performance RAG serving.
Chameleon consists of three primary components.
Firstly, \textit{ChamVS} is a distributed and accelerated vector search engine. It consists of several disaggregated memory nodes, each containing a shard of quantized database vectors in DRAM, a near-memory retrieval accelerator prototyped on FPGA, and a hardware TCP/IP stack. 
Secondly, \textit{ChamLM} is a multi-GPU LLM inference engine. It produces query vectors and generates texts using the retrieved information.
Lastly, a CPU coordinator server orchestrates the network communication between the retrieval and LLM accelerators.

I evaluate Chameleon with various LLM architectures, model sizes, database sizes, and retrieval frequencies. 
For large-scale vector search, ChamVS achieves up to 23.72$\times$ latency reduction compared to the optimized CPU baselines while consuming 5.8$\sim$26.2$\times$ less energy. 
For RAG serving, Chameleon achieves up to 2.16$\times$ and 3.18$\times$ speedup in latency and throughput compared to the hybrid CPU-GPU architecture. 
I further illustrate that the optimal balance between the two types of accelerators varies significantly across different RAG configurations, making disaggregation essential for achieving both flexibility and high accelerator utilization rates.

The chapter makes the following \textbf{contributions:}

\begin{itemize}[leftmargin=*]
    \item I present Chameleon, an efficient RAG serving system designed around two proposed principles: accelerator heterogeneity and disaggregation. 
    \item I design and implement ChamVS, a distributed engine for large-scale vector search, which includes:
    \begin{itemize}
	\item Near-memory accelerators for vector search, including a novel resource-efficient top-K selection architecture.
        \item A GPU-based index scanner to prune search space.
    \end{itemize}
    \item I evaluate Chameleon on various RAG configurations and showcase its remarkable performance and efficiency.
\end{itemize}

\section{Motivation}
\label{sec_chameleon:background}




An efficient RAG serving engine should meet the following \textbf{system requirements}:

\begin{itemize}[leftmargin=*]
    \item Both the LLM inference and the large-scale vector search components should be fast and resource-efficient. 
    \item The system should be flexible enough to accommodate diverse RAG configurations, spanning various combinations model sizes, database sizes, and retrieval frequencies.
\end{itemize}

However, little effort has been devoted to developing efficient RAG systems that meet the above requirements. This is likely because RAG has been an emerging topic within the machine learning community~\cite{borgeaud2022improving, khandelwal2019generalization, lewis2020retrieval, izacard2022few, izacard2020leveraging}, with their prototype implementations exhibiting the following problems:

\underline{\textbf{(P1)}} Each research RAG system focuses on \textit{being able to run} one or a small number of RAG configurations, paying little attention to latency, throughput, resource efficiency, and system flexibility to serve diverse RAG configurations. 

\underline{\textbf{(P2)}} While hardware accelerators for LLMs, such as GPUs, are advancing rapidly, less attention has been paid to the vector search aspect, which, as the evaluations will demonstrate, can become the performance bottleneck in RAG serving. 

\underline{\textbf{(P2.1)}} CPUs are slow in scanning PQ codes during query time.
This is due to the frequent cache accesses (for each byte of PQ code, load the code and use it as an address to load a distance) and the instruction dependencies between operations  (distance lookups depend on PQ codes and distance accumulations depend on the lookup values). 
Even with the state-of-the-art SIMD-optimized CPU implementation~\cite{faiss}, the throughput peaks at roughly 1 GB/s per core when scanning PQ codes (1.2 GB/s on Intel Xeon Platinum 8259CL @ 2.50GHz).
Within a CPU-memory-balanced server, the PQ code scanning process significantly underutilizes the available memory bandwidth, as about 16 cores are required to saturate the bandwidth of a single memory channel (around 20 GB/s).

\underline{\textbf{(P2.2)}} GPUs suffer from two major limitations for large-scale vector search. 
Firstly, the limited memory capacity of each GPU makes large-scale searches on GPU clusters cost-prohibitive. 
For instance, accommodating only 1 TB of PQ codes necessitates at least 16 NVIDIA A100 GPUs (cost 300K USD as of March 2024), each with 80 GB of memory, given that a portion of memory should be reserved for intermediate search states. 
Although an alternative solution is to adopt a hybrid CPU-GPU architecture where the GPU fetches vectors from CPU's memory, the inter-processor bandwidth is way lower than the GPU memory bandwidth. Even for NVIDIA Grace Hopper, with the latest high-performance CPU-GPU interconnect, the single-direction bandwidth of 450 GB/s is only 15\% of the GPU's bandwidth. 
Secondly, the throughput for PQ code scanning on GPUs is considerably lower than the GPU's bandwidth, only around 50\% of the bandwidth even with large batch sizes (evaluated on NVIDIA A100), due to the multiple passes of memory accesses to write and read intermediate results at each search step~\cite{johnson2019billion}. 

\section{Chameleon: System Overview}

I design and implement Chameleon, an efficient, flexible, and performant RAG serving system built around the following principles:



    
\begin{itemize}[leftmargin=*]
    \item Chameleon employs heterogeneous hardware to accelerate both LLM inference and vector search efficiently.
    \item Chameleon disaggregates the accelerators, enabling independent scaling for each type of hardware, thus supporting various RAG configurations efficiently.
    \item The modular design of Chameleon allows flexible hardware upgrades, such as integrating more powerful LLM inference accelerators or ASIC-based ChamVS accelerators in the future.
\end{itemize}

\begin{figure*}[t]
	\centering
  \includegraphics[width=1.0\linewidth]{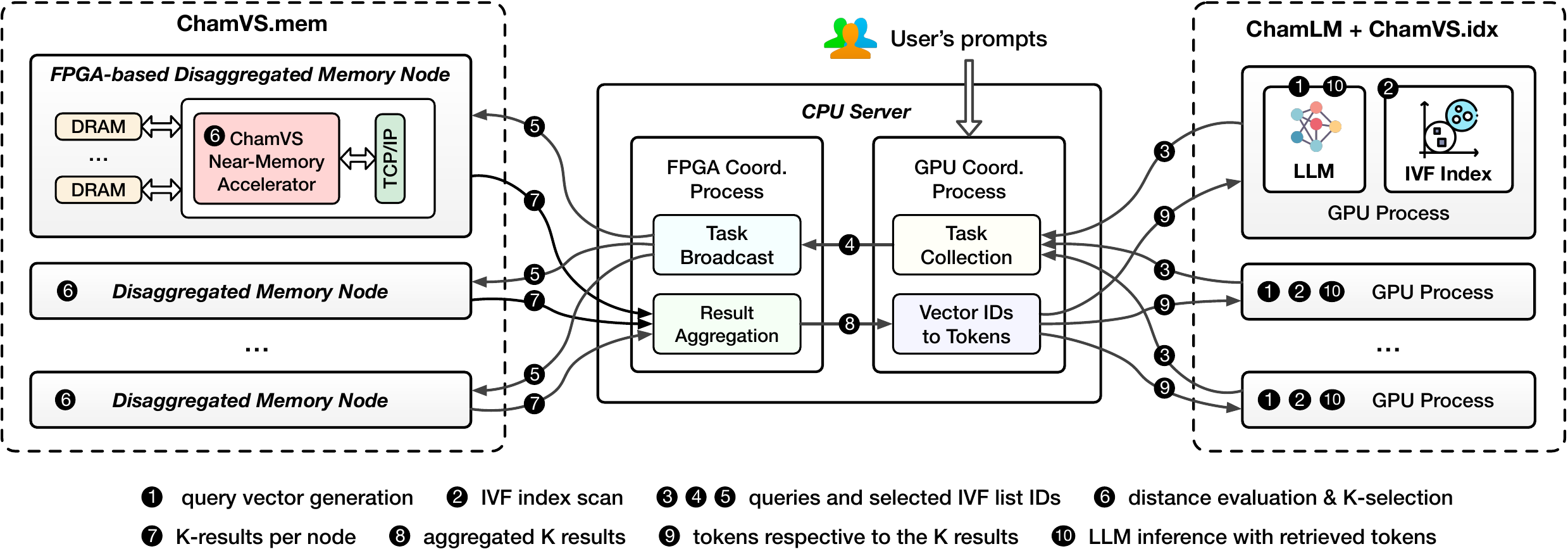}
  \caption{Chameleon is a heterogeneous and disaggregated accelerator system for efficient RAG serving.}
  \label{fig_chameleon:overview}
\end{figure*}

\textbf{Figure~\ref{fig_chameleon:overview} overviews the Chameleon architecture}, which primarily consists of the following components.

\textit{Firstly, ChamVS is a distributed accelerator engine for low-latency vector search.}
On the one hand, ChamVS.idx is a GPU-based IVF index scanner colocated with the ChamLM GPUs (right side of Figure~\ref{fig_chameleon:overview}). {While Chameleon also supports index scan on CPUs}, GPUs are generally more favorable for handling this embarrassingly parallel workload due to their superior memory bandwidth and computational capability.
Given that GPUs are already integrated into Chameleon, no additional devices are required. 
{The only overhead is a slight increase in GPU memory usage, as the index sizes are small relative to the database vectors. For example, assuming 1KB per vector and one thousand vectors per IVF list, a single GB of index can support one million IVF lists, enough for a large database containing one billion vectors.}
On the other hand, ChamVS.mem is responsible for querying quantized database vectors. ChamVS.mem contains one or multiple disaggregated memory nodes, each with a partition of the database vectors and a near-memory retrieval accelerator prototyped on FPGA for query processing (left side of Figure~\ref{fig_chameleon:overview}). 


\textit{Secondly, ChamLM is a multi-GPU LLM inference engine}, as shown on the right side of Figure~\ref{fig_chameleon:overview}. 
Each GPU, managed by an independent GPU process, can reside on the same or different servers. 
Currently, ChamLM assigns each GPU a full copy of the LLM, as RAG can achieve high generation quality even with smaller LLMs~\cite{borgeaud2022improving, lewis2020retrieval}. Future larger models could be accommodated by extending ChamLM to support tensor or pipeline parallelism~\cite{narayanan2019pipedream, shoeybi2019megatron, rajbhandari2020zero} across GPUs.
{Once a retrieval request is sent, a GPU pauses inference to wait for results. While one potential solution to avoid such GPU idleness is to split the generation into two sub-batches --- one executes inference when the other one is waiting for retrieved contents --- this approach does not necessarily improve performance. This is because (a) using sub-batches reduces inference throughput, and (b) retrieval latency may not align with inference latency.
 }

\textit{Thirdly, the CPU serves as the cluster coordinator, managing the lightweight communication between the GPUs and FPGAs.} After receiving search requests from the GPU processes, it dispatches them to the FPGA-based disaggregated memory nodes, aggregates the per-partition results returned by the FPGAs, converts the K nearest neighbor vector IDs into their corresponding texts, and sends the retrieved tokens back to the GPUs. Since each query only requires less than ten KBs of network data transfer, the communication latency is negligible compared to vector search and LLM inference. 

\textbf{Token generation workflow.}
For each token generation step, the procedure diverges depending on whether the retrieval is invoked. 
Without retrieval, the GPUs infer the next token as in regular LLMs.
With retrieval, the first step is to generate a contextual query vector~\ballnumber{1}, either by using the hidden state of the current context~\cite{khandelwal2019generalization, khandelwal2020nearest} or encoding the query tokens through another model~\cite{borgeaud2022improving}. 
Following this, the IVF index residing on the same GPU is scanned to select the $nprobe$ most relevant IVF lists~\ballnumber{2}.
The query vector and the list IDs are then transmitted to the GPU coordinator process running on the CPU node via the network~\ballnumber{3}. After recording the association between queries and GPU IDs, the query and list IDs are forwarded to the FPGA coordination process~\ballnumber{4}, which broadcasts them to the FPGA-based disaggregated memory nodes~\ballnumber{5}.
The ChamVS near-memory processor on each node then uses the query vectors to construct distance lookup tables for each IVF list, computes the distances between the query and quantized database vectors, and collects the K nearest neighbors~\ballnumber{6}. 
Subsequently, the result vector IDs and distances from all memory nodes are sent back to the CPU server~\ballnumber{7}, which aggregates the results~\ballnumber{8} and returns the tokens of the nearest neighbors to the originating GPU~\ballnumber{9}.
Finally, the GPU predicts the next token based on both the context and the retrieved tokens~\ballnumber{10}.

\section{ChamVS Near-Memory Accelerator}

\begin{figure}[t]
  \centering
  \includegraphics[width=0.8\linewidth]{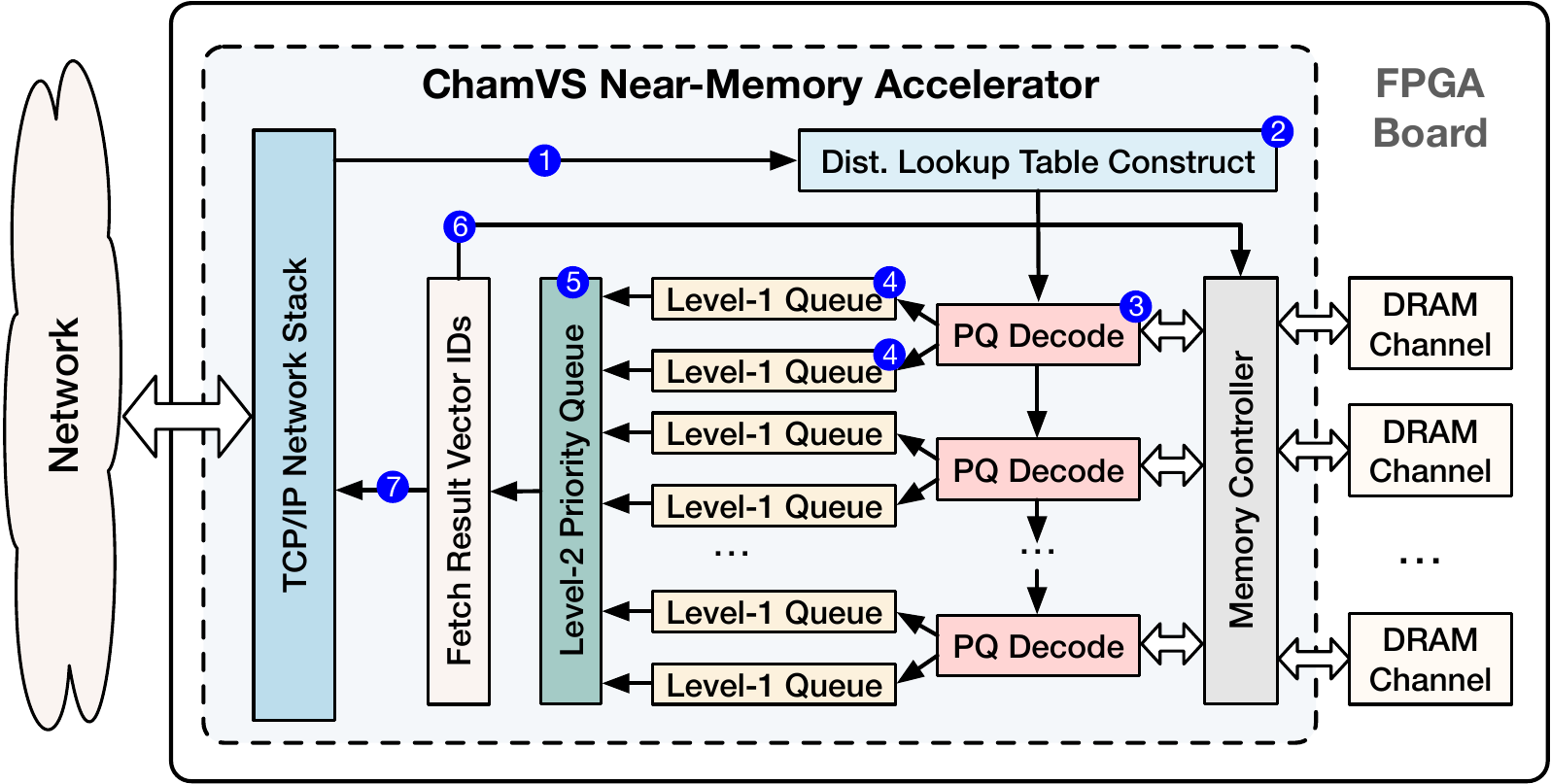}
  \caption{The ChamVS near-memory retrieval accelerator.}
  \label{fig_chameleon:fpga-design}
\end{figure}

ChamVS enables high-performance, large-scale vector search by pairing each disaggregated memory node with a near-memory retrieval accelerator. 
As shown in Figure~\ref{fig_chameleon:fpga-design}, the accelerator comprises a distance lookup table construction unit, several PQ decoding units for distance evaluations between query vectors and quantized database vectors, a group of systolic priority queues for parallel $K$-selection, and multiple memory channels.

\subsection{PQ Decoding Units}

As shown in Figure~\ref{fig_chameleon:fpga-design}~\blueballnumber{3}, each ChamVS accelerator contains multiple PQ decoding units to fully utilize the memory bandwidth. These units read database vectors (PQ codes) from DRAM and compute their distances to query vectors using a distance lookup table.

\begin{figure}[t]
  \centering
  \includegraphics[width=0.65\linewidth]{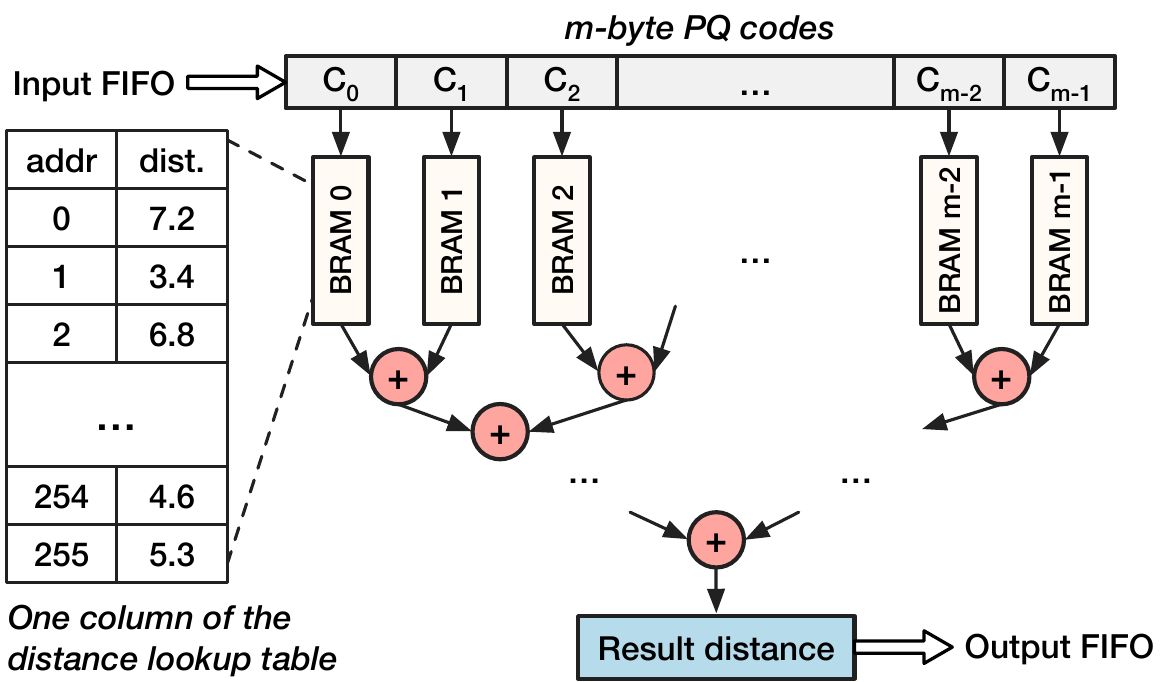}
  \caption{The architecture design of a PQ decoding unit.}
  \label{fig_chameleon:PE_scan}
\end{figure}

\textbf{The design of a PQ decoding unit involves both operator and pipeline parallelisms, enabling a high throughput of producing one result distance every clock cycle.}
As shown in Figure~\ref{fig_chameleon:PE_scan}, the decoding steps --- including data ingestion, distance lookups, computation, and output egestion --- are fully pipelined, similar to that of~\cite{jiang2023co, lee2022anna}. 
The unit also parallelizes the operators within the distance lookup and computation steps.

\textit{Decoding procedure.} For each IVF list to scan, the unit first stores the input distance lookup table in BRAM (on-chip SRAM in FPGAs). 
The shape of the lookup table is $m\times256$ for the typical 8-bit PQ codes ($2^8=256$), where $m$ is the number of bytes per quantized vector. Different table columns are stored in separate BRAM slices, facilitating parallel distance lookups. 
Subsequently, the PQ codes are loaded from DRAM to the unit via an $m$-byte-wide FIFO, with each byte serving as an address to retrieve a value from the corresponding column of the table. 
Finally, an adder tree sums up the retrieved values to produce the approximate distance between the query vector and the quantized database vector.

\subsection{Efficient $K$-Selection Module}

{The $K$-Selection module in ChamVS selects the $K$ nearest neighbors from distances computed by the PQ decoding units.}
Designing an efficient $K$-selection microarchitecture is challenging, because it has to handle multiple incoming elements per cycle due to the high throughput of PQ decoding units. 
I propose approximate hierarchical priority queue (AHPQ), a high-throughput, resource-efficient architecture for parallel $K$-selection in hardware.

\subsubsection{Primitive: Systolic Priority Queue}


The systolic priority queue facilitates high-throughput input ingestion on hardware accelerators~\cite{huang2014scalable, leiserson1979systolic}, consuming \textit{one input element every two clock cycles}.
In short, it is a register array equipped with compare-swap units between the registers, thus \textit{the hardware resource consumption of the queue increases linearly with its length.} 



A natural approach to implement $K$-selection in ChamVS is to instantiate a group of systolic priority queues in a hierarchical structure, as shown in Figure~\ref{fig_chameleon:fpga-design}~\blueballnumber{4}\blueballnumber{5}.
Since a systolic priority queue can only ingest one input every two cycles, two queues, termed as level-one (L1) queues, should be paired with one PQ decoding unit, as it can produce one output per cycle. 
For each query, each L1 queue collects a subset of the $K$ nearest neighbors, and the level-two (L2) queue subsequently selects the final $K$ results.


Unfortunately, a straightforward implementation of the hierarchical priority queue can consume excessive hardware resources, making the solution unaffordable even on high-end FPGAs. 
For example, given 32 instantiated PQ decoding units and $K=100$, the accelerator would necessitate 64 L1 queues of length 100, an amount that already exceeds the total the total available FPGA resources.

\subsubsection{Approximate Hierarchical Priority Queue (AHPQ)} 
\textbf{I propose the AHPQ architecture for high-performance and resource-efficient $K$-selection.} Recognizing that ANN search is inherently approximate, I relax the $K$-selection objective from selecting the $K$ smallest distances in all queries to collecting precise results in the vast majority of cases, such as in 99\% of the queries.

\begin{figure}[t]
  \centering
  \includegraphics[width=0.9\linewidth]{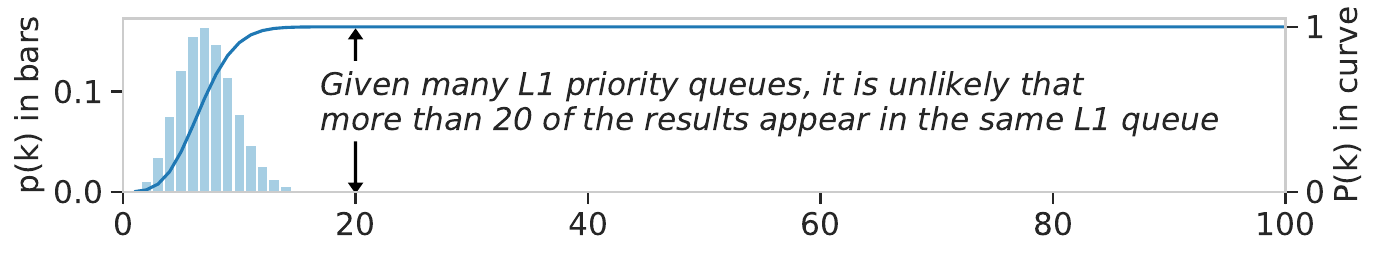}
  \caption{The probability distribution that one out of the 16 L1 priority queues holds k out of the 100 nearest neighbors.}
  \label{fig_chameleon:single-queue-prob}
\end{figure}

The intuition behind AHPQ is simple: \textit{it is unlikely that all $K$ results are produced by a single PQ decoding unit.}
For example, given 16 level-one queues of length $K$=100, the average number of the results in a queue is only $100/16=6.25$.
Specifically, the probability that one queue holds $k$ of the $K$ results is $p(k) = C_{K}^{k} * (\frac{1}{num_{queue}})^{k} * (1-\frac{1}{num_{queue}})^{K-k}$, where $C_{K}^{k}$ represents the number of combinations selecting $k$ out of $K$ items.
The cumulative probability that a queue contains no more than $k$ of the $K$ results is $P(k) = \sum_{i=0}^{k}p(i)$.
Figure~\ref{fig_chameleon:single-queue-prob} shows the probability distribution of $p$ and $P$ given different $k$ in bars and curve: it is almost impossible that a queue holds more than 20 out of the \textit{K=100} results. Thus, the lengths of the L1 queues can be truncated to 20 while producing almost the same results.  



The design aims to reduce the size of the L1 queues while ensuring that the results for 99\% of queries remain identical to those obtained with an exact $K$-selection module. 
Specifically, for 99\% of the queries, none of the L1 queues will omit any result that is supposed to be returned to the user.

\begin{figure}[t]
  \centering
  \includegraphics[width=0.8\linewidth]{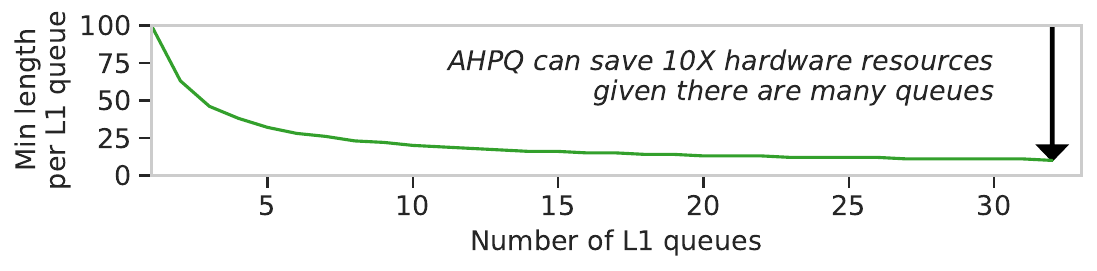}
  \caption{The proposed approximate hierarchical priority queue can save hardware resources by an order of magnitude.}
  \label{fig_chameleon:queue-len}
\end{figure}

Figure~\ref{fig_chameleon:queue-len} shows the resource savings achieved by applying the approximate hierarchical priority queue. 
As the number of L1 queues increases, the queue sizes can be reduced by an order of magnitude while still retaining 99\% of identical results, leading to a corresponding decrease in hardware resource consumption.

\subsection{Memory Management and Load Balancing}
\label{sec_chameleon:memory_management}


{The memory management mechanism of ChamVS balances workloads across memory nodes and channels.
In the current implementation, vectors within each IVF list are evenly partitioned among memory nodes, with these sub-lists further distributed across memory channels to ensure workload balance.
For potential scenarios where IVF lists are too small to be partitioned, each list may reside on different nodes or channels, which could lead to load imbalances, especially with small query batches. Such imbalances can be mitigated by larger batches, as it is less likely that all queries happen to hit the same node or channel. 
Additionally, for the case of uneven access frequencies across IVF lists, adjusting their placement based on these frequencies can help achieve better load balancing~\cite{chen2021spann}.
}

\section{Implementation}
\label{sec_chameleon:implementation}

{Chameleon is implemented in 11K lines of code, including 3K lines of Vitis HLS C/C++ for the ChamVS near-memory accelerator, 1.4K lines of C++ for the CPU coordinator, 3.5K lines of Python for ChamLM, and 3.2K lines of Python for various evaluations. 
}
{ 
Referring to existing RAG research~\cite{khandelwal2019generalization, khandelwal2020nearest}, I build ChamLM on Fairseq~\cite{ott2019fairseq}, a PyTorch-based LLM toolkit.
ChamLM extends Fairseq to support multi-GPU inference, initiating retrieval requests, integrating the retrieved tokens into generation processes, and network communication between the retrieval engines and GPU processes.
}
{ 
ChamVS.idx uses Faiss~\cite{johnson2019billion} for index scanning on GPUs or CPUs. 
ChamVS.mem integrates an FPGA TCP/IP stack~\cite{100gbps}. 
}
{ 
The CPU coordinator process for query broadcasting and result aggregation is implemented in C++ using the socket library. The simple messages in RAG allow us to avoid higher-level abstractions like RPCs, minimizing performance overhead.
}

\section{Evaluation}
\label{sec_chameleon:eval}

I evaluate Chameleon to answer the following questions:

\begin{itemize}[leftmargin=*]
    \item How much performance and energy benefits can ChamVS attain in large-scale vector search? \S~\ref{sec_chameleon:eval_chamvs}
    \item How does Chameleon perform across different RAG configurations by introducing heterogeneous accelerators? \S~\ref{sec_chameleon:eval_e2e}
    \item Is accelerator disaggregation necessary? \S~\ref{sec_chameleon:eval_e2e}
\end{itemize}

\subsection{Experimental Setup}

\textbf{LLMs.} I evaluate models of similar sizes to those in existing RAG research~\cite{borgeaud2022improving, li2022decoupled, sachan2021end, yogatama2021adaptive, izacard2020leveraging}, up to several billions of parameters.
I evaluate both smaller (S) and larger (L) decoder-only (Dec) and encoder-decoder (EncDec) models. Table~\ref{tab_chameleon:models} summarizes the four RAG configurations for evaluation, including input dimensionalities, numbers of layers and attention heads, model sizes, retrieval intervals, and neighbor numbers. For EncDec models, I follow~\cite{borgeaud2022improving} to use a two-layer shallow encoder and a deeper decoder, and set different retrieval intervals. 
For all the models, I use a vocabulary size of 50K and let them generate 512 tokens per sequence. 

\textbf{Vector datasets.} 
Table~\ref{tab_chameleon:datasets} summarizes the four evaluated vector datasets. 
The SIFT and Deep datasets are popular benchmarks for billion-scale ANN. 
Due to the lack of available datasets for RAG, I create two synthetic datasets by replicating each SIFT vector to the models' dimensionalities (512 and 1024). 
As a common practice, I set $nlist$, the number of clusters in the IVF index, to approximately the square root of the number of dataset vectors (\textit{nlist=32K}). I set $nprobe$ as 32 to scan 0.1\% of database vectors per query, for which high recall can be achieved on both real-world datasets (93\% on Deep and 94\% on SIFT for 100 nearest neighbors). I quantize the SIFT and Deep datasets to 16-byte PQ codes, while the two synthetic datasets adopt 32 and 64-byte PQ codes, respectively.

\textbf{Software.}
For vector search, I use \textit{Faiss}~\cite{faiss} developed by Meta, known for its optimized PQ implementations for both CPUs and GPUs. Due to its vector-only nature, Faiss's ANN search performance surpasses vector data management systems that support additional relational data functionalities~\cite{pan2023survey}. 
For LLM inference, I extend Fairseq~\cite{ott2019fairseq} to support RAG as introduced in \S\ref{sec_chameleon:implementation}.


\textbf{Hardware.} 
I instantiate the ChamVS near-memory accelerator on AMD Alveo U250 FPGAs (16 nm) equipped with 64 GB of DDR4 memory (4 channels x 16 GB) and set the accelerator frequency to 140 MHz.
For a fair comparison, each ChamVS memory node is compared to a CPU-based vector search system with equivalent memory capacity (64 GB) and an 8-core AMD EPYC 7313 processor (7 nm) with a base frequency of 3.0 GHz. 
I evaluate NVIDIA RTX 3090 GPUs (8nm) with 24 GB GDDR6X memory. 

\begin{table}
  \begin{center}
    \caption{Various RAG configurations in the evaluation.}
    \label{tab_chameleon:models}
    \scalebox{0.9} {
    \begin{tabular}{L{4.5em} R{2.5em} R{2.5em} R{2.5em} R{2.5em} R{3em} R{2em}} 
      \toprule
       & \multicolumn{1}{c}{Dim.}  & \multicolumn{1}{c}{Layers} & 
 \multicolumn{1}{c}{Heads} &  \multicolumn{1}{c}{Param.}  &  \multicolumn{1}{c}{Interval}  &  \multicolumn{1}{c}{$K$} \\
      \midrule
    Dec-S & 512 & 24 & 8 & 101M & 1 & 100 \\
    Dec-L &  1024 & 96 & 16 & 1259M & 1 & 100 \\
    EncDec-S &  512 & 2,24 & 8 & 158M & 8/64/512 & 10 \\
    EncDec-L &  1024 &  2,96 & 16 & 1738M & 8/64/512 & 10\\
      \bottomrule
    \end{tabular}
    } 
  \end{center}
\end{table}

\begin{table}
  \begin{center}
    \caption{The vector datasets used in the evaluation.}
    \label{tab_chameleon:datasets}
    \scalebox{0.9} {
    \begin{tabular}{L{11em} R{4em} R{4em} R{4em} R{4em}} 
      \toprule
       & \multicolumn{1}{c}{Deep} & \multicolumn{1}{c}{SIFT} &  \multicolumn{1}{c}{SYN-512} & \multicolumn{1}{c}{SYN-1024} \\
      \midrule
      \#vec & 1E+9 & 1E+9 & 1E+9 & 1E+9 \\
      $m/D$ & 16/96 & 16/128  & 32/512 & 64/1,024  \\
      \textit{nprobe/nlist} & 32/32K & 32/32K & 32/32K & 32/32K \\
      Raw vectors (GB) & 384 & 512 & 2,048 & 4,096 \\
      PQ and vec IDs (GB) & 24 & 24 & 40 & 72 \\
      \bottomrule
    \end{tabular}
    } 
  \end{center}
\end{table}

\begin{figure*}[t]
  
  \centering

  \begin{subfigure}[b]{0.4\linewidth}
    \includegraphics[width=\linewidth]{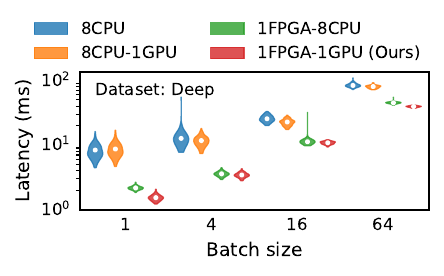}
  \end{subfigure}
  \begin{subfigure}[b]{0.4\linewidth}
    \includegraphics[width=\linewidth]{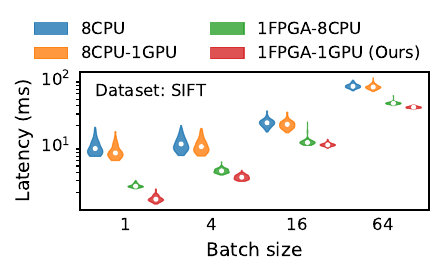}
  \end{subfigure}
  \begin{subfigure}[b]{0.4\linewidth}
    \includegraphics[width=\linewidth]{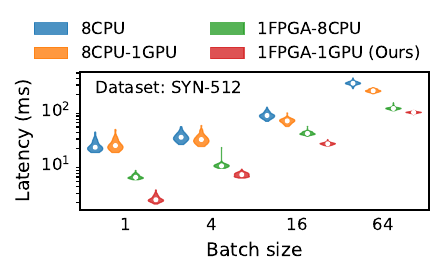}
  \end{subfigure}
  \begin{subfigure}[b]{0.4\linewidth}
    \includegraphics[width=\linewidth]{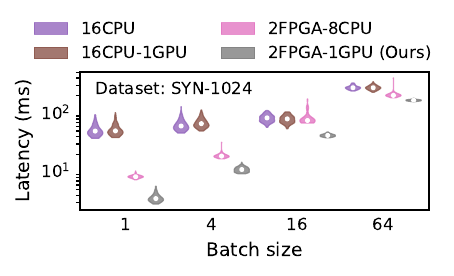}
  \end{subfigure}
  \hfill
  
  \caption{ChamVS achieves significantly lower search latency than CPUs and GPUs.}
  \label{fig_chameleon:chamvs_latency}
\end{figure*}

  

  
  


\subsection{Large-Scale Vector Search on ChamVS}
\label{sec_chameleon:eval_chamvs}

\textbf{Search performance.} 
I compare ChamVS with baseline systems using four hardware setups. PQ codes can be processed on CPU/FPGA while the IVF index can be scanned on CPU/GPU, leading to four hardware configurations: CPU, CPU-GPU, FPGA-CPU, and FPGA-GPU.
To report the best baseline performance, the CPU and CPU-GPU systems are monolithic, while the FPGA-CPU and FPGA-GPU systems are disaggregated over the network. 
Figure~\ref{fig_chameleon:chamvs_latency} compares the latency distributions of the four solutions. Each white dot in the violin plots denotes a median latency. The number of CPU cores and the number of accelerators used are listed in the plot legends. There are two primary observations from the experiments:


\textit{Firstly, the near-memory accelerator in ChamVS significantly lowers vector search latency.} Across different datasets and batch sizes (Figure~\ref{fig_chameleon:chamvs_latency}), the FPGA-CPU solution achieves 1.36$\sim$6.13$\times$ speedup compared to the CPU baseline, and the FPGA-GPU solution shows even higher speedup (2.25$\sim$23.72$\times$). This is because the ChamVS near memory accelerator can (a) decode PQ codes in parallel, (b) pipeline the decoding, distance calculation, and K-selection, such that each quantized vector can be processed by the pipeline rapidly.

\textit{Secondly, scanning the IVF index on GPU allows further latency improvements compared to the FPGA-CPU solution.} 
As shown in Figure~\ref{fig_chameleon:chamvs_latency}, the FPGA-GPU approach achieves 1.04$\sim$3.87$\times$ speedup compared to the FPGA-CPU solution. This is because the IVF index scan procedure can easily leverage the massively parallelism and the high memory bandwidth of GPUs. 
In contrast, the hybrid CPU-GPU solution shows little or even negative improvements compared to the CPU-only solution (0.91$\sim$1.42$\times$), because the search performance is limited by the slow PQ code scan process on CPU.

\begin{figure}[t]
  \centering
  \includegraphics[width=0.75\linewidth]{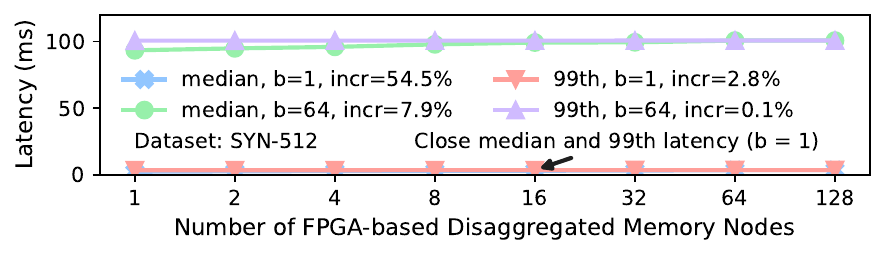}
  \caption{{The performance scalability of ChamVS.}}
  \label{fig_chameleon:scalability}
\end{figure}

\textbf{Scalability.}
I extrapolate query latency beyond the limited number of accelerators available in the evaluation. Considering the one-GPU and $N$-FPGA setup, I estimate the latency distribution by summing up accelerator and network latencies. Each query latency number is the maximum of $N$ randomly sampled latency numbers from the 1-FPGA setup. For network latency, I assume a 100 Gbps bandwidth for the CPU server and apply the LogGP model~\cite{alexandrov1995loggp, culler1993logp}, which assumes a tree topology for broadcast and reduce communications, setting the latency between two endpoints as 10.0 $\mu$s (a conservative number compared to 6.0 $\mu$s reported in~\cite{hoefler2007low, hoefler2014energy}). 
Figure~\ref{fig_chameleon:scalability} presents the median and the 99th percentile latencies for different batch sizes on the SYN-512 dataset. The tail latencies remain almost identical to those in the one-node setup due to the negligible network latency compared to the query. As for the median latencies, there is only a 7.9\% increase for a batch size of 64, while for the case without batching, the latency increases by 54.5\% as the accelerator latency is determined by the slowest one.

\textbf{Energy consumption.} \textit{ChamVS achieves 5.8$\sim$26.2$\times$ energy efficiency compared to the CPU.} Table~\ref{tab_chameleon:energy} summarizes the average energy consumption to serve a single query across different systems. I measure CPU, GPU, and FPGA energy consumption using Running Average Power Limit (RAPL) and NVIDIA System Management Interface, and Vivado, respectively. For ChamVS, I report the energy per query by measuring the power consumption times latency for scanning index on GPU and scanning PQ codes on FPGAs, respectively, and summing the two parts up.

\begin{table}
\begin{small}
  \begin{center}
    \caption{Average energy consumption per query (in mJ) on ChamVS and CPUs using various batch sizes (1$\sim$16).}
    \label{tab_chameleon:energy}
    \begin{small}
    \scalebox{0.9} {
    \begin{tabular}{L{5em}     M{3em} M{0em} M{2em} M{0em} M{2em} M{0em}      M{2em} M{0em} M{2em} M{0em} M{2em}   }\\
      \toprule
      & \multicolumn{5}{c}{CPU} & \phantom{}& \multicolumn{5}{c}{ChamVS (FPGA + GPU)} \\
        \cmidrule{2-6} \cmidrule{8-12}
      & b=1 & \phantom{}& b=4 & \phantom{}& b=16 & \phantom{}&   b=1 & \phantom{}& b=4 & \phantom{}& b=16  \\
      \midrule
SIFT &  950.3 && 434.0 && 143.3 && 53.6 && 28.2 && 21.5 \\
Deep &  929.5 && 412.9 && 141.9 && 52.3 && 26.9 && 20.5 \\
SYN-512 &  1734.9 && 957.8 && 372.5 && 95.6 && 55.0 && 41.1 \\
SYN-1024 &  4459.9 && 2315.0 && 918.5 && 170.1 && 107.8 && 85.2 \\
      \bottomrule
    \end{tabular}
    } 
    \end{small}
  \end{center}
\end{small}
\end{table}



{\textbf{Recall}. 
\textit{ChamVS, with approximate hierarchical priority queues (AHPQ), delivers results nearly identical to those of the software.}
Table~\ref{tab_chameleon:recall} shows the recall given various AHPQ lengths (8$\sim$32) when searching for the $K=100$ nearest neighbors. Here, R1@100 indicates the percentage of queries where the top nearest neighbor is within the results, while R@100 represents the percentage of overlap between the true 100 nearest neighbors and the 100 results returned.
Compared to software, AHPQ only decreases recall by up to 0.06\%. 
Interestingly, on the Deep dataset, reducing the queue lengths to eight does not necessarily result in lower recall than using a length of 32. This is likely due to the nature of PQ approximation --- a higher distance indicated by PQ does not always mean that the original vector is actually farther from the query.
}

\begin{table}
\begin{small}
  \begin{center}
    \caption{{Recall of ChamVS using approximate queues.}}
    \label{tab_chameleon:recall}
    \begin{small}
    \scalebox{0.9} {
    \begin{tabular}{L{8em}    R{8em} R{8em}   R{8em}  R{8em} }\\
      \toprule
      & CPU (len=100) & AHPQ (len=8) &   AHPQ (len=16) & AHPQ (len=32)  \\
      \midrule

R1@100 (Deep) &  92.88\% & 92.85\%  &  92.84\% & 92.84\% \\
R@100 (Deep) &  45.54\% & 45.49\% &  45.49\% &  45.48\% \\
R1@100 (SIFT) &  94.21\% & 94.20\% &  94.21\% & 94.21\% \\
R@100 (SIFT) &  48.68\% & 48.66\% &  48.67\% & 48.67\% \\

      \bottomrule
    \end{tabular}
    } 
    \end{small}
  \end{center}
\end{small}
\end{table}

\subsection{End-to-end RAG serving on Chameleon}
\label{sec_chameleon:eval_e2e}

I evaluate RAG serving performance on Chameleon with different models and retrieval intervals, using the SYN-512 and SYN-1024 datasets for the smaller and larger models, respectively. 

\textbf{RAG performance.} 
I evaluate system performance when generating a 512-token sequence using a single GPU for LLM inference .
For the latency evaluation, I disable batching, while the throughput evaluation uses the maximum allowed batch sizes given the GPU's memory capacity (64 for Dec-S and EncDec-S; 8 for Dec-L and EncDec-L).
For vector search in RAG, I use the FPGA-GPU solution for ChamVS and the CPU-only solution as the baseline, as CPU-GPU vector search can be even slower using small batches.

\begin{figure}[t]  
  \centering
    \includegraphics[width=0.9\linewidth]{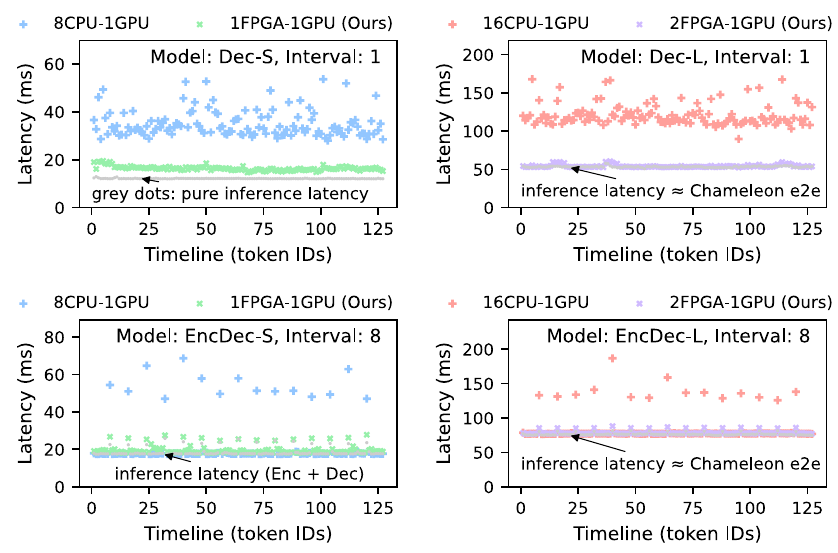}
  \caption{{Latency of RAG serving given different LLM configurations and retrieval intervals.}}
  \label{fig_chameleon:chameleon_latency_over_time}
\end{figure}

\textit{Chameleon significantly outperforms the CPU-GPU baseline system in latency for inference steps involving vector search.}
{Figure~\ref{fig_chameleon:chameleon_latency_over_time} visualizes the RAG serving latency of Chameleon and the baseline system (CPU-GPU) for the first 128 generated tokens. 
Inference latency is represented by the grey dots, while retrieval latency accounts for the remaining portion of the end-to-end latency. The time spent on coordinator and index scanning is not marked in the figure, as their latencies of hundreds of microseconds are negligible compared to up to tens of milliseconds for inference and retrieval. }
{Figure~\ref{fig_chameleon:chameleon_latency_over_time} shows that ChamVS significantly reduces the latency at the token generation steps requiring retrieval, as the retrieval latency of Chameleon is almost negligible compared to the inference latency executed on GPUs. }
Specifically, the speedup provided by Chameleon at retrieval-based inference steps (retrieval + inference) ranges from 1.94$\sim$4.11$\times$, 1.71$\sim$3.02$\times$, 1.76$\sim$3.41$\times$, and 1.29$\sim$2.13$\times$ for Dec-S, EncDec-S, Dec-L, and EncDec-L, respectively.

\begin{figure}[t]
  \centering

  \begin{subfigure}[b]{0.49\linewidth}
    \includegraphics[width=\linewidth]{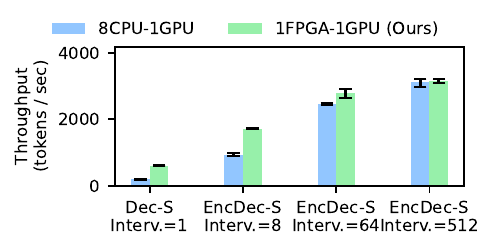}
  \end{subfigure}
  \begin{subfigure}[b]{0.49\linewidth}
    \includegraphics[width=\linewidth]{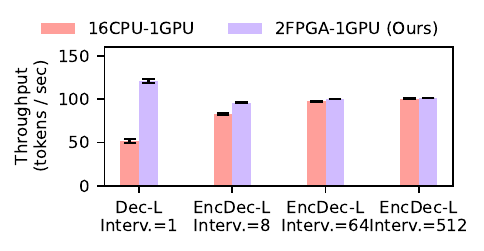}
  \end{subfigure}
  \hfill
  
  \caption{Throughput of RAG serving given different LLM configurations and retrieval intervals.}
  \label{fig_chameleon:chameleon_throughput}
\end{figure}

\textit{Chameleon achieves up to 3.18$\times$ throughput compared to the CPU-GPU baseline.} Figure~\ref{fig_chameleon:chameleon_throughput} shows that the lower the retrieval interval, the more throughput advantage Chameleon offers, with the speedup being 3.18$\times$ and 2.34$\times$ for Dec-S and Dec-L that require retrieval per token generation (interval=1). 
Chameleon attains greater speedup in batched inference than single-sequence inference (as in latency experiments), because, as the batch size grows, the latency increase for LLM inference is not as significant as that of vector search, due to the many-core parallelism that GPUs offer. 

\begin{figure}[t]
  \centering
  \includegraphics[width=0.75\linewidth]{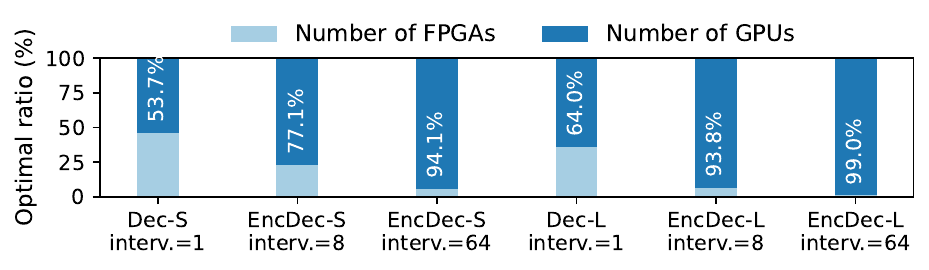}
  \caption{{Disaggregation is essential as the optimal accelerator ratio varies significantly across RAG configurations.}}
  \label{fig_chameleon:accelerator_ratio}
\end{figure}

{\textbf{The need for resource disaggregation.} 
Accelerator disaggregation allows Chameleon to adjust the ratio between the two types of accelerators across RAG configurations. 
I model the overall system throughput, measured by generated tokens per second, across various accelerator ratios using a total of 1,000 accelerators, assuming the cost for an inference accelerator and a retrieval accelerator is equivalent.  Given retrieval interval \(i\), batch size \(b\), number of inference and retrieval accelerators \(N_I\) and \(N_R\), latency per batch for inference and retrieval \(L_I(b)\) and \(L_R(b)\), the system throughput is determined by the minimum of the inference and retrieval throughput: \( Th_{system} = \min(Th_{I}, Th_{R}) \), where \( Th_{I} = \frac{i \cdot b \cdot N_I}{i \cdot L_I(b) + L_R(b)}\) and \( Th_{R} = \frac{i \cdot b \cdot N_R }{L_R(b)} \).
}
{
Figure~\ref{fig_chameleon:accelerator_ratio} shows that the optimal ratio of accelerators to achieve the highest throughput varies significantly, ranging from 53.7\%$\sim$99.0\% across RAG configurations.}

\begin{figure}[t]
  \centering
  \includegraphics[width=0.9\linewidth]{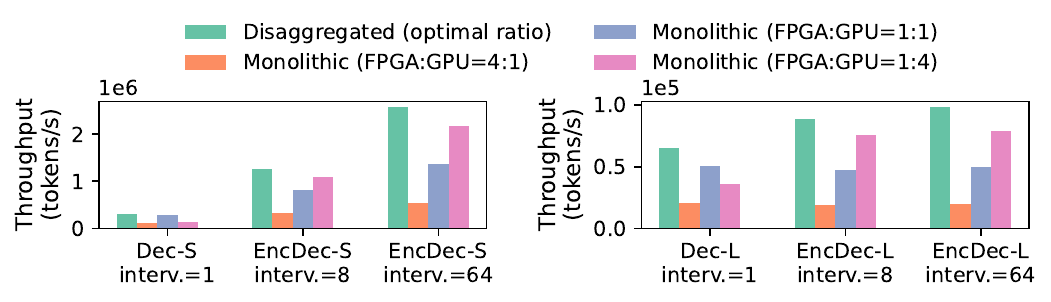}
  \caption{{The disaggregated design consistently outperforms the monolithic ones using fixed accelerator ratios.}}
  \label{fig_chameleon:disaggregated_vs_monolithic}
\end{figure}

{
\textit{The disaggregated design, using the optimal accelerator ratio, consistently outperforms the monolithic ones with fixed ratios, as shown in Figure~\ref{fig_chameleon:disaggregated_vs_monolithic}.}
Given the impracticality of adjusting the ratio for each RAG configuration in a monolithic design, the performance of a monolithic design can only match that of Chameleon on a limited set of RAG configurations.
}

\section{Related Work}
\label{sec_chameleon:related_work}

Chameleon represents the first endeavor to improve RAG serving performance by using heterogeneous accelerator systems. I now introduce related research about vector search on modern hardware.
Lee et al.~\cite{lee2022anna} study ASIC designs for IVF-PQ, and a couple of works~\cite{jiang2023co, zhang2018efficient} implement IVF-PQ on an FPGA, but their designs are constrained by either the limited HBM capacity or the slow CPU-FPGA interconnect. 
In contrast, Chameleon disaggregates IVF-PQ, with the index on GPUs and PQ codes on FPGA-based memory nodes, and employs the innovative hardware priority queue design to achieve high performance with little hardware resources. 
While graph-based vector search accelerators can achieve low latency~\cite{jiang2024accelerating, zeng2023df}, the memory consumption is high, requiring up to one TB of memory for only one billion SIFT vectors, in contrast to 24 GB in PQ evaluated in this chapter. 
Apart from accelerators, researchers also study memory and storage for vector search. 
One can leverage non-volatile memory~\cite{ren2020hm} and CXL~\cite{jang2023cxl} to scale up graph-based ANN, while on-disk ANN has to be more careful with I/O cost~\cite{chen2021spann, jayaram2019diskann, lejsek2008nv}. 
Hu et al.~\cite{hu2022ice} further push down distance evaluation into NAND flash to reduce data movement.

\section{Conclusion}
\label{sec_chameleon:discussion}


I present Chameleon, a heterogeneous and disaggregated accelerator system for efficient RAG serving. 
Given the rapidly evolving algorithms, software, and hardware related to RAG, Chameleon can be potentially upgraded in the following ways.
For LLM inference, ChamLM could be enhanced by supporting low precision~\cite{fastertransformer}, continuous batching~\cite{yu2022orca}, paged-attention~\cite{kwon2023efficient}, and disaggregated prompt computation and token generation~\cite{patel2023splitwise}. 
While currently supporting PQ, ChamVS could potentially be replaced by graph-based ANN accelerators~\cite{zeng2023df, peng2021optimizing}. 
ChamVS could also be extended to support index updates~\cite{xu2023spfresh} and relational features~\cite{zhang2023vbase}.

\chapter{PipeRAG: Fast Iterative RAG via Adaptive Pipeline Parallelism} 
\label{chap:piperag}

As identified in Chapter~\ref{chap:rago}, iterative retrieval can degrade RAG serving performance due to inference accelerator idleness and high retrieval latency.  
This chapter addresses the serving efficiency of RAG with iterative retrieval by co-designing algorithms and systems, overlapping retrieval and generation latency by approximate data prefetching.

\section{Introduction}
\label{sec_piperag:intro}


While one retrieval prior to the generation process can be enough when generating short sequences~\cite{lewis2020retrieval, izacard2020leveraging}, a more general RAG approach involves periodic retrievals throughout the generation~\cite{borgeaud2022improving, norlund2023generalization, ram2023context, jiang2023active, trivedi2022interleaving}. This necessity arises due to the potential shift in the generation context, such as changes in topics.  Therefore, periodic retrievals ensure the retrieved content remains relevant to the latest context of the generation. 
A popular example of this category is \textsc{Retro}~\cite{borgeaud2022improving}, which tailors the transformer neural network architecture to support the integration of retrieved content at regular intervals.

However, periodic retrievals on large databases, potentially comprising trillions of tokens~\cite{borgeaud2022improving}, can significantly slow down the sequence generation. \textit{A natural research question is: can we optimize the system performance of RAG while preserving or even improving generation quality?} 

\begin{figure*}[t]
	\centering
  \includegraphics[width=\linewidth]{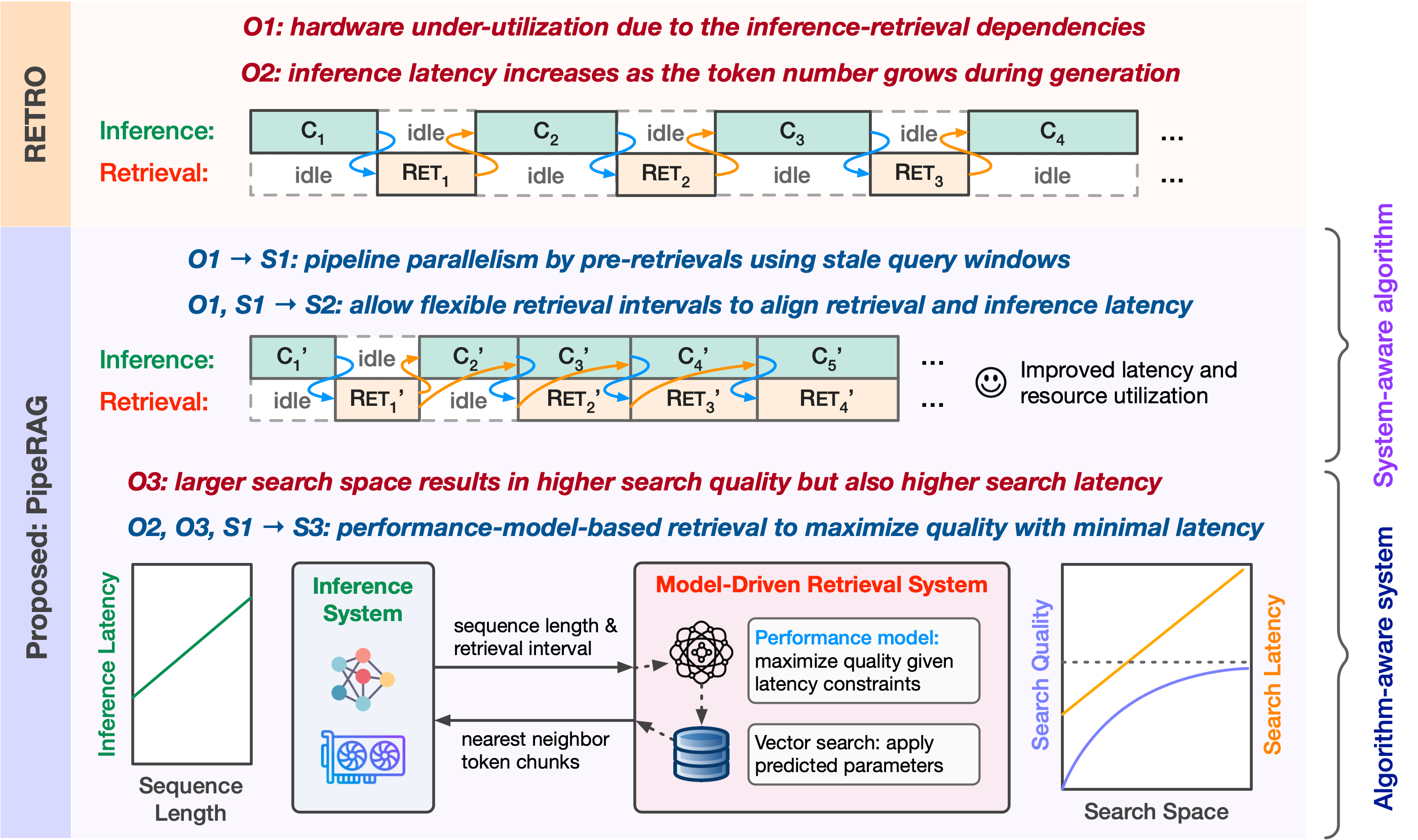}
  \caption{Based on three performance-centric observations (O1$\sim$O3), PipeRAG combines a system-aware algorithm integrating pipeline parallelism (S1) with flexible retrieval intervals (S2) and an algorithm-aware retrieval system guided by a performance model (S3).}
  \label{fig_piperag:overview}
\end{figure*}

 
I propose PipeRAG, a pioneering approach to improve RAG efficiency via a collaborative algorithm-system co-design --- including a system-aware RAG algorithm and an algorithm-aware retrieval system as overviewed in Figure~\ref{fig_piperag:overview}. 

The foundation of PipeRAG is established on three observations centered on performance. Firstly, the dependencies between retrievals and LLM inferences lead to hardware underutilization, with either the inference or retrieval system being idle at any given time during the generation process~(\textbf{O1}). Secondly, the inference latency per token increases with sequence lengths, due to the growing workloads of the attention mechanism in transformer neural networks~(\textbf{O2}). Lastly, the retrieval process, particularly the approximate nearest neighbor search, exhibits a trade-off between search latency and search quality~(\textbf{O3}).

\textit{The key idea of PipeRAG is to prefetch content from databases to facilitate pipeline parallelism between the inference and retrieval systems.} This solution reduces end-to-end generation latency by allowing simultaneous inference and retrievals, effectively addressing the hardware inefficiencies identified in O1~(\textbf{S1}). 
I then enhance this key idea with two additional solutions.
On the model side, PipeRAG modifies \textsc{Retro}'s attention mechanism to support flexible retrieval intervals (invoke a retrieval after generating a certain number of tokens), because the intervals must be carefully tuned to capitalize the efficiency of pipeline parallelism~(\textbf{S2}). 
On the system side, the retrieval system adopts a performance model informed by O2 and O3 to dynamically adjust the retrieval search space (the amount of database vectors to scan per query) according to the latency expectation of the upcoming token inferences in the pipeline, thereby optimizing search quality without increasing end-to-end generation latency~(\textbf{S3}).

The evaluation of PipeRAG, involving various evaluation datasets and using large databases based with up to 200 billion tokens, clearly illustrates its efficiency in both generation performance (latency) and generation quality (perplexity). Specifically, the quality-performance Pareto frontier of PipeRAG significantly outperforms that of \textsc{Retro}: PipeRAG can achieve up to 2.6$\times$ speedup in latency without compromising perplexity; alternatively, maintaining the same latency allows PipeRAG to reduce perplexity by as much as 0.93 compared to \textsc{Retro}. These encouraging results highlight the importance of algorithm-system co-design in retrieval-augmented generation, paving the way for deploying PipeRAG in future RAG systems.

\textbf{Contributions:} I propose PipeRAG, an algorithm-system co-design approach aimed at improving retrieval-augmented generation efficiency. Specifically:
\begin{itemize}
\item I design a system-aware RAG algorithm that leverages pipeline parallelism, whose efficiency is further improved by supporting flexible retrieval intervals.
\item I propose an algorithm-aware retrieval system that uses performance models to dynamically balance search quality and performance.
\item I showcase the impressive performance of PipeRAG in various datasets, demonstrating the importance of algorithm-system co-design in optimizing RAG.
\end{itemize}

\section{Background and Motivation}
\label{sec_piperag:background}



\begin{figure}
  \centering
  \includegraphics[width=0.65\linewidth]{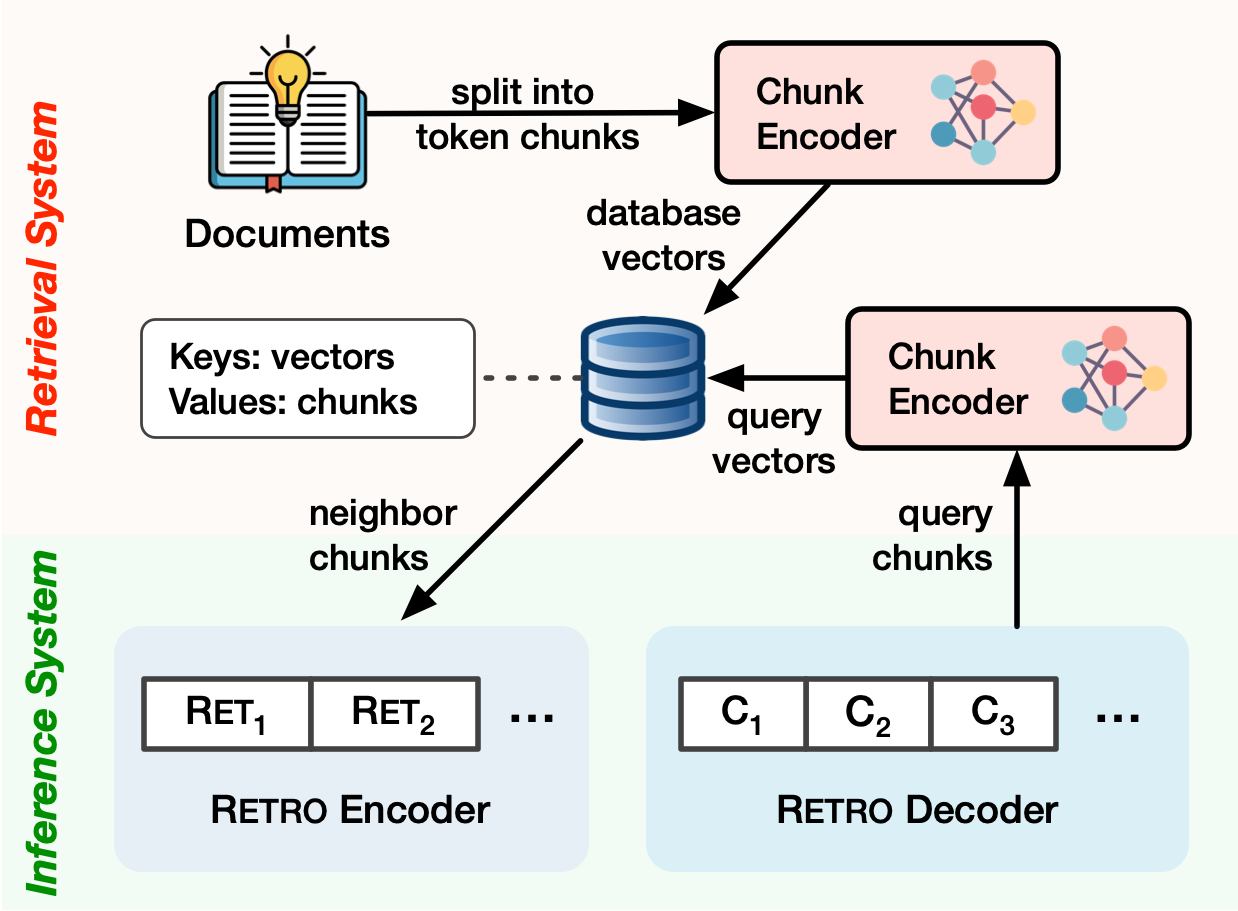}
  \caption{Retrieval augmentation with \textsc{Retro}.}
  \label{fig_piperag:retro}
\end{figure}

Sequence generation quality of LLMs can be improved through periodically retrieving from large token databases~\cite{borgeaud2022improving, norlund2023generalization, ram2023context}. Here, periodic retrievals, instead of retrieving only once, are essential in handling potential contextual shifts during generation, such as topic changes, ensuring alignments between the retrieved content and the evolving generation context. 
\textsc{Retro} is a representative model in this category~\cite{borgeaud2022improving}. 
As illustrated in Figure~\ref{fig_piperag:retro}, \textsc{Retro} integrates a retrieval system with an inference system for token generation. It employs an encoder for incorporating retrieved tokens and a decoder for token generation.

\textbf{Database construction.} A \textsc{Retro} database comprises a large collection of documents segmented into \( n \) chunks of tokens \( S = (S_1, \ldots, S_n) \), where each chunk \( S_i \) spans \( m \) tokens. These token chunks are each converted into vector representations \( R(S) \). The database is then structured as a key-value store, with keys being the vector representations \( R(S) \) and values corresponding to the original token chunks \( S \), along with \( F \), in which \( F_i \) representing the immediately following token chunks of each chunk \( S_i \). Given a query vector \( q \), the database performs an approximate nearest neighbor (ANN) search to retrieve \( k \) closest token chunks and their immediately following chunks. 

\textbf{Retrieval process.}  \textsc{Retro} performs retrievals at regular intervals during the generation phase. Specifically, when generating a sequence of $t$ tokens \( X = (x_1, \ldots, x_t) \), \textsc{Retro} partitions \( X \) into \( l \) chunks \( (C_1, \ldots, C_l) \), each consisting of \( m \) tokens. Consequently, token \( x_{i} \) belongs to chunk \( C_{\lceil \frac{i}{m} \rceil} \). For the generation of chunk \( C_i \), \textsc{Retro} employs the preceding chunk \( C_{i-1} \) as the query to retrieve \( k \) nearest neighbors \( \textsc{Ret}(C_{i-1}) \) from the database.

\textbf{Attention mechanisms.} 
\textsc{Retro} involves both decoder-to-encoder and encoder-to-decoder attention mechanisms. 
The decoder within \textsc{Retro} utilizes chunked cross-attention to integrate the retrieved information encoded by the encoder. To preserve causality, the generation of a chunk $C_i$ incorporates the retrieved tokens \( \textsc{Ret}(C_{i-1})\) by integrating the encoder states \( \textsc{Enc}(\textsc{Ret}(C_{i-1}))\).
On the other hand, the \textsc{Retro} encoder states \( \textsc{Enc}(\textsc{Ret}(C_{i-1}))\) integrates the decoder's states of the \( \textsc{Dec}(C_{i-1}) \) via a standard cross-attention (CA) mechanism, such that the encoder can blend the retrieved information with the generation context.
Because both decoder-to-encoder and encoder-to-decoder attention mechanisms operate on a chunk-wise basis, \textsc{Retro} avoids the excessive computational demands of attending to all previous retrieval and generation states.


\textbf{Motivation: improving RAG efficiency.}
Although periodically retrieving tokens from a large database can effectively improve the generation quality of LLMs, frequent retrievals can account for a considerable portion of the total generation time, thereby significantly slowing down the end-to-end generation process.

In this chapter, I ask the following question: \textbf{is it possible to further enhance the efficiency of retrieval augmented generation?} Here, I conceptualize \textit{RAG efficiency} as a Pareto frontier considering two objectives: \textit{generation quality} and \textit{system performance}. 
Specifically, given a quality requirement (achieving certain perplexity), can we optimize RAG's system performance (reducing generation latency)? On the other hand, given a system performance requirement, can we improve the quality of generation?

\section{Solution: PipeRAG}
\label{sec_piperag:approach}

I propose PipeRAG, a novel retrieval augmented generation approach to improve the performance-quality Pareto frontier through an in-depth algorithm-system co-design.
The development of PipeRAG stems from performance-centric observations revealing (1) the \textit{fundamental} system inefficiencies in existing RAG algorithms and (2) the distinct performance characteristics of LLM inference and retrieval systems. 
Based on these observations, PipeRAG includes (1) a system-aware RAG algorithm to address the system inefficiencies and (2) an algorithm-aware retrieval system to dynamically balance retrieval quality and latency. 

\subsection{Performance-Centric Observations in RAG}
 
\textbf{O1: Hardware inefficiency due to RAG dependencies.} A conventional RAG process introduces dependencies between retrievals and inferences: the current generation context is used as a query to retrieve relevant token chunks stored in the database; the inference process must wait for the retrieval to finish before it can continue generating a few more tokens, until the next retrieval is triggered.

A RAG system typically comprises two sub-systems: the retrieval system and the inference system, each hosted on separate hardware platforms.
AI accelerators such as GPUs and TPUs are the ideal hardware platforms for LLM inference due to the high demands for computation and memory bandwidth during inference. 
On the other hand, the retrieval systems consisting of large databases are usually not based on GPUs. This is because (1) the limited memory capacity of individual GPUs (GPUs adopt high-bandwidth memory that is fast but limited in capacity) makes the hosting of large databases cost-prohibitive, necessitating the setup comprising many GPUs, and (2) the communication bandwidth between the CPU and GPU is significantly lower compared to GPU's device memory bandwidth, thus the CPU-GPU solution, in which database vectors are stored in CPU-side memory and then transferred to GPUs at query time, could be exceedingly slow.
Given the capacity requirements, the retrieval system is typically CPU-based~\cite{borgeaud2022improving, lewis2020retrieval}, with the database either held in substantial main memory (DRAM), or, in more budget-friendly setups, stored on disks. 

Given that the two systems are based on separate hardware, the dependencies between retrievals and inferences in RAG result in significant underutilization of hardware resources. Figure~\ref{fig_piperag:overview} illustrates this inefficiency using RETRO as a representative example: due to the dependencies, either the inference or retrieval system is idle at any given time during the generation process, leading to hardware inefficiencies.

\textbf{O2: Increasing inference time with sequence length.} 
In a standard transformer neural network~\cite{vaswani2017attention}, the cost of generating each new token correlates with the sequence length, rather than remaining a constant. This is due to the attention mechanism in transformers: although the workload of the fully-connected layers remains constant throughout the generation process, the cost of attention layers increases with the sequence length~\cite{beltagy2020longformer}. Specifically, for each new token generated, the query states (Q) of the most recent token are compared against the key states (K) of all preceding tokens to calculate relevance scores. These scores are then utilized for a weighted sum over the value states (V) (note that the queries, keys, and values mentioned here under the context of transformers are distinct from those terms in RAG systems).
Consequently, the inference cost per token can be approximated as a linear function to sequence length.

\textbf{O3: Trade-offs between retrieval quality and latency.} Large-scale vector search in RAG employs approximate nearest neighbor (ANN) search instead of exact nearest neighbor search due to the latter's prohibitive cost on large databases. In ANN search, database vectors are indexed, with popular choices including clustering-based inverted-file (IVF) indexes~\cite{IVF} and graph-based indexes~\cite{malkov2014approximate, malkov2018efficient}. Optionally, database vectors may also be compressed via product quantization (PQ)~\cite{PQ} to shrink database sizes and reduce memory bandwidth usage at query time at the expense of search accuracy. During a search, a query vector is only compared against a subset of database vectors selected by the index. 

Regardless of the index types, there exists a \textit{fundamental} trade-off between search quality and latency in ANN search. Typically, the index first directs the search towards those database vectors that are most likely to be the nearest neighbors of the query vector, and then gradually expands the search space. The number of database vectors scanned per query can be directly or indirectly controlled by ANN search hyper-parameters. Expanding the search space would enhance the probability of finding the query vector's true nearest neighbors in the database (improved search quality), but also would also lead to higher latency (lower search performance) due to the greater number of comparisons between query vectors and database vectors.

Figure~\ref{fig_piperag:overview} visualizes the relationship between search quality and latency~\cite{PQ}. As the search space expands (number of scanned database vectors), the search quality (recall of the retrieval) gradually improves until reaching a plateau where the nearest neighbors are likely found. Simultaneously, the search cost (latency) grows linearly with the search space, with an initial cost of scanning the index (which could be zero in some graph-based indexes). 

\subsection{Algorithm-System Co-deisgn in PipeRAG}

\begin{figure}[t]
	\centering
  \includegraphics[width=0.7\linewidth]{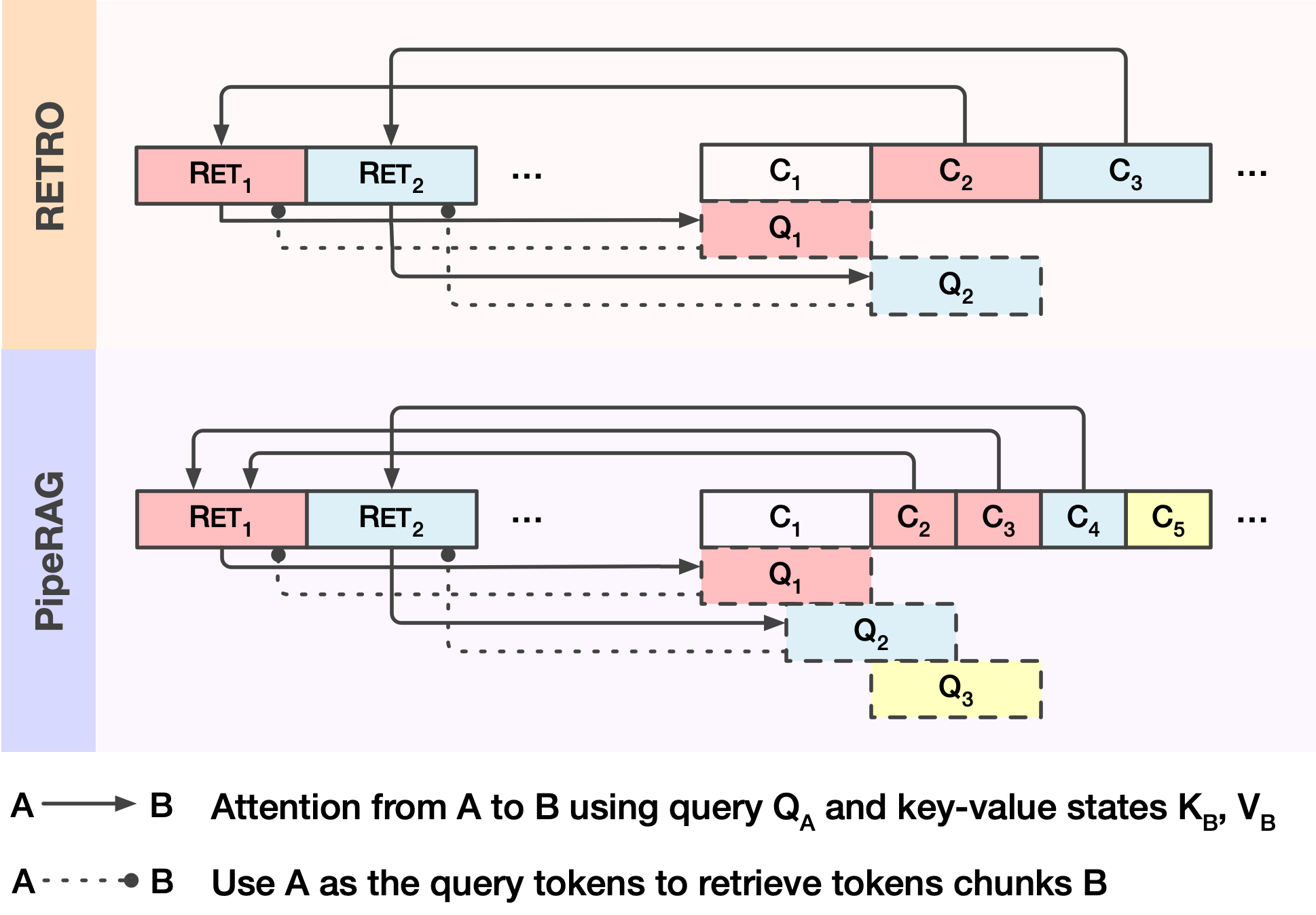}
  \caption{PipeRAG's attention mechanism.}
  \label{fig_piperag:attention}
\end{figure}
Given the aforementioned performance-centric observations, I propose PipeRAG, an algorithm-system co-design approach aimed at enhancing RAG's performance-quality Pareto frontier. 
PipeRAG addresses the \textit{fundamental} issue of hardware inefficiency (O1) by employing pipeline parallelism (S1) and allowing flexible retrieval intervals (S2). Leveraging the distinct performance characteristics of the inference and retrieval sub-systems (O2, O3), PipeRAG further offers an option to enable automatic search space selection within the retrieval system, facilitating high-quality generation without introducing additional generation latency.

\textbf{S1: Pipeline parallelism across RAG sub-systems.} 
Because the hardware under-utilization issue in RAG is caused by dependencies between retrievals and inferences, my first solution is about revisiting RAG algorithms to enable pipeline parallelism: the retrievals and inferences should be executed concurrently, thus overlapping their execution latency and improving hardware utilization. 

To facilitate pipeline parallelism, I relax the RAG dependencies as illustrated in Figure~\ref{fig_piperag:overview}: instead of depending on the content retrieved using the query representing the most recent generation context (the latest generated tokens),
the inference process can utilize a slightly older, or \textit{stale}, query window to \textit{prefetch} content from the database.
The intuition here is that if the stale query window closely aligns with the latest generation context, it is likely to retrieve content similar to that obtained using the most recent query tokens.
Once the dependency constraint is relaxed, retrievals can be proactively initiated to \textit{prefetch} content from the database, thus enabling pipeline parallelism as shown in Figure~\ref{fig_piperag:overview}.

Formally, when generating token chunk \( C_{j+1} \), PipeRAG does not use the immediately preceding chunk as the query \( Q = C_j = (x_{jm}, \ldots, x_{jm + m - 1}) \) to retrieve \( \textsc{Ret}(Q) \). Instead, it opts for a stale token window \( \hat{Q} = (x_{jm - s}, \ldots, x_{jm + m - 1 - s}) \) as an approximate query, offset by \( s \) tokens from the latest query window. Subsequently, \( \hat{\textsc{Ret}(Q)} = \textsc{Shift}(\textsc{Ret}(\hat{Q}), s) \) serves as the approximation of \( \textsc{Ret}(Q) \). Given that the stale query is \( s \) tokens behind the most recent generation context, the retrieved results \( \textsc{Ret}(\hat{Q}) \) are correspondingly left-shifted by \( s \) tokens. This shift ensures that the first \( s \) retrieved tokens, which are likely less relevant for the upcoming generation due to staleness, are excluded while maintaining the overall length of retrieval tokens. Note that the concept of stale query windows does not apply for the initial retrieval, which is conducted using the first chunk \( C_{1} \), as illustrated in Figure~\ref{fig_piperag:overview}.


\textbf{S2: Flexible retrieval intervals.} 
\textsc{Retro} utilizes a fixed retrieval interval of \( m=64 \), aligning with the generation chunk size, database token chunk size, and query window size. 
However, the effectiveness of pipeline parallelism (S1) is maximized when the retrieval and inference subsystems have similar latencies --- generating \( m=64 \) tokens does not always consume similar time as one retrieval. 

In order to improve the effectiveness of pipeline parallelism, PipeRAG supports alternative retrieval intervals \( m' \) and modifies \textsc{Retro}'s attention mechanism accordingly.
Here, \( m' \) remains constant during a single generation process but can vary from the default value of 64. When using shorter intervals, such as \( m'=32 \), the staleness of queries is also reduced (\( s=32 \), thereby improving the quality of the retrieved content to more closely resemble that obtained from a non-stale query.
Figure~\ref{fig_piperag:attention} illustrates the differences in retrievals and attention mechanisms between \textsc{Retro} and PipeRAG, taking \( m'=32 \) as an example. As shown in the figure, while a query \( Q_i \) still has a window size of \( m=64 \) tokens, the retrieval interval is halved. This necessitates adjustments in the attention regions to align with these modified intervals. For encoder-to-decoder attention, the attention is directed from the retrieved chunk to the query window whose position is different from that of \textsc{Retro}. For decoder-to-encoder attention, the generation of chunk \( C_{j+1} \) of length \( m' \) applies chunked cross-attention on \( \textsc{Ret}(Q_{j-1}) \).

\textbf{S3: Performance-model-driven retrievals.} 
PipeRAG has the potential to match the generation latency of LLMs that do not introduce retrievals, especially when the retrievals and inferences are completely overlapped in the pipeline.
However, achieving this ideal overlap is challenging because of the distinct performance characteristics of the retrieval and inference systems as introduced in O2 and O3. 

To address this, I propose a performance-model-driven retrieval system to automatically enable perfectly overlapped pipeline windows.
In this context, a performance model refers to any model (not limited to neural networks) designed to predict the performance characteristics of a system.
Specifically, the retrieval system takes the generation states as inputs and automatically adjusts the search space using performance models, ensuring that the retrieval latency can be hidden by the generation latency of the next token chunk. By maximizing the search space under the latency constraint, the retrieval quality is also maximized without incurring extra generation latency.

The inference performance can be modeled as follows. 
The time required to generate a token chunk can be represented by \( T_{C} = T_{\textsc{Enc}} + T_{\textsc{Dec}} \). The latency of encoder inference is related to the number of retrieved neighbors and the number of tokens per neighbor, while the decoder inference latency depends on the current sequence length and the chunk size (O2). 

On the other hand, retrieval latency can be represented modeled as \( T_{\textsc{Ret}} = T_{Network} + T_{EncQuery} + T_{ScanIndex} + T_{ScanVec}\), encompassing the time spent on network communications, encoding the query tokens as vectors, scanning the vector index, and scanning a subset of database vectors. 
In this chapter, I apply the widely-adopted IVF-PQ vector search algorithm~\cite{PQ} that combines a clustering-based inverted-file (IVF) index with product quantization (PQ). The IVF index clusters the database to \( nlist \) IVF lists. At query time, \( nprobe \) out of the \( nlist \) IVF lists are selected to scan (database vectors within the selected lists are compared to the query vectors). 

As the performance of both retrievals and inferences are related to hardware, I measure and model their performance on the deployment hardware. I record the time consumption of both encoder and decoder inferences with various input sequence lengths. For retrieval, I model the relationship between \( nprobe \) and search latency using linear regression, given that \( nprobe \) is approximately proportional to the number of scanned database vectors.

The retrieval system then leverages these performance models to predict the maximal search space, indicated by $nlist$, given the latency constraint for generating the next token chunk, ensuring that \( T_{\textsc{Ret}} \leq T(C) \). 
Since the \( T(C) \) can be easily obtained from the recorded performance numbers, we can then derive the maximal \( nprobe \) during the search based on the retrieval performance model.

While an alternative approach to achieve a perfectly overlapped pipeline is adjusting the retrieval intervals in the inference system, I rule out this option due to generalizability concerns. In future deployment scenarios, a retrieval system may serve multiple inference systems. Thus, the retrieval performance is impacted by the number of concurrent queries being processed. In this case, it could be challenging for the inference system to accurately predict the retrieval latency, as it lacks the information about the retrieval system's workload at the moment. Therefore, it is the retrieval system, instead of the inference system, that should be responsible for constructing a perfectly overlapped pipeline via performance modeling.

\section{Evaluation}
\label{sec_piperag:evaluation}

I evaluate PipeRAG in various aspects, showing its effectiveness in both generation quality and generation latency. 

\subsection{Experimental Setup}



\textbf{Database.}
The token database was constructed from the C4 corpus with deduplicated English documents. I did not choose the Pile dataset as previous work~\cite{borgeaud2022improving} due to its current copyright issues. Adhering to~\cite{borgeaud2022improving}, I segmented the documents into chunks of \( m=64 \) tokens, yielding a total of three billion chunks. Following~\cite{norlund2023generalization}, I transformed each token chunk into a 384-dimensional vector using a sentence transformer\cite{reimers2019sentence} checkpoint \textit{all-MiniLM-L6-v2}.

\textbf{Model.}
I developed PipeRAG based on the \textsc{Retro} checkpoint with 582M parameters provided by~\cite{norlund2023generalization}, the only available pre-trained \textsc{Retro} model when I conducted the experiments.

\textbf{Evaluation Set.}
To evaluate language modeling quality, I used the Wikipedia dataset~\cite{wikipedia}, the RealNews subset of the C4 dataset, and C4's English document subset~\cite{dodge2021documenting, raffel2020exploring}.

\textbf{Software.}
The implementation of the PipeRAG model is based on a \textsc{Retro} baseline obtained from~\cite{norlund2023generalization}, which is built on top of PyTorch.  To enhance inference performance, I supported the caching of key-value states in the transformer and converted the model to ONNX format, enabling model inference by ONNX runtime. With the above optimizations, the inference latency on GPU is improved by around 3$\times$ over the original Pytorch implementation. I maintained the fp32 (32-bit floating point) precision of the model.
For the retrieval system, I used the Faiss library~\cite{johnson2019billion}, which is known for its efficient product-quantization-based vector search implementation. 
I adopted the IVF-PQ vector search algorithm, setting the number of IVF list centroids to \( nlist=16384 \) and quantizing each 384-dimensional vector into 64 bytes of PQ code. 
During retrievals, I set the number of nearest neighbors as \(k=2 \).
The communication between the inference and retrieval systems was managed via the gRPC library.

\textbf{Hardware.}
For model inference, I utilized an NVIDIA A100 GPU (40 GB). The retrieval process was handled by a server equipped with dual-socket Intel(R) Xeon(R) Platinum 8259CL CPUs @2.50GHz (48 cores and 96 threads) and 384 GB memory. 
The retrieval and inference servers were interconnected through a network, with a round-trip time of around 1 ms.

\subsection{Perplexity Evaluation}

I report the language modeling quality of PipeRAG as the main quality metric, because most QA benchmarks only contain short answers, which cannot exhibit the latency benefit of PipeRAG when generating longer sequences.

\begin{figure*}
\centering    
\begin{subfigure}[b]{0.55\linewidth}
    \centering
    \includegraphics[width=\linewidth]{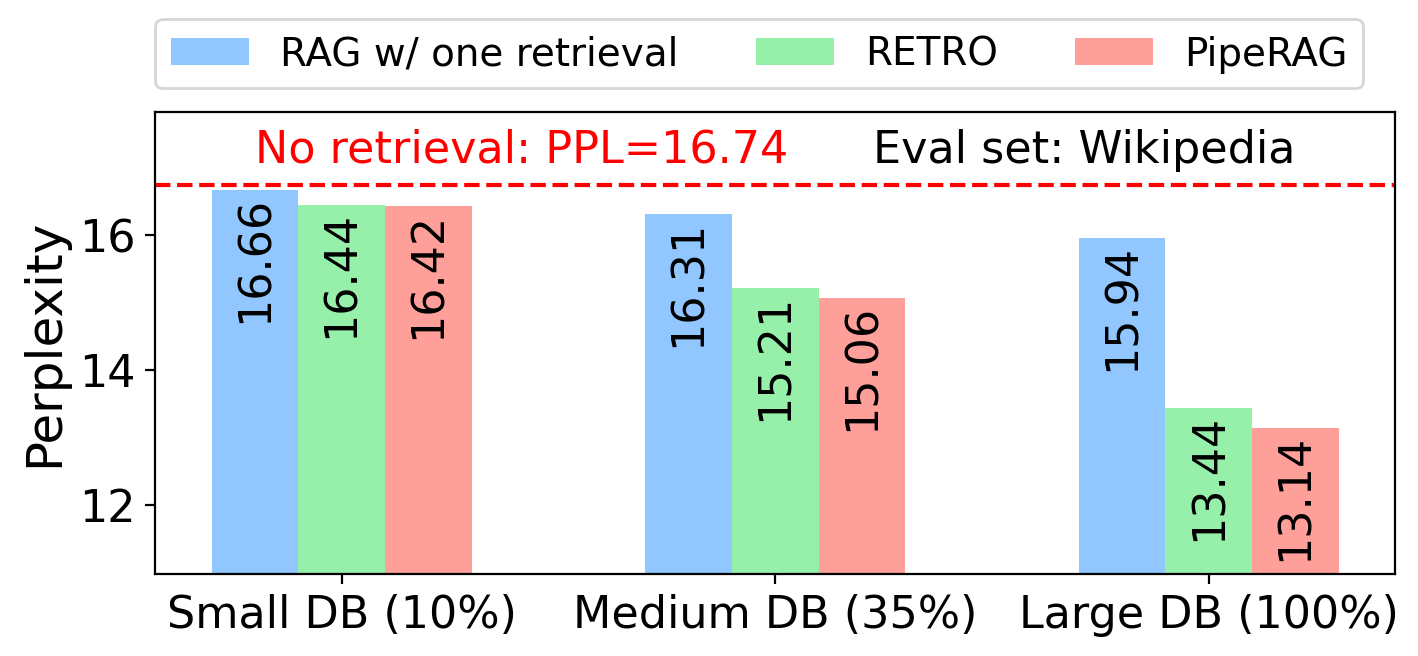}
\end{subfigure}

\begin{subfigure}[b]{0.55\linewidth}
    \centering
    \includegraphics[width=\linewidth]{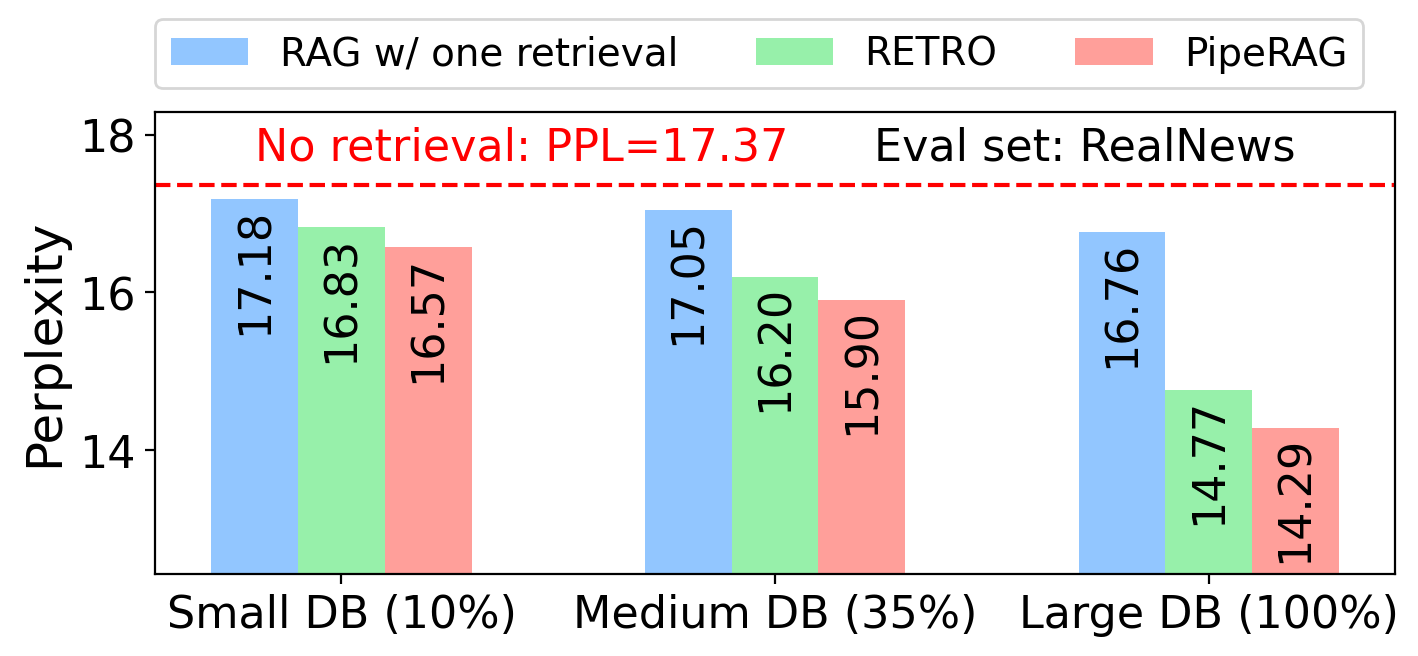}
\end{subfigure}

\begin{subfigure}[b]{0.55\linewidth}
    \centering
    \includegraphics[width=\linewidth]{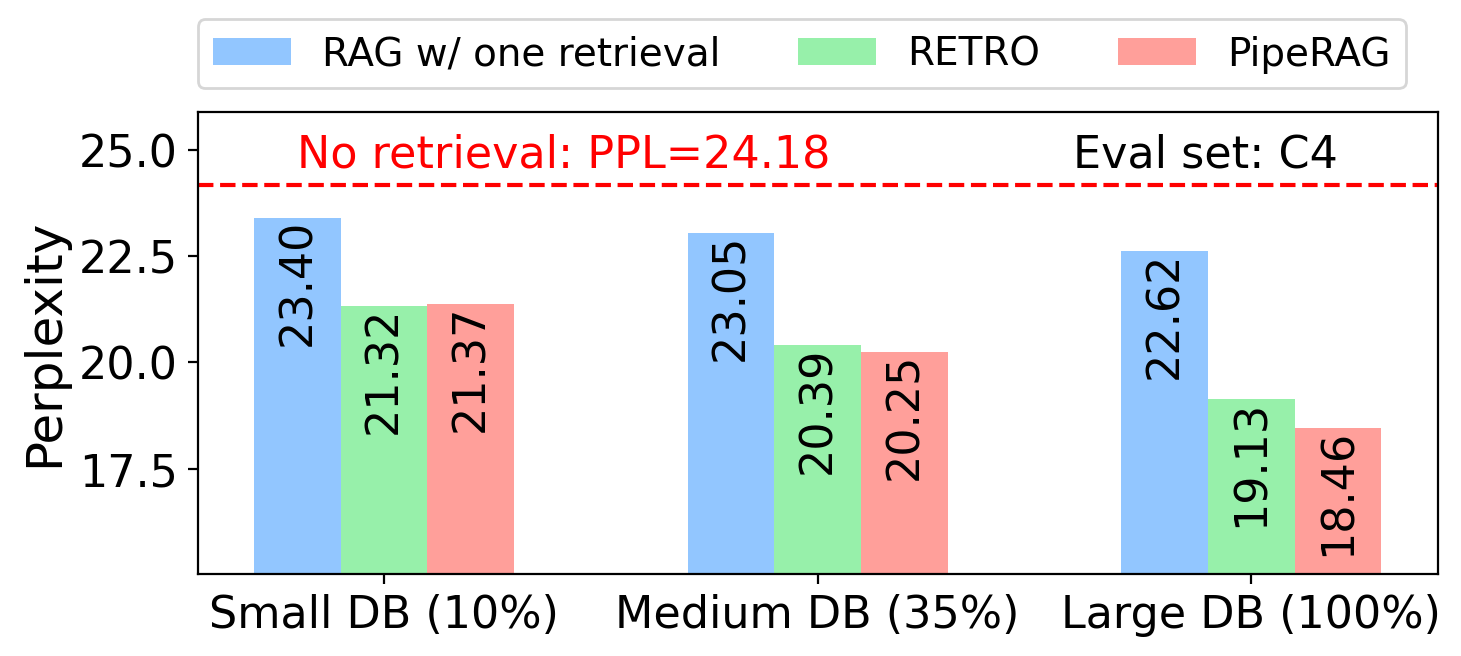}
\end{subfigure}

  \caption{The effect of database sizes and retrieval strategies on language modeling perplexity (lower is better).}
  \label{fig_piperag:eval_dbsize}
\end{figure*}

\begin{figure*}
\centering    

\begin{subfigure}[b]{0.55\linewidth}
    \centering
    \includegraphics[width=\linewidth]{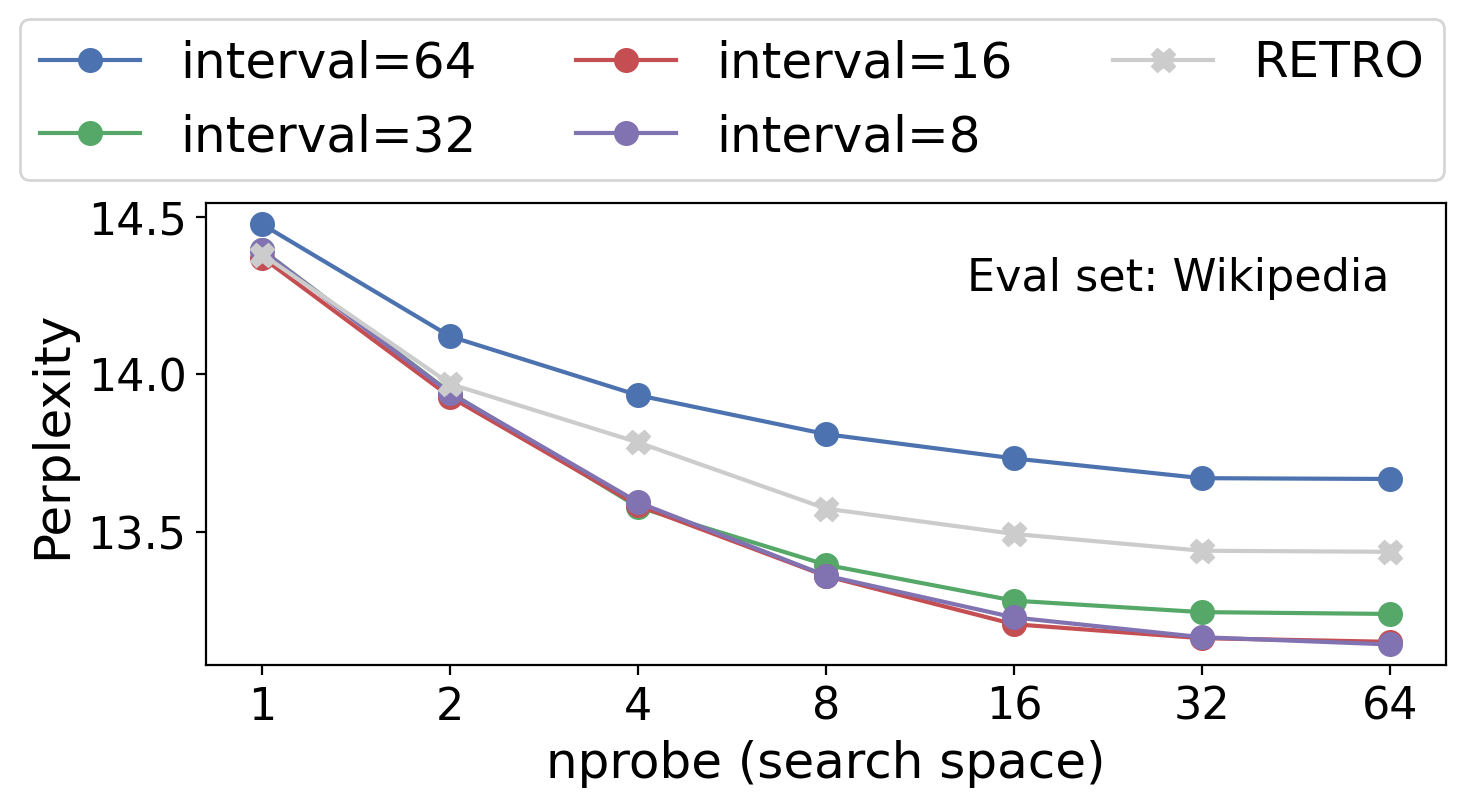}
\end{subfigure}

\begin{subfigure}[b]{0.55\linewidth}
    \centering
    \includegraphics[width=\linewidth]{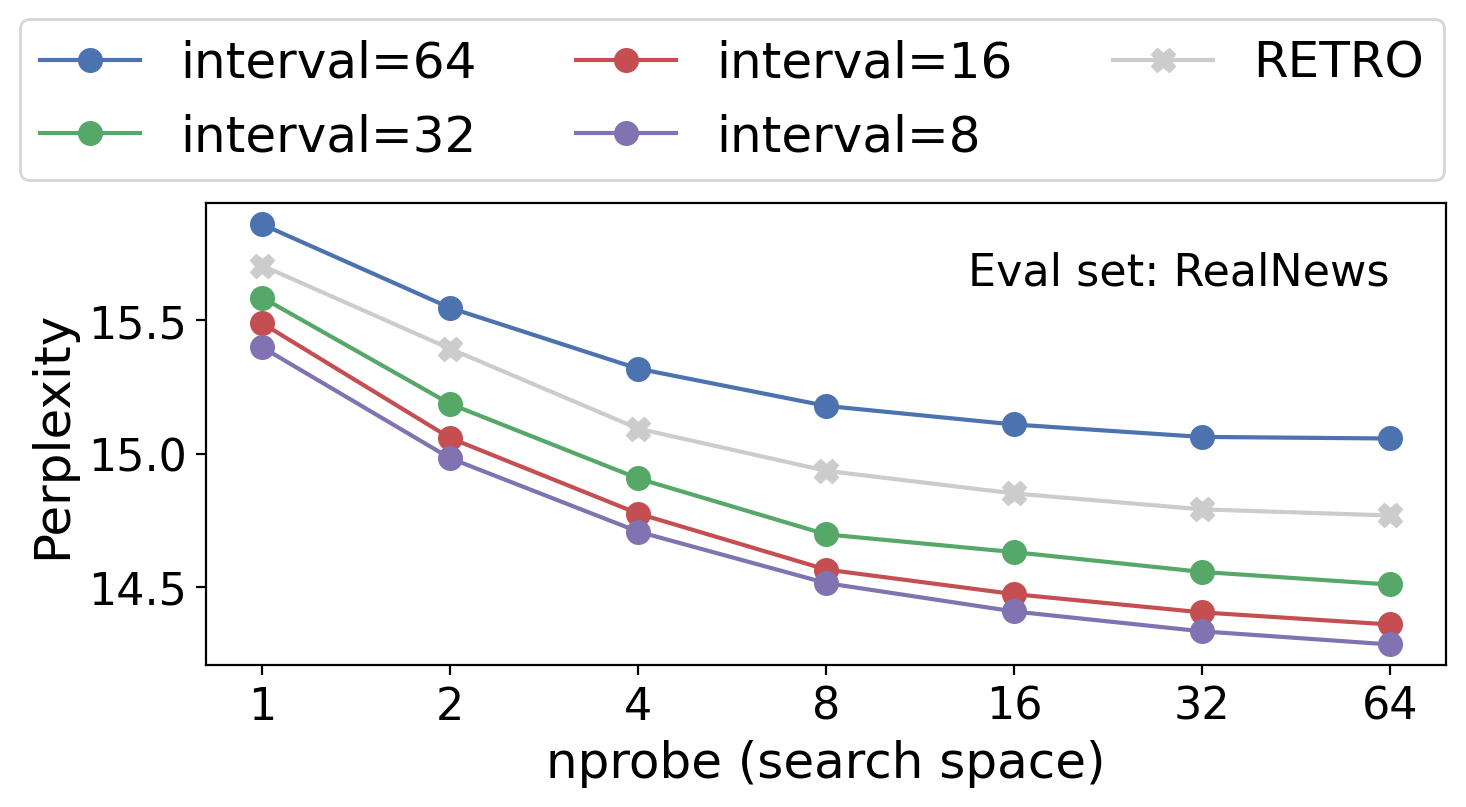}
\end{subfigure}

\begin{subfigure}[b]{0.55\linewidth}
    \centering
    \includegraphics[width=\linewidth]{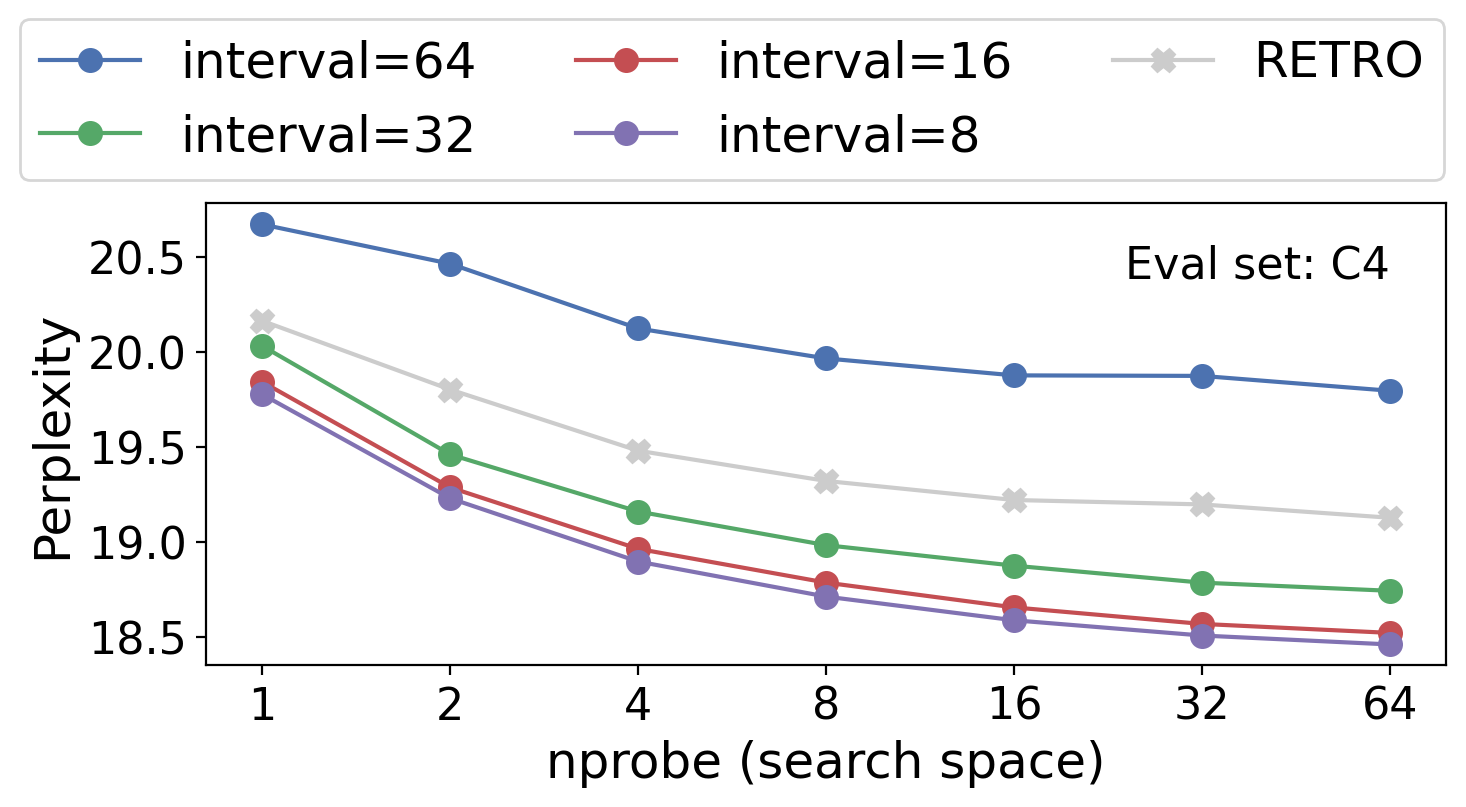}
\end{subfigure}

  \caption{Perplexity of RAG when applying various retrieval intervals and search space configurations (\( nprobe \)).}
  \label{fig_piperag:eval_ppl}
\end{figure*}

Figure~\ref{fig_piperag:eval_dbsize} shows the impact of various retrieval strategies across different database sizes. This comparison includes PipeRAG, \textsc{Retro}, retrieval-augmented generation with only one retrieval at the beginning of generation, and generation without retrieval. For the last two strategies, \textsc{Retro} still serves as the base model. 
As indicated in the figure, retrieval, especially on large databases, plays a crucial role in improving generation quality (lower perplexity is better). Across all evaluated datasets, generation without retrieval performs the worst, followed by only retrieving once, showing the effectiveness of periodic retrieval in \textsc{Retro}. Additionally, perplexity decreases as the dataset size increases, highlighting the importance of comprehensive content coverage in the databases. Notably, when pairing with the largest database, PipeRAG outperforms \textsc{Retro} in generation quality, as I will analyze in greater detail later on.


\textit{From now on, I report results in generation quality and performance based on the full (largest) database}, as using subsets significantly compromises generation quality.

Figure~\ref{fig_piperag:eval_ppl} compares the perplexity between PipeRAG and \textsc{Retro} across various retrieval configurations. I assess PipeRAG with different retrieval intervals, setting the search space through \( nprobe \), which represents the number of scanned vector lists per query in the IVF index. As shown in Figure~\ref{fig_piperag:eval_ppl}, both PipeRAG and \textsc{Retro} show reduced perplexity with an expanded search space, which leads to better search quality (O3).

\begin{tcolorbox}[
    enhanced,
    arc=2mm, 
    outer arc=2mm, 
    boxrule=0.8pt, 
    colframe=black, 
    colback=white, 
    boxsep=0pt, 
    drop shadow southeast, 
]

\textbf{Takeaway 1:} The quality of retrieval-augmented generation benefits from higher retrieval quality achieved by expanding the search space during vector search.
\end{tcolorbox}

\begin{figure*}[t]

\centering    
\begin{subfigure}[b]{0.55\linewidth}
    \centering
    \includegraphics[width=\linewidth]{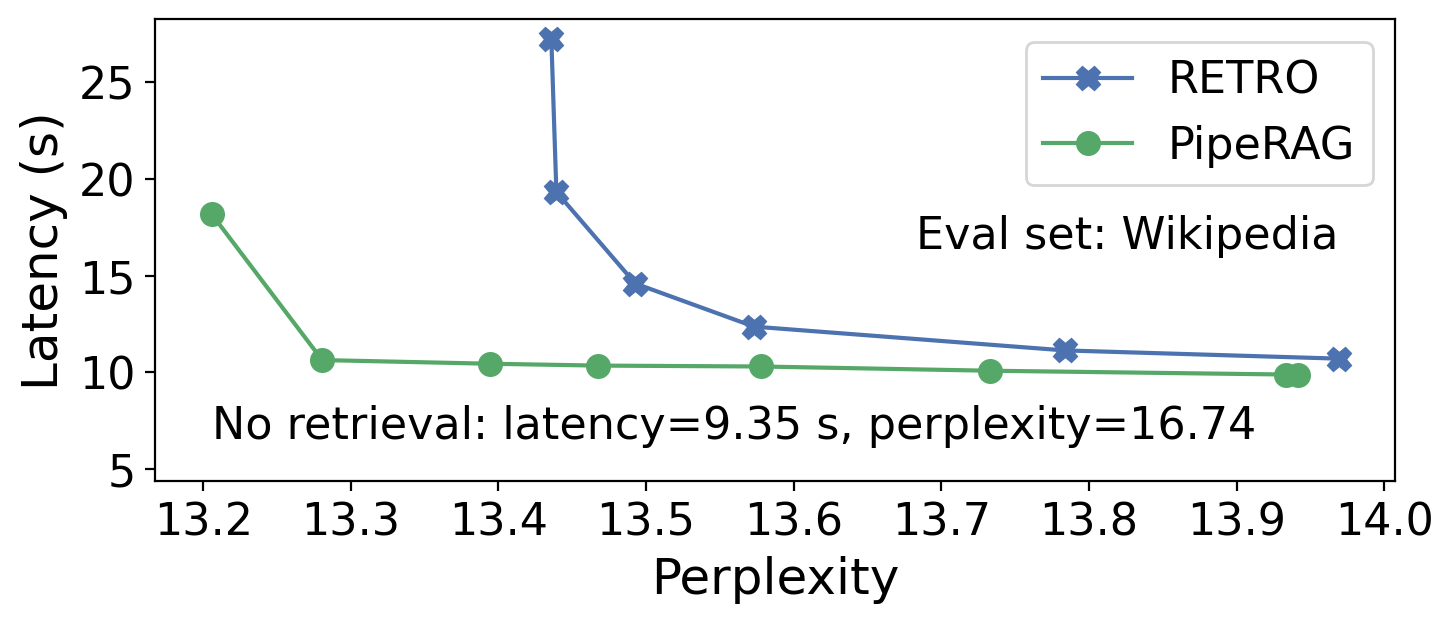}
\end{subfigure}

\begin{subfigure}[b]{0.55\linewidth}
    \centering
    \includegraphics[width=\linewidth]{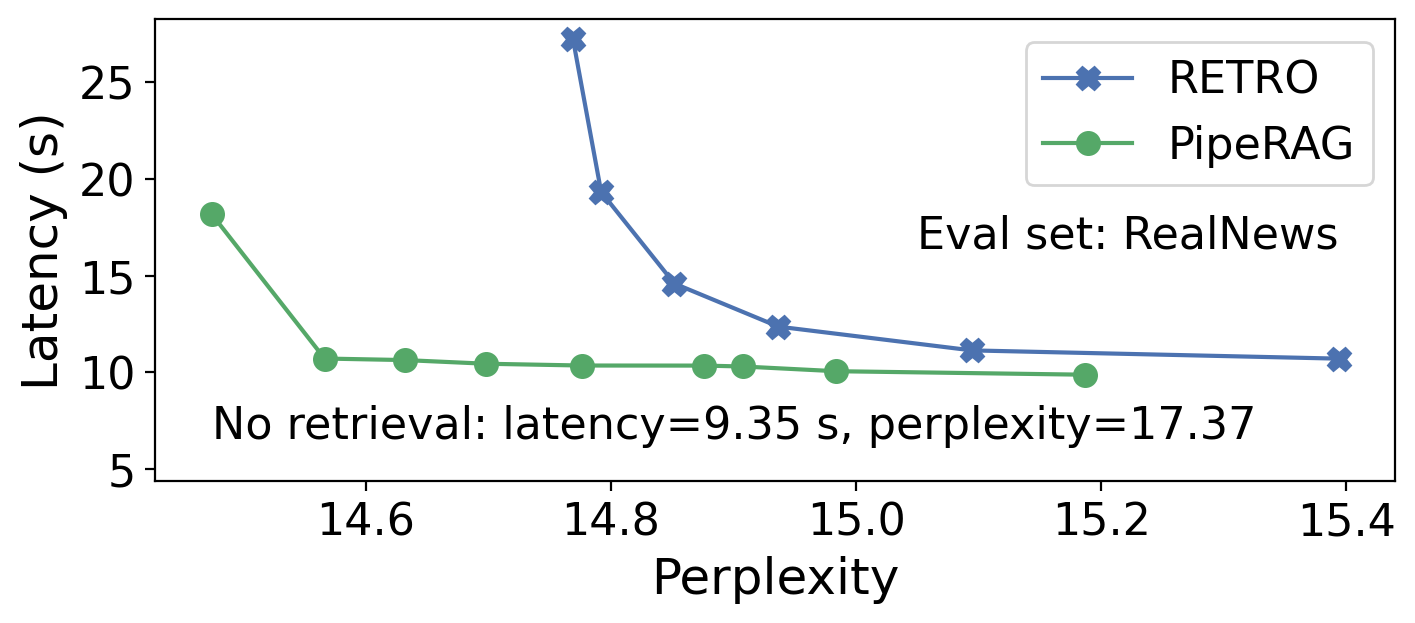}
\end{subfigure}

\begin{subfigure}[b]{0.55\linewidth}
    \centering
    \includegraphics[width=\linewidth]{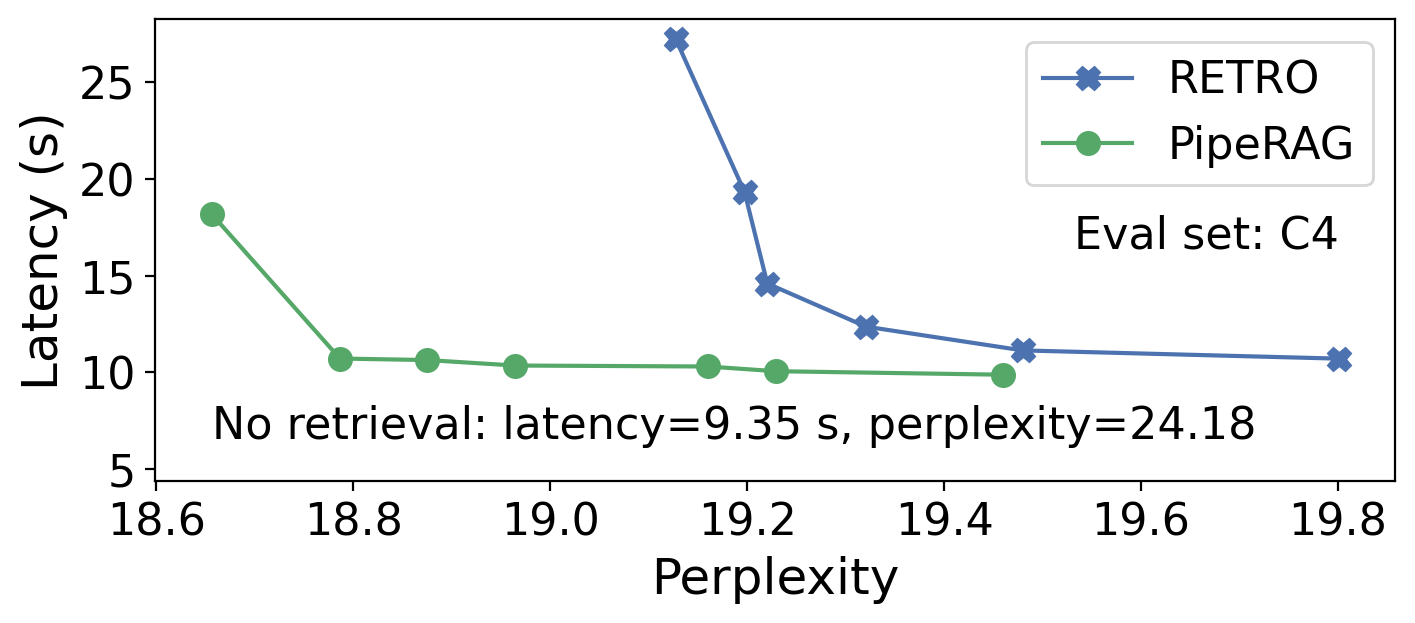}
\end{subfigure}

  \caption{PipeRAG significantly outperforms \textsc{Retro} on latency-perplexity trade-offs (lower latency and perplexity are better).}
  \label{fig_piperag:eval_e2e}
\end{figure*}

\begin{table*}
\centering 

\caption{Performance-driven retrieval (S3) facilitates latency comparable to non-retrieval models while reducing perplexity. Values in parentheses indicate the difference to the baseline model without retrieval (lower latency and perplexity are better).} 

\scalebox{0.8}{
\begin{tabular}{
L{4em} 
M{5em} M{6em} M{6em} 
M{0em}
M{5em} M{6em} M{6em} 
}\toprule
\multirow{2}{*}{Eval Set} & \multicolumn{3}{c}{Latency (s)} & \phantom{}& \multicolumn{3}{c}{Perplexity} \\
\cmidrule{2-4} \cmidrule{6-8}
 & No retrieval & RETRO & PipeRAG (S3) &\phantom{}& No retrieval & RETRO & PipeRAG (S3) \\
\midrule
Wikipedia & 9.35 & 14.59 (+5.23) & 10.34 (\textbf{+0.99})  && 16.74 & 13.49 (-3.25) & 13.47 (\textbf{-3.28})    \\ 
RealNews & 9.35 & 12.36 (+3.00) & 10.58 (\textbf{+1.22})  && 17.37 & 14.94 (-2.43) & 14.87 (\textbf{-2.50})   \\ 
C4 & 9.35 & 11.13 (+1.78) & 10.58 (\textbf{+1.22})  && 24.18 & 19.48 (-4.70) & 19.36 (\textbf{-4.82})   \\ 

\bottomrule 
\end{tabular} }

\label{tab_piperag:dynamic_nprobe} 

\end{table*}

Furthermore, PipeRAG demonstrates superior generation quality over \textsc{Retro}, particularly when using shorter retrieval intervals of no more than 32 (Figure~\ref{fig_piperag:eval_ppl}). This advantage is attributed to PipeRAG's revised attention mechanism. Shorter intervals not only reduce query staleness (equivalent to the interval) but improve the content integration frequency, in contrast to \textsc{Retro} with a fixed interval of 64. The increased retrieval frequency in PipeRAG does not necessarily add to generation latency thanks to the pipeline parallelism, a point I will further elaborate on.

\begin{tcolorbox}[
    enhanced,
    arc=2mm, 
    outer arc=2mm, 
    boxrule=0.8pt, 
    colframe=black, 
    colback=white, 
    boxsep=0pt, 
    drop shadow southeast, 
]

\textbf{Takeaway 2:} PipeRAG can surpass \textsc{Retro} in generation quality when using shorter retrieval intervals backed by PipeRAG's attention mechanism.
\end{tcolorbox}

\subsection{Performance-Quality Pareto Frontier}

In this section, I assess the efficiency of PipeRAG. The primary performance metric is the end-to-end latency to generate a 1024-token sequence, which I reported by taking the median latency of five individual runs. 


Figure~\ref{fig_piperag:eval_e2e} compares the Pareto frontiers of the performance-quality (latency-perplexity) trade-offs between PipeRAG and \textsc{Retro}.
For \textsc{Retro}, I manipulate the search space by tuning \( nprobe \).
For PipeRAG, I explore a range of retrieval intervals in conjunction with either a fixed search space or the performance-model-driven search space selection (S3).  
Across all datasets, the Pareto frontier of PipeRAG demonstrates significant advantages over \textsc{Retro}, as shown in Figure~\ref{fig_piperag:eval_e2e}. For example, PipeRAG can attain up to a 2.6$\times$ reduction in latency while maintaining or reducing perplexity relative to \textsc{Retro}; alternatively, under the same latency constraint, PipeRAG can achieve lower perplexity of as much as 0.93 points compared to \textsc{Retro}.

\begin{tcolorbox}[
    enhanced,
    arc=2mm, 
    outer arc=2mm, 
    boxrule=0.8pt, 
    colframe=black, 
    colback=white, 
    boxsep=0pt, 
    drop shadow southeast, 
]
\textbf{Takeaway 3:} PipeRAG shows impressive efficiency, achieving up to 2.6$\times$ speedup in latency over \textsc{Retro} without compromising generation quality.
\end{tcolorbox}

Table~\ref{tab_piperag:dynamic_nprobe} demonstrates the effectiveness of the proposed performance-model-driven retrieval system. The objective of the performance model is to dynamically maximize search quality while minimizing additional performance costs.
To evaluate this, I compare the generation latency and quality of PipeRAG applying performance-model-driven retrievals to that of \textsc{Retro} as well as the same base \textsc{Retro} model without invoking retrievals.
As shown in Table~\ref{tab_piperag:dynamic_nprobe}, PipeRAG achieves a notable reduction in perplexity (2.50$\sim$4.82) with a minor increase in performance overhead (merely 10.6\%$\sim$13.2\% in latency overhead), outperformance \textsc{Retro} in both latency and perplexity. 
This slight increase in latency is attributed to the extra computational workload of the cross-attention mechanism when integrating the retrieved content from the encoder. 


\begin{tcolorbox}[
    enhanced,
    arc=2mm, 
    outer arc=2mm, 
    boxrule=0.8pt, 
    colframe=black, 
    colback=white, 
    boxsep=0pt, 
    drop shadow southeast, 
]

\textbf{Takeaway 4:} 
Leveraging performance-model-driven retrievals, PipeRAG can achieve comparable latency to models without retrievals while significantly improving generation quality. 
\end{tcolorbox}

While the model checkpoint we used~\cite{norlund2023generalization} was relatively small and was not fine-tuned on QA datasets, we still conducted QA experiments to show the effectiveness of PipeRAG over the baseline.
Specifically, we conducted QA experiments on the open-domain version of the Natural Questions dataset. However, the ground truth answers are typically short (less than five tokens). In order to compare not only the QA quality but also the generation latency between PipeRAG and the baseline, we extended the model's output to sequences of 256 tokens, involving multiple retrievals. We evaluated several configurations: (1) no retrieval, which resulted in low latency, (2) RETRO with retrieval, which showed high latency, and (3) PipeRAG with retrieval, achieving low latency. For PipeRAG, the staleness was set as 64 tokens. 

\begin{table}[h]
\centering
\caption{Summary of recall and latency for different retrieval settings in QA experiments.}
\label{tab_piperag:qa_experiments}
\scalebox{0.9}{
\begin{tabular}{lcc}
\toprule
Setting & Average Recall & Latency (ms) \\
\midrule
No retrieval & 0.098 & 1859.2 \\
RETRO & 0.150 & 3237.16 \\
PipeRAG & 0.148 & 1920.6 \\
\bottomrule
\end{tabular}
}
\end{table}

Table~\ref{tab_piperag:qa_experiments} summarizes the recall and latency results. Retrieval-augmented settings, such as those employed by PipeRAG and the baseline, demonstrated significant improvements in generation quality compared to the no-retrieval configuration. Notably, even with a staleness of 64 tokens, PipeRAG achieved recall comparable to RETRO while delivering 1.64$\times$ lower latency.

\subsection{Serving Performance on Various Hardware}

PipeRAG is versatile across various deployment scenarios, which may involve a wide range of models, database scales, search algorithms, retrieval intervals, and hardware configurations. This versatility is essential, as it allows PipeRAG to adapt to environments where the latency balance between retrieval and generation may differ significantly. 

To estimate PipeRAG's efficiency on various hardware, I model its performance using hypothetical hardware with enhanced inference and/or retrieval performance. I enable this by scaling the current latencies meansured in Table~\ref{tab_piperag:retrieval_latency_wide} and \ref{tab_piperag:generation_latency_wide}, which show the latencies for retrievals (with different search spaces indicated by $nprobe$, the number of lists to scan per query) and generation (where per-token inference latency increases with the number of tokens generated).

\begin{table}[t]
\centering
\caption{Average retrieval latency in milliseconds with varying search spaces.}
\label{tab_piperag:retrieval_latency_wide}
\scalebox{0.9}{
\begin{tabular}{lcccccccc}
\toprule
$nprobe$ & 1 & 2 & 4 & 8 & 16 & 32 & 64 & 128 \\
\midrule
Latency (ms) & 26.06 & 48.12 & 92.59 & 178.43 & 344.49 & 666.43 & 1298.52 & 2544.44 \\
\bottomrule
\end{tabular}
}
\end{table}

\begin{table}[t]
\centering
\caption{Average generation latency per token as the sequence length (number of tokens) increases, with merged intervals.}
\label{tab_piperag:generation_latency_wide}
\scalebox{0.9}{
\begin{tabular}{lcccc}
\toprule
Token ID & $0 \sim 127$ & $128 \sim 255$ & $256 \sim 383$ & $384 \sim 511$ \\
\midrule
Latency (ms) & 7.24 & 7.29 & 7.71 & 8.34 \\
\midrule
Token ID & $512 \sim 639$ & $640 \sim 767$ & $768 \sim 895$ & $896 \sim 1023$ \\
\midrule
Latency (ms) & 9.31 & 10.47 & 11.84 & 13.19 \\
\bottomrule
\end{tabular}
}
\end{table}


For \textsc{Retro}, the end-to-end latency is the sum of inference and retrieval time. In PipeRAG, due to the parallelism, the latency for generating a chunk of tokens is determined by the maximum value of the inference and retrieval latency of that chunk, except for the first chunk where the pipeline is not yet active (see Figure~\ref{fig_piperag:overview}).

I then input the measured performance of inference and retrievals into the performance model. This allows us to simulate performance scaling, such as a 4$\times$ improvement in retrieval or a 16$\times$ enhancement in inference. The result generation latency as well as the respective conclusions are included in Section~\ref{sec_piperag:evaluation}. The model's accuracy is then verified by comparing these projected results against actual experimental data, with deviations found to be within a reasonable range (the median difference is only 5.7\%).



\begin{figure}
	\centering
  \includegraphics[width=0.8\linewidth]{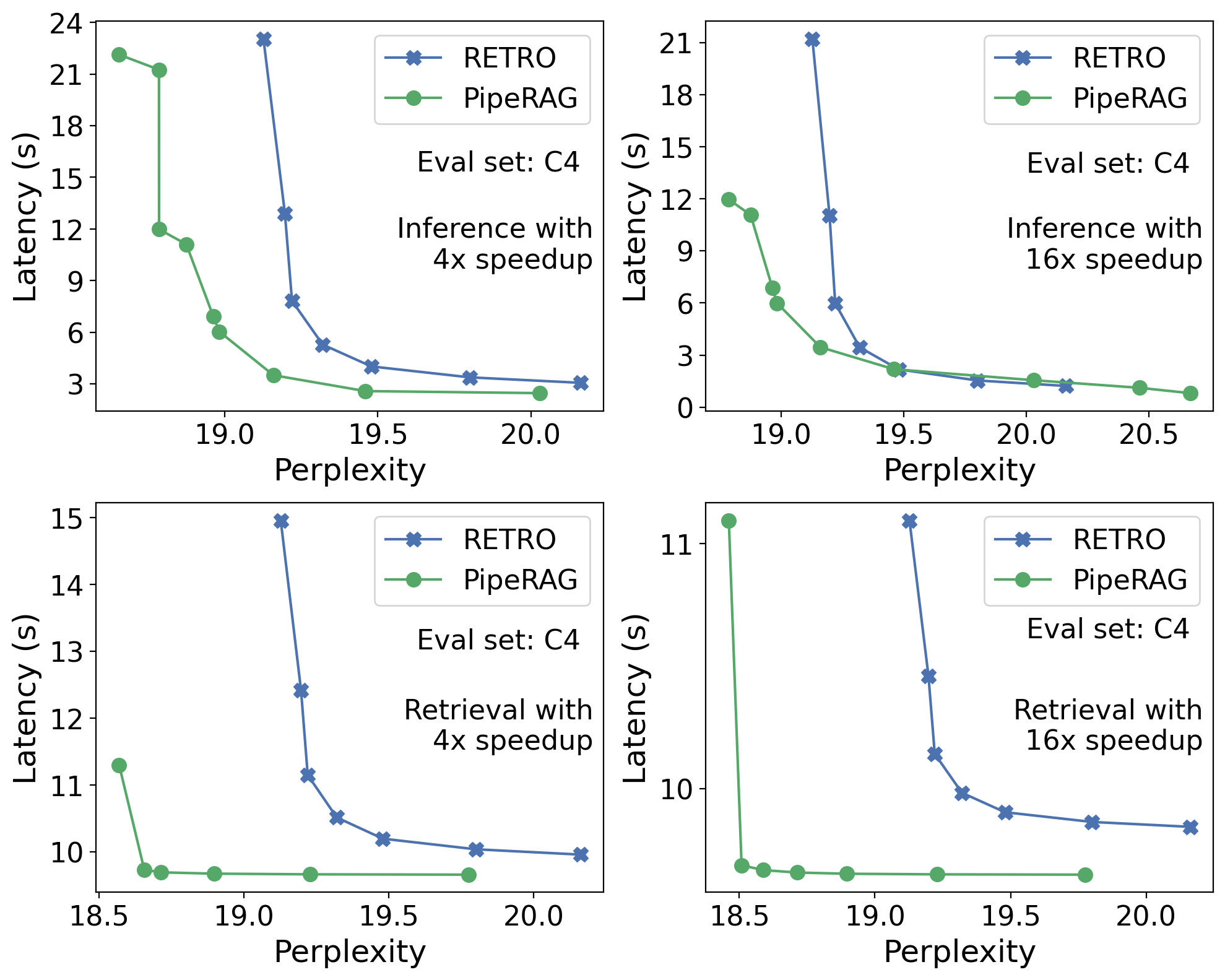}
  \caption{Trends in PipeRAG efficiency when deployed on future hardware that enables faster retrieval or inference.}
  \label{fig_piperag:different_performance}
\end{figure}

\begin{table*}[t]
\centering 

\caption{Cosine similarity between content retrieved by stale and non-stale queries. The results indicate that stale queries are still highly effective in identifying relevant token chunks from the database.} 

\scalebox{0.8}{
\begin{tabular}{
L{4em} M{5em} M{0em}
M{4.1em} M{4.1em} M{4.1em} M{4.1em} M{4.1em} M{4.1em} M{4.1em} 
}\toprule
& \multirow{2}{*}{No staleness} & \phantom{}& \multicolumn{7}{c}{Staleness (number of stale tokens in the query)} \\
 \cmidrule{4-10}
 &&& 1 & 2 & 4 & 8 & 16 & 32 & 64 \\
\midrule
Wikipedia & 1.0000 && 0.9262 & 0.9204 & 0.9138 & 0.9062 & 0.8990 & 0.8921 & 0.8875 \\
RealNews & 1.0000 && 0.9219 & 0.9147 & 0.9073 & 0.8996 & 0.8925 & 0.8850 & 0.8794 \\
C4 & 1.0000 && 0.9323 & 0.9263 & 0.9193 & 0.9127 & 0.9052 & 0.8980 & 0.8929 \\

\bottomrule 
\end{tabular} }
\label{tab_piperag:stale_query_results_similarity}
\end{table*}

\begin{figure*}[t]

\centering    
\begin{subfigure}[b]{0.5\linewidth}
    \centering
    \includegraphics[width=\linewidth]{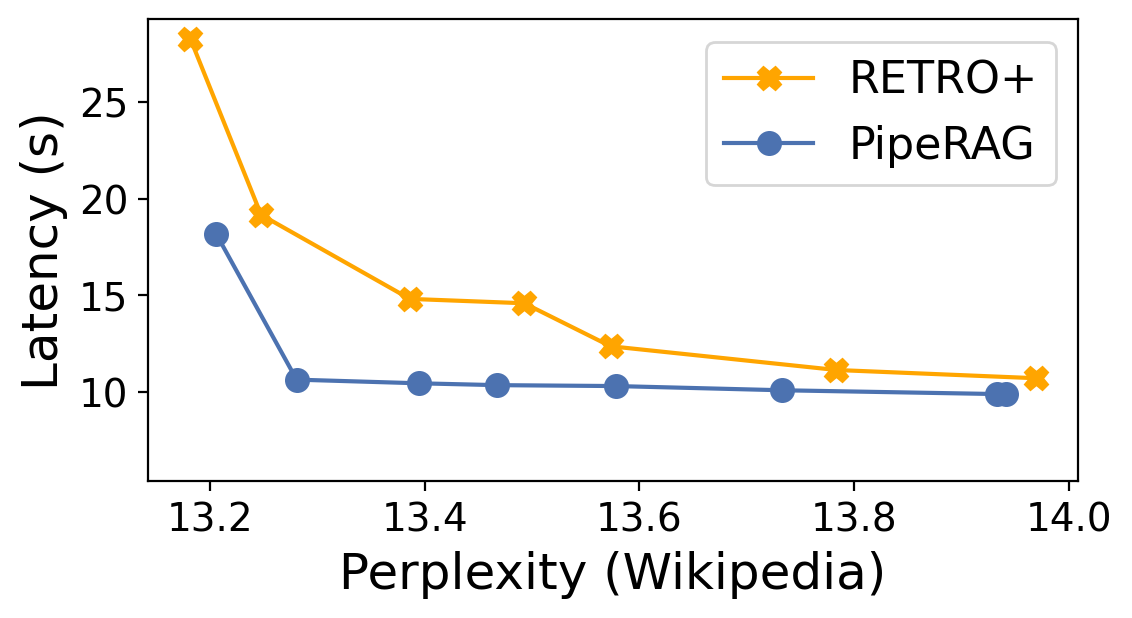}
\end{subfigure}

\begin{subfigure}[b]{0.5\linewidth}
    \centering
    \includegraphics[width=\linewidth]{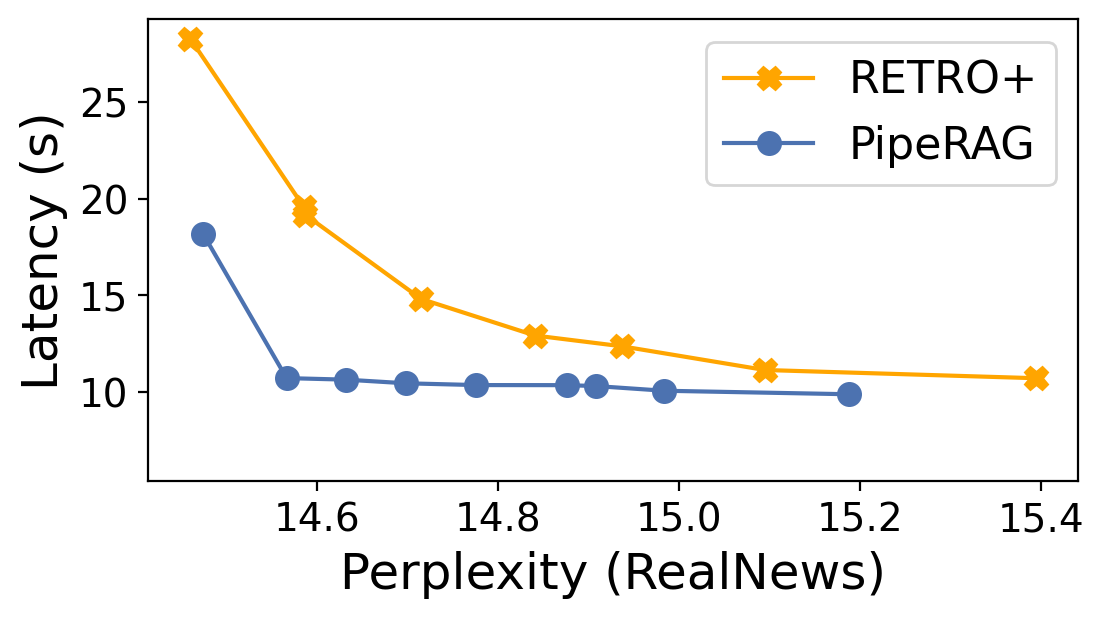}
\end{subfigure}

\begin{subfigure}[b]{0.5\linewidth}
    \centering
    \includegraphics[width=\linewidth]{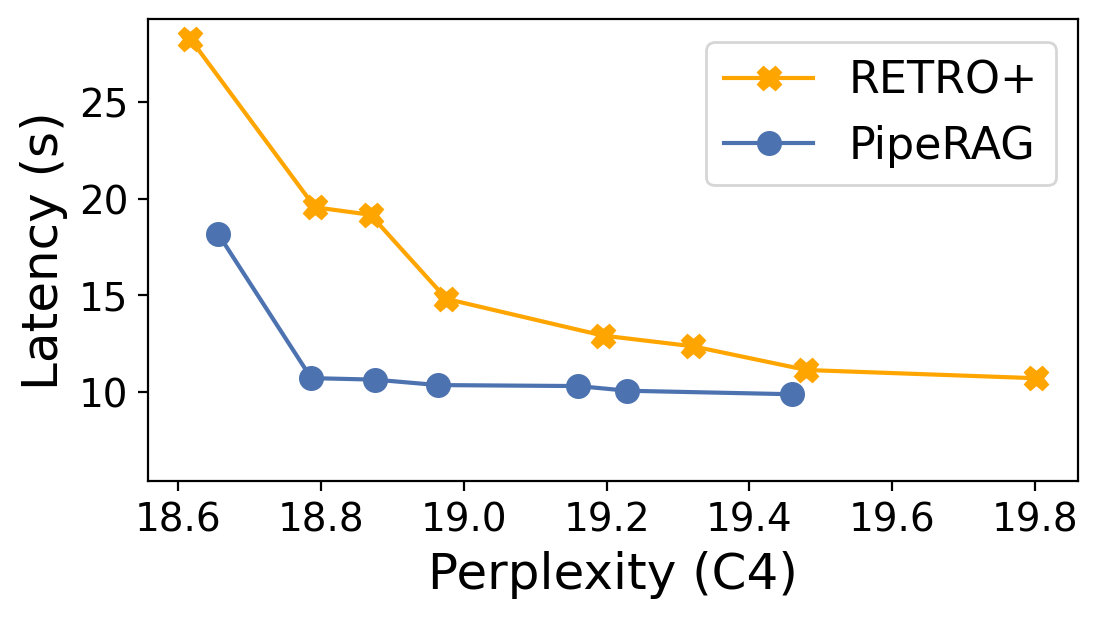}
\end{subfigure}

  \caption{Even if the baseline model supports flexible retrieval intervals (\textsc{Retro+}), PipeRAG still significantly outperforms it in efficiency thanks to the proposed pipeline parallelism.}
  \label{fig_piperag:eval_e2e_retro_flexible_interval}
\end{figure*}

Figure~\ref{fig_piperag:different_performance} illustrates the projected performance trends of PipeRAG across a range of system and hardware configurations. Considering the rapid advancements in hardware accelerators, I expect shifts in performance of both retrieval and inference systems over years. To analyze PipeRAG's effectiveness on future hardware, I model the latency of PipeRAG and \textsc{Retro} when using faster retrieval or inference systems. 
The first row of Figure~\ref{fig_piperag:different_performance} demonstrates the generation latency when the inference system becomes 4$\times$ and 16$\times$ faster, while the second row examines the effects of accelerated retrieval. Across all scenarios, PipeRAG achieves superior efficiency compared to \textsc{Retro}. When either system experiences an order of magnitude speedup (e.g., 16$\times$), however, the benefits of applying PipeRAG become less significant. This trend aligns with my expectations, as the effectiveness of pipeline parallelism peaks when both system components have comparable latencies and diminishes when one component significantly outpaces the other.

\begin{tcolorbox}[
    enhanced,
    arc=2mm, 
    outer arc=2mm, 
    boxrule=0.8pt, 
    colframe=black, 
    colback=white, 
    boxsep=0pt, 
    drop shadow southeast, 
]

\textbf{Takeaway 5:} PipeRAG outperforms \textsc{Retro} in efficiency across different hardware, though the extent of improvements depends on sub-system performance.
\end{tcolorbox}

\subsection{Ablation Study}

\textbf{The effectiveness of retrievals using stale queries.}
I investigate the fundamental applicability of prefetching content using stale queries. For this purpose, I compare the \( k=1 \) nearest neighbors retrieved by non-stale queries in the evaluation set with their staleness versions. Same as Section~\ref{sec_piperag:evaluation}, I use the largest C4 database, which consists of three billion token chunks, and set \( nprobe=64 \) to ensure high retrieval quality. I then employ the \textit{msmarco-bert-base-dot-v5} checkpoint from sentence transformers~\cite{reimers2019sentence} to evaluate the cosine similarity between contents retrieved by stale and non-stale queries.

Table~\ref{tab_piperag:stale_query_results_similarity} presents the retrieval quality using stale queries. Here, I use different degrees of staleness, ranging from 1 token to 64 tokens, while maintaining a consistent retrieval interval of \( m=64 \). The results indicate that, despite the staleness, the retrieved content closely resembles that obtained through non-stale queries, with around 90\% cosine similarity across datasets. As expected, this similarity shows a gradual decline as the staleness increases.

\textbf{Enable flexible retrieval interval in \textsc{Retro} baseline.}
Since PipeRAG not only introduces pipeline parallelism but also modifies \textsc{Retro}'s attention mechanism to maximize the effectiveness of pipelining, it is natural to ask how a baseline model would perform if it integrates the same attention mechanism. To illustrate the effectiveness of pipeline parallelism itself, I compare PipeRAG with an enhanced variant of \textsc{Retro}, named \textsc{Retro+}, which also supports flexible retrieval intervals by integrating PipeRAG's attention mechanism.

Figure~\ref{fig_piperag:eval_e2e_retro_flexible_interval} compares the performance-quality Pareto-frontier between PipeRAG and \textsc{Retro+}. 
Both models use retrieval intervals ranging from 8 to 64.
While \textsc{Retro+}, benefiting from flexible intervals, matches PipeRAG in perplexity, PipeRAG consistently achieves lower latency given the same perplexity. This is attributed to the proposed pipeline parallelism: PipeRAG effectively hides the retrieval latencies by overlapping them with generation latencies, whereas for \textsc{Retro+}, more frequent retrievals lead to increased total generation latency. 


\begin{tcolorbox}[
    enhanced,
    arc=2mm, 
    outer arc=2mm, 
    boxrule=0.8pt, 
    colframe=black, 
    colback=white, 
    boxsep=0pt, 
    drop shadow southeast, 
]

\textbf{Takeaway 6:} Pipeline parallelism is essential for RAG efficiency, as PipeRAG outperforms \textsc{Retro+} that supports flexible retrieval intervals using PipeRAG's attention mechanism.
\end{tcolorbox}

\section{Discussion}
\label{sec_piperag:discussion}

\subsection{Broader Applicability of PipeRAG}

\textit{The idea of improving RAG efficiency through pipeline parallelism is broadly applicable across various RAG configurations, as long as they include periodic retrievals.} 
In this chapter, I have focused on improving RAG efficiency based on the \textsc{Retro} model and evaluated generation performance using specific hardware and software setups described in Section~\ref{sec_piperag:evaluation}. 
In the future, RAG can evolve in several ways: models may adopt a decoder-only transformer architecture~\cite{radford2018improving, brown2020language} although the high cost of periodically appending the retrieved content has to be addressed~\cite{ram2023context, jiang2023active}; retrieval engines could incorporate LLM-based or BM25-based result reranking~\cite{nogueira2019passage, macavaney2019cedr, doostmohammadi2023surface}, instead of solely relying on vector-level similarity; and hardware may evolve to include dedicated retrieval accelerators~\cite{jiang2023co, jiang2023chameleon, jiang2024accelerating}. 
However, regardless of these potential advancements in algorithms and hardware, the dependencies between retrievals and inferences in RAG systems --- especially when retrievals are periodic --- remains a \textit{fundamental} obstacle to fully leveraging hardware resources and achieving maximal inference efficiency. Thus, whenever the time consumption of one retrieval and multiple steps of inferences are on a similar scale, pipeline parallelism by \textit{prefetching} content from databases, which can also be combined with retrieval caches~\cite{jin2024ragcache, zhang2024accelerating}, should be a great option to improve generation efficiency.

\textit{Prefetching content from databases using stale queries is applicable regardless of the specific models used for generation.} To demonstrate this, I show that using a stale query window can retrieve content very similar to that obtained via a regular query window. 
These findings address a potential limitation in the evaluation, as the experimentation with PipeRAG was conducted using the \textsc{Retro} checkpoint provided by~\cite{norlund2023generalization}, which was the only available \textsc{Retro} checkpoint at the time of this research project.

\subsection{Factors Influencing Retrieval and Inference Performance}

\textbf{Retrieval performance} depends on the following factors:

\begin{itemize}
    \item \textbf{Hardware.} The memory bandwidth and computational capacity of the hardware used for retrieval are key factors influencing performance. It is worth noticing that there are emerging hardware accelerators that are specialized for retrievals~\cite{jiang2023co} and integrated into RAG systems~\cite{jiang2023chameleon}, offering impressive retrieval performance as well as cost efficiency.
    \item \textbf{Document numbers.} The total number of documents, along with encoding granularity as introduced below, determines the vector count in the database.
    \item \textbf{Encoding granularity.} Documents can be encoded in various granularities by LLMs, ranging from one vector per document~\cite{huang2013learning, karpukhin2020dense} to one vector per passage~\cite{dai2019deeper, reimers2019sentence} or even per token~\cite{khattab2020colbert, santhanam2021colbertv2}.
    \item \textbf{Dimensionality.} The dimensionality of the database vectors, as well as the compression ratio when employing product quantization, are critical to retrieval performance.
    \item \textbf{Indexes.} The selection of indexes, such as IVF or graph-based ones, and their parameter configurations are crucial for retrieval efficiency.
    \item \textbf{Reranking.} Optionally, the retrieved content can be reranked using LLMs, which often yields better ranking quality than relying solely on vector similarity~\cite{nogueira2019passage}.
\end{itemize}

\textbf{LLM inference performance} is influenced by the following factors:

\begin{itemize}
    \item \textbf{Hardware.} The performance of inference is heavily dependent on the hardware, particularly its memory bandwidth and computational capacity. LLM accelerators such as GPUs are evolving rapidly in these metrics. 
    \item \textbf{Software.} The choice of software for inference also plays a significant role. For instance, PyTorch's eager execution mode might not fully exploit hardware accelerators due to the slow execution speed of Python programs. In such cases, software overhead could exceed the GPU kernel execution time.
    \item \textbf{Quantization}. Quantizing models to lower precisions can markedly reduce inference time, thanks to reduced memory footprint and bandwidth usage. For instance, converting models to 3-bit precision can lead to a 3$\sim$5$\times$ speedup compared to 16-bit floating point formats~\cite{frantar2022gptq}.
    \item \textbf{Sparsity}. Techniques like mixture-of-experts allow for scaling LLMs without proportionate increases in computational costs~\cite{fedus2022switch, du2022glam}, because only a small subset of neurons are activated during inference. 
\end{itemize}

\section{Related Work}

PipeRAG is a pioneer work to enhance RAG efficiency through an in-depth algorithm-system co-design, diverging from existing RAG research that mainly focuses on improving generation quality. We now briefly introduce these related works.

Since knowledge is primarily retrieved rather than encoded in the LLM's parameters, RALMs, even with LLMs of one to two orders of magnitude fewer parameters, can achieve superior or comparable performance to conventional LLMs on various natural language processing (NLP) tasks~\cite{lewis2020pre, izacard2022few, komeili2021internet, guu2020retrieval}. While the generation may only involve a single passage retrieval at the beginning~\cite{lewis2020retrieval, izacard2020leveraging, sachan2021end}, the generated sequence may gradually diverge from the initially retrieved contents as the sequence grows longer. Thus, a more general RAG approach involves multiple retrievals during text generation to improve token generation quality~\cite{ram2023context, borgeaud2022improving}.

Another line of RAG research emphasizes token-level retrievals, exemplified by kNN-LM~\cite{khandelwal2019generalization} and subsequent works~\cite{khandelwal2020nearest, meng2021fast, xu2023nearest}. In these models, during each token generation step, the hidden state of the last layer is used as a query to retrieve contextually similar passages as well as their subsequent tokens (with a retrieval interval of one). The next token of the current context is then predicted by interpolating the model's next-token probability distribution with that of the retrieved contents. There are also arguments suggesting that token-level content integration may not be as effective as integrating longer passages~\cite{wang2023knn}.
\section{Conclusion}
\label{sec_piperag:conclusion}

Iterative retrieval-augmented generation presents both opportunities and efficiency challenges, due to the overheads of retrieval on large databases. 
I propose PipeRAG, which improves generation efficiency by adopting pipeline parallelism, allowing flexible retrieval intervals, and dynamically adjusting retrieval quality via performance modeling. PipeRAG achieves up to 2.6$\times$ speedup over \textsc{Retro} without compromising generation quality. This not only establishes a solid foundation for integrating pipeline parallelism in future RAG systems but also showcasing future research opportunities in optimizing RAG through algorithm-system co-design.

\part{Algorithm-Hardware Co-Design for Vector Search}
\label{part:retrieval}

\chapter{FANNS: Accelerating Quantization-Based Vector Search}
\label{chap:fanns}

This chapter introduces hardware specialization for vector search, which serves as the foundation of the Chameleon project presented in Chapter~\ref{chap:chameleon}.  
In particular, this chapter focuses on algorithm-hardware co-design for product-quantization-based algorithms, a widely used vector search paradigm introduced in Chapter~\ref{chap:prelim:retrieval}.  
The next chapter will shift focus to graph-based retrieval, another major category of vector search algorithms.

\section{Introduction}
\textbf{Target algorithm: IVF-PQ.}
The algorithm uses an inverted file (IVF) index to group vectors into many vector lists by clustering. It then applies product quantization (PQ) to compress high-dimensional vectors into a series of byte codes, reducing memory consumption and accelerating the similarity evaluation process. 
When a query arrives, IVF-PQ goes through six search stages to retrieve similar vectors. The main stages include comparing the vector with all the vector list centroids to identify a subset of relevant lists, scanning the quantized vectors within the selected lists, and collecting the topK most similar vectors. 


\textbf{The benefit of hardware-algorithm co-design.} 
Maximizing the performance of an IVF-PQ accelerator is challenging because one needs to carefully balance the design choices of both the hardware and the algorithm.
Given a target chip size, there are many valid designs to implement IVF-PQ: how should we choose the appropriate microarchitecture for each of the six IVF-PQ search stages? How should we allocate the limited hardware resources to the six stages? 
From the algorithm's perspective, the multiple parameters in IVF-PQ can significantly influence performance bottlenecks and recall.
Due to the vast design space, hardware specialization tailored to a specific set of algorithm parameters can achieve better performance-recall trade-offs: as I will show in the experiments, the accelerators without algorithm parameter awareness are 1.3$\sim$23.0$\times$ slower than the co-designed accelerators given the same recall requirement.

\begin{figure}[t]

	\centering
  \includegraphics[width=0.65\linewidth]{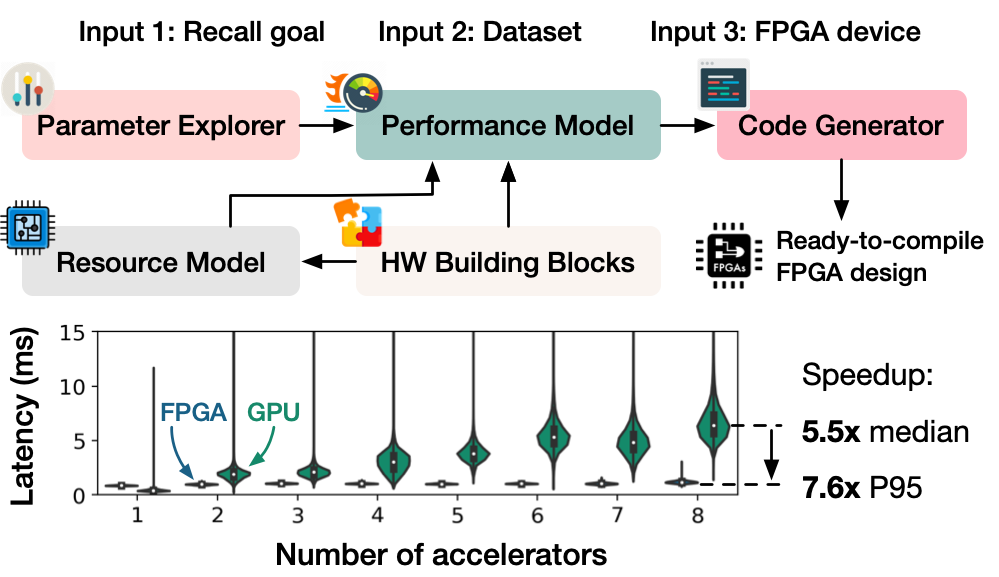}
  \caption{{By co-designing hardware and algorithm, \textit{FANNS} significantly outperforms GPUs in scale-out vector search.}}
  \label{fig_fanns:first_page_overview}
\end{figure}

\textbf{Proposed solution.} 
Considering the numerous design possibilities for an IVF-PQ accelerator, I exploit the reconfigurability of FPGAs to examine various design points. \textbf{I propose \textit{FANNS} (FPGA-accelerated Approximate Nearest Neighbor Search), an end-to-end accelerator generation framework for IVF-PQ-based vector search, which automatically co-designs algorithm and hardware to maximize the accelerator performance for target datasets and deployment recall requirements.} 
Figure~\ref{fig_fanns:first_page_overview} illustrates the FANNS workflow. 
Provided with a dataset, a deployment recall requirement, and a target FPGA device, \textit{FANNS} automatically (a) identifies the optimal combination of parameter settings and hardware design and (b) generates a ready-to-deploy accelerator. 
Specifically, \textit{FANNS} first evaluates the relationship between IVF-PQ parameters and recall on the given dataset. 
It also lists all valid accelerator designs given the FPGA hardware resource budget.
Then, the \textit{FANNS} performance model predicts the queries-per-second (QPS) throughput of all combinations between algorithm parameters and hardware designs.
Finally, using the best combination determined by the performance model, the \textit{FANNS} code generator creates the corresponding FPGA code, which is compiled into an FPGA bitstream. 
Besides a single-accelerator solution, \textit{FANNS} can also scale out by instantiating a hardware TCP/IP stack in the accelerator.

\textbf{Results.} 
Experiments conducted on various datasets demonstrate the effectiveness of the hardware-algorithm co-design: the accelerators generated by \textit{FANNS} achieve up to 23.0$\times$ speedup over fixed FPGA designs and up to 37.2$\times$ speedup compared to a Xeon CPU. While a GPU may outperform an FPGA due to its higher flop/s and bandwidth, FPGAs exhibit superior scalability compared to GPUs thanks to the stable hardware processing pipeline. As shown in Figure~\ref{fig_fanns:first_page_overview}, experiments on eight accelerators show that the FPGAs achieve 5.5$\times$ and 7.6$\times$ speedup over GPUs in median and 95\textsuperscript{th} percentile (P95) latency, respectively. 

\textbf{Contributions}
\begin{itemize}
        \item I identify a major challenge in designing accelerators for the IVF-PQ-based vector search algorithm: handling the shifting performance bottlenecks when applying different algorithm parameters.
	\item I show the benefit of co-designing hardware and algorithm for optimizing large-scale vector search performance.
        \item I propose \textit{FANNS}, an end-to-end accelerator generation framework for IVF-PQ, maximizing accelerator performance for target datasets and recall requirements. FANNS includes:
	\begin{itemize}
		\item A collection of hardware building blocks for IVF-PQ.
		\item An index explorer that captures the relationship between algorithm parameters and recall. 
		\item A hardware resource consumption model that returns all accelerator designs on a given FPGA device.
		\item A performance model to predict the accelerator QPS of arbitrary combinations of algorithm parameters and accelerator designs.
		\item A code generator that creates ready-to-compile FPGA code given arbitrary accelerator designs. 
	\end{itemize}
	\item{I demonstrate the impressive performance and scalability of \textit{FANNS}, achieving 7.6$\times$ P95 latency speedup over GPUs when utilizing eight accelerators.}
	
\end{itemize}

\section{Hardware-Algorithm Design Space}
\label{sec_fanns:design_space}

The main challenge in designing compelling IVF-PQ accelerators is to find the optimal option in a huge algorithm-hardware design space, as summarized in Table~\ref{tab_fanns:design_space}. 
From the algorithm's perspective, multiple parameters in IVF-PQ can significantly influence recall and performance bottlenecks. 
From the hardware's perspective, there are many valid designs to implement IVF-PQ.

\subsection{The Six Search Stages at Query Time} 

IVF-PQ contains six search stages for query serving. 
\textit{First}, if OPQ is involved, transform the query vector by the OPQ matrix (\ul{Stage OPQ}).
\textit{Second}, evaluate the distances between a query vector and all Voronoi cell centroids (\ul{Stage IVFDist}). 
\textit{Third}, select a subset of cells that are closest to the query vector to scan (\ul{Stage SelCell}). 
\textit{Fourth}, in order to compare distances between PQ codes and a query vector efficiently, construct a distance lookup table per Voronoi cell (\ul{Stage BuildLUT}). More specifically, this step divides the query vector into $m$ sub-vectors and computes the distances between the normalized query vector and all centroids of the sub-quantizer.
\textit{Fifth}, approximate the distances between a query vector and the PQ codes (\ul{Stage PQDist}) by Equation~\ref{eq:adc}, in which $d^{2}(x_i,\hat{y_i}$ only requires looking up the distance tables constructed in Stage BuildLUT. This lookup-based distance computation process is also known as asymmetric distance computation (ADC). 
\textit{Finally}, collect the $K$ vectors closest to the query (\ul{Stage SelK}). 

\begin{equation}
\label{eq:adc}
  \hat{d}^{2}(x,y)=d^{2}(x,\hat{y})
  =\sum_{i=1}^{m}d^{2}(x_i,\hat{y_i})
\end{equation}

\begin{table}

  \begin{center}
    \caption{The list of choices during design space exploration.}
    \label{tab_fanns:design_space}
    \scalebox{0.82}{
    \begin{tabular}{L{5em} L{40em}} 
      \toprule
      \multicolumn{2}{c}{Algorithm parameter space} \\
      \midrule
      \textit{nlist} & The total Voronoi cell number. \\
      \textit{nprobe} & The number of cells to be scanned per query. \\
      \textit{K} & The number of most similar vectors to return. \\
      \textit{OPQ\textsubscript{enable}} & Whether to apply OPQ. \\
      \midrule
      \multicolumn{2}{c}{Hardware design space} \\
      \midrule
      \textit{Design\textsubscript{s}} & The microarchitecture design of stage $s$. \\
      \textit{\#PE\textsubscript{s}} & The number of processing elements in stage $s$. \\
      \textit{Cache\textsubscript{s}} & Cache index on-chip or store it off-chip for stage $s\in$ \{Stage IVFDist, Stage BuildLUT\}. \\
      \bottomrule
    \end{tabular}
    }
  \end{center}
   
\end{table}

\begin{figure*}[t]
  \centering
  
  \begin{subfigure}[b]{0.49\linewidth}
    \includegraphics[width=\linewidth]
    {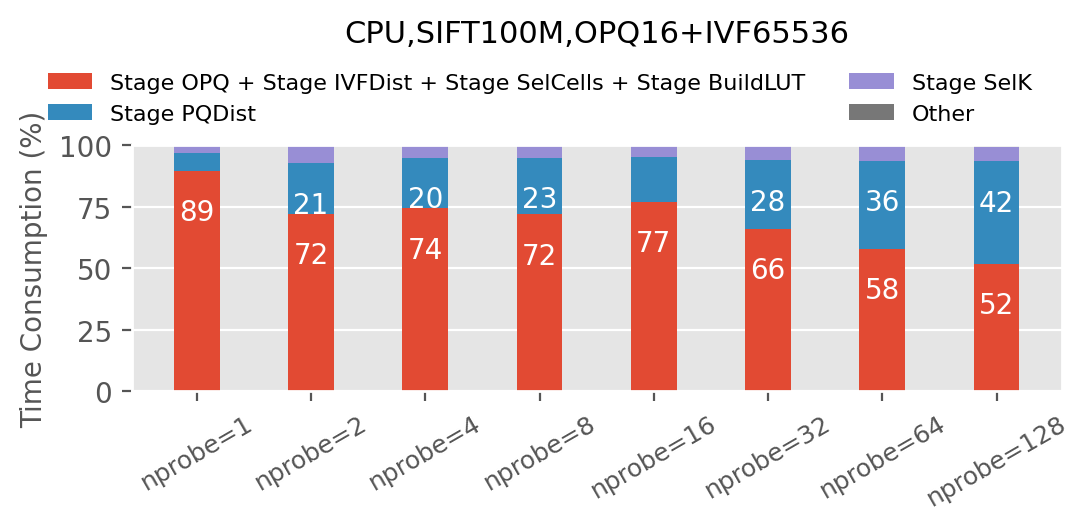}
  \end{subfigure}
  \hfill
  \begin{subfigure}[b]{0.49\linewidth}
    \includegraphics[width=\linewidth]
    {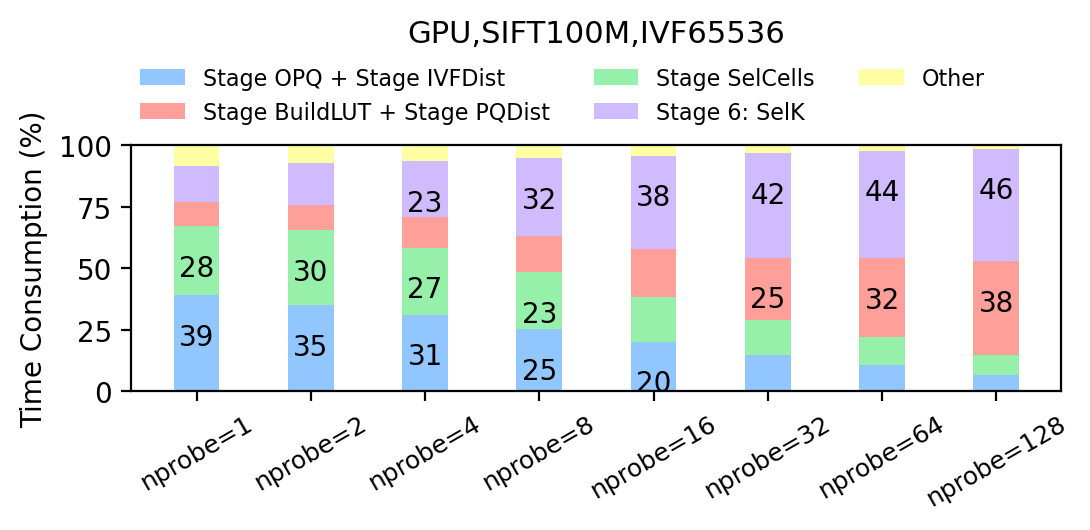}
  \end{subfigure} 
  
  \begin{subfigure}[b]{0.49\linewidth}
    \includegraphics[width=\linewidth]
    {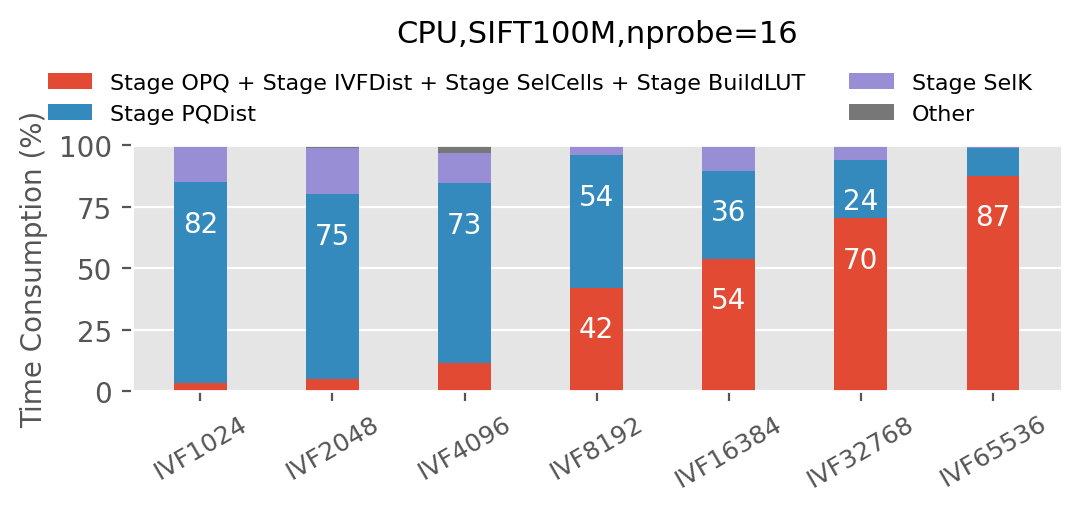}
  \end{subfigure}
  \hfill
  \begin{subfigure}[b]{0.49\linewidth}
    \includegraphics[width=\linewidth]
    {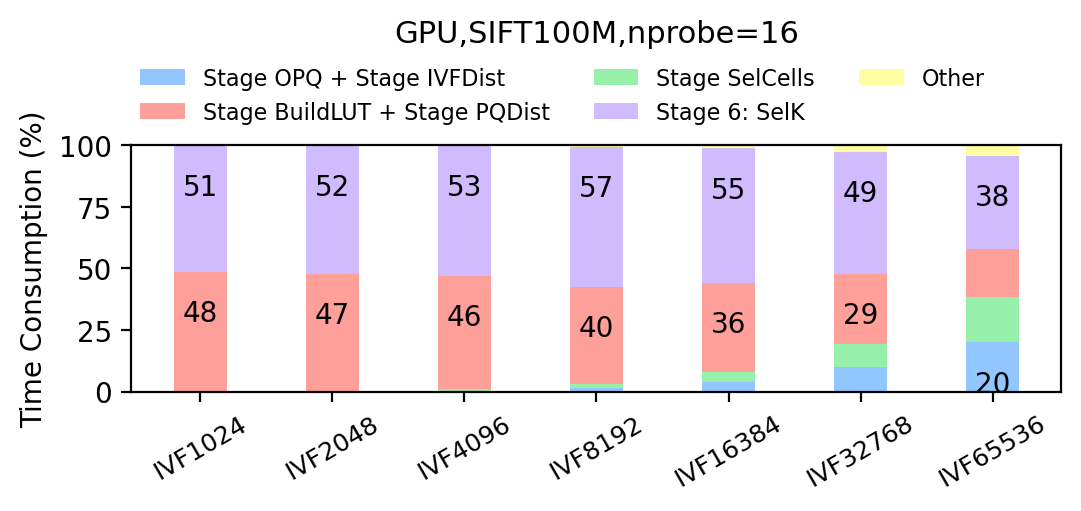}
  \end{subfigure}

  \begin{subfigure}[b]{0.49\linewidth}
    \includegraphics[width=\linewidth]
    {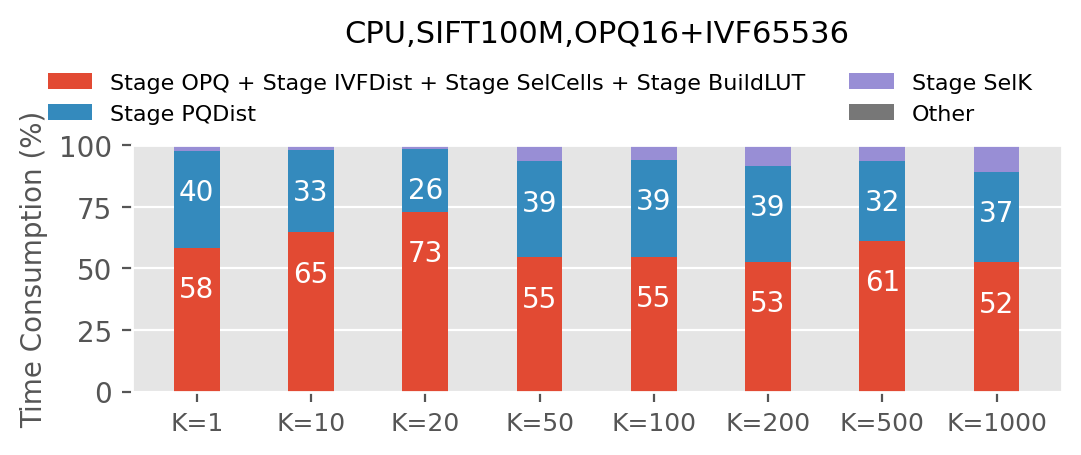}
  \end{subfigure}
  \hfill
  \begin{subfigure}[b]{0.49\linewidth}
    \includegraphics[width=\linewidth]
    {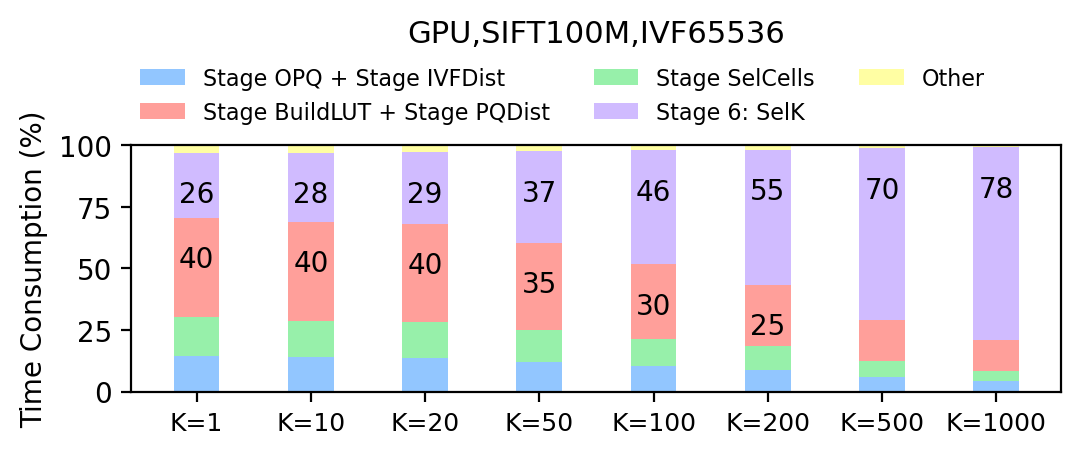}
  \end{subfigure}
  
  \caption{IVF-PQ bottleneck analysis on CPU (1st column) and GPU (2nd column). By tuning \textit{nprobe} (1st row), \textit{nlist} (2nd row), and $K$ (3rd row), we can find that the bottlenecks shift across different algorithm parameters.}
  \label{fig_fanns:cpu_gpu_profiling}

\end{figure*}

\subsection{Algorithm Parameter Space}

\textbf{To achieve a certain recall requirement, there are many options for selecting algorithm parameters.} 
For example, as I will present in the experiments, all the indexes I evaluated can achieve a target recall of R@100=95\% by different \textit{nprobe}. 
It is hard to tell which set of parameters I should deploy on the accelerator.

\textbf{Parameter selections can change the performance bottleneck drastically, which must be considered during the accelerator design phase.} 
I profile the search process on CPUs and GPUs and break down the time consumption per search stage in Figure~\ref{fig_fanns:cpu_gpu_profiling}.
Unlike many applications with a single outstanding bottleneck, the bottlenecks of IVF-PQ shift between the six search stages when different parameters are used.  
However, a specialized accelerator cannot handle shifting bottlenecks because it contains a certain number of dedicated processing elements (PE) for each stage. 
Thus, the accelerator should either target to achieve acceptable performance running arbitrary algorithm parameters or to achieve optimal performance on a certain parameter setting. 
I now break down the IVF-PQ bottlenecks:

\textit{The performance effect of \textit{nprobe}.} 
I fix the index and tune \textit{nprobe}.
I use the indexes that can achieve the highest QPS of R@100=95\% on the SIFT100M dataset on CPU and GPU, respectively.
As shown in the first column of Figure~\ref{fig_fanns:cpu_gpu_profiling}, increasing the number of cells to scan results in more time consumption in Stage PQDist and Stage SelK, regardless of hardware platforms. 
The time consumption of these two stages, on GPUs for example, can increase from 20\% to 80\% as \textit{nprobe} grows.

\textit{The performance effect of \textit{nlist}.} 
By contrast to the first experiment, I now observe the effect of the total number of clusters of the index by fixing the number of clusters to scan (\textit{nprobe=16}). As shown in the second column of Figure~\ref{fig_fanns:cpu_gpu_profiling}, higher \textit{nlist} results in more time consumption on Stage IVFDist to evaluate distances between the query vector and cluster centroids. The consumption is more significant on CPUs due to their limited flop/s compared with GPUs, while the main bottlenecks of GPUs are still in later stages even if \textit{nlist} is reasonably large.

\textit{The performance effect of $K$.}
I fix the index per hardware as in the $\mathrm{nprobe}$ experiment. As shown in the third column of Figure~\ref{fig_fanns:cpu_gpu_profiling}, the time consumption on Stage SelK on GPUs increases significantly as $K$ grows, while the phenomenon is unobvious on CPUs as the bottlenecks are in other stages.

\subsection{Hardware Design Space}

\textbf{There are many ways to implement an IVF-PQ accelerator, and the design choices are summarized in Table~\ref{tab_fanns:design_space}.} 


{The \textit{first} choice is the microarchitecture per search stage}. Not only does the processing element (PE) design differ between stages, there are multiple valid designs per stage. For example, Stage SelK collects $K$ nearest neighbors from a series of distance values, which can either be implemented by a hierarchical priority queue consisting of systolic compare-swap units or by a hybrid design involving sorting network and priority queues, as I will show in Section~\ref{sec_fanns:hardware_design_space}.

{The \textit{second} choice is chip area allocation across the six search stages, i.e., choosing PE numbers per stage}.  
Due to the finite transistors within a chip, this is a zero-sum game: increasing the number of PEs in one stage implies reducing them in another. 

{The \textit{third} decision is about index caching}. Though storing them in off-chip DRAM is the only option for larger IVF indexes, we can decide whether to cache smaller indexes in on-chip SRAM. Caching index guarantees low accessing latency and high bandwidth but increases hardware resource consumptions.


\subsection{How Does One Choice Influence Others?}

The choices of algorithm parameters will influence the optimal hardware design and vice versa.
Since the relationship between the design choices is intricate, I only convey the intuition here with a couple of examples, while the quantitative model will be presented in later sections. 
First, tuning a single parameter can affect the optimal accelerator design.
Increasing $nlist$ results in more workload in comparing the distances between query vectors and IVF centroids. As a result, more PEs in Stage IVFDist should be instantiated to handle the increasing workload, while fewer PEs can be instantiated in other stages due to the limited chip size. Besides, if the $nlist$ is large enough, caching the IVF index on-chip is not a choice at all, while caching small indexes can be beneficial at the cost of consuming on-chip memory that other PEs could have taken. 
Second, a specific accelerator design has its favorable parameter settings.
Assume the accelerator has a lot of Stage IVFDist PEs, while other stages are naturally allocated with fewer resources.
Such design naturally favors a parameter setting of high $nlist$ and low $nprobe$: the reverse case (low $nlist$ and high $nprobe$) will underutilize the Stage IVFDist PEs yet overwhelming the limited Stage PQDist PEs, resulting in low QPS.




\section{\textit{FANNS} Framework Overview}
\label{sec_fanns:system_overview}

\begin{figure*}
  \centering
  \includegraphics[width=1.0\linewidth]{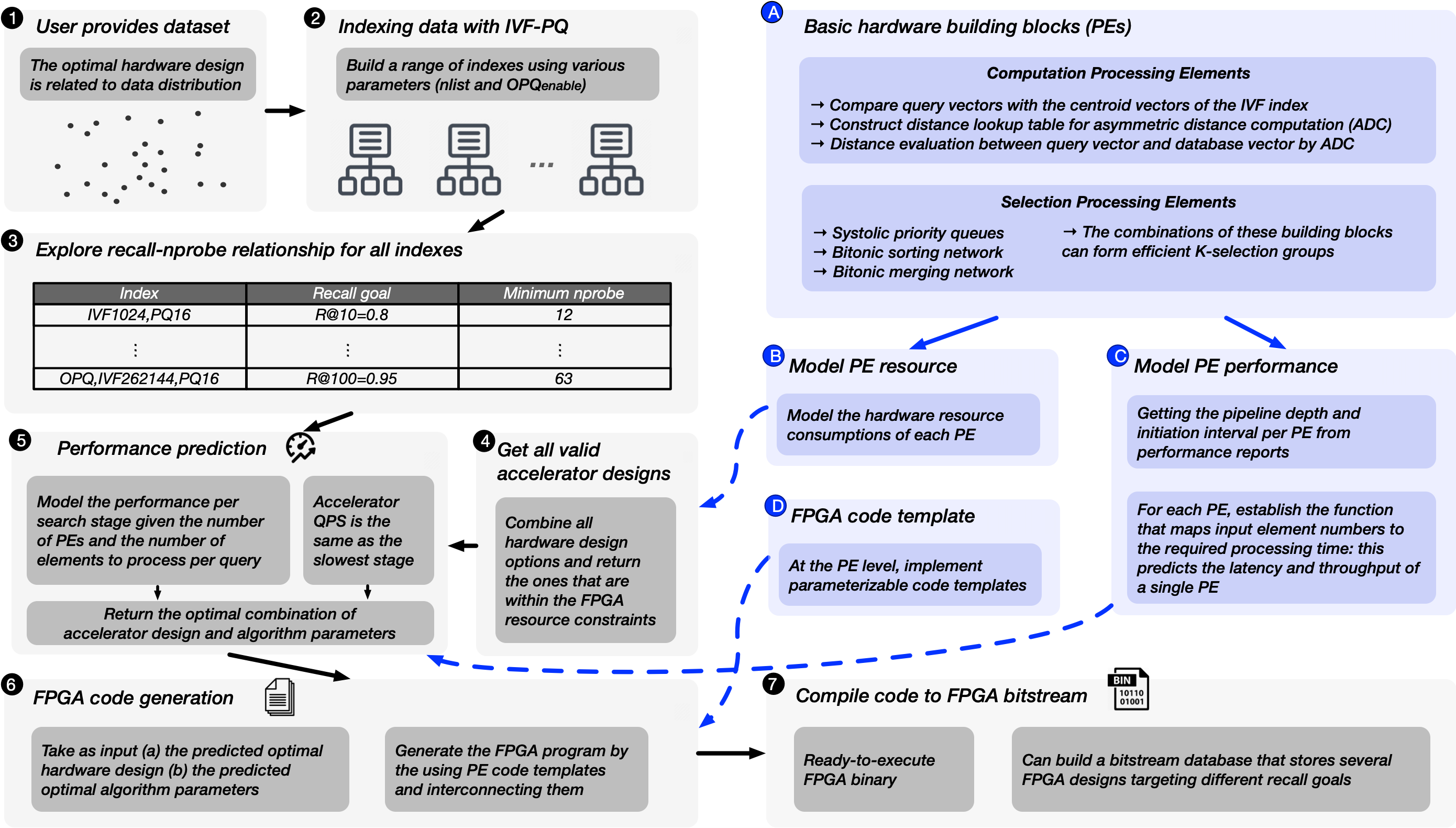}
  \caption{
  The workflow of \textit{FANNS}. The letter-labeled blue blocks are the framework building blocks independent of user requirements, while the digit-labeled gray blocks are the automatic accelerator generation steps. }
  \label{fig_fanns:workflow}
\end{figure*}


\begin{figure*}
  \centering
  \includegraphics[width=1.0\linewidth]{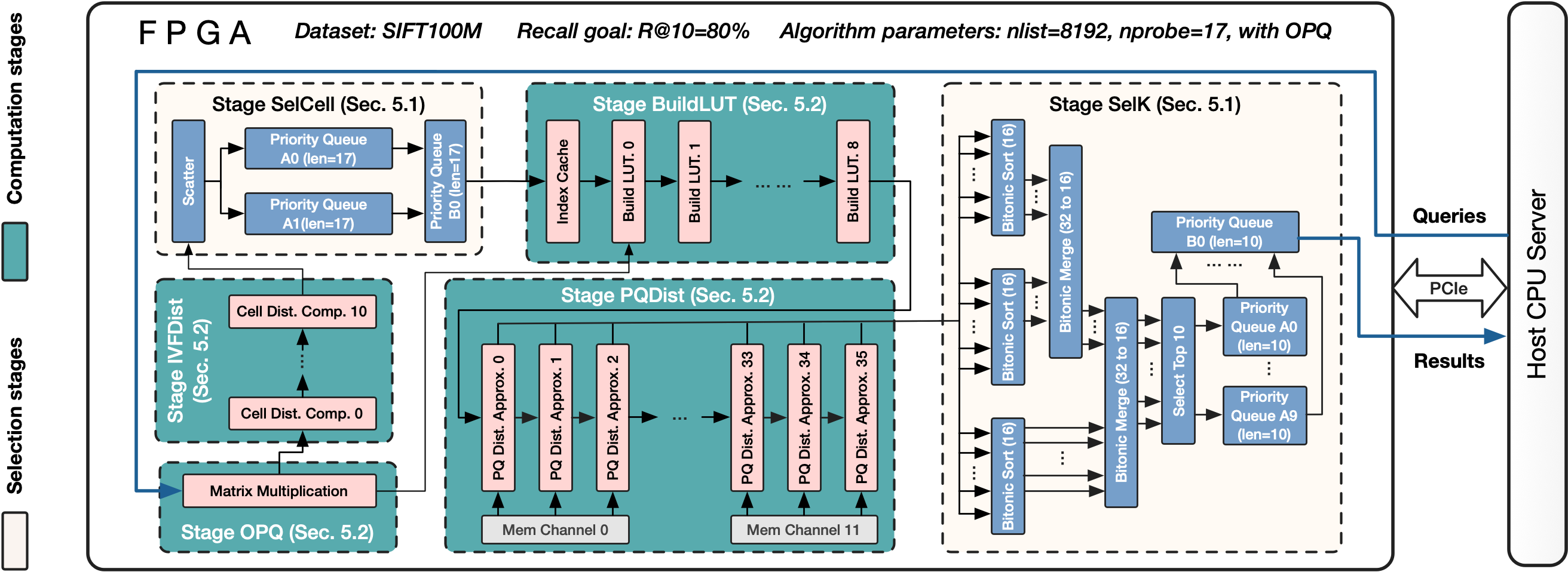}
  \caption{An example accelerator design generated by \textit{FANNS}.}
  \label{fig_fanns:fpga-sample-design}
\end{figure*}

I present \textit{FANNS} (FPGA-accelerated Approximate Nearest Neighbor Search), an end-to-end vector search framework by hardware-algorithm co-design. 
\textit{FANNS} targets the deployment scenario where the user (deployer) has a target recall goal (e.g., for all queries, achieve 80\% recall for top 10 results on average) on a given dataset and a given hardware device. 
Due to the many design options for an IVF-PQ accelerator as introduced above, I leverage the reconfigurability of FPGAs to implementvarious designs designs.
In this case, \textit{FANNS} can automatically figure out the optimal combination of algorithm parameters and hardware design, and generate the specialized FPGA accelerator for the combination.
\textit{FANNS} also supports scale-out by instantiating a hardware network stack~\cite{100gbps} in the accelerator.

Figure~\ref{fig_fanns:workflow} overviews the \textit{FANNS} workflow. 

\textbf{Framework building blocks~(\blueballnumber{A}$\sim$~\blueballnumber{D}).}
To build an IVF-PQ accelerator, I first build a set of PEs for all six search stages~\blueballnumber{A}. 
These building blocks are independent to user requirements.
I design multiple PEs per stage when there are several valid microarchitecture solutions. 
Given the designed PEs, I naturally know their hardware resource consumptions~\blueballnumber{B}.
I can model the PE performance in both latency and throughput~\blueballnumber{C}: knowing the pipeline depth and initiation interval per PE, one can establish the relationship between the number of input elements to process and the respective time consumption in clock cycles. 
Finishing the PE design step, I also have a set of PE code templates~\blueballnumber{D}. 

\textbf{Automatic accelerator generation workflow~(\ballnumber{1}$\sim$\ballnumber{7}).}
The gray blocks in Figure~\ref{fig_fanns:workflow}~presents the automatic workflow that customizes the hardware per user recall requirement.
The inputs of the framework are the user-provided dataset and recall goal~\ballnumber{1}. 
Given the dataset, \textit{FANNS} trains a number of indexes using a range of parameters~\ballnumber{2}. 
Then, for each index, \textit{FANNS} evaluates the relationship between \textit{nprobe} and recall~\ballnumber{3}. 
On the other hand, \textit{FANNS} returns all valid hardware designs whose resource consumption is under the constraint of the given FPGA device~\ballnumber{4}.
Subsequently, \textit{FANNS} uses a performance model to predict the optimal combination of parameter setting and accelerator design~\ballnumber{5}. The performance model takes two input sources: (a) the set of all possible accelerator designs by combining different hardware-level options summarized in Table~\ref{tab_fanns:design_space} and (b) the minimal \textit{nprobe} per index given the recall requirement. For each combination of the hardware-level and parameter-level choices, \textit{FANNS} performance model can predict QPS based on per-PE performance. 
Given the predicted optimal design, \textit{FANNS} code generator outputs the ready-to-compile FPGA code by instantiating the respective PEs and interconnecting them~\ballnumber{6}. 
Finally, the FPGA code is compiled to bitstream (FPGA executable)~\ballnumber{7}. 
Table~\ref{tab_fanns:time} breaks down the time consumption of the \textit{FANNS} workflow.

\begin{table}[t]
  \begin{center}
    \caption{Time consumption of the \textit{FANNS} workflow.}
    \label{tab_fanns:time}
    \scalebox{0.9}{
    \begin{tabular}{L{14em} L{14em}} 
      \toprule
      \multicolumn{1}{c}{Step} & \multicolumn{1}{c}{Time consumption} \\
      \midrule
      Build Indexes & Several hours per index. \\ 
      Get recall-nprobe relationship & Up to minutes per index. \\
      Predict optimal design & Up to one hour per recall goal. \\
      FPGA code generation & Within seconds. \\
      FPGA bitstream generation  & Around ten hours per design. \\
      \bottomrule
    \end{tabular}
    }
  \end{center}
   
\end{table}

\textbf{Framework deployment.} Given its ability to optimize accelerator performance based on specific datasets and recall objectives, FANNS is well-suited for integration into production vector search systems. 
Such systems often manage dynamic datasets, subject to regular insertions and deletions. This is accomplished through the maintenance of a primary IVF-PQ index for a specific dataset snapshot, an incremental (usually graph-based) index for new vectors added since the last snapshot, and a bitmap to track deleted vectors. These two indexes are periodically merged, e.g., once a week, into a new primary index~\cite{adb-v}. In this scenario, FANNS targets optimizing performance for the main index, thus also periodically redesigning accelerators for the new dataset snapshot and, if applicable, the new recall goal. When building the accelerator for the new snapshot, the existing accelerator and the CPU's incremental index continue to process queries. As such, the time taken to build the new accelerator is effectively concealed by the ongoing operation of the older system, barring the initial build. This setup also allows FANNS to always target a static dataset snapshot. The algorithm explorer, therefore, does not need to handle any shifts in data distribution, allowing accurate performance modeling.

\textbf{Example FPGA design.}
Figure~\ref{fig_fanns:fpga-sample-design} shows a generated accelerator targeting R@10=80\% on the SIFT100M dataset. 
In this single-accelerator-search scenario, the FPGA communicates with the host CPU through PCIe to receive query requests and return results.
\textit{FANNS} processes queries in a deeply pipelined fashion: there can be multiple queries on the fly in different stages in order to maximize throughput. 
Each stage of processing is accomplished by a collection of PEs.
The arrows connecting those PEs are FIFOs: a PE loads values from the input FIFO(s), processes the inputs, and pushes the results to the output FIFO(s). 
A stage can contain homogeneous PEs such as Stage IVFDist or heterogeneous PEs such as Stage SelK which involves sorting networks, merging networks, and priority queues. 
The PE numbers are typically irregular (11 in Stage IVFDist, 9 in Stage BuildLUT, etc.) as they are calculated by the performance model, unlike being restricted to the exponential of two which human designers favor. 
I will specify the hardware design per search stage in the following section.

\section{Hardware Processing Elements}
\label{sec_fanns:hardware_design_space}

I present the accelerator hardware processing elements and the design choices. I group the six search stages into selection stages and computation stages to explain related concepts together. 

\subsection{Designs for the Selection Stages}

Two stages need selection functionality. Stage SelCells selects the closest Voronoi cells to the query vector, given a set of input distances. Stage SelK collects the $K$ smallest distances between the query vector and database vectors, given the many approximated distances output by Stage PQDist every clock cycle. Since there can be multiple PEs producing inputs to the two stages, the selection hardware should support multiple input streams. 

\subsubsection{$K$-Selection Primitives}
Bitonic sort networks and systolic priority queues are the building blocks for $K$-selection. 



\textit{Bitonic Sort.} Bitonic sort is a parallel sorting algorithm that takes several input elements in parallel, performs a certain series of compare-swap operations, and outputs the sorted array. Bitonic sort exhibits high sorting throughput, and its parallelism aligns very well with FPGAs~\cite{batcher1968sorting, mueller2012sorting, papaphilippou2018flims, song2016parallel, salamat2021nascent, papaphilippou2020adaptable}.
As a result, the latency of sorting an array is $\sum_{i=1}^{log_{2}l}i=\frac{log_{2}l * (1 + log_{2}l)}{2}$ clock cycles where $l$ is the width of the sorting network.




\begin{figure}[t]
  \centering
  \includegraphics[width=0.7\linewidth]{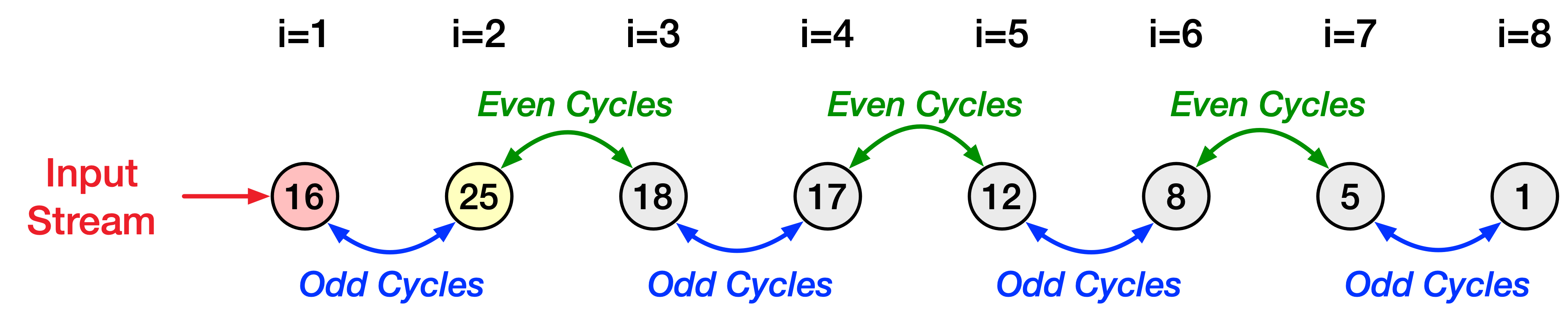}
  \caption{A hardware systolic priority queue.}
  \label{fig_fanns:priority-queue}
\end{figure}

\textit{Systolic Priority Queue.} 
While software-based priority queues support enqueue, dequeue, and replace operations, FANNS only needs the replace operation: if the input is smaller than the current root, dequeue the root and enqueue the input.
Figure~\ref{fig_fanns:priority-queue} shows in the implemented systolic priority queue~\cite{huang2014scalable, leiserson1979systolic} that supports such minimal required functionality while consuming the least hardware resources. 
It is a register array interconnected by compare swap units, supporting one replace operation every two clock cycles. 
In the first cycle, the leftmost node is replaced with a new item, and all the even entries in the array are swapped with the odd entries. 
In the second cycle, all the odd entries are swapped with the even entries.
During this process, the smallest elements are gradually swapped to one side of the queue.

\subsubsection{$K$-Selection Microarchitecture Design}
Parallel $K$-selection collects the $s$ smallest numbers per query ($s=nprobe$ in Stage SelCells; $s=K$ in Stage SelK) out of $z$ input streams given that each stream produces $v$ values per query. I propose two design options for this task with different trade-offs:

\textbf{Option 1: hierarchical priority queue (HPQ).} I propose HPQ as a straightforward way for parallel selection. 
The first level of HPQ contains $z$ queues to collect $s$ elements from each stream.
The second level takes the $zs$ elements collected in the first level and selects the $s$ results.
The HPQ allows $z/2$ input elements per cycle since each replace operation in a priority queue requires two cycles. As a result, if an input stream generates one element per cycle, one should split it into two substreams and match it with two priority queues in the first level.

\begin{figure}
  \centering
  \includegraphics[width=0.9\linewidth]{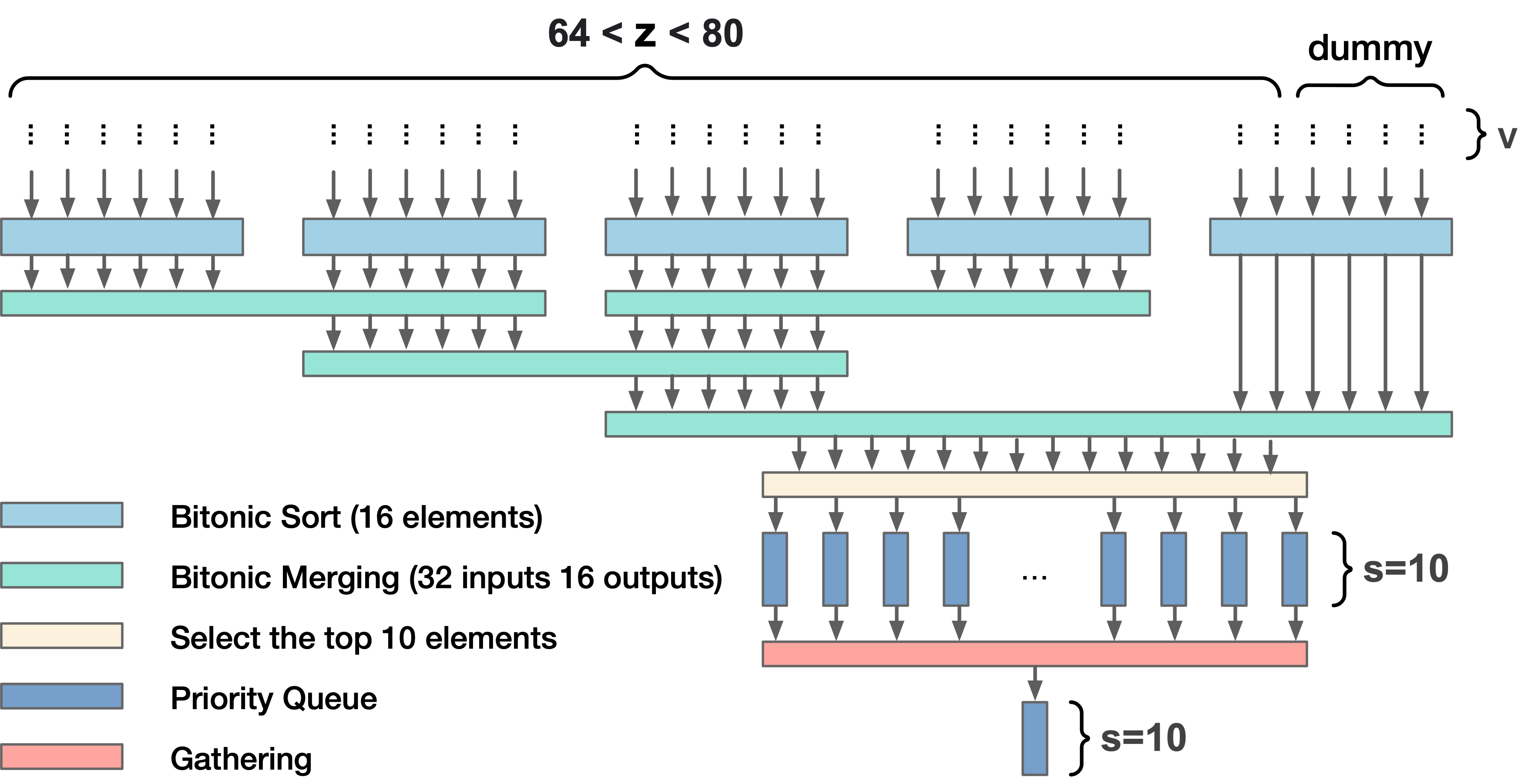}
  \caption{An example of hybrid bitonic sorting, merging, and priority queue architecture that selects the top 10 elements out of up to 80 input streams.}
  \label{fig_fanns:hybrid-topk}
\end{figure}

\textbf{Option 2: hybrid sorting, merging, and priority queue group (HSMPQG).} 
The key idea is to collect the $s$ results per clock cycle before inserting them into the priority queues, such that the number of required queues can be significantly reduced.
Figure~\ref{fig_fanns:hybrid-topk} shows an example of such design ($64<z\leq80$ and $s=10$). 
The first step is to sort every 16 elements since 16 is the minimum bitonic sort width greater than $s=10$. Handling up to 80 inputs per cycle requires five bitonic sort networks. Some dummy streams are added as the input for the last sorting network. 
The second step is to merge the sorted elements by several bitonic partial mergers. Each bitonic merger outputs the top 16 elements out of the two input sorted arrays. After several merging steps, one has the sorted top 16 elements per cycle.
Afterward, the $s=10$ elements per cycle are picked out of the 16 and inserted into a hierarchical priority queue, which outputs the $s$ results per query.
Note that one can configure the number of bitonic sort and parallel merge networks for different workloads. For example, if $16<z\leq32$, two sorting and one merging modules are required; one will need three sorting and two merging networks when $32<z\leq48$.

\textbf{Intuition behind different $K$-selection microarchitecture.} 
The HPQ design suits the situation when the input stream number $z$ is small, because the few priority queues instantiated will not consume many resources. This design is also the only option when $s\geq{z}$, for which the second option cannot filter out unnecessary elements per cycle at all.
The HSMPQG design targets to collect a small result set over many input streams. It could save hardware resources by significantly reducing the number of priority queues compared with the first option. However, the bitonic sorting and merging networks also count for resource consumption, thus the second option is not always better even if $s<z$.


\subsection{Designs for the Computation Stages}

Computation stages include Stage OPQ, Stage IVFDist, Stage BuildLUT, and Stage PQDist.
In this section, I first specify the Stage PQDist PEs to convey the compute PE design principles on FPGAs, and then introduce the PE interconnection topology.

\subsubsection{Stage PQDist.}

As shown in Figure~\ref{fig_fanns:fpga-sample-design}, there are many Stage PQDist PEs working in parallel, approximating distances between the query vector and the quantized database vectors.

\begin{figure}
  \centering
  \includegraphics[width=0.9\linewidth]{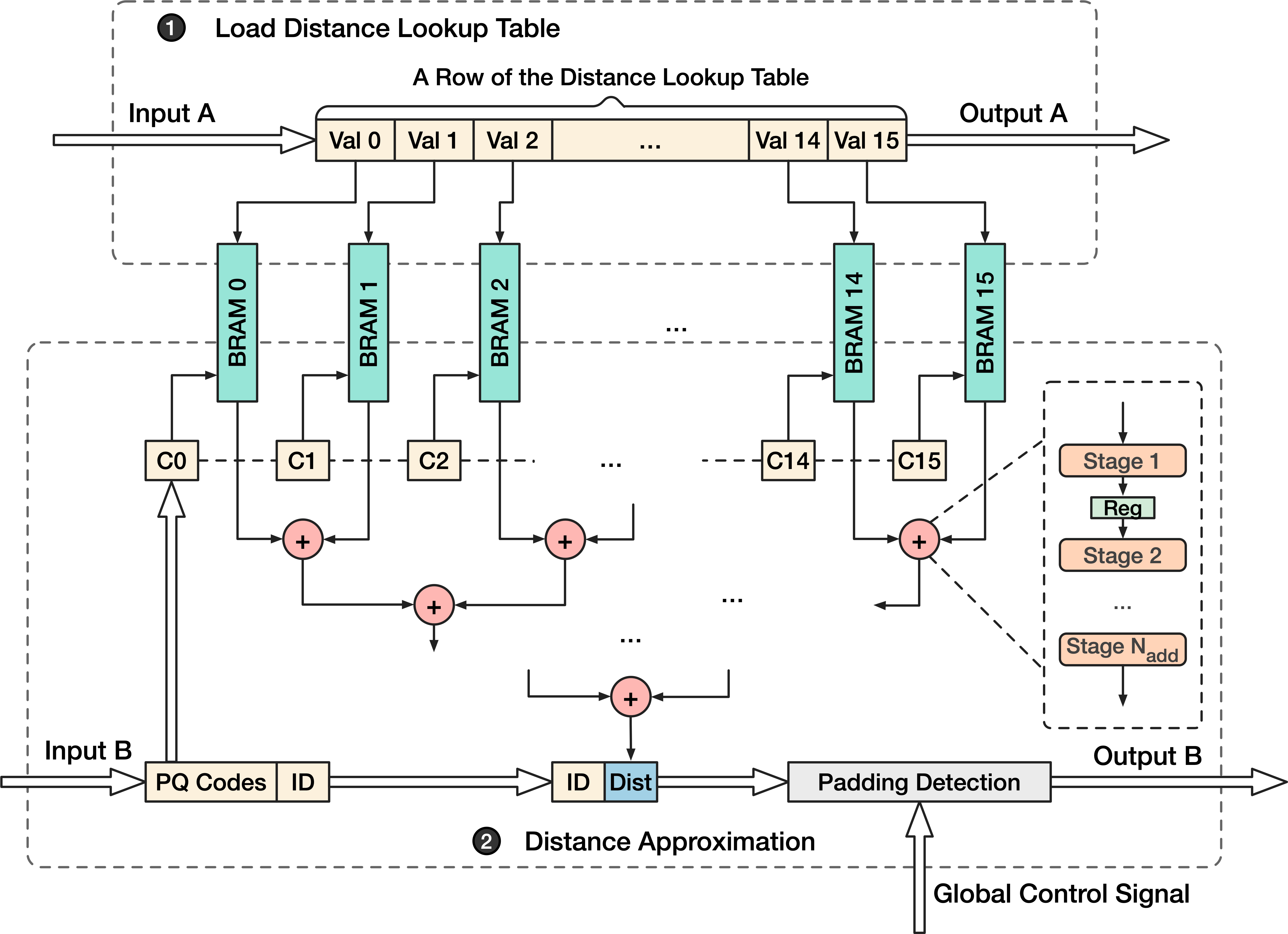}
  \caption{The PE hardware design for Stage PQDist.}
  \label{fig_fanns:PE_PQDist}
\end{figure}

\textbf{PE design.} 
Figure~\ref{fig_fanns:PE_PQDist} presents the PE design for decoding 16-byte PQ codes.
The PE takes two inputs: the distance lookup tables produced by Stage BuildLUT and the PQ codes stored in off-chip memory channels. 
For a single query, a PE repeats two major steps $nprobe$ times.
The first step is reading a distance lookup table of size of $km$.
I use BRAM, a type of fast on-chip SRAM, to cache the tables. 
In order to provide memory access concurrency for the computing step, I assign $m$ BRAM slices per PE --- each slice stores a column of a table. 
The second step is approximating the distances between the query vector and the database vectors by asymmetric distance computation.
Each PQ code (1-byte) of a database vector is used as the lookup index for one column of a distance table, and $m$ distances are retrieved from the BRAM slices in parallel per cycle. 
These partial distances are then fed into an add tree that produces the total distance.
In order to maximize the computation throughput, each logical operator (addition) in the tree is composed of several DSPs (for computation) and FFs (as registers), such that the computation is fully pipelined, allowing the add tree to consume $m$ input distances and to output one result per clock cycle.
During the last iteration of scanning a cell, the PE performs padding detection and overwrites the output by a large distance for the padded case. 
The meta-info about padding is streamed into the PE by the accelerator's global controller.

\textbf{PE size.} 
In principle, a PE with more computation logic is usually more efficient in terms of performance delivered per hardware resource unit. This is because each PE has some surrounding logic as the interface to other PEs --- the smaller each PE, the more significant the total overhead. However, it is hard for the FPGA compiler to map a huge chunk of logic to the FPGA successfully~\cite{de2020flexible}.
Thus, I experiment with several PE sizes and select the largest one that can be successfully compiled. 

\subsubsection{PE interconnection Topology.}

Within a computation stage, I adopt a 1-D array architecture to forward data between the homogeneous PEs.
For example, the second PE consumes the query vector and the results of the first PE, appends its local results, and sends the query vector as well as the aggregated results to the third PE. 
Another design choice, which I do not adopt, is the broadcasting/gather topology.
The advantage of the 1-D array architecture over the broadcasting/gather one is the minimized wire fan-out: too many long wires connected to a single source can lead to placement \& routing failure during FPGA compilation~\cite{de2020transformations}.
For communication between stages and within a selection stage, the FIFO connections are straightforward because there is no input sharing as in computation stages.

\section{End-to-End Hardware Generation}
\label{sec_fanns:e2e}

This section illustrates the end-to-end accelerator generation flow of \textit{FANNS}, as visualized in Figure~\ref{fig_fanns:workflow} (\ballnumber{1}$\sim$\ballnumber{7}). I implement the end-to-end generation flow using a set of Python scripts, while the hardware processing elements are implemented in Vitis HLS.

\vspace{-1em}
\subsection{Explore Algorithm Parameters}
\label{sec_fanns:design_space_exploration_nprobe}

\textbf{Given a dataset, \textit{FANNS} first captures the relationship between the algorithm parameters and recall, which will be used for accelerator QPS prediction.}
\textit{FANNS} trains a number of IVF indexes trying different \textit{nlist}~\ballnumber{2}. 
Each index is trained both with and without OPQ. 
Given the user-provided sample query set, \textit{FANNS} evaluates the minimum \textit{nprobe} that can achieve the user-specified recall goal on each index~\ballnumber{3} (e.g., 80\% of average recall for top 10 results). 
The result of this step is a list of index-\textit{nprobe} pairs that serve as the inputs of the FPGA performance model.

\vspace{-1em}
\subsection{List Valid Accelerator Designs}
\label{sec_fanns:resource_model}

\textbf{\textit{FANNS} lists all valid accelerator designs on a given FPGA device by resource consumption modeling}~\ballnumber{4}.
Specifically, \textit{FANNS} combines all hardware choices in Table~\ref{tab_fanns:design_space} to form accelerators and returns the valid ones whose consumptions of all types of resources (BRAM, URAM, LUT, FF, and DSP) are under the device constraint. 
Consuming all resources on the FPGA is unrealistic because such designs will fail at the placement \& routing step during FPGA compilation. 
But it is also impossible to predict the maximum resource utilization per design because the EDA algorithms for FPGA compilation are nondeterministic. 
As a result, I set the resource utilization rate as a constant for all accelerators, e.g., a conservative value of 60\% in the experiments.

\begin{equation}
\label{eq:resource_model}
\scalebox{0.8}{
$
\begin{gathered}
    \sum_{i} C_{r}(PE_i) + \sum_{i} C_{r}(FIFO_i) + C_{r}(infra) \leq Constraint_{r} , \\
    \forall r\in\{BRAM, URAM, LUT, FF, DSP\}
\end{gathered}
$
} 
\end{equation}

\textit{FANNS} models an accelerator's resource consumption by summing up several components as in Equation~\ref{eq:resource_model} ($C_r$ denotes the consumption of resource $r$).
The first part is of all the PEs. The consumption of a PE is known once I finish designing and testing the PE. For priority queues of variable lengths, I employ a linear consumption estimation model since the numbers of registers and compare-swap units in a priority queue are linear to the queue length. 
The second part is the FIFOs connecting PEs, which can be modeled by measuring the consumption of a single FIFO and counting the required FIFO numbers.
The final component is the infrastructure surrounding the accelerator kernel, such as the memory controller, which consumes constant resources.

\vspace{-1em}
\subsection{Model Accelerator Performance}
\label{sec_fanns:design_space_exploration_perf_modeling}

\textbf{The \textit{FANNS} performance model predicts the QPS of all combinations of algorithm parameters and accelerator designs, then returns the optimal one.}
Given the large design space, it is unrealistic to evaluate QPS by compiling all accelerators and testing all parameters.
Thus, one needs an effective performance model to predict the QPS per combination~\ballnumber{5}. By using the following modeling strategy, \textit{FANNS} can evaluate all (millions of) combinations given a recall requirement within an hour. I now introduce the model in a top-down manner. 

\textit{Model the performance of an accelerator.} As the six search stages of IVF-PQ are pipelined, the throughput of the entire accelerator is that of the slowest stage, as in Equation~\ref{eq:performance_all}. 
\begin{equation}
\label{eq:performance_all}
\scalebox{0.95}{$
\begin{gathered}
    QPS_{accelerator} = min(QPS_{s})\mathrm{, where~} s\in \mathrm{\{Stages}\}
\end{gathered}
$}
\end{equation}

\textit{Model the performance of a search stage.} 
A search stage typically consists of multiple PEs functioning concurrently.
If these PEs share the same amount of workload, the time consumption per query of the stage is the same as the time consumption for a single PE to handle its own workload.
If the workloads are imbalanced per PE, the performance of the stage is decided by its slowest PE.

\textit{Model the performance of a PE.} 
Inspired by de Fine Licht et al.~\cite{de2020transformations}, I estimate the throughput of a single PE by predicting the number of clock cycles it takes to process a query ($CC$). 
For a single query, I suppose that the PE processes $N$ input elements. 
The pipeline initiation interval is $II$, which represents the number of cycles that must pass before a new input can be accepted into the pipeline. 
The pipeline has a latency $L$, which is the number of cycles it consumes for an input to propagate through the pipeline and arrive at the exit.
$L$ and $II$ are known constants after implementing the hardware. $N$ can be either a constant or a variable. For example, after deciding the algorithm parameters and the accelerator design, the number of distances evaluated per PE in Stage IVFDist is a constant (\textit{N=nlist/PENum}). By contrast, the number of PQ codes scanned in Stage PQDist differs for every query due to the imbalanced number of codes per cell. In this case, I estimate $N$ by taking the expected scanned entries per query (assume the query vector distribution is identical to the database vectors, such that cells containing more vectors are more likely to be scanned). 
Given the numbers of $L$, $II$ and $N$, I can estimate the consumed clock cycles as $CC=L+(N-1)*II$. The QPS of the PE can then be derived by Equation~\ref{eq:performance_pe} where \textit{freq} is the accelerator frequency. 
Similar to predicting the maximum resource utilization rate, it is impossible to know the operational frequency of an accelerator before compilation.
Thus, I assume the frequency to be a constant for all accelerators.

\vspace{-1em}
\begin{equation}
\label{eq:performance_pe}
\begin{gathered}
    QPS_{PE} = freq/(L+(N-1)*II)
\end{gathered}
\end{equation}

\vspace{-1em}
\subsection{Generate FPGA Programs}
\label{sec_fanns:design_space_exploration_codegen}

{\textit{FANNS} code generator takes as inputs the optimal combination of parameter setting and hardware design and produces the ready-to-compile FPGA code.}
Refering to the inputs, the code generator instantiates the given numbers of PEs using the PE code templates, the respective on-chip memory for index caching, the FIFOs interconnecting the aforementioned components, and the off-chip memory interfaces between the accelerator kernel and the FPGA shell~\ballnumber{6}.
Since the code generation step does not involve complex logic, it only consumes seconds to return the FPGA program, which will be further compiled to the bitstream~\ballnumber{7}.

\section{Evaluation}
\label{sec_fanns:evaluation}

This section shows the effectiveness and necessity of algorithm-hardware co-design to achieve the optimal vector search performance on FPGAs. 
I also integrate the FPGAs with network stacks to show their great scalability.

\subsection{Experimental Setup}
\label{sec_fanns:target_platform}

\textbf{Baseline.}
I compare \textit{FANNS} with CPU, GPU, and FPGA baselines. 
The CPU and GPU baselines run Faiss (version 1.7.0), a popular ANN library developed by Meta known for its efficient IVF-PQ implementation.
The FPGA baseline uses the same set of hardware building blocks as \textit{FANNS} but without being parameter-aware.

\textbf{Hardware Setup.}
I choose CPUs, GPUs, and FPGAs that are manufactured in the generation of technology. 
I use an m5.4xlarge CPU server on AWS, which contains 16 vCPUs of 16 vCPUs of Intel(R) Xeon(R) Platinum 8259CL @ 2.50GHz (Cascade Lake, 14nm technology) and 64 GB of DDR4 memory.
I use NVIDIA V100 GPUs (CUDA version 11.3) manufactured by the TSMC 12 nm FFN (FinFET NVIDIA) technology (5,120 CUDA cores; 32 GB HBM).
I use Xilinx Alveo U55c FPGA fabricated with TSMC's 16nm process. It contains 1.3M LUTs, 9K DSPs, 40MB on-chip memory, and 16 GB HBM. I develop the accelerators using Vitis HLS 2022.1. 

\textbf{Benchmark.}
I evaluate \textit{FANNS} on standard and representative vector ANN benchmarks: the SIFT and Deep datasets.
The SIFT dataset contains 128-dimensional vectors, while the Deep dataset consists of 96-dimensional vectors.
For both datasets, I adopt the 100-million-vector size scale as they can fit into the FPGA memory after product quantization. 
Both datasets contain 10,000 query vectors and respective ground truths of nearest neighbor search for recall evaluation.
I set various recall goals on each dataset. 
As recalls are related to $K$ (the more results returned, the more likely they overlap with true nearest neighbors) and the data distribution, I set one recall goal per K per dataset, i.e., R@1=30\%, R@10=80\%, and R@100=95\% on the SIFT dataset and R@1=30\%, R@10=70\%, and R@100=95\% on the Deep dataset.

\textbf{Parameters.} 
I explore a range of algorithm parameters and set a couple of constant factors for \textit{FANNS} performance model.
On the algorithm side, I trained a range of indexes with different numbers of Voronoi cells (\textit{nlist} ranges from $2^{10}$ to $2^{18}$ ) for each dataset, so as to achieve the best QPS for not only FPGA but the CPU and GPU baselines. 
Per \textit{nlist}, I trained two indexes with and without OPQ to compare the performance.
I quantize the vectors to 16-byte PQ codes ($m=16$) for all indexes and all types of hardware.
The primary consideration is to fit the dataset within FPGA's device memory while achieving high recall. 
On the \textit{FANNS} side, I set the maximum FPGA resource utilization rate as 60\% to avoid placement \& routing failures. I also set the target accelerator frequency as 140MHz based on the design experience with the U55c FPGA device.

\subsection{\textit{FANNS}-Generated Accelerators}

This section presents the \textit{FANNS} generated accelerators. I show that the optimal designs shift with parameter settings. I then present the fully customized accelerator designs under target recall and compare them against the parameter-independent FPGA baseline design.

\begin{figure*}[t]
  \centering
  
  \begin{subfigure}[b]{0.55\linewidth}
    \includegraphics[width=\linewidth]
    {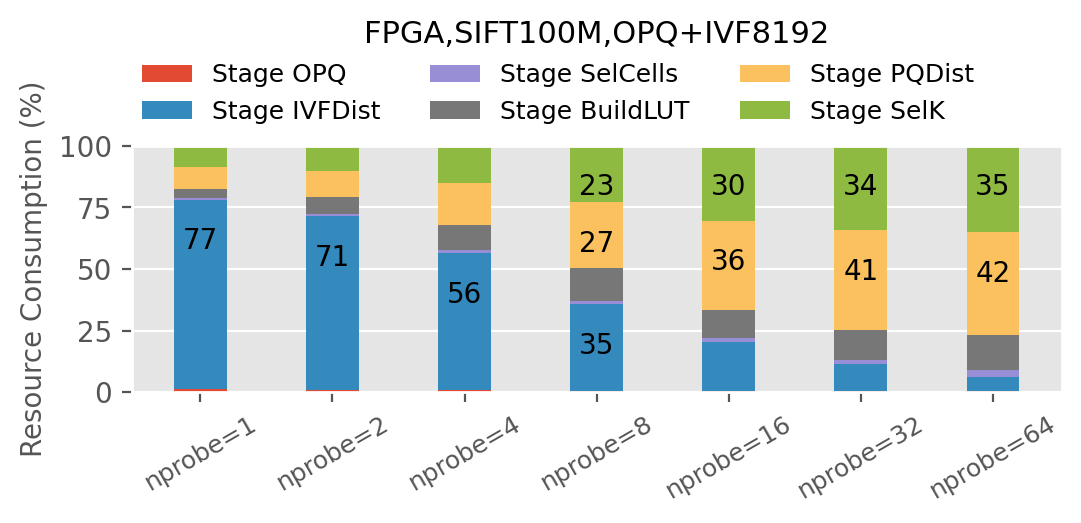}
  \end{subfigure}
  
  \begin{subfigure}[b]{0.55\linewidth}
    \includegraphics[width=\linewidth]
    {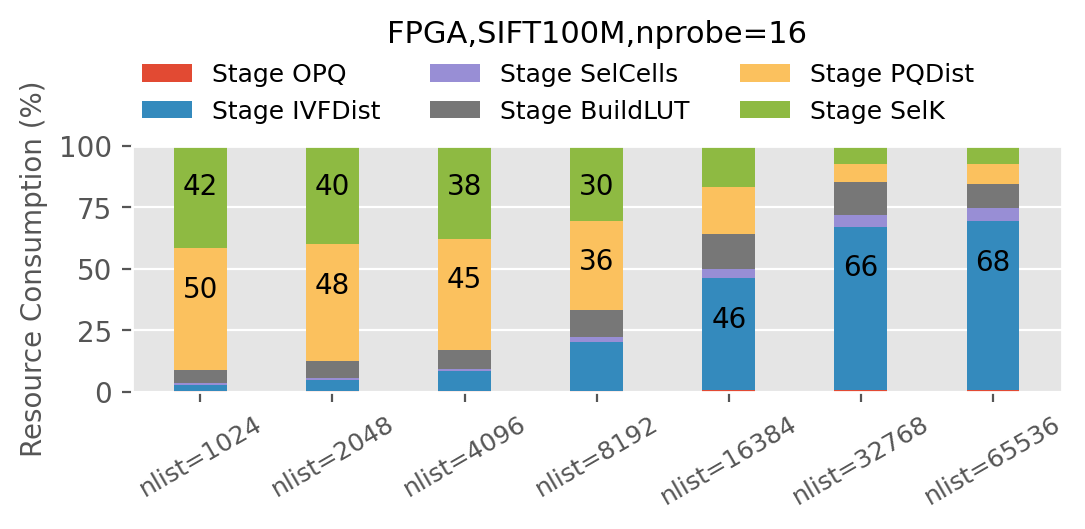}
  \end{subfigure}
  
  \begin{subfigure}[b]{0.55\linewidth}
    \includegraphics[width=\linewidth]
    {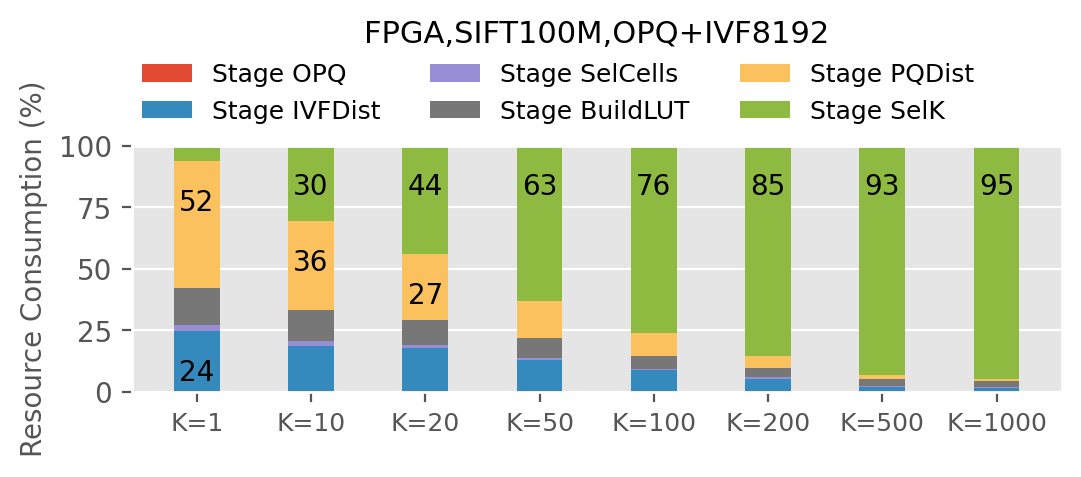}
  \end{subfigure}

  \caption{The optimal FPGA designs shift with various algorithm parameters (\textit{nprobe}, \textit{nlist}, and $K$.)}
  \label{fig_fanns:FPGA_design_shift}
\end{figure*}

\subsubsection{The Effect of Algorithm Parameters on Hardware Designs}

\textbf{Optimal accelerator designs shift significantly with algorithm parameters, as shown in Figure}{~\ref{fig_fanns:FPGA_design_shift}}. In this experiment, I assign the parameters to the \textit{FANNS} performance model, which predicts the optimal hardware design under the parameter constraints. To draw high-level conclusions from the designs, I only visualize the resource consumption ratio per search stage, omitting the microarchitecture-level choices.
First, I observe the effect of \textit{nprobe} on the designs. 
As the number of PQ codes to scan increases, more hardware resources are dedicated to Stage PQDist and Stage SelK, while most of the resources are allocated to Stage IVFDist when $nprobe$ is small.
Then, I fix \textit{nprobe} and observe the designs when shifting \textit{nlist}. As \textit{nlist} raises, more PEs are instantiated for Stage IVFDist so as to improve the performance of centroid distance computation. 
Finally, as $K$ increases, the resources spent on Stage SelK surge because the resource consumption of hardware priority queues is linear to the queue length $K$.

\subsubsection{The Optimal Accelerator Designs of Given Recall Goals}

Table~\ref{tab_fanns:final_designs}~summarizes the \textit{FANNS}-generated designs per recall and compares them with the baseline designs. It shows that: 

\textbf{First, \textit{FANNS} picks different parameters per recall.} For the three recall requirements, \textit{FANNS} adopts different indexes and $nprobe$ to maximze the performance.

\textbf{Second, \textit{FANNS} generates different hardware designs for each recall requirement.} Stage SelK, for example, applies the two different microarchitecture designs (HPQ and HSMPQG) and invests different amounts of hardware resources (2.9\%$\sim$31.7\% LUT) for the three recall requirements. Even for the stages using the same microarchitecture, e.g., Stage IVFDist, the PE numbers of these accelerators can also be different.

\subsubsection{Parameter-independent Accelerator Designs}

{I design a set of parameter-independent ANNS accelerators that can serve queries on arbitrary indexes as the FPGA baseline. }
I design three parameter-independent accelerators for different $K$ requirements (1, 10, 100) as shown in Table~\ref{tab_fanns:final_designs}. 
Each accelerator roughly balances resource consumption across stages such that the accelerator should perform well on a wide range of algorithm settings. 
Saying this, I do not simply allocate 1/6 resources to each of the six stages due to the following facts. 
First, the number of PEs between Stage PQDist and Stage SelK should be proportional, as more distance computation PEs should be matched with more priority queues to balance the performance between the two stages.
Second, Stage OPQ performs a lightweight vector-matrix multiplication that consumes few resources.

\begin{landscape} 
\begin{table*}
  \caption{
  Comparison between human-crafted design and \textit{FANNS}-generated designs (for the SIFT100M dataset), including index selection, architectural design, resource consumption (LUT), and predicted QPS.}
  \label{tab_fanns:final_designs}
\scalebox{0.5}{ 
\begin{tabular}{
@{} l l c 
l c 
l l c 
l l l c 
l l l c 
l l l c 
l l c 
l l l c 
l @{}} 
    \toprule
    \phantom{} & \multirow{2}{2em}{Index} & \phantom{} & 
    \multirow{2}{2em}{nprobe} & \phantom{} & 
    \multicolumn{2}{c}{Stage OPQ} & \phantom{} 
    & \multicolumn{3}{c}{Stage IVFDist} & \phantom{} 
    & \multicolumn{3}{c}{Stage SelCells} & \phantom{} 
    & \multicolumn{3}{c}{Stage BuildLUT} & \phantom{} 
    & \multicolumn{2}{c}{Stage PQDist} & \phantom{} 
    & \multicolumn{3}{c}{Stage SelK} & \phantom{} 
    & \multirow{2}{4.5em}{Pred. QPS (140 MHz)} \\
    \cmidrule{6-8} 
    \cmidrule{9-11} 
    \cmidrule{13-15} 
    \cmidrule{17-19} 
    \cmidrule{21-22}
    \cmidrule{24-26}  
    & & & 
    & & 
    \#PE & LUT.(\%) &&
    \#PE & Index store & LUT.(\%) &&
    Arch. & \#InStream & LUT.(\%) &&
    \#PE & Index store & LUT.(\%) &&
    \#PE & LUT.(\%) && 
    Arch. & \#InStream & LUT.(\%) && 
     \\
    \midrule 
    
    K=1 (Baseline) & N/A && 
    N/A && 
    1 & 0.2 &&
    10 & HBM & 6.9 && 
    HPQ & 2 & 6.4 && 
    5 & HBM & 6.9 && 
    36 & 15.2 && 
    HPQ & 72 & 1.8 && 
    N/A  \\
    
    K=10 (Baseline) & N/A &&  
    N/A && 
    1 & 0.2 &&
    10 & HBM & 6.9 &&
    HPQ & 2 & 6.4 &&
    4 & HBM & 6.3 &&
    16 & 6.7 &&
    HPQ & 32 & 5.7 && 
    N/A \\
    
    K=100 (Baseline) & N/A && 
    N/A && 
    1 & 0.2 &&
    10 & HBM & 6.9 &&
    HPQ & 2 & 6.4 &&
    4 & HBM & 6.3 &&
    4 & 1.7 &&
    HPQ & 8 & 15.0 &&
    N/A \\
    
    \midrule 
    K=1 (\textit{FANNS})  & IVF4096 &&  
    5 && 
    0 & 0 &&
    16 & on-chip & 11.0 &&
    HPQ & 2 & 0.3 &&
    5 & on-chip & 2.6 &&
    57 & 24.0 &&
    HPQ & 114 & 2.9 &&
    31,876 \\
    
    K=10 (\textit{FANNS})  & OPQ+IVF8192 && 
    17 && 
    1 & 0.2 &&
    11 & on-chip & 7.6 &&
    HPQ & 2 & 0.9 &&
    9 & on-chip &5.2 &&
    36 & 15.2 &&
    HSMPQG & 36 & 12.7 &&
    11,098 \\
    
    K=100 (\textit{FANNS}) & OPQ+IVF16384 && 
    33 && 
    1 & 0.2 &&
    8 & on-chip & 5.5 &&
    HPQ & 1 & 0.6 &&
    5 & on-chip & 3.6 &&
    9 & 3.8 &&
    HPQ & 18 & 31.7 &&
    3,818 \\
    
    \bottomrule
  \end{tabular}
  } 
\end{table*}
\end{landscape}

\subsection{Performance Comparison}
\label{sec_fanns:overall_performance}

















\begin{figure*}
  \centering
  
  \begin{subfigure}[b]{0.49\linewidth}
    \includegraphics[width=\linewidth]{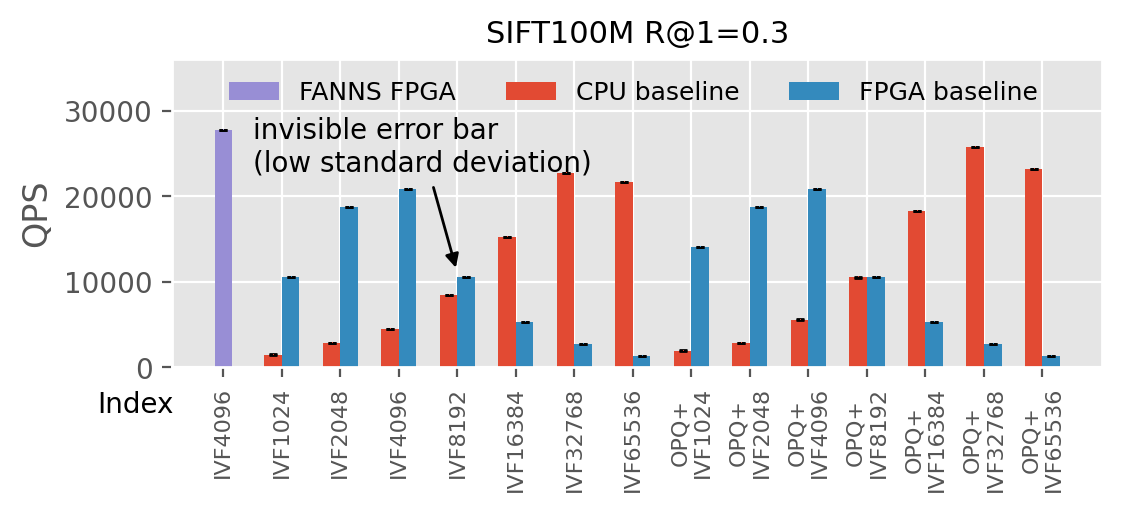}
  \end{subfigure}
  \begin{subfigure}[b]{0.49\linewidth}
    \includegraphics[width=\linewidth]{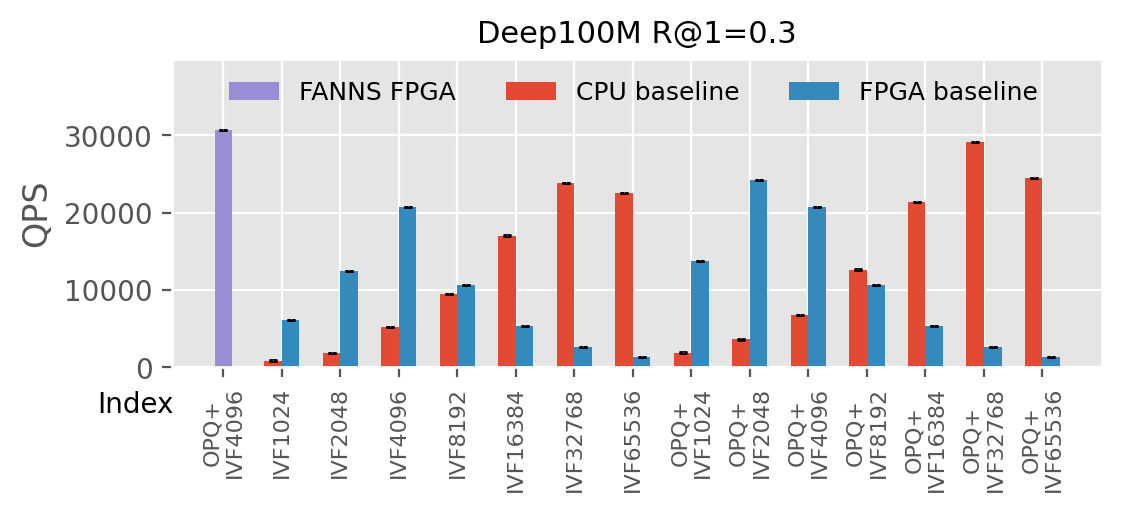}
  \end{subfigure}

  \begin{subfigure}[b]{0.49\linewidth}
    \includegraphics[width=\linewidth]{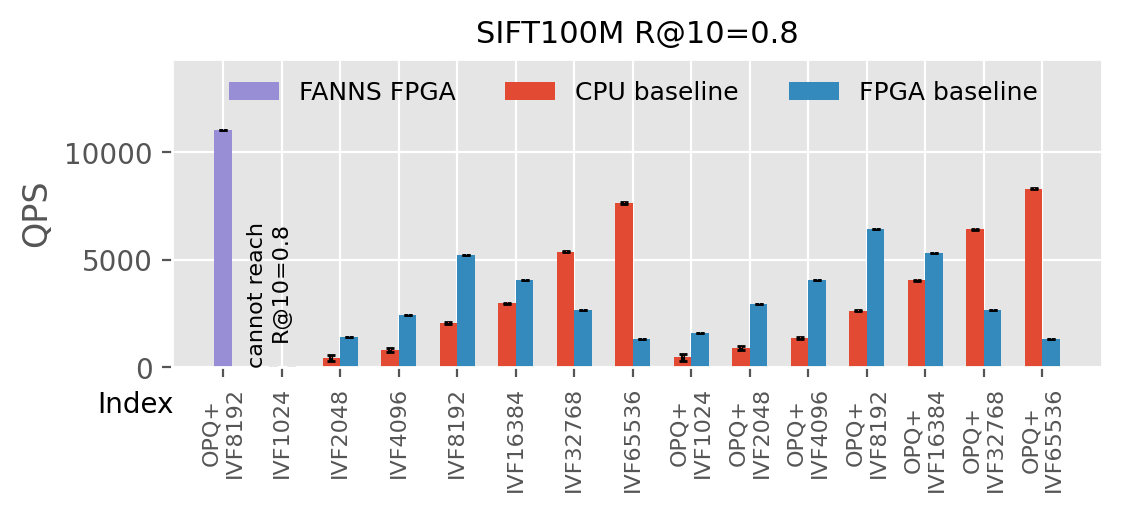}
  \end{subfigure}
  \begin{subfigure}[b]{0.49\linewidth}
    \includegraphics[width=\linewidth]{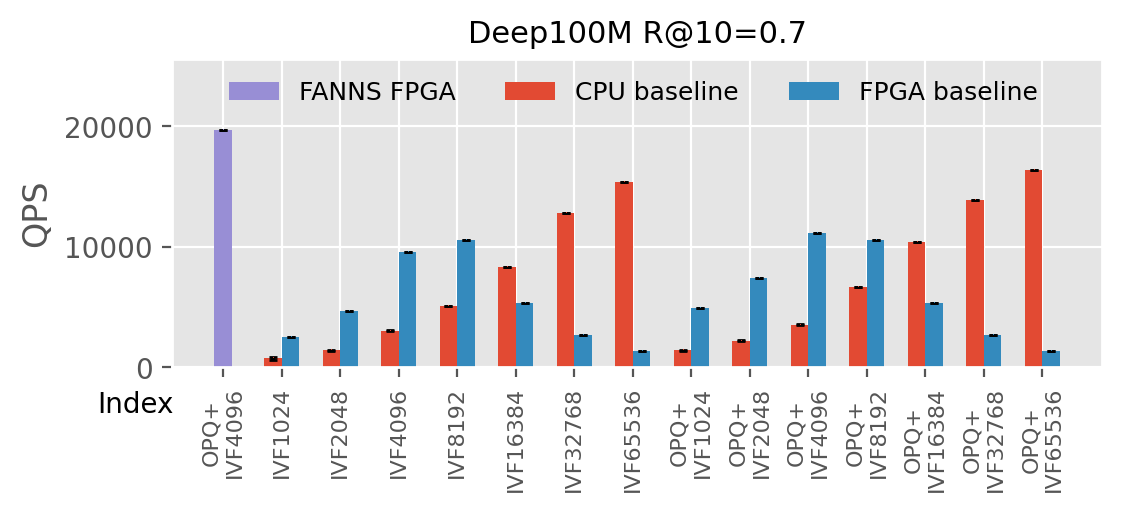}
  \end{subfigure}
  
  \begin{subfigure}[b]{0.49\linewidth}
    \includegraphics[width=\linewidth]{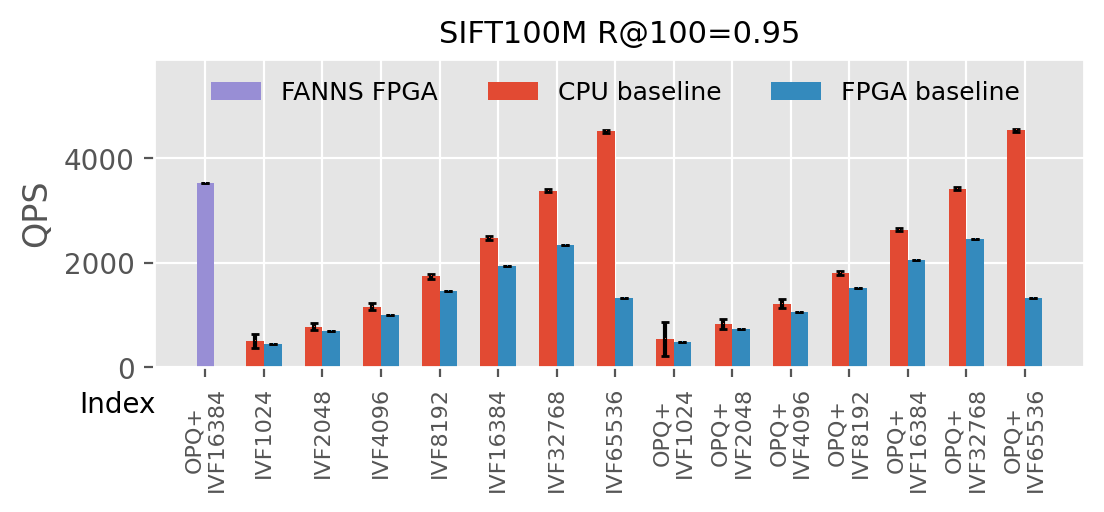}
  \end{subfigure}
  \begin{subfigure}[b]{0.49\linewidth}
    \includegraphics[width=\linewidth]{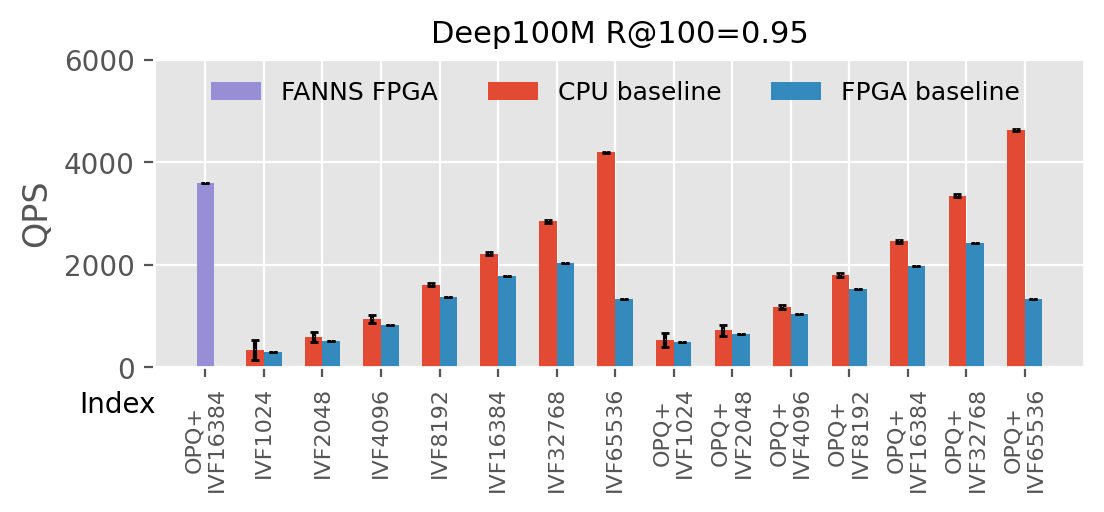}
  \end{subfigure}
  
  \caption{The throughput comparison between \textit{FANNS}-generated accelerators and CPU/FPGA baselines on the SIFT dataset (first column) and the Deep dataset (second column) under various recall requirements (three rows). }
  \label{fig_fanns:fpga_cpu_throughput}
\end{figure*}

\begin{figure}
  \centering

  \begin{subfigure}[b]{0.4\linewidth}
    \includegraphics[width=\linewidth]{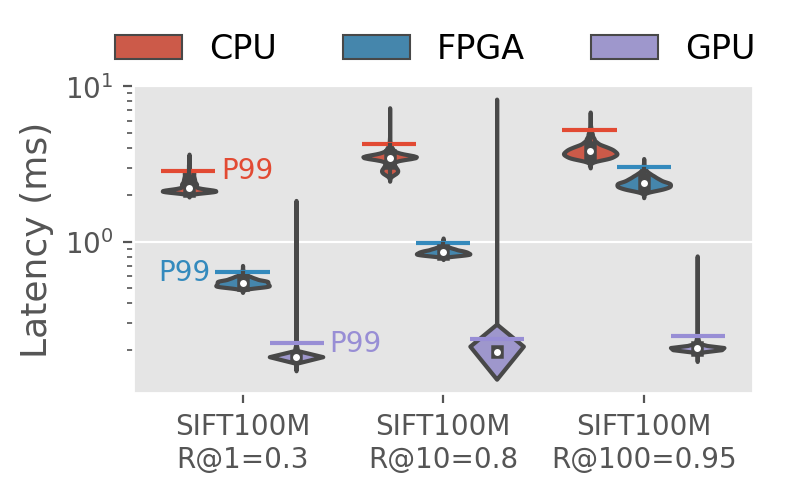}
  \end{subfigure}
  \begin{subfigure}[b]{0.4\linewidth}
    \includegraphics[width=\linewidth]
    {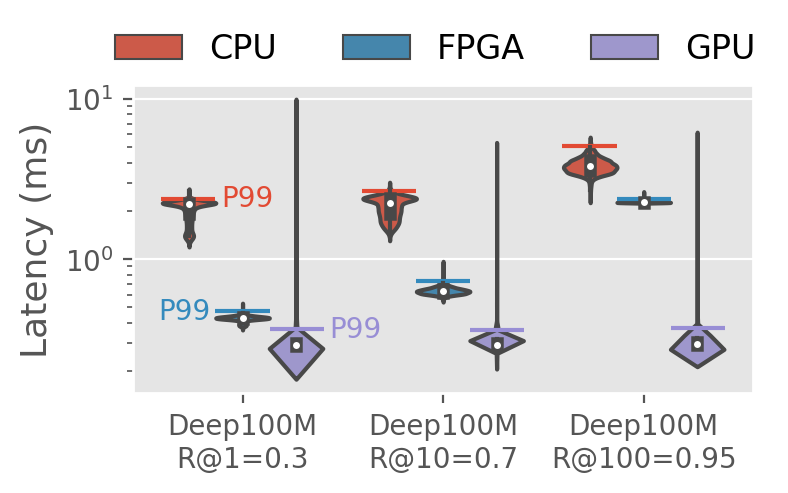}
  \end{subfigure}

  \caption{Latency of single-node CPU, GPU, and FPGA.}
  \label{fig_fanns:fpga_cpu_latency}
\end{figure}

\subsubsection{Offline Batch Processing}
\hfill \\
I first compare the throughput (QPS) between \textit{FANNS} and the CPU/FPGA baselines in Figure~\ref{fig_fanns:fpga_cpu_throughput}.
The throughput experiments have no latency constraints, thus allowing query batching (size = 10K) to report the highest QPS.
\textit{FANNS} reports $1.3\sim23.0\times$ QPS as the baseline FPGA designs and $0.8\sim37.2\times$ as the CPU. 
As FPGAs have two orders of magnitude lower flop/s than GPUs, GPU still achieves significantly higher QPS than FPGAs ($5.3\sim22.0\times$), although the FPGAs show comparable latency and better scalability, as I will present later.
Several observations from the throughput experiments include:

\textbf{First, customizing the FPGA per use case is essential to maximize performance.} Although I have done the best to design the parameter-independent FPGA baseline, the \textit{FANNS}-generated accelerators are customized for a target recall requirement on a given dataset, thus showing significant QPS improvements and latency reductions compared with the baseline designs.

\textbf{Second, the performance model can effectively predict the accelerator performance.} 
By comparing the actual FPGA performance in Figure~\ref{fig_fanns:fpga_cpu_throughput} and the \textit{FANNS}-predicted performance in all experiments, I find the actual QPS can reach 86.9\%$\sim$99.4\% of the predicted performance.
In the case when the generated accelerators can achieve the target frequency, the actual performance is virtually the same as the predicted one.
When the target frequency cannot be met due to the nondeterministic FPGA placement and routing algorithm, the achieved performance drops almost proportionally with the frequency.

\textbf{Third, FPGA performance is closely related to $K$, as instantiating longer priority queues consumes a lot of resources.} 
To match the performance of Stage PQDist that contains many compute PEs, \textit{FANNS} needs to instantiate many hardware priority queues in Stage SelK. 
But the resource consumption per queue is roughly linear to the queue size $K$. As $K$ grows, more resource consumption on queues results in fewer resources for other stages and leads to overall lower performance. This explains why the FPGA performance is slightly surpassed by the CPU when $K=100$.

\textbf{Fourth, picking appropriate algorithm parameters is essential for performance, regardless of hardware platforms.}
The performance numbers of the CPU and the baseline FPGA designs show that the QPS difference can be as significant as one order of magnitude with different parameters. 

\subsubsection{Online Query Processing and Scalability}

To support low-latency online query processing, I integrate \textit{FANNS} with a hardware TCP/IP stack~\cite{100gbps}, such that clients can query the FPGA directly, bypassing the host server. I also compare system scalability of GPUs and FPGAs in this scenario. As the network stack also consumes hardware resources, I rerun the \textit{FANNS} performance model to generate the best accelerators. I assume the queries already arrive at the host server for CPU and GPU baselines, while for FPGAs, the measurements include the network latency (around five $\mu$s RTT). 

\textbf{FPGA achieves 2.0$\sim$4.6$\times$ better P95 latency than the best CPU baseline.} Figure~\ref{fig_fanns:fpga_cpu_latency} captures the latency distributions~\cite{hoefler2015scientific} of each type of hardware. Although showing high tail latency, GPUs still achieve lower median and P95 latency than FPGAs and CPUs due to the much higher flop/s and bandwidth.
The FPGA shows much lower latency variance than its counterparts, thanks to the fixed accelerator logic in FPGAs.

\textbf{FPGAs achieves 5.5$\times$ and 7.6$\times$ speedup over GPUs in median and P95 latency in an eight-accelerator setup, as shown in Figure}~\ref{fig_fanns:first_page_overview}.
I run the prototype scale-out experiments on a cluster of eight FPGAs. 
Each FPGA or GPU holds a 100-million vector partition, running the same index (\textit{nlist}=8192, $m$=16) to achieve R@10=80\%.
For FPGAs, I use a CPU server that sends queries to all FPGAs and aggregates the results. 
For GPUs, Faiss natively supports multi-GPU workload partitioning. 
FPGAs achieve better scalability thanks to their stable latency distribution, as shown in Figure~\ref{fig_fanns:fpga_cpu_latency}. In contrast, GPUs experience long tail latencies, thus a multi-GPU query is more likely to be constrained by a slow run.

\begin{figure}[t]
	\centering
  \includegraphics[width=0.6\linewidth]{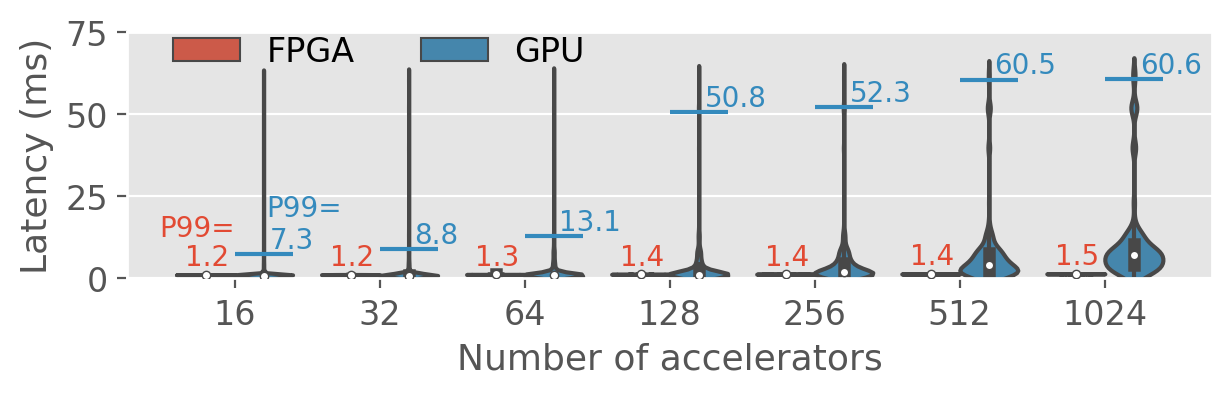}
  \caption{{Estimated latency on large-scale deployments.}}
  \label{fig_fanns:latency_simulation}
\end{figure}

\textbf{FPGAs are expected to exhibit increasing speedup over GPUs as the search involves more (hundreds or thousands of) accelerators}.
To extrapolate latency trends beyond eight accelerators, I estimate the latency distribution of large-scale vector search using the following method. The query latency consists of search and network components. 
I record search latencies of 100K queries on a single FPGA/GPU using the same parameters as the above paragraph. For a distributed query, I randomly sample \textit{N\textsubscript{accelerator}} latency numbers from the latency history and use the highest number as the search latency. 
I assume the implementation of broadcast/reduce communication collectives follows a binary tree topology. Subsequently, I apply LogGP~\cite{culler1993logp, alexandrov1995loggp} to model the network latency, using previously reported values measured for InfiniBand using MPI~\cite{hoefler2014energy, hoefler2007low}: the maximum communication latency between two endpoints is 6.0 $\mu$s; the constant CPU overhead for sending or receiving a single message is 4.7 $\mu$s; and the cost per injected byte at the network interface is 0.73 ns/byte. I assume merging partial results from two nodes takes 1.0 $\mu$s.
As shown in Figure~\ref{fig_fanns:latency_simulation}, FPGA's P99 latency speedup over GPUs increases from 6.1$\times$ with 16 accelerators to 42.1$\times$ with 1024 accelerators, thanks to the low search latency variance on FPGAs. 

\section{Related Work}


\textit{FANNS} is the first hardware-algorithm co-design framework for vector search. 
I now introduce related works about vector search on modern hardware.
The most popular GPU-accelerated ANN library so far is Faiss developed by Meta~\cite{johnson2019billion}. 
The academia has also built several GPU-based ANNS systems~\cite{wieschollek2016efficient, chen2019robustiq, chen2019vector}. 
Google researchers accelerate exact nearest neighbor search on TPUs and show great performance on one-million-vector datasets~\cite{chern2022tpu}.
\citet{lee2022anna} propose a fixed ASIC design for PQ supporting arbitrary algorithm parameters.
\citet{zhang2018efficient} implements a variation of the PQ algorithm on FPGAs and focuses on compressing large IVF indexes to fit it to BRAM. 
\citet{ren2020hm} stores full-precision vectors in non-volatile memory to scale up graph-based ANNS, while on-disk ANNS has to be careful with I/O cost~\cite{chen2021spann, jayaram2019diskann, lejsek2008nv}.

\section{Conclusion}

Commercial search engines are driven by a large-scale vector search system operating on a massive cluster of servers.
I introduce \textit{FANNS}, a scalable FPGA vector search framework that co-designs hardware and algorithm.
The eight-FPGA prototype demonstrates 7.6$\times$ improvement in P95 latency compared to eight GPUs, with the presented performance model indicating that this advantage will only increase as more accelerators are employed. 
The remarkable performance of \textit{FANNS} lays a robust groundwork for future FPGA integration in data centers, with potential applications spanning large-scale search engines, LLM training, and scientific research in fields such as biomedicine and chemistry.


\chapter{Falcon: Delayed-Synchronization Traversal for Graph-based Search}
\label{chap:falcon}

This chapter focuses on algorithm-hardware co-design for graph-based vector search, another major category of vector search algorithms, complementing the quantization-based retrieval acceleration discussed in Chapter~\ref{chap:fanns}.

\section{Introduction}
\label{sec_falcon:intro}


Among various ANN search algorithms, \textit{graph-based vector search (GVS)} algorithms are particularly popular due to their high search performance and quality~\cite{li2019approximate, malkov2018efficient, fu2017fast}, with the latter measured by recall, the percentage of true nearest neighbors correctly identified by the search. The key idea of GVS is to construct a proximity graph on database vectors: each vector is a node, and similar vectors are linked by edges. During a search, the query vector is compared to a subset of database vectors by iteratively traversing the graph using best-first-search (BFS), which greedily selects the best candidate node to evaluate for each search iteration.

Given the rising adoption of ANN search in online systems, an ideal GVS system should \textit{achieve low search latency for real-time query batches}, while being cost- and energy-efficient. 
For example, in a RAG system, the LLM serving engine may perform on-demand retrievals in the middle of the generation process~\cite{borgeaud2022improving, khandelwal2019generalization, jiang2023active, jeong2024adaptive}. These retrievals typically involve small query batches or even individual queries because (a) the sequence batch sizes are constrained by accelerator memory capacity~\cite{yu2022orca, kwon2023efficient}, and (b) these sequences can trigger retrievals asynchronously due to their different generation contexts~\cite{jiang2023active, jeong2024adaptive, trivedi2022interleaving}.
Consequently, high search latency not only prolongs the overall generation time but also leads to idleness of the inference accelerators such as GPUs and TPUs, which have to wait for search results before proceeding~\cite{jiang2025rago, jiang2024piperag, zhang2024accelerating}.

However, reducing GVS latency remains challenging due to limitations imposed by both existing hardware architectures (CPUs and GPUs) and inherent difficulty of parallelizing graph traversals.
CPUs and GPUs operate on a time-multiplexed basis, executing GVS operations --- such as database vector fetching, distance computation, and result insertion --- sequentially, with only limited overlap between them, even if data prefetching is applied.
Thus, given the classic BFS traversal algorithm~\cite{malkov2018efficient, fu2017fast}, query latency accumulates over multiple iterations as the search progresses through each operator.  
While improving throughput of queries per second (QPS) is straightforward by parallelizing execution across a large batch of queries, reducing search latency for a single query is significantly more challenging.
This is because, when implementing intra-query parallelization, the synchronization overhead among CPU cores or GPU streaming multi-processors~\cite{lagrone2011set, zhang2020study} is disproportionately high relative to a single iteration of graph traversal, which typically takes only microseconds and involves just dozens of distance computations.

While previous research has explored hardware accelerator designs for GVS based on FPGA prototyping~\cite{zeng2023df, peng2021optimizing}, these approaches have three main limitations.
Firstly, they only support the Hierarchical Navigable Small World (HNSW) graph.
While HNSW is widely used today, more efficient graph construction algorithms are emerging~\cite{malkov2014approximate, malkov2018efficient, fu2017fast, zhao2023towards, zuo2023arkgraph, lu2021hvs, peng2023efficient}. For example, the Navigating Spreading-out Graph (NSG)~\cite{fu2017fast}, with additional time invested in index construction, can achieve better recall than HNSW.
Secondly, directly implementing the software-oriented BFS algorithm on these accelerators results in sub-optimal search latency, because it significantly under-utilizes the accelerators, as I will further explain in conjunction with the hardware designs.
Thirdly, existing architectures are mainly throughput-oriented and either do not support~\cite{peng2021optimizing} or suboptimally support intra-query parallelism for low-latency search~\cite{zeng2023df}.

\textit{To achieve low-latency GVS while supporting various graphs, both algorithm-level and hardware-level optimizations are essential.}
To this end, I propose a hardware-algorithm co-design solution including \textit{Falcon}, a specialized GVS accelerator, and \textit{delayed-synchronization traversal (DST)}, an accelerator-optimized graph traversal algorithm designed to simultaneously improve accelerator search performance and recall.

\textit{Falcon is an in-memory GVS accelerator with four key features.}
Firstly, Falcon involves fast distance computations and sorting units, and minimizes off-chip memory accesses by using an on-chip Bloom filter to track visited nodes. 
Secondly, Falcon supports both intra-query parallelism, utilizing all compute and memory resources to process a single query, and across-query parallelism, handling multiple queries through separate processing pipelines.
Thirdly, Falcon supports general GVS, allowing it to leverage emerging algorithms offering better recall and performance.
Finally, Falcon functions as a networked service with an integrated TCP/IP stack, thus reducing end-to-end service latency by bypassing the accelerator's host server from the communication path. 
%


\textit{Delayed-synchronization traversal (DST) relaxes the greedy graph traversal order to improve accelerator utilization.}
The design of the algorithm is motivated by two key observations.
First, from a system performance perspective, the synchronous and greedy nature of the software-oriented best-first search (BFS) limits the amount of parallelism the accelerator can exploit and thus leads to significant accelerator under-utilization. 
Second, from a traversal-pattern perspective, I found that relaxing the order of candidate evaluations does not compromise recall. 
Building on these observations and drawing inspiration from label-correcting algorithms for parallel shortest path computation on graphs~\cite{bertsekas1993simple, meyer2003delta}, DST relaxes synchronizations that enforce the greedy traversal order, thereby increasing the amount of parallel workloads that Falcon can handle.
Consequently, DST both reduces search latency by improving accelerator utilization and improves recall by allowing the exploration of search paths that the greedy BFS would otherwise overlook.

I prototype Falcon on FPGAs and evaluate it on various vector search benchmarks across different types of graphs. 
In combination with DST, Falcon achieves up to 4.3$\times$ and 19.5$\times$ lower online search latency and up to 8.0$\times$ and 26.9$\times$ better energy efficiency compared to CPU and GPU-based GVS systems, respectively.
Besides, the proposed DST algorithm outperforms the classic BFS by 1.7$\sim$2.9$\times$ in terms of latency on Falcon and simultaneously improves recall.

The chapter makes the following \textbf{contributions}:
\begin{itemize}
    \item I identify the hardware primitives essential for efficient GVS, design Falcon, a specialized GVS accelerator, prototype it on FPGAs, and expose it as a networked service.
    \item I analyze the graph traversal patterns of best-first search and propose DST, an accelerator-optimized graph traversal algorithm that reduces GVS latency by relaxing traversal order.
    \item I evaluate Falcon and DST across diverse graphs and datasets, demonstrating their high performance and energy efficiency.
\end{itemize}

\section{Background and Motivation}
\label{sec_falcon:background}

In this section, I introduce GVS algorithms~(\S\ref{sec_falcon:background_gvs}), and discuss the limitations of existing processors for online GVS~(\S\ref{sec_falcon:background_limitataions}).


\subsection{Best-first Search (BFS) for Query Processing.}
\label{sec_falcon:background_gvs}

While various graph construction algorithms exist~\cite{malkov2014approximate, malkov2018efficient, fu2017fast, zhao2023towards, zuo2023arkgraph, lu2021hvs}, they all handle ANN queries using the classic best-first search (BFS) algorithm. 

\textit{BFS traverses a graph by greedily evaluating the best candidate node in each search iteration.}
As illustrated in Algorithm~\ref{algo:bfs}, BFS begins by adding the typically fixed entry node \(p\) to the candidate queue \(C\), which stores nodes for potential exploration; the result queue \(R\), which holds the nearest neighbors found so far; and the visited set \(Visited\), which tracks nodes that have already been visited. It then searches on the graph iteratively as long as there is at least one candidate that is reasonably close to the query \(q\). Here, reasonably close means that the minimum distance from the candidates in \(C\) to \(q\) is less than the maximum distance of the nodes currently in \(R\). The algorithm then pops and evaluates the best candidate \(c\) by visiting all of its neighbors. Each neighbor that has not been visited is added to the visited set, the candidate queue, and the result queue, ensuring that no node is processed more than once. Following the exploration of neighbors, \(R\) is adjusted to maintain only the closest $l$ elements. 

The maximum size of the result queue \(l\) (\(k \leq l\)) controls the trade-off between search performance and quality. A larger \(l\) increases the threshold distance for considering a candidate, thereby expanding the number of candidate nodes evaluated during the search. Although visiting more nodes increases the likelihood of finding the true nearest neighbors, it also leads to higher search latency.

\begin{algorithm}[t]
\caption{Best-First Search (BFS)}
\label{algo:bfs}
\begin{algorithmic}[1]
\Require graph $G$, entry node $p$, query vector $q$, maximum result queue size $l$, number of results to return $k$ ($k \leq l$)
\Ensure $k$ approximate nearest neighbors of query $q$

\State $C \gets \{p\}, R \gets \{p\}, Visited \gets \{p\}$ 
\While{$C \neq \emptyset$ \textbf{and} $\Call{Min}{C.dist} \leq \Call{Max}{R.dist}$}
    \State $c \gets \Call{Extract-Min}{C}$ \Comment{pop the nearest candidate}
    \For{all neighbors $n$ of $c$}
        \If{$n \notin Visited$}
            \State $dist \gets  \Call{Compute-Dist}{q, n} $
            \State $Visited.\text{add}(n), C.\text{add}(n,dist), R.\text{add}(n,dist)$ 
        \EndIf
    \EndFor
    \State $R.\text{resize}(l)$ \Comment{keep only the closest $l$ elements}
\EndWhile
\State \textbf{return} $\Call{Sort}{R}[:k]$ \Comment{return the first $k$ elements}
\end{algorithmic}
\end{algorithm}

\subsection{Limitations of Existing Processors for GVS}
\label{sec_falcon:background_limitataions}

Existing GVS systems have been mostly CPU-based, and recent research has explored their deployments on GPUs and FPGAs. 
However, current solutions remain sub-optimal for latency-sensitive online vector search.

\subsubsection{Search on CPU} 
CPUs have several limitations in online GVS systems.
Firstly, CPUs operate on a time-multiplexing basis, executing GVS operators such as fetching, computing, and insertion sequentially, with only limited timeline overlaps due to data prefetching. This sequential processing leads to cumulative search latency for each operator, in contrast to Falcon's design as I will introduce in this chapter.
Secondly, software implementations typically employ a byte array to track visited nodes for each query~\cite{malkov2018efficient, fu2017fast}, resulting in additional read and write operations per visited node.
Thirdly, CPUs struggle with random memory accesses to fetch vectors, which are typically less than 1 KB, and to update the visited arrays (one byte per read or write).




\subsubsection{High-throughput GVS on GPUs}
GPUs are known for their massive parallelism, featuring thousands of cores grouped into many streaming multi-processors~\cite{choquette20213}. 
Thus, GPUs are well-suited for high-throughput GVS applications, as evidenced by recent studies~\cite{groh2022ggnn, zhao2020song}. 
However, GPUs exhibit two shortcomings for online GVS.
Firstly, GPUs show much higher GVS latency than CPUs as shown in the evaluation, because the limited amount of workload per search iteration makes it infeasible to effectively parallelize one query across multiple streaming multi-processors.
Secondly, the scale of graphs that GPUs can efficiently serve is constrained by memory capacity.
GPUs typically use either HBM or GDDR memory, which offers high bandwidth but less capacity compared to DDR memory. Although utilizing CPU-side memory is a potential option, search performance remains a concern: the throughput of fast CPU-GPU interconnects like the NVLink in NVIDIA Grace Hopper~\cite{gracehopper} is still an order of magnitude lower than that of GPU memory.


\subsubsection{Specialized GVS Accelerators}
Two recent studies~\cite{zeng2023df, peng2021optimizing} implemented HNSW, a popular GVS algorithm, on FPGAs. Peng et al.~\cite{peng2021optimizing} present a first implementation, and Zeng et al.~\cite{zeng2023df} further optimized the design by supporting data prefetching and multi-FPGA search. However, they are still not optimal for online GVS for the following reasons. 

Firstly, supporting only one type of graph (HNSW) may be inadequate given the rapid emergence of efficient GVS algorithms~\cite{malkov2014approximate, malkov2018efficient, fu2017fast, zhao2023towards, zuo2023arkgraph, lu2021hvs}. 
For example, NSG~\cite{fu2017fast}, given longer graph construction time, can achieve better performance-recall trade-offs than HNSW. 
Specializing the accelerator for HNSW~\cite{peng2021optimizing, zeng2023df} restricts the accelerator's flexibility in supporting various types of graphs: HNSW has a unique multi-level architecture, while the vast majority of graphs in GVS do not incorporate a leveled structure.

Secondly, applying the software-friendly BFS on the accelerators leads to sub-optimal search performance. This is because BFS can cause significant under-utilization of the accelerators, as I will specify in \S\ref{sec_falcon:dst}.

Thirdly, although Zeng et al.~\cite{zeng2023df} supports intra-query parallelism, an improvement over Peng et al.~\cite{peng2021optimizing}, the parallel strategy remains suboptimal. 
Specifically, the method of partitioning the graph into several sub-graphs and searching all sub-graphs in parallel~\cite{zeng2023df} leads to significantly more nodes being visited per query compared to traversing a single, larger graph, as I will explain further in \S\ref{sec_falcon:accelerator_intra_inter}.


\section{Falcon: Accelerator Design}
\label{sec_falcon:accelerator}

\begin{figure*}[t]
  \centering
  \includegraphics[width=1.0\linewidth]{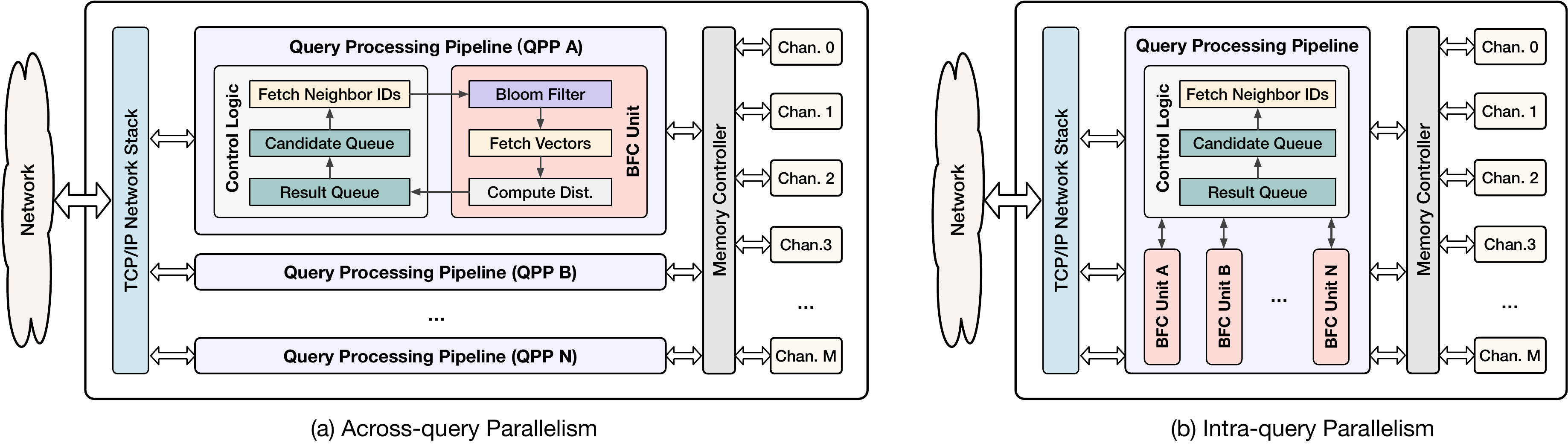}
  \caption{Falcon overview. It has two architecture variants supporting across-query and intra-query parallelisms.}
  \label{fig_falcon:accelerator_overview}
\end{figure*}

I present Falcon, a low-latency GVS accelerator that I prototype on FPGAs but also applicable to ASICs~(\S\ref{sec_falcon:accelerator_overview}).
Falcon consists of various high-performance hardware processing elements~(\S\ref{sec_falcon:accelerator_pe}).
It has two variants supporting across-query and intra-query parallelisms, optimized for processing batches of queries and individual queries, respectively~(\S\ref{sec_falcon:accelerator_intra_inter}). 
The accelerator is directly accessible as a networked service and supports various types of graphs~(\S\ref{sec_falcon:accelerator_network}).

\subsection{Design Overview}
\label{sec_falcon:accelerator_overview}

\textbf{Accelerator components.} Figure~\ref{fig_falcon:accelerator_overview} shows Falcon, a spatial dataflow accelerator for GVS. Each \textit{query processing pipeline (QPP)} handles one query at a time, containing both control logics and \textit{Bloom-fetch-compute (BFC) units}. Falcon is composed of various processing elements (PEs) interconnected via FIFOs, including systolic priority queues for storing candidate nodes and search results, Bloom filters to avoid revisiting nodes, and compute PEs for efficient distance calculations between query vectors and database vectors.

\textbf{Parallel modes.} Falcon has two variants that support \textit{across-query parallelism} and \textit{intra-query parallelism}, as shown in Figure~\ref{fig_falcon:accelerator_overview}(a) and (b), respectively. 
Across-query parallelism processes different queries across QPPs, while intra-query parallelism minimizes per-query latency by utilizing all compute and memory resources (multiple BFC units) to process one query at a time. 

\textbf{Differences compared to existing accelerators.} Falcon distinguishes itself from previous GVS accelerators~\cite{zeng2023df, peng2021optimizing} in four aspects.
Firstly, Falcon utilizes on-chip Bloom filters to manage the list of visited nodes, thereby minimizing memory accesses~(\S\ref{sec_falcon:accelerator_pe}). 
Secondly, Falcon's intra-query parallel design utilizes all compute and memory resources to traverse a single graph rather than partitioned sub-graphs~(\S\ref{sec_falcon:accelerator_intra_inter}). 
Thirdly, Falcon supports various GVS algorithms, rather than being limited to a specific one such as HNSW, allowing it to benefit from emerging algorithms that offer improved search quality and performance~(\S\ref{sec_falcon:accelerator_network}).
Finally, Falcon employs the proposed accelerator-optimized traversal algorithm that significantly reduces vector search latency~(\S\ref{sec_falcon:delayed_sync_traversal}).

\subsection{Hardware Processing Elements}
\label{sec_falcon:accelerator_pe}

I now introduce the main types of PEs in the order of their appearance in Algorithm ~\ref{algo:bfs}.

\subsubsection{Priority Queues}

\begin{figure}[t]
	\centering
  \includegraphics[width=0.65\linewidth]{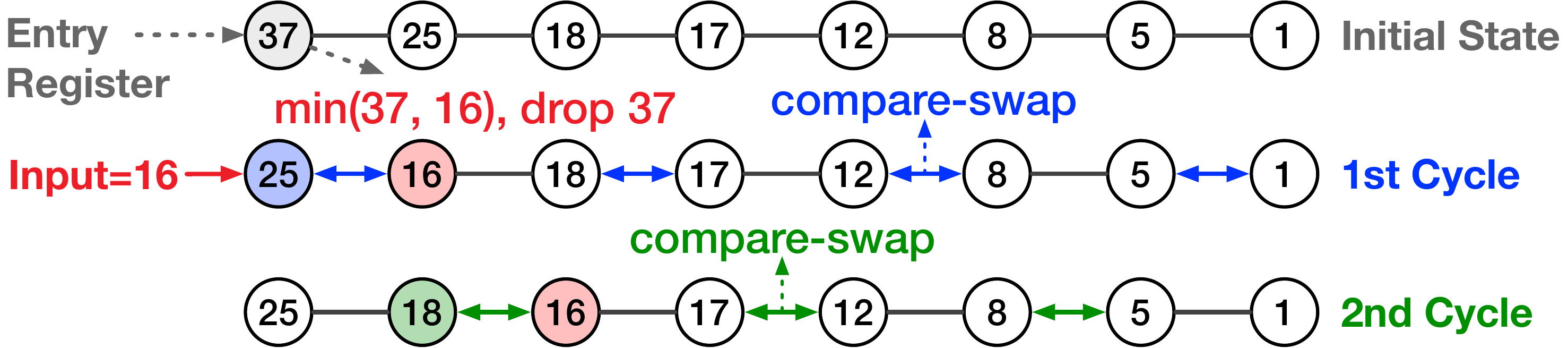}
  \caption{A systolic priority queue with $s=8$ elements.}
  \label{fig_falcon:priority-queue}
\end{figure}

I implement the systolic priority queue architecture~\cite{huang2014scalable, leiserson1979systolic} for the candidate and result queues in Algorithm~\ref{algo:bfs}. As shown in Figure~\ref{fig_falcon:priority-queue}, a systolic priority queue is a register array of $s$ elements interconnected by $s-1$ compare-swap units. It enables high-throughput input ingestion of one insertion per two clock cycles by comparing and swapping neighboring elements in parallel in alternating odd and even cycles. The queue can be sorted in $s-1$ cycles. 


\subsubsection{Bloom Filters}

\begin{figure}[t]
	\centering
  \includegraphics[width=0.7\linewidth]{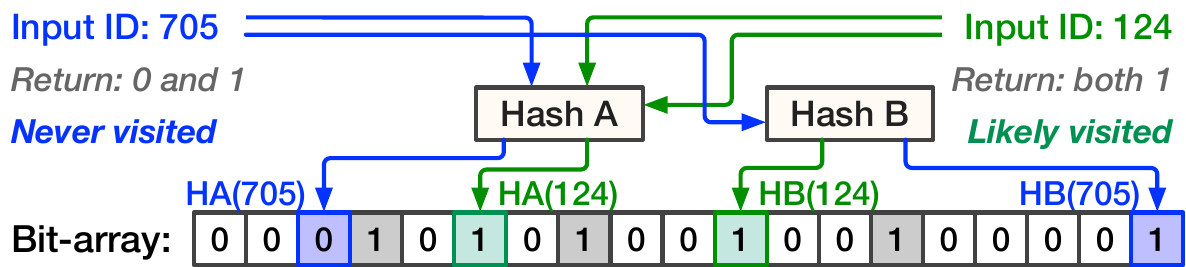}
  \caption{A Bloom filter for visited nodes filtering with $h=2$.}
  \label{fig_falcon:bloom}
\end{figure}

Once the candidate queue pops a candidate to be explored, the next step is to check whether each of the candidate's neighbors is already visited. 

Previous software and specialized hardware implementations either maintain a visited array or a hash table, but neither is ideal for Falcon.
For example, software-based implementations~\cite{malkov2018efficient, fu2017fast} maintain an array with a length as the number of nodes in the graph. Node IDs are used as the array addresses to access the visited tags. However, this approach leads to extra memory accesses, requiring one read operation per check and one extra write operation to update the array for unvisited nodes.
Zeng et al.~\cite{zeng2023df} developed on-chip hash tables as part of the accelerators to track the visited nodes to avoid off-chip memory accesses. Each entry of the hash table stores up to four visited node IDs. However, given the limited on-chip SRAM, it is unlikely to instantiate large hash tables, and thus collisions would appear during the search. A collision would not only lead to redundant node visits, but those visited nodes will be inserted into the candidate and result queues repetitively, thus eventually degrading recall.  

Falcon, in contrast to existing solutions, adopts on-chip Bloom filters to track visited nodes. A Bloom filter is a space-efficient probabilistic data structure designed to test whether an element is a member of a set, e.g., determining whether a node has been visited based on its ID. As shown in Figure~\ref{fig_falcon:bloom}, a Bloom filter uses multiple ($h$) hash functions to map each input to several positions in a $b$-bit array. To check if a node has been visited, the same hash functions are used to check the status of these specific positions: if any of the bits are not set, the node is definitely not visited; if all are set, the node is highly likely visited (but not guaranteed, a scenario known as false positive). Given $m$ inserted elements, the false positive rates can be calculated by \( \left(1 - e^{-\frac{{hm}}{{b}}}\right)^h \)~\cite{bloom1970space}.

\textit{Compared to hash tables, Bloom filters are significantly more space efficient for identifying visited nodes}. For example, instantiating a hash table with 1K slots for 4-byte node IDs requires 32Kbit SRAM. 
Using a chaining strategy to resolve hash collisions~\cite{mehta2004handbook}, where collided elements are moved to DRAM, the collision probability for a new incoming node ID is as high as 63.2\% when 1K nodes have already been visited.
In contrast, using the same amount of SRAM, a Bloom filter can provide 32K slots. With an equivalent number of nodes visited, the false positive rate for a new node ID is only 3.0\% and 0.07\% using a single hash function and three hash functions, respectively.
As I will show in the evaluation, the very few false positives, meaning that an unvisited node is reported as visited, would not visibly degrade recall. This is because a well-constructed graph typically offers multiple paths from the query vector to the nearest neighbors, mitigating the effects of these very few false positives.

Falcon implements Bloom filters in the following manner. Both the number of hash functions and the size of the Bloom filters are configurable. Currently, Falcon uses three Murmur2 hashes~\cite{murmur} per filter. These hash functions are computed in parallel, and each hash function pipeline can yield a hash code every clock cycle. The size of the bitmap is set to 256Kbit, which translates to low false positive rates --- only one in 600K for 1K visited nodes.


\subsubsection{Fetching Vectors}
Upon identifying nodes to visit, the next step is reading the vectors for each node.

Falcon optimizes bandwidth utilization by pipelining vector fetches. Rather than waiting for the first vector to return before issuing a second read, each fetch unit pipelines up to 64 read requests (configurable), thus improving read throughput by hiding the latency associated with memory and the memory controller. The data width of the FIFO connecting a fetch unit to the memory controller is set to 64 bytes.

\subsubsection{Distance Computations} 
Each vector fetch unit is connected to a compute PE that calculates L2 distances between queries and database vectors. A compute PE instantiates multiple multipliers and adders and pipelines different compute stages, such that the compute throughput can match the maximum read throughput of a vector fetch unit.

\subsection{Intra-query and Across-query Parallelism}
\label{sec_falcon:accelerator_intra_inter}

While across-query parallelism for batched queries can be straightforwardly implemented by instantiating multiple query processing pipelines (QPP) on the accelerator, there are two design choices for intra-query parallelism, which aim to minimize latency for individual queries.
One option involves adopting the architecture of across-query parallelism by partitioning the dataset into multiple subsets, querying each subset with an individual QPP, and aggregating the results, as Zeng et al.~\cite{zeng2023df} described.


Alternatively, the choice of this work is to \textit{speed up the traversal of a single graph} by instantiating multiple BFC units in a single QPP to utilize all the compute and memory resources for a single query (Figure~\ref{fig_falcon:accelerator_overview}(b)).
This decision stems from the observation that traversing several sub-graphs significantly increases the total amount of workload per query compared to traversing a single graph.

\begin{figure*}[t]
  \centering
  \includegraphics[width=0.5\linewidth]{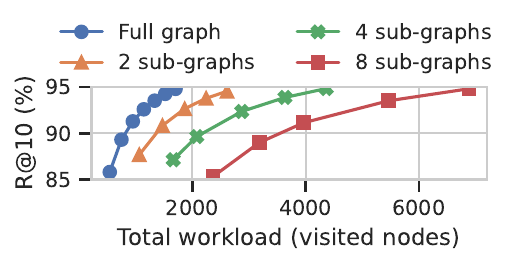}
  \caption{Traversing one graph versus several sub-graphs.}
  \label{fig_falcon:subgraph} 
\end{figure*}

Figure~\ref{fig_falcon:subgraph} shows that, to achieve a recall of  $R@10=90\%$ on the SPACEV natural language embedding dataset~\cite{spacev}, the total number of visited nodes per query when using eight subgraphs is 4.2$\times$ of that for a single graph. Thus, the maximum speedup (assuming perfect load balancing) that eight partitions and eight QPPs can achieve is only 1.9$\times$ that of traversing a single graph with one QPP.

When traversing a single graph using intra-query parallelism, Falcon leverages its direct message-passing mechanism via FIFOs to enable low-overhead, fine-grained task dispatching among different BFC units.
This is a significant architectural advantage compared to CPUs and GPUs, where synchronization overhead among CPU cores or GPU streaming processors~\cite{lagrone2011set, zhang2020study} is too high compared to a single iteration of graph traversal, which only takes microseconds typically involving dozens of distance computations.

\begin{figure*}[t]
  \centering
  \includegraphics[width=1.0\linewidth]{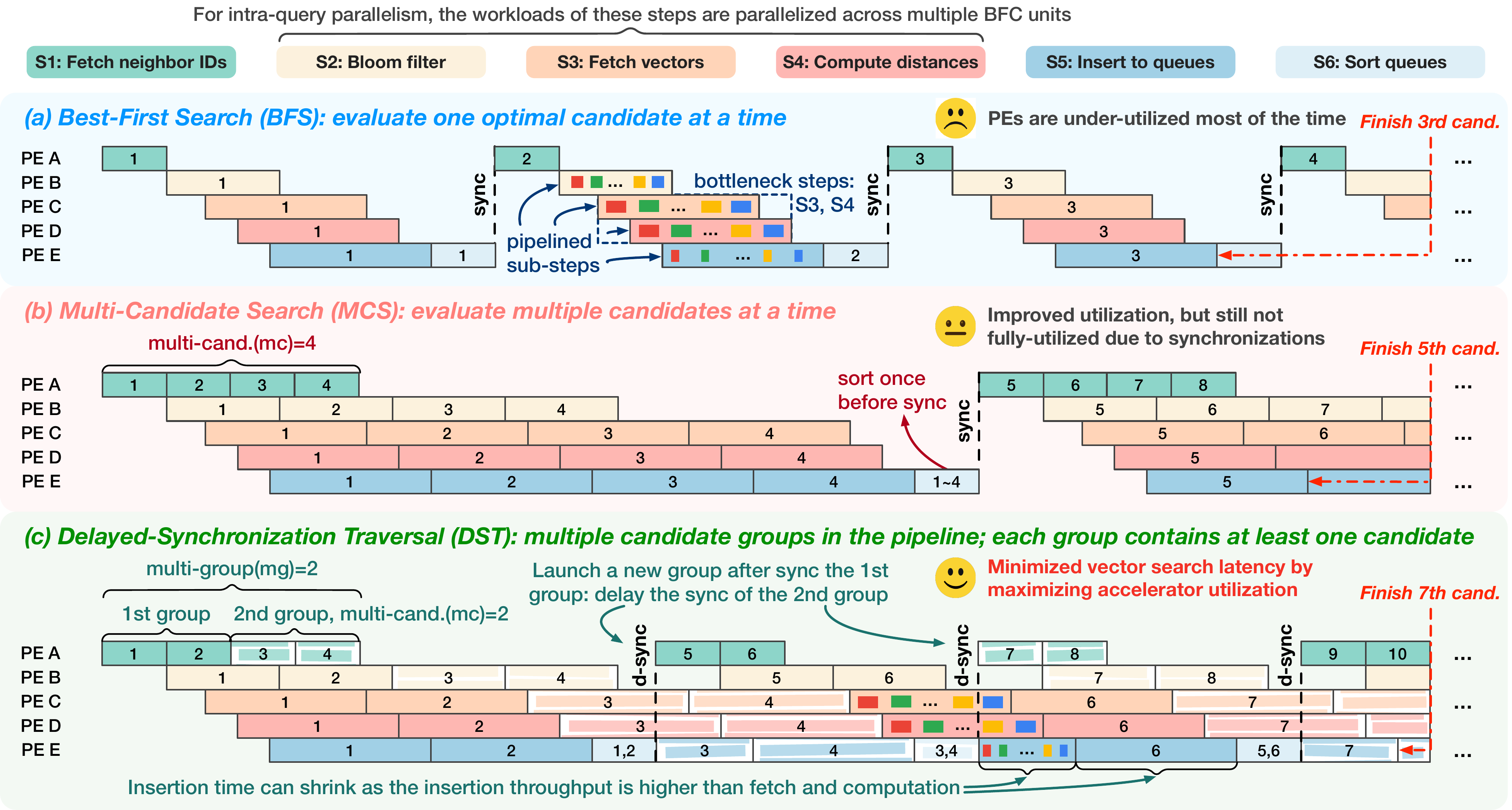}
  \caption{The proposed Delayed-Synchronization Traversal (DST) reduces vector search latency by maximizing accelerator utilization. It delays synchronizations and allows multiple candidates to be evaluated simultaneously in the processing pipeline.}%
  \label{fig_falcon:timeline}
\end{figure*}

\subsection{Accelerator-as-a-Service}
\label{sec_falcon:accelerator_network}

\subsubsection{Network Stack Integration} 
Vector search systems are typically wrapped as services for real-time LLM serving or recommender systems. To minimize service latency, I integrate a TCP/IP network stack~\cite{100gbps} into Falcon, as shown in Figure~\ref{fig_falcon:accelerator_overview}. This integration allows Falcon to function as a networked accelerator service in data centers~\cite{putnam2014reconfigurable, fowers2018configurable}, facilitating direct communication with clients. This approach differs from common setups where the accelerator operates as a PCIe-based operator offloading engine, which involves additional latency including CPU handling requests from the network, accelerator kernel invocation, and data copying between the CPU and the accelerator. 

Compared to CPU and GPU-based services, Falcon can partially overlap communication and query latency: for a batch of queries, it begins processing the first query upon its arrival rather than waiting for the entire batch to be received.

\subsubsection{Supporting Various Graphs} 
Falcon supports arbitrary graphs by representing them with a unified graph format, accommodating common graph elements including nodes, edges, entry nodes, and degrees.
This approach is naturally compatible with the vast majority of graphs~\cite{malkov2014approximate, fu2017fast, zhao2023towards, zuo2023arkgraph}, except for HNSW~\cite{malkov2018efficient} that has a unique multiple-layer structure. The upper layers of HNSW are designed to identify a high-quality entry point into the base layer, which contains all the database vectors --- thus the base layer is comparable to the entire graph in other GVS algorithms~\cite{malkov2014approximate, fu2017fast}. Instead of customizing the accelerator for this case, I prioritize the Falcon's versatility by initiating searches from a fixed entry point on the base layer of HNSW. I found that this approach, without starting from the optimal entry node for each query, would not compromise recall, although more hops might be necessary to reach the nearest neighbors, a finding also supported by existing research~\cite{wang2021comprehensive, lin2019graph}.





\section{Delayed-Synchronization Traversal}
\label{sec_falcon:delayed_sync_traversal}

Realizing the inefficiencies of BFS on Falcon~(\S\ref{sec_falcon:bfs_inefficiency}), I investigate its graph traversal patterns~(\S\ref{sec_falcon:goal_improve_util}) and propose DST, an accelerator-optimized traversal algorithm applicable for both intra- and across-query parallelisms~(\S\ref{sec_falcon:dst}).

\subsection{Inefficiency of BFS on Accelerators}
\label{sec_falcon:bfs_inefficiency}


Figure~\ref{fig_falcon:timeline}(a) visualizes the timeline of BFS on Falcon, where each unique color represents one of the six search steps (S1$\sim$S6), and each PE handles a specific step, except for the priority queues that manage two steps, including distance insertions and sorting (S5 and S6).
Some steps must wait for the previous step to complete: sorting only begins after all distances are inserted into the queues. 
Other steps like filtering, fetching vectors, computing distances, and insertions can partially overlap because these PEs pipeline the execution of sub-steps, where each sub-step involves one of the neighbors of the candidate being evaluated. 
Between search iterations, an \textit{implicit synchronization} between all of the PEs ensures that the queues are sorted, such that the best candidate can be popped for evaluation in the next iteration.

Unfortunately, directly implementing the software-oriented BFS on a GVS accelerator like Falcon can lead to sub-optimal search performance due to under-utilization of the accelerator.
As shown in Figure~\ref{fig_falcon:timeline}(a), only a fraction of the PEs are utilized simultaneously because of the inherently greedy nature of BFS, which processes only one candidate at a time, offering little opportunity for parallelization.


\subsection{Goal: Improving Accelerator Performance through Traversal Algorithm Redesign}
\label{sec_falcon:goal_improve_util}



A natural idea to optimize accelerator performance is to \textit{maximize accelerator utilization by minimizing PE  idleness}. 
Given the imbalanced workloads across different search steps, this approach does not necessitate all PEs to be always active but rather focuses on keeping those PEs involved in bottleneck steps consistently busy. 
In the context of GVS, the bottleneck steps usually include fetching neighbor vectors (S3) and calculating their distances relative to the queries (S4). 

\begin{figure}[t] 
  \centering
  \includegraphics[width=0.7\linewidth]{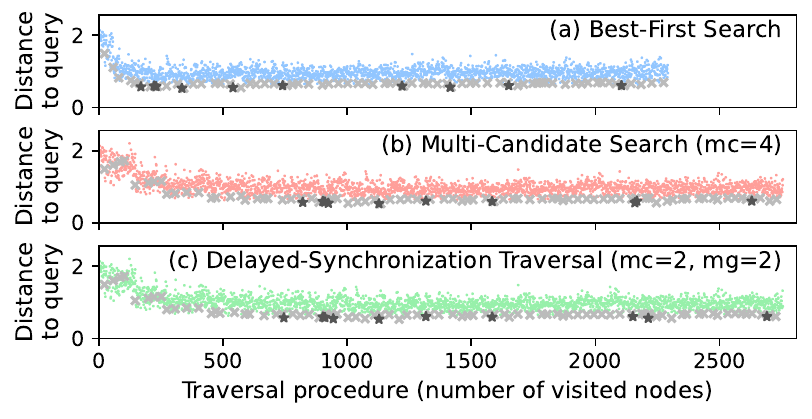}
  \caption{Traversal procedures of BFS, MCS, and DST. Each cross is an evaluated candidate, each dot a visited neighbor node, and each star one of the ten nearest neighbors.}
  \label{fig_falcon:dist_trend}
\end{figure}

\subsubsection{Algorithm-specific Observations.} 
Given the critical role of accelerator utilization in search performance, \textit{is it necessary to strictly follow the BFS traversal order and synchronization pattern to achieve high search quality?}

To answer this question, I examine the traversal patterns of GVS. 
Figure~\ref{fig_falcon:dist_trend}(a) shows the BFS traversal procedure for a sample query on the Deep1M dataset~\cite{babenko2016efficient} using HNSW. Each grey cross represents an evaluated candidate node, colored dots denote its neighbor nodes, and black stars mark the ten nearest neighbors.
Notably, while the node distances to the query decrease at the beginning of the traversal, most subsequent candidates maintain similar distances rather than showing a monotonically decreasing trend --- an observation consistent across queries and datasets.

This observation suggests that \textbf{traversals in GVS do not have to adhere to a strictly greedy approach --- relaxing the traversal order of different candidate nodes should result in comparable search quality}, assuming the same or a similar set of candidates is evaluated.

\subsubsection{Naive Solution: MCS}
Leveraging the intuition above, one straightforward way to improve accelerator utilization is increasing the number of candidates evaluated per iteration, a strategy I term \textit{multi-candidate search (MCS)}. As illustrated in Figure~\ref{fig_falcon:timeline}(b), each iteration evaluates $\mathit{mc}=4$ candidates instead of just the closest one, because the second to the fourth best candidates per iteration may also be close to the query and could be on the search path of BFS.

However, the PE utilization is not yet optimal due to the synchronization required between iterations, where the candidate queue must be sorted before evaluating the next $\mathit{mc}$ nearest candidates.
While increasing$\mathit{mc}$ could push PE utilization rates towards 100\%, this approach can potentially degrade end-to-end search performance as I will show in the evaluation, because evaluating many candidates per iteration means potentially processing irrelevant candidates. 

\subsection{Low-latency GVS via DST}
\label{sec_falcon:dst}

To maximize accelerator utilization with minimal overhead (the number of extra nodes visited), I propose \textit{Delayed-Synchronization Traversal (DST)}, a parallel, low latency graph traversal algorithm for GVS. 
\textbf{The key idea of DST is to allow on-the-fly processing of multiple groups of candidates within the query processing pipeline by delaying synchronizations between search iterations.} Each candidate group can contain one or multiple candidate nodes.

\subsubsection{DST Procedure.} Figure~\ref{fig_falcon:timeline}(c) demonstrates how DST enhances accelerator utilization. In this example, there are two candidate groups (\(mg=2\)), each with two candidates (\(mc=2\)), thus allowing four candidates to be processed simultaneously in the pipeline, mirroring the MCS setup (\(mc=4\)) in Figure~\ref{fig_falcon:timeline}(b). 
Unlike MCS, DST introduces \textit{delayed synchronization}: as the evaluation of the candidate group containing the 5th and 6th candidates begins, only the first group, containing the 1st and 2nd candidates, has been fully evaluated --- the delayed synchronization sorts the existing results, while the synchronization of the second group (with 3rd and 4th candidates) is deferred. This strategy ensures that the processing pipeline remains filled and that the bottleneck-step PEs for fetching vectors and computing distances are fully utilized, thereby avoiding the periods of idleness around synchronizations as shown in Figure~\ref{fig_falcon:timeline}(a) and (b). When applying DST to intra-query parallelism, steps S2$\sim$S4 can be parallelized across multiple BFC units, unlike across-query parallelism, which utilizes one BFC unit per QPP.

Algorithm~\ref{algo:dst} details the procedure of DST from the accelerator controller's perspective. DST starts by evaluating the entry node as the first candidate group. As soon as a candidate group is evaluated, DST tries to fill the accelerator pipeline by launching the evaluation of additional candidate groups, where both the number of groups in the pipeline (\(mg\)) and the number of candidates per group (\(mc\)) can be set by the user. DST terminates when there are no active groups in the pipeline and there are no more valid candidates.

\begin{algorithm}[t]
\caption{Delayed-Synchronization Traversal (DST)}
\label{algo:dst}
\begin{algorithmic}[1]
\Require graph $G$, entry node $p$, query vector $q$, result queue size $l$,  number of candidate groups $mg$, number of candidates per group $mc$, number of results $k$ ($k \leq l$)
\Ensure $k$ approximate nearest neighbors of query $q$

\State $C \gets \{p\}, R \gets \{p\}, Visited \gets \{p\}$ 
\State \Call{Launch-Eval-Non-Block}{$\{p\}$}, $GroupCnt \gets 1$

\While{$GroupCnt > 0$ \textbf{or} $\Call{Min}{C.dist} \leq \Call{Max}{R.dist}$} \Comment{stop if no active groups and qualified candidates}

    \If{\Call{Earliest-Eval-Done}{}} \Comment{check task status}
        \State $GroupCnt \gets GroupCnt - 1$ 
        \While{$GroupCnt < mg$} \Comment{fill the pipeline}
            \State $threshold \gets \Call{Max}{R.dist}$
            \State $Group \gets \Call{Extract-Min}{C, mc, threshold}$ 
            \If{$\Call{Size}{Group} > 0$}
                \State \Call{Launch-Eval-Non-Block}{$Group$}
                \State $GroupCnt \gets GroupCnt + 1$
            \EndIf
        \EndWhile
    \EndIf
\EndWhile
\State \textbf{return} $\Call{Sort}{R}[:k]$ \Comment{return the first $k$ elements}
\end{algorithmic}
\end{algorithm}

\subsubsection{Performance Benefits.} 
DST achieves significantly higher throughput than BFS and MCS in terms of the number of candidates processed per unit of time.
Figure~\ref{fig_falcon:timeline} marks the count of processed candidates by the end of the timeline on the right side.
In this example, BFS completes only three candidates, meaning that the results for the 3rd candidate have been inserted into the candidate queue. 
MCS shows improved throughput, managing to finish processing five candidates in the same time frame.
DST, given an equivalent number of candidates in the pipeline as MCS (four), achieves the highest throughput by completing seven candidates by the end of the timeline. 
Notably, DST fully utilizes the critical PEs for vector fetching and distance computations, thanks to the delayed-synchronization mechanism.

\subsubsection{Search Quality.} Given the algorithmic relaxations in DST compared to BFS, one might immediately question: \textit{Will the reordered traversal in DST degrade recall?} Contrary to this concern, DST can actually improve recall while lowering search latency as the experiments will demonstrate (Figure~\ref{fig_falcon:dst}) for the following reasons. 
On one hand, BFS traverses the graph in a greedy manner, striving to avoid visiting nodes that are not sufficiently close to the query. On the other, DST, by delaying synchronizations and allowing multiple candidates to be processed in the pipeline, relaxes the threshold for node evaluation. 
Considering that the termination condition remains consistent with BFS (when there is no qualified candidate left), DST likely evaluates the high-quality candidates on the search path of BFS and additionally explores other potentially relevant candidates. Thus, the evaluation of these extra sub-optimal candidates (a) does not prevent the evaluation of better candidates close to the queries and (b) may uncover extra paths leading to the nearest neighbors, thereby potentially improving recall.

Figure~\ref{fig_falcon:dist_trend} compares the search convergence of BFS, MCS, and DST. All of them find the nearest neighbors in this example, with DST and MCS visiting more nodes than BFS. 

\subsubsection{Parameter Configuration.} DST introduces two additional runtime configurable parameters compared to BFS: the number of candidate groups in the pipeline (\(mg\)) and candidates per group (\(mc\)). The optimal configuration depends on several factors, including vector dimensionalities, data distributions, and degrees (number of neighbors per node). 
I found it challenging to determine the optimal parameters by performance modeling due to (a) the significant variance in node degrees and (b) the unpredictable proportion of visited nodes as traversal progresses.
Thus, to ensure optimal search performance, it is advisable to perform an empirical parameter search using a set of sample queries before system deployment. Typically, this process only takes minutes, as the search space is relatively small, with both \(mg\) and \(mc\) usually not exceeding ten according to the experiments.

\section{Evaluation}
\label{sec_falcon:eval}

The evaluation aims to answer the following questions:

\begin{itemize}[leftmargin=*]
    \item How does Falcon's search performance and energy efficiency compare to that of CPUs and GPUs? \S~\ref{sec_falcon:eval_e2e}
    \item How much speedup and recall improvement can DST achieve on Falcon over BFS? \S~\ref{sec_falcon:eval_dst}
    \item Where is the performance cross-over point between intra-query and across-query parallelism? \S~\ref{sec_falcon:eval_inter_intra}
\end{itemize}


\begin{figure*}[t]
  \centering
  \begin{subfigure}[b]{1.0\linewidth}
    \includegraphics[width=\linewidth]{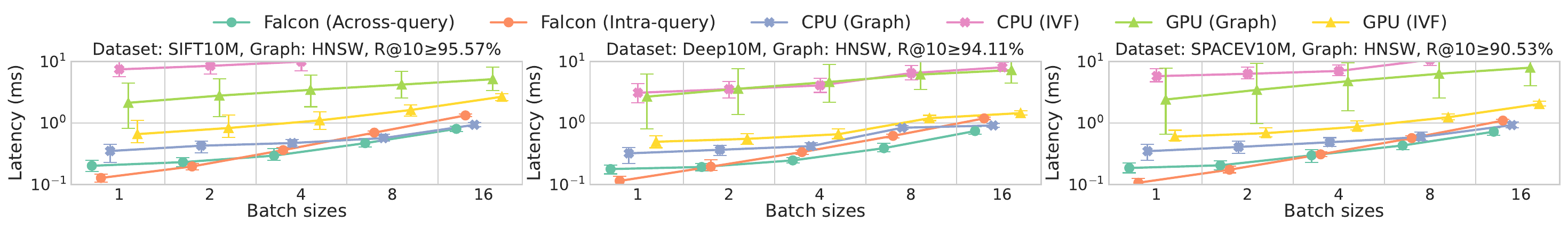}
  \end{subfigure}
  \hfill
  \begin{subfigure}[b]{1.0\linewidth}
    \includegraphics[width=\linewidth]{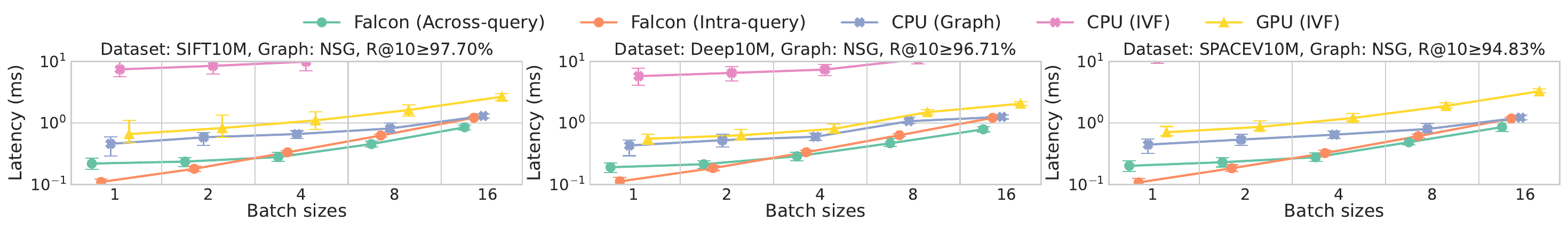}
  \end{subfigure}
  
  \caption{End-to-end GVS latency distribution of CPU, GPU, and Falcon across various graphs (rows) and datasets (columns). 
  The error bar shows the range within which 95\% of query latencies fall; CPU latency with IVF may surpass the $y$-axis limit. }
  \label{fig_falcon:latency}
\end{figure*}

\begin{figure}[t]
  \centering
  \includegraphics[width=0.7\linewidth]{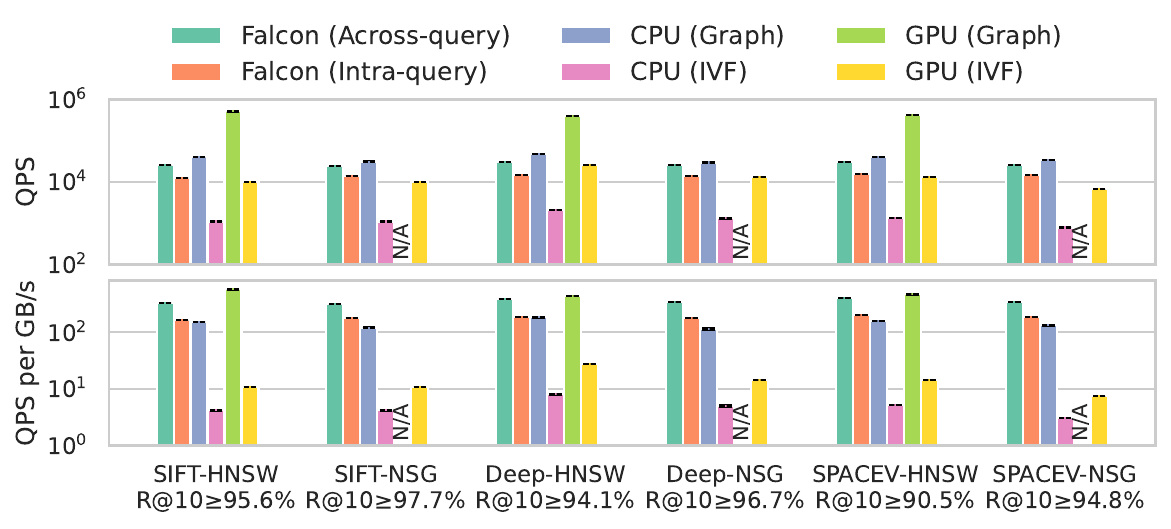}
  \caption{Throughput in queries-per-second (QPS) of different processors and indexes given large batch sizes (10K).}
  \label{fig_falcon:throughput}
\end{figure}

\subsection{Experimental Setup}
\label{sec_falcon:experiment_setup}

\textbf{Baseline systems.}
For CPUs, I evaluate two popular graphs, namely HNSW~\cite{malkov2014approximate} and NSG~\cite{fu2017fast}, using their official implementations.
For GPUs, I evaluate GGNN~\cite{groh2022ggnn}, an approximate version of HNSW optimized for GPU architectures.
Additionally, I evaluate the inverted-file (IVF) index~\cite{IVF}, a clustering-based index, using the Faiss library~\cite{faiss} for both CPUs and GPUs.
As the previous FPGA GVS implementations~\cite{zeng2023df, peng2021optimizing} are not open-sourced, I mainly compare their traversal strategies with DST based on Falcon in \S\ref{sec_falcon:eval_dst}.

\textbf{Hardware.} 
I use server-class hardware manufactured in similar generations of technology (12$\sim$16 nm), where the CPU and GPU hold advantages over the FPGA in terms of bandwidth. 
I develop Falcon using Vitis HLS 2022.1, instantiate it on the AMD Alveo U250 FPGA (16 nm) with 64 GB of DDR4 memory (four channels x 16 GB, 77 GB/s in total), and set the accelerator frequency to 200 MHz.
I use a CPU server with 48 cores of Intel Xeon Platinum 8259CL operating at 2.5 GHz and 384 GB DDR4 memory (12 channels, 256 GB/s).
GPU evaluations are performed on NVIDIA V100 with 16 GB HBM2 memory (900 GB/s).


\textbf{Datasets.} 
I use the SIFT~\cite{SIFT}, Deep~\cite{babenko2016efficient}, and SPACEV~\cite{spacev} datasets, containing 128, 96, and 100-dimensional vectors, respectively, thus covering both vision features (SIFT and Deep) and text embeddings (SPACEV).
I evaluate their subsets of the first ten million vectors, such that the constructed graphs can fit within the GPU and FPGA memory. 

\textbf{Algorithm settings.}
Unless specified otherwise, I set the maximum degree of the graphs to 64, balancing between graph size and search quality.
I set the candidate queue size as 64, which ensures at least 90\% recall for ten nearest neighbors across datasets. 
Falcon uses the best-performing DST parameters unless otherwise specified.
For IVF indexes, I set the number of IVF lists as 4096, approximately the square root of the number of vectors as a common practice.


\subsection{End-to-end Performance and Efficiency}
\label{sec_falcon:eval_e2e}

I compare Falcon with baseline systems on the six combinations between datasets and graphs.  
The software recall of these experiments is noted in Figure~\ref{fig_falcon:latency}: NSG consistently achieves better recall than HNSW. Falcon always achieves better recall than software because DST explores more search paths per query than BFS, as I will analyze in \S\ref{sec_falcon:eval_dst}.

\subsubsection{End-to-end Online Search Latency.} 
For online search, I treat all systems as a service where both the client and the server are connected to the same network switch. The network transmission time between CPU servers and between CPUs and FPGAs are similar --- around 50$\mu$s given a batch size of one, only a tiny fraction of the end-to-end query latency. 
Figure~\ref{fig_falcon:latency} shows the distributions of vector search latency for various batch sizes across six graph-dataset combinations.
I set the IVF-based index parameters for each scenario to achieve at least the same recall as GVS.

\textit{Falcon consistently outperforms all baselines in median latency, achieving speedups of up to 4.3$\times$ over CPU with graphs, 19.5$\times$ over GPU with graphs, 102.1$\times$ over CPU with IVF, and 6.5$\times$ over GPU with IVF.} 
Falcon achieves the lowest search latency among the compared systems, with its intra-query and across-query parallel modes preferable for different batch sizes as I will discuss in \S\ref{sec_falcon:eval_inter_intra}. 
For CPUs, GVS outperforms the IVF index as the latter requires more database vectors to scan to achieve comparable recall~\cite{gao2023high, li2019approximate}. 
As batch sizes increase, CPU GVS latency becomes closer to that of Falcon, mainly benefiting from the CPU server's 3.3$\times$ higher bandwidth than the FPGA, whose bandwidth is saturated at a batch size of four. 
On GPUs, the embarrassingly parallel scan pattern of IVF results in better latency than GVS.
Despite their high bandwidth and numerous cores, GPUs struggle to efficiently handle queries with small batch sizes due to the GPU’s throughput-oriented architecture, which prioritizes parallel processing of many queries but results in high latency for individual queries. 


\subsubsection{Throughput without Latency Constraints.} 
Figure~\ref{fig_falcon:throughput} presents search throughput in queries-per-second (QPS) without latency constraints by setting the batch size as 10K.

\textit{Without latency constraints, GVS throughput on accelerators becomes a contest of memory bandwidth.} 
For both CPUs and GPUs, graph-based indexes outperform IVF, which necessitates scanning more database vectors to reach the same recall~\cite{gao2023high, li2019approximate}.
For GVS, the GPU exhibits superior throughput thanks to its 12$\times$ memory bandwidth over the FPGA, as shown in the upper half of Figure~\ref{fig_falcon:throughput}.
Upon normalization by bandwidth (Figure~\ref{fig_falcon:throughput} lower), the performance of Falcon and GPUs becomes comparable, with GPUs showing a slight edge for SIFT. This is because the GPU adopts the greedy BFS algorithm, whereas Falcon uses DST that trades off additional nodes to visit for reduced latency, as I will analyze in \S\ref{sec_falcon:eval_dst}. 
The CPU performs the worst in QPS per unit bandwidth due to additional memory accesses required to check and update the visit status array.


\begin{figure*}[t]
  \centering
  \begin{subfigure}[t]{0.48\linewidth}
    \includegraphics[width=\linewidth]{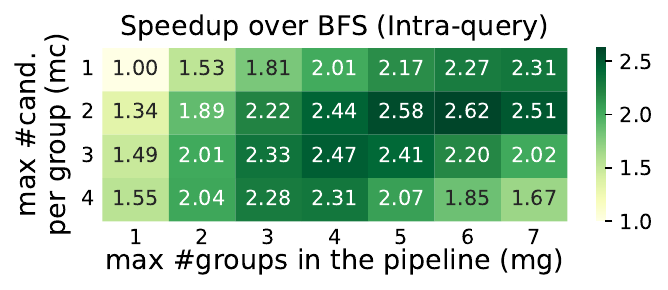}
  \end{subfigure}
  \begin{subfigure}[t]{0.48\linewidth}
    \includegraphics[width=\linewidth]{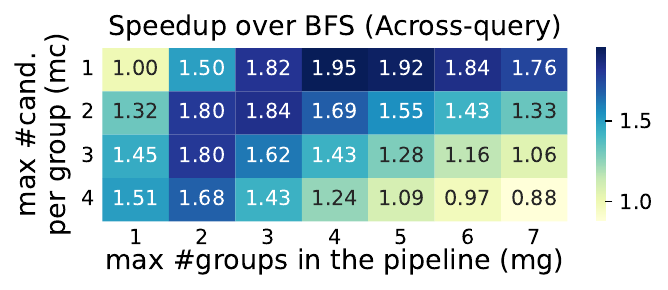}
  \end{subfigure}
  
  \begin{subfigure}[t]{0.48\linewidth}
    \includegraphics[width=\linewidth]{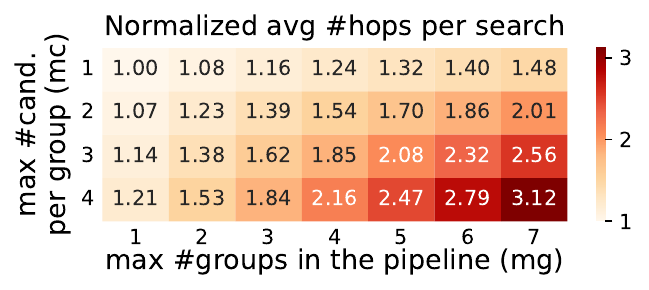}
  \end{subfigure}
  \begin{subfigure}[t]{0.48\linewidth}
    \includegraphics[width=\linewidth]{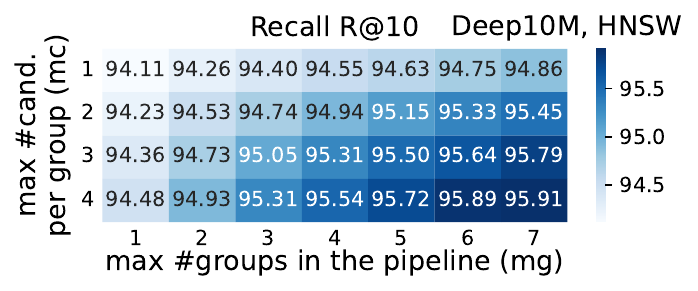}
  \end{subfigure}

  \caption{The performance, recall, and amount of evaluated candidates given different DST parameters~(\(mg\) and \(mc\)). }
  %
  \label{fig_falcon:dst}
\end{figure*}

\subsubsection{Energy Efficiency.} 
I measure the power consumption (in Watt) of CPU, GPU, and Falcon using \textit{Intel RAPL}, \textit{NVIDIA System Management Interface}, and \textit{AMD's Vitis Analyzer}. 
The energy consumption per query batch (in Joule) is calculated by multiplying power with batch latency. 

\textit{Falcon is energy efficient, achieving up to 8.0$\times$, 26.9$\times$, 231.1$\times$, and 5.5$\times$ better energy efficiency than CPU graph, GPU graph, CPU IVF, and GPU IVF, respectively.} 
For online GVS with batch sizes up to 16, the power consumption of CPU, GPU, and Falcon ranges from 136.9$\sim$209.2W, 183.4$\sim$324.2W, and 55.2$\sim$62.3W, respectively. Considering energy consumption per batch, Falcon achieves 2.2$\sim$8.0$\times$ and 11.9$\sim$26.9$\times$ better energy efficiency than CPUs and GPUs. 
For offline GVS without latency constraints (using batch size of 10K), Falcon still achieves 1.9$\sim$3.9$\times$ energy efficiency over CPUs, but is outperformed by GPUs by 5.3$\sim$11.1$\times$, indicating that GPUs remain the preferred option for scenarios requiring high-throughput thanks to their superior memory bandwidth.

\begin{figure}[t]
  \centering
  \includegraphics[width=0.75\linewidth]{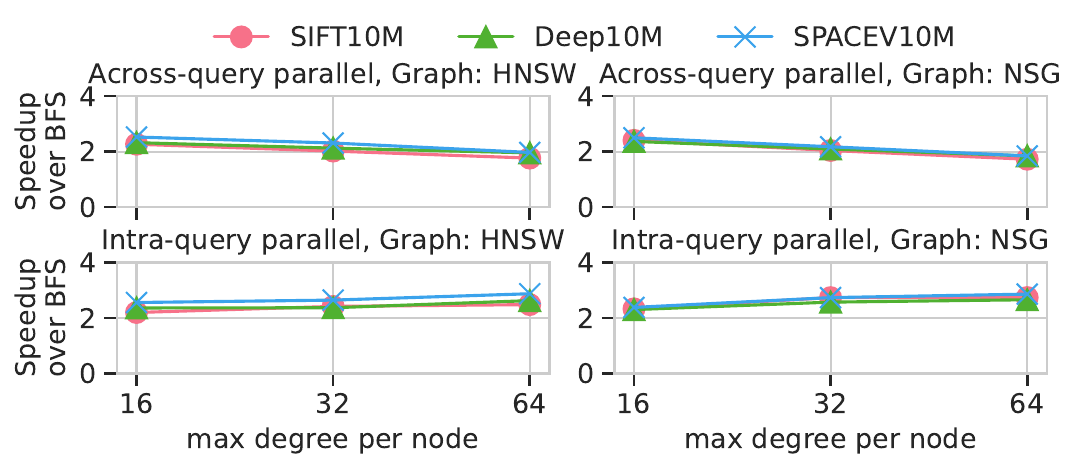}
  \caption{DST consistently outperforms BFS across various datasets, graph configurations, and parallel modes.}
  \label{fig_falcon:dst_speedup_across_settings}
\end{figure}

  

\subsection{DST Efficiency on Accelerators}
\label{sec_falcon:eval_dst}

\subsubsection{Performance Benefits}
I now discuss the speedup achieved with different DST parameters and the maximum speedup across various experimental setups.

\textbf{The impact of DST configurations on performance.} 
I evaluate the impact of the numbers of candidate groups in the pipeline (\(mg\)) and candidates per group (\(mc\)) on DST performance. 
Figure~\ref{fig_falcon:dst} shows the throughput speedup achieved by DST compared to BFS on the Deep10M dataset with HNSW, across both the intra-query and across-query parallel versions of Falcon.
BFS is equivalent to \(mg=1, mc=1\) (upper-left corner), whereas MCS, evaluating multiple candidates per iteration without delayed synchronization, is shown in the first column \((mg=1, mc\geq 1)\). All the other setups are considered as DST.
Note that previous FPGA designs~\cite{peng2021optimizing, zeng2023df} adopts BFS, with Zeng et al.~\cite{zeng2023df} implementing a prefetching strategy on BFS that, at best (zero miss rate), matches the performance of MCS with \(mc=2\).


\textit{The optimal configuration for DST varies across use cases, with intra-query parallelism typically requiring higher parameter values than across-query parallelism.} In Figure~\ref{fig_falcon:dst}, the optimal parameters are \(mg=6, mc=2\) for intra-query parallelism and \(mg=4, mc=1\) for across-query parallelism. This is because the intra-query version parallelizes the distance computations, thus achieving a higher throughput of workload processing per query, leading to a higher throughput of processing nodes and thus necessitating a greater workload intensity to fully utilize the accelerator. However, higher \(mg=6\) and \(mc=2\) also lead to a greater amount of query-wise workloads as more hops are needed before the search terminates, as shown in Figure~\ref{fig_falcon:dst}. Thus, the maximum speedup is determined by the balance between accelerator utilization and the number of extra hops per query.

\textbf{Maximum speedup in various experimental setups.} Figure~\ref{fig_falcon:dst_speedup_across_settings} shows the speedup of DST over BFS across various settings, including parallel modes, datasets, graph types, and the maximum degrees of each graph. 


\textit{DST consistently outperforms BFS across all setups, achieving speedups from 1.7$\sim$2.9$\times$.} 
DST is particularly advantageous in intra-query parallelism: with a maximum degree size of 64, it achieves speedups of 2.5$\sim$2.9$\times$ over BFS for intra-query parallelism, compared to 1.7$\sim$2.5$\times$ for across-query parallelism. This is because intra-query parallelism utilizes more BFC units for a single query, thus benefits more from increased workloads in the pipeline using DST. 

\subsubsection{Recall Benefits.} The rightmost heatmap in Figure~\ref{fig_falcon:dst} shows the improvements in search quality achieved by DST.

\textit{In general, larger numbers of candidates in the processing pipeline (higher \(mg\) and \(mc\)) lead to increased recall.} This is due to the evaluation of a broader range of candidates. Although some candidates may not be on the optimal search path, they could still lead to paths that reach the nearest neighbors.

\textit{DST consistently achieves better recall than BFS across all experiments.}
In Figure~\ref{fig_falcon:dst}, employing the performance-optimal DST configurations enhances R@10 from 94.11\% to 94.55\% and 95.33\% for across-query and intra-query parallelism, respectively. 
Given various experimental setups as in Figure~\ref{fig_falcon:dst_speedup_across_settings}, the R@10 improvements range from 0.14\% to 4.93\%.

\subsection{Across-query and Intra-query Parallelism}
\label{sec_falcon:eval_inter_intra}

\begin{figure}[t]
  \centering
  \begin{subfigure}[b]{0.4\linewidth}
    \includegraphics[width=\linewidth]{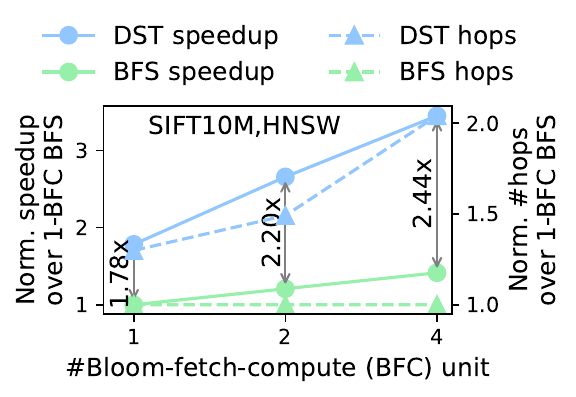}
  \end{subfigure}
  \begin{subfigure}[b]{0.4\linewidth}
    \includegraphics[width=\linewidth]{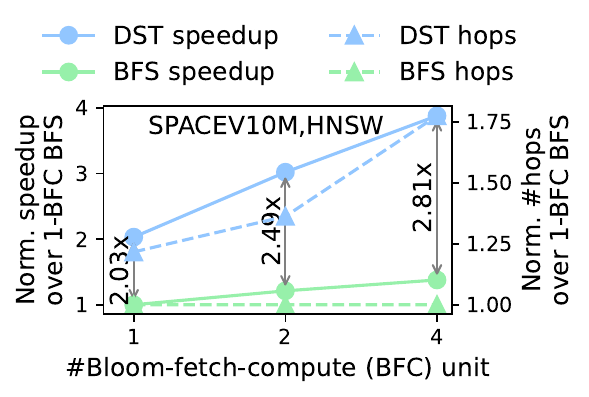}
  \end{subfigure}
  \caption{The scalability of DST and BFS for intra-query parallelism across various numbers of BFC units.}
  \label{fig_falcon:intra_query_scalability}
\end{figure}

\subsubsection{Scalability of Intra-query Parallelism}
Figure~\ref{fig_falcon:intra_query_scalability} compares the scalability of DST and BFS given various numbers of Bloom-fetch-compute (BFC) units across datasets, with all units sharing a common control unit to form a query processing pipeline (QPP). For DST, I use \(mc\) and \(mg\) that achieve the highest performance.

\textit{DST demonstrates better performance scalability than BFS.}
For example, on the SIFT dataset (left side of Figure~\ref{fig_falcon:intra_query_scalability}), the speedup of DST over BFS increases from 1.78$\times$ to 2.44$\times$ as the number of BFC units grows from one to four. 
BFS, with four BFC units, achieves only a speedup of 1.41$\times$ over the single BFC version. This limited scalability of BFS stems from its greedy traversal pattern, which processes only one candidate at a time, resulting in minimal parallelizable workloads per iteration while the control overhead associated with the queues remains constant. 
In contrast, DST expands the workloads in the pipeline, ensuring that each BFC unit has sufficient workload to work with.

\subsubsection{Performance Trade-offs between Intra-query and Across-query Parallelism}
Figure~\ref{fig_falcon:latency} compares the performance of the two types of parallelism, with each accelerator containing four BFC units forming one QPP (intra-query parallel) or four QPPs (across-query parallel).

\textit{The optimal choice of parallel mode is related to batch sizes.} As shown in as shown in Figure~\ref{fig_falcon:latency}, intra-query parallelism is always advantageous for a query size of one. However, since the latency speedup from intra-query parallelism does not scale linearly with the number of BFC units (Figure~\ref{fig_falcon:intra_query_scalability}), across-query parallelism performs better for queries with batch sizes at least equal to the number of QPPs (four in this case). For batch sizes that fall between these two scenarios, the preferred parallel mode depends on the dataset, vector dimensionality, and graph construction parameters.

\section{Discussion}
\label{sec_falcon:discussion}


I have shown the performance advantages of Falcon and DST over CPUs and GPUs through FPGA-prototyping. 
I now discuss potential future extensions of the prototype to enable broader deployments, including adding more functionalities, supporting larger-scale searches, and achieving even higher efficiency.

\textbf{Handling insertions and updates.}
To support data insertions, deletions, or updates in Falcon, one could refer to the designs of software vector search systems. They typically manage a primary index for a dataset snapshot, an incremental (smaller) index for newly added vectors since the last snapshot, and a bitmap marking deleted vectors~\cite{adb-v}. These two indexes are merged periodically, e.g., daily, into a new primary index. 
Falcon can adopt this approach by focusing on serving the primary index, while the incremental index remains small enough to be efficiently managed by CPUs.

\textbf{Scale-out the system.}
I have not yet scaled out Falcon due to the limited number of FPGAs available.
However, I expect the scale-out design to be similar to software-based GVS systems~\cite{doshi2020lanns}. Specifically, the dataset is partitioned into subsets, each associated with a graph managed by a separate Falcon node. Queries are then directed to one or several of these partitions, with the results subsequently aggregated.

\textbf{Extensions for ASICs.}
Both Falcon's architecture and DST are applicable to ASICs. 
The remaining decision involves choosing between prioritizing memory capacity or bandwidth --- opting for DDR to serve larger graphs or HBM to process smaller datasets more rapidly.
Based on the data ingestion speed measured for each BFC units and the total memory bandwidth, the number of PEs to be instantiated on the ASIC accelerator can then be calculated.




\section{Conclusion}
\label{sec_falcon:conclusion}


To meet the surging demands of online GVS, I propose Falcon, a high-performance GVS accelerator, and DST, an accelerator-optimized traversal algorithm.
Evaluated across various graphs and datasets, they shows up to 4.3$\times$ and 19.5$\times$ speedup in online search latency compared to CPUs and GPUs, while being up to 8.0$\times$ and 26.9$\times$ more energy efficient.
These compelling results show the potential for Falcon and DST to become the standard solutions for GVS acceleration.



\part{Vector Table Management in Recommender Systems} 
\label{part:dlrm}


\chapter{MicroRec: Efficient DLRM on Heterogeneous Memory Systems}
\label{chap:microrec}

This chapter and the next focus on recommender systems, another essential use case of vector data systems in machine learning.  
In Part I, which addresses RAG performance, I show that RAG is a heterogeneous system in terms of both its components and underlying hardware. 
In the case of recommender systems, heterogeneity can exist even within a single model, including embedding table lookups and DNN inference.
This chapter addresses recommender model serving efficiency from both hardware and data structure perspectives.

\section{Introduction}
\label{sec_microrec:intro} 
 
Personalized recommendations are widely used to improve user experience and increase sales. Nowadays, deep learning has become an essential building block in such systems. For example, Google deploys wide-and-deep models for video and application recommendations~\cite{wide_and_deep_app_store, wide_and_deep_MT_youtube}; Facebook uses different kinds of deep models for a range of social media scenarios~\cite{facebook_benchmark}; and Alibaba combines attention mechanism with DNNs and RNNs for online retail recommendations~\cite{din_alibaba_attention_fc, dien_alibaba_attention_rnn}. Due to the popularity of DNN-based recommendation models, they can comprise as much as 79\% of the machine learning inference workloads running in data centers~\cite{facebook_benchmark}.

\begin{figure*}[t]
  \centering
  \includegraphics[width=0.9\linewidth]{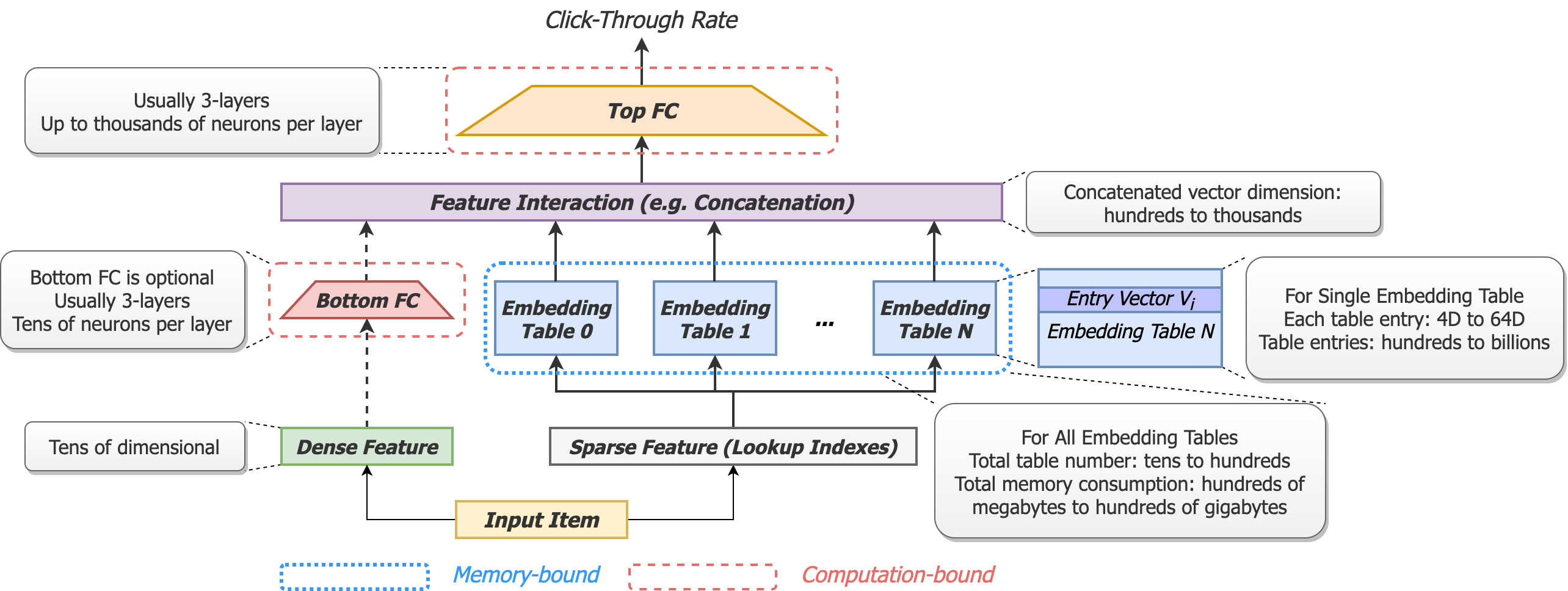}
  \caption{A typical deep recommendation model and it's workload specification.}
  \label{fig_microrec:embedding}
\end{figure*}

\begin{figure*}[t]
  \centering
  \includegraphics[width=\linewidth]{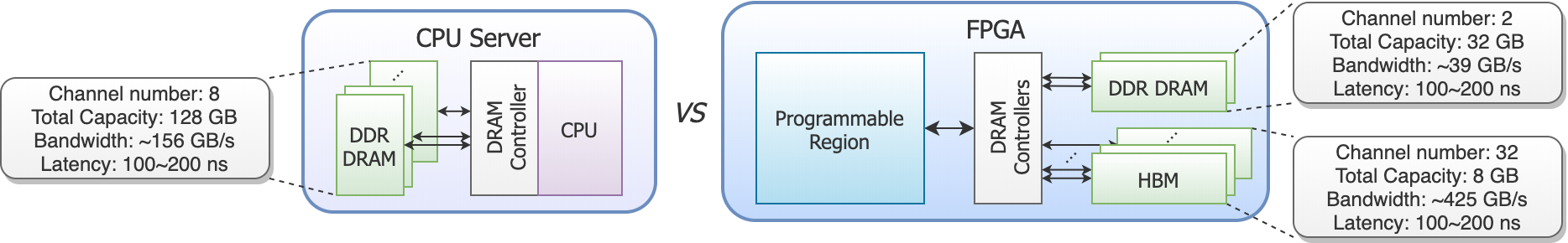}
  \caption{Two hardware choices for recommendation inference. Left: a typical CPU server on which models are stored in DDR DRAM (memory channel number varies from server to server) and computation is done in CPU. Right: an FPGA accelerator where embedding tables are distributed over many memory channels and fast inference is supported by reprogrammable circuit.}
  \label{fig_microrec:fpga_cpu_system}
\end{figure*}

\paragraph*{Deep Recommendation Models.} I first briefly introduce deep recommendation models to provide the necessary context to discuss the challenges, solutions, and contributions.
Figure~\ref{fig_microrec:embedding} illustrates a classical deep recommendation model for \textit{Click-Through Rate} (CTR) prediction~\cite{facebook_benchmark, wide_and_deep_app_store} and summarize its workload characteristics. An input feature vector consists of dense features (e.g., age and gender) and sparse features (e.g., location and advertisement category). Over the dense feature
vector, some systems apply a neural feature extractor that consists of multiple fully connected (FC) layers~\cite{facebook_benchmark, tensordimm_kaist}, while some design~\cite{wide_and_deep_app_store} does not contain the bottom FC layers. Over the sparse feature vector, the system translates each feature into a dense feature embedding by looking up it in an embedding table. These features are then combined (e.g., concatenated) and fed to a neural classification model consisting of multiple fully connected layers.

\paragraph*{Challenges in a CPU-based System.} 
When deploying recommendation systems on typical CPU servers (left half of Figure~\ref{fig_microrec:fpga_cpu_system}), embedding tables are stored in DDR DRAM, and the cores are responsible for the computation. 
There are two system bottlenecks in such deployments.

\textit{First}, embedding table lookups are costly because they induce massive random DRAM accesses on CPU servers.
Production recommendation models usually consist of at least tens of embedding tables, thus each inference requires the corresponding lookup operations. Due to the tiny size of each embedding vector, the resulting DRAM accesses are nearly random rather than sequential. Since CPU servers have only a few memory channels, these random DRAM accesses are expensive. 

\textit{Second}, both embedding lookups and computation can be expensive if one resorts to ML frameworks such as TensorFlow and PyTorch. For \textit{TensorFlow Serving} which is optimized for inference, the embedding layer involves 37 types of operators (e.g., concatenation and slice) and these operators are invoked multiple times during inference,
resulting in significant time consumption especially in small batches. 
Similarly, the throughput of neural network computation can also be restricted when using small batches.
Unfortunately, small batch sizes are usually required in CPU-based recommendation engines to meet the latency requirements of tens of milliseconds, thus the framework overhead is non-negligible.

Not surprisingly, there has been a range of work trying to accelerate deep recommendation models. \citet{tensordimm_kaist} and \citet{facebook_benchmark} observed the main system bottleneck of substantial random memory accesses. \citet{tensordimm_kaist} and \citet{ke2020recnmp} thus proposed to redesign DRAM in micro-architectural level; however, it would take years to put such new DRAM chips in production even if they are adopted.
\citet{gupta2020deeprecsys} suggested GPUs could be useful in recommendation for large batches, but the memory bottleneck still remains and GPUs suffer from high latency. Similarly, \citet{hwang2020centaur} implemented an FPGA accelerator for recommendation but without removing the memory bottleneck.
In this chapter, I ask: \textit{Can we
accelerate deep recommendation models, at industrial scale, with \textit{practical} yet \textit{efficient} hardware acceleration?}

\paragraph*{Solution.}
Based on careful analysis of two production-scale models from Alibaba, I design and implement MicroRec, a low-latency and high-throughput recommendation inference engine. The speedups are rooted in two sources. 
First, I employ more suitable hardware architecture for recommendation with (a) hybrid memory system containing High Bandwidth Memory (HBM), an emerging DRAM technology, for highly concurrent embedding lookups; and (b) deeply pipelined dataflow on FPGA for low-latency neural network inference. 
Second, I revisit the data structures used for embedding tables to reduce the number of memory accesses. By applying Cartesian products to combine some of the tables, the number of  DRAM accesses required to finish the lookups are significantly reduced .

The contributions of this chapter include:
\begin{enumerate}[wide = 0pt]
	\item I show how to use high-bandwidth memory to scale up the concurrency of embedding lookups. This introduces 8.2$\sim$11.1$\times$ speedup over the CPU baseline.
	\item I propose to reduce the number of random memory accesses in deep recommendation systems by data structure design. I show that applying Cartesian Products between embedding tables further improves the lookup performance by 1.39$\sim$1.69$\times$ with marginal storage overhead (1.9$\sim$3.2\%). 
	\item To optimize performance with low storage overhead, I propose a heuristic algorithm to combine and allocate tables to the hybrid memory system on the FPGA.
    \item I implement MicroRec on FPGA and test it on two production models from Alibaba  (47 tables, 1.3 GB; 98 tables, 15.1 GB). The end-to-end latency for a single inference only consumes 16.3$\sim$31.0 microseconds, 3 to 4 orders of magnitude lower than common latency requirements for recommender systems. In terms of throughput, MicroRec achieves 13.8$\sim$14.7$\times$  speedup on the embedding layer, and 2.5$\sim$5.4$\times$ speedup on the complete inference process compared to the baseline (16 vCPU; 128 GB DRAM with 8 channels; AVX2-enabled). 
\end{enumerate}

\section{Background}
\label{sec_microrec:deep_rec}

Personalized recommendation systems are widely deployed by YouTube~\cite{deep_youtube, wide_and_deep_MT_youtube}, Netflix~\cite{gomez2015netflix}, Facebook~\cite{park2018deep}, Alibaba~\cite{din_alibaba_attention_fc, dien_alibaba_attention_rnn}, and a number of other companies~\cite{underwood2019use, xie2018personalized, chui2018notes}. 
In this section, I review their basic properties and analyze their performance to identify the main bottlenecks.

\subsection{Embedding Table Lookups}
\label{sec_microrec:embedding}

Embedding table lookup is the key difference between deep recommendation models and regular DNN workloads, and it shows the following traits. First, the embedding tables contribute to the majority of storage consumption in deep recommendation models. Large embedding tables at industry scale can contain up to hundreds of millions of entries, consuming tens or even hundreds of gigabytes of storage. 
Second, the size of the tables varies wildly between a few hundred (e.g., countries or ``province ID'') to hundreds of millions of entries (e.g., ``user account ID'').

Embedding table lookup is problematic from a performance perspective. Due to the traits mentioned above, most tables are held in main memory, inducing many random memory accesses during inference. \citet{ke2020recnmp} proves this point by showing that high cache miss rates are common in deep recommendation inference.

\begin{figure}[t]
	\centering
  \includegraphics[width=0.8\linewidth]{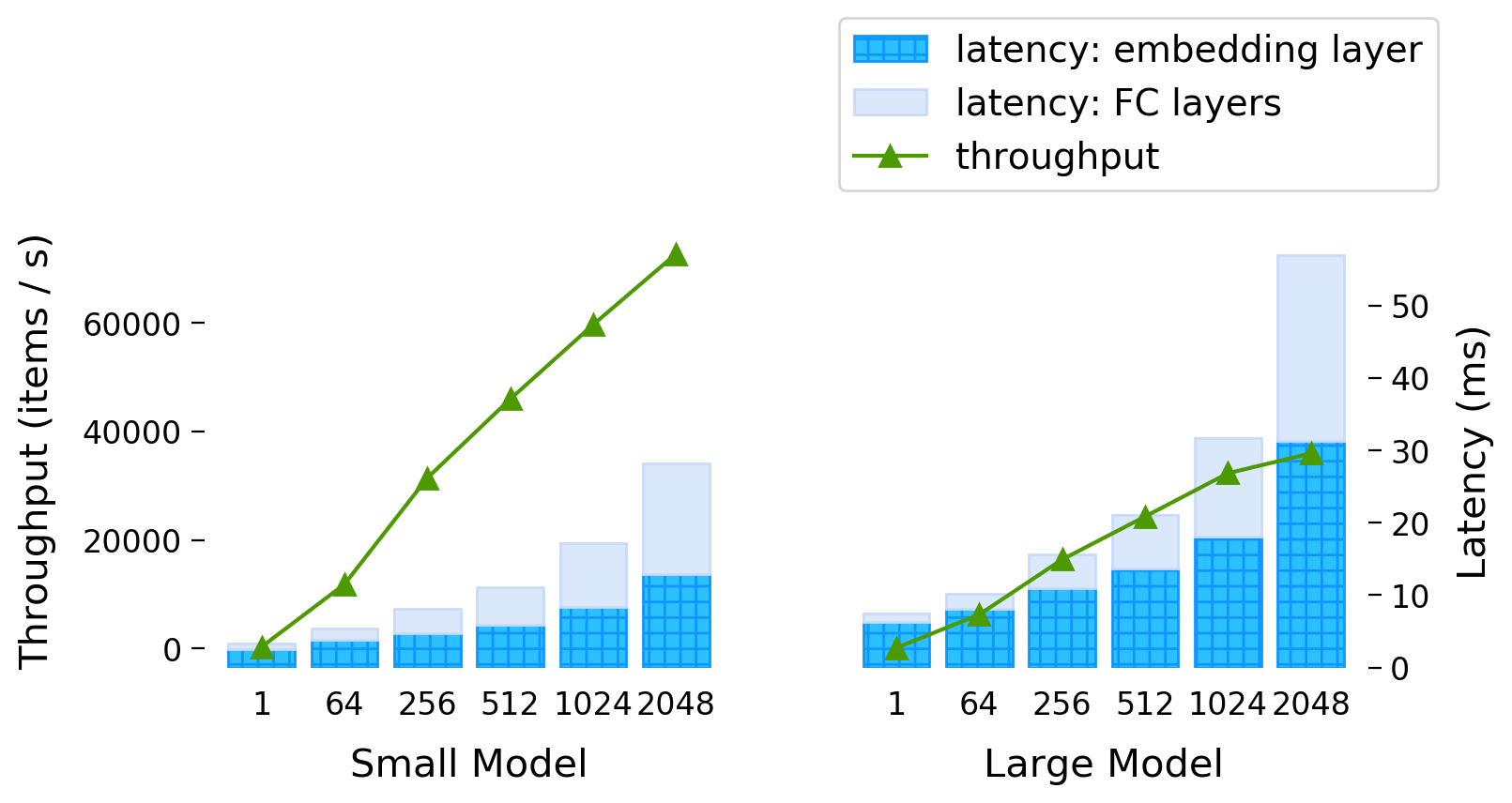}
  \caption{The embedding layer is expensive during inference.}
  \label{fig_microrec:cpu_perf_breakdown}
\end{figure}

\subsection{Performance Analysis}
\label{sec_microrec:cpu_perf}

I chose CPUs as the hardware platform for baseline experiments. 
Although GPUs are popular for neural network training, they have not shown clear advantages over CPUs for deep recommendation inference. As reported by \citet{gupta2020deeprecsys}, GPUs can only outperform CPUs when (a) the model is computation-intensive (less embedding lookups), and (b) very large batch sizes are used. 

Figure \ref{fig_microrec:cpu_perf_breakdown} shows the cost of the embedding layer during inference on two models from Alibaba (models specified in Table \ref{tab_microrec:model_size}) .
As a side effect of the massive number of memory accesses, the many related operators also lead to significant overhead. According to our observation on \textit{TensorFlow Serving}, an optimized ML framework for inference, 37 types of operators are involved in the embedding layer (e.g., slice and concatenation), and these operators are invoked many times during inference. The close latency to infer small batches (size of 1 and 64) illustrates the expense of operator-calls. 
Larger batch sizes can lead to better throughput, yet SLA (latency requirement) of tens of milliseconds must be met, thus extremely large batches are not allowed for recommendations.

\section{MicroRec}
\label{sec_microrec:microrec}

I present MicroRec, an FPGA-enabled high-performance recommendation inference engine which involves both hardware and data structure solutions to reduce the memory bottleneck caused by embedding lookups. On the hardware side, the FPGA accelerator features highly concurrent embedding lookups on a hybrid memory system (HBM, DDR DRAM, and on-chip memory). On the data structure side, I apply Cartesian products to combine tables so as to reduce random memory accesses. Putting them together, I show how to find an efficient strategy to combine tables and allocate them across hybrid memory resources.

\begin{figure}[t!]
	\centering
  \includegraphics[width=1.0\linewidth]{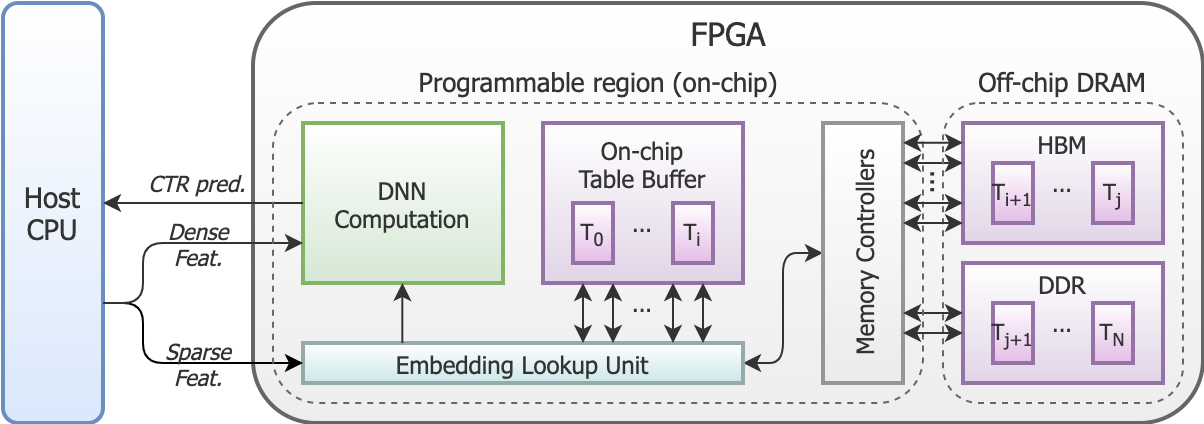}
  \caption{System overview of MicroRec.
  }
  \label{fig_microrec:system_overview}
\end{figure}

\subsection{System Overview}
\label{sec_microrec:microrec_overview}

Figure \ref{fig_microrec:system_overview} overviews the hardware design of MicroRec. Embedding tables are distributed over both on-chip memory (BRAM and URAM) and off-chip memory (HBM and DDR). Neural network inference is taken cared by the DNN computation units which contain both on-chip buffers storing weights of the model and computation resources for fast inference. 
To conduct inference, the host server first streams dense and sparse features to the FPGA\footnote{\label{foot_CPU_input}The Vitis hardware development platform does not yet support streaming from the host server to a Xilinx U280 FPGA, thus I have prototyped the design by caching the input features on FPGA.}. Then, the embedding lookup unit translates the sparse features to dense vectors by looking up embedding tables from both on-chip and off-chip memory. Finally, the computation unit takes the concatenated dense vector as input and finishes inference before returning the predicted CTR to the host. 

\subsection{Boost Emebdding Lookup Concurrency by Increased Memory Channels}
\label{sec_microrec:microrec_memory}

The tens of embedding table lookup operations during inference can be parallelized when multiple memory channels are available. MicroRec resorts to high-bandwidth memory as the main force supporting highly concurrent embedding lookups. Besides that, I also take advantage of other memory resources on FPGA, i.e., DDR4 DRAM and on-chip memory, to further improve lookup performance. 

\paragraph{High-Bandwidth Memory}
\label{sec_microrec:microrec_HBM}

I resort to HBM to parallelize embedding lookups. As an attractive solution for high-performance systems, HBM offers improved concurrency and bandwidth compared to conventional DRAMs~\cite{jun2017hbm, o2014highlights}. In this chapter, I use a Xilinx Alveo U280 FPGA card~\cite{xilinx_u280} equipped with 8 GBs of HBM which provides a bandwidth of up to 425 GB/s~\cite{zeke2020benchmark_hbm}. More specifically, the HBM system on U280 consists of 32 memory banks, which can be accessed concurrently by independent pseudo-channels. Thus, embedding tables can be distributed to these banks so that each bank only contains one or a few tables, and up to 32 tables can be looked up concurrently.

\paragraph{Hybrid Memory System on FPGA}
\label{sec_microrec:microrec_hybrid_memory}

The Xilinx Alveo U280 FPGA involves multiple types of memory resources, including on-chip memory (BRAM and URAM) and off-chip memory (DDR4 DRAM and HBM), which exhibit different traits. 
HBM and DDR show close access latency of a couple of hundreds of nanoseconds given the memory controller generated by Vitis~\cite{vitis}, but have different concurrency-capacity trade-off (HBM: 32 channels, 8GB; DRAM: 2 channels, 32 GB).
Besides HBM and DDR, FPGAs also equip a few megabytes of on-chip memory that plays a similar role as CPU cache (small yet fast memory to cache frequently-accessed data or intermediate results).
Without read initiation overhead as in DRAM, the latency to access on-chip memory only consists of control logic and sequential read. According to the experiments, finish retrieving an embedding vector from an on-chip memory bank only consumes up to around 1/3 time of DDR4 or HBM.

\subsection{Reduce Memory Accesses by Cartesian Products}
\label{sec_microrec:microrec_cartesian}

\begin{figure}
  \centering
    \includegraphics[width=0.8\columnwidth]{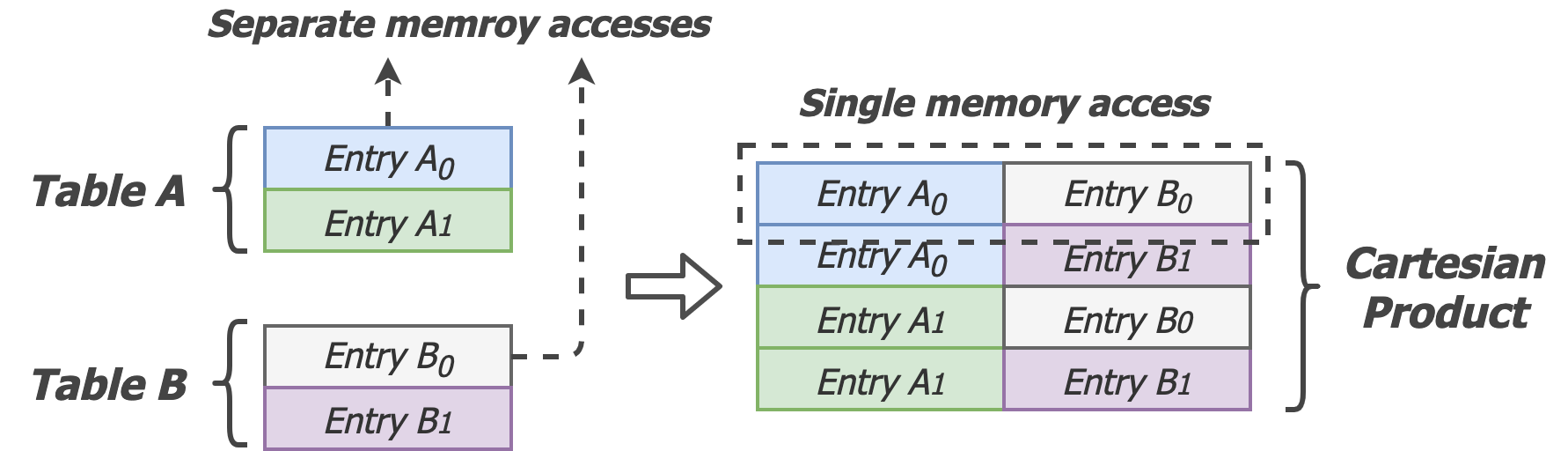}
    \caption{Cartesian product of two embedding tables. Each entry of the product concatenates an entry from table A and another from B: one memory access retrieves two embedding vectors.}
    \label{fig_microrec:cartesian}
\end{figure}

I reduce the number of memory accesses by combining tables so that each memory access can retrieve multiple embedding vectors. As shown in Figure \ref{fig_microrec:cartesian}, two embedding tables can be joined into a single larger one through a relation Cartesian Product. Since tables A and B in Figure \ref{fig_microrec:cartesian} have two entries, the product table ends up with four entries: each of them is a longer vector obtained by concatenating an entry from table A and another from table B.
Using such a representation, the number of memory accesses is reduced by half: instead of two separate lookup operations now only one access is needed to retrieve the two vectors. 

By applying a Cartesian product, the latency to lookup two tables is reduced by almost half. Embedding tables in deep recommendation models usually contain short entry vectors (with between 4 to 64 elements in most cases). Although the entry vectors of the product are longer, i.e., the sum of two individual entries, they are still not long enough to fully take advantage of the spatial locality within DRAM. To retrieve a vector up to a few hundreds of bytes, a DRAM spends most of the time initiating the row buffer, while the following short sequential scan is less significant in terms of time consumption. As a result, reducing the memory accesses by half can lead to a speedup of almost 2$\times$.

Though Cartesian products lead to higher storage consumption, this overhead is comparatively small. This may sound counter-intuitive, 
however, most deep recommendation models contain tables of different size scales, so applying Cartesian products on small tables is almost for free compared to some of the largest tables in the model. 
According to the observations of real-world deployments, while some tables only consist of 100 4-dimensional embedding vectors, large tables can contain up to hundreds of millions of entries with a vector length of 64 due to the reasons discussed in section \ref{sec_microrec:embedding}. In this case, a Cartesian product of two small tables requires only tens of kilobytes (assume 32-bit floating-point storage): almost negligible compared to a single large table of tens or hundreds of gigabytes.

Cartesian products can help balancing the workload on off-chip DRAM (DDR and HBM). 
For example, suppose there are 34 off-chip memory channels (32 for HBM and 2 for DDR), and 40 tables should be allocated on them. 
In this case, some banks have to store two embedding tables while others only hold one. 
When retrieving one vector from each table, the lookup performance is bound by the channels holding two tables, as the lookup latency on them is potentially 2$\times$ that of those containing only one table. 
Using Cartesian products, the total number of tables can be reduced from 40 to 34. This allows us to balance the workload on each memory channel resulting in potentially 2$\times$ speedup compared to an unbalanced workload situation. 

\subsection{Putting Everything Together: A Rule-based Algorithm for Table Combination and Allocation}
\label{sec_microrec:microrec_algorithm}

The goal of this work is to minimize embedding lookup latency given the memory constraints discussed in section \ref{sec_microrec:microrec_hybrid_memory}, i.e., available capacity and channels of each type of memory. To achieve this, an algorithm is required to explore solutions of combining tables through Cartesian products and deploying the result on memory banks.

\paragraph{Brute-force Search}
\label{sec_microrec:microrec_brute_force}

A straightforward way to achieve this objective is to explore all possibilities in a brute-force manner and choose the best solution. 
First, one would list all possibilities of using tables as Cartesian product candidates. Then, for each one of these options, all possible combinations of Cartesian products would be calculated (including joining more than two tables). Based on the combinations of tables available, the single and combined tables are allocated to memory banks (solutions exceeding the memory capacity of a bank can be dropped) minimizing the latency. For ties in latency, the solution with the least storage overhead is chosen.

However, applying brute-force search is unrealistic because of the large exploration space. 
For example, selecting $n$ of out $N$ total tables as Cartesian candidates is a combinatorial problem with a time complexity of \(O(\frac{N!}{n!(N-n)!})\). Then, it costs \(O(n!)\) to explore any Cartesian products combinations of the candidates. Each outcome, including Cartesian products and original tables, are then allocated to memory banks at the cost of \(O(N)\). Using a parameter to control how many tables are selected for Cartesian products, the overall time complexity of the brute-force search is \(O(\sum_{n=1}^{N}N\frac{N!}{(N-n)!})\), making brute-force searching infeasible as the number of tables grows up.

\paragraph{Heuristic-rule-based Search}
\label{sec_microrec:microrec_heuristic}
To optimize embedding lookup latency, I propose a heuristic search algorithm that can efficiently search for near-optima solutions with a low time complexity of $\mathcal{O}(N^2)$. Besides, this algorithm can be generalized to any FPGAs, no matter whether they are equipped with HBM, and no matter how many memory channels they have. Due to the memory traits introduced in section \ref{sec_microrec:microrec_hybrid_memory}, the algorithm simply regards HBM as additional memory channels: designers can adjust the memory channel number and bank capacities in the algorithm according to the available hardware.

Four heuristics are applied in the algorithm to reduce the search space where the optimal solution is unlikely to appear\footnote{The rules can be expanded, modified, or removed to adpat different models since these rules are table-size-dependent.}. Consequently, the algorithm can return near-optimas with low time complexity. The first three rules are designed to explore Cartesian combinations efficiently, while the fourth rule is for memory allocation.

Heuristic rule 1: large tables are not considered for Cartesian products. Tables are sorted by size and only the $n$ smallest tables should be selected for Cartesian products, otherwise products of large tables can lead to heavier storage overhead.

Heuristic rule 2: Cartesian products for table pairs of two. Although Cartesian products of the three smallest tables may only consume tens of megabytes storage (still small compared to a single large table of several or tens of gigabytes), the overall solution could be sub-optimal because this method consumes too many small tables at once while they are appropriate candidates to pair with larger tables.

Heuristic rule 3: within the product candidates, the smallest tables are paired with the largest tables for Cartesian products. This rule avoids terrible solutions where a Cartesian product is applied between two large tables. 

Heuristic rule 4: cache smallest tables on chip. 
After applying Cartesian products, all tables are sorted by size to determine the number of small tables to store on the chip.
Two constraints must be considered during this process. First, the size of selected tables should not exceed assigned on-chip storage. Second, if multiple tables are co-located in the same on-chip bank, the total lookup latency should not exceed off-chip (DDR or HBM) lookups, otherwise caching tables on-chip is meaningless.

Algorithm~\ref{algo:heuristic} sketches the heuristic-rule-based search for table combination and allocation. It starts by iterating over the number of tables selected as Cartesian product candidates. Within each iteration, the candidates are quickly combined by applying the first three heuristic rules ($\mathcal{O}(N)$). All tables are then allocated to memory banks efficiently by rule 4 ($\mathcal{O}(N)$). The algorithm ends by returning the searched solution that achieves the lowest embedding lookup latency. Considering the outer loop iterating over Cartesian candidate numbers, the total time complexity of the heuristic algorithm is as low as $\mathcal{O}(N^2)$.

\begin{algorithm}[t!]
\caption{Heuristic Search}
\label{algo:heuristic} 

\textbf{Input:} \\
\hspace{1em} $N$: total number of embedding tables \\
\hspace{1em} $n$: number of tables selected for Cartesian products \\
\hspace{1em} $c$: candidate tables for Cartesian products \\
\hspace{1em} $p$: all tables after applying Cartesian products \\
\vspace{-1em}

\textbf{Output:} \\
\hspace{1em} $current\_best$: best solution found, including table number, sizes, and bank allocations \\

\vspace{-2em}

\begin{tabbing}
    \hspace{2em} \= \hspace{2em} \= \hspace{2em} \= \kill
    \textbf{for} $n \in \{1...N\}$ \textbf{do} \+ \\
        $c \gets \texttt{select\_tables}(n, N)$ \` // Heuristic Rule 1 \\
        $p \gets \texttt{Cartesian\_product}(c)$ \` // Heuristic Rule 2 \& 3 \\
        $\texttt{solution} \gets \texttt{allocate\_to\_banks}(p)$ \` // Heuristic Rule 4 \\
        \textbf{if} $\texttt{solution}$ is better than $\texttt{current\_best}$ \textbf{then} \+ \\
            $\texttt{current\_best} \gets \texttt{solution}$ \- \\
    \textbf{end for} \\
\end{tabbing}
\vspace{-2em}
\textbf{Return:} $\texttt{current\_best}$

\end{algorithm}

\section{FPGA Implementation}

In this section, I describe the implementation of MicroRec on an FPGA with an emphasis on its low inference latency.

\subsection{Reduce Latency by Deeply Pipelined Dataflow}
\label{sec_microrec:design_pipeline}

\begin{figure}[t]
  \centering
  \includegraphics[width=0.6\linewidth]{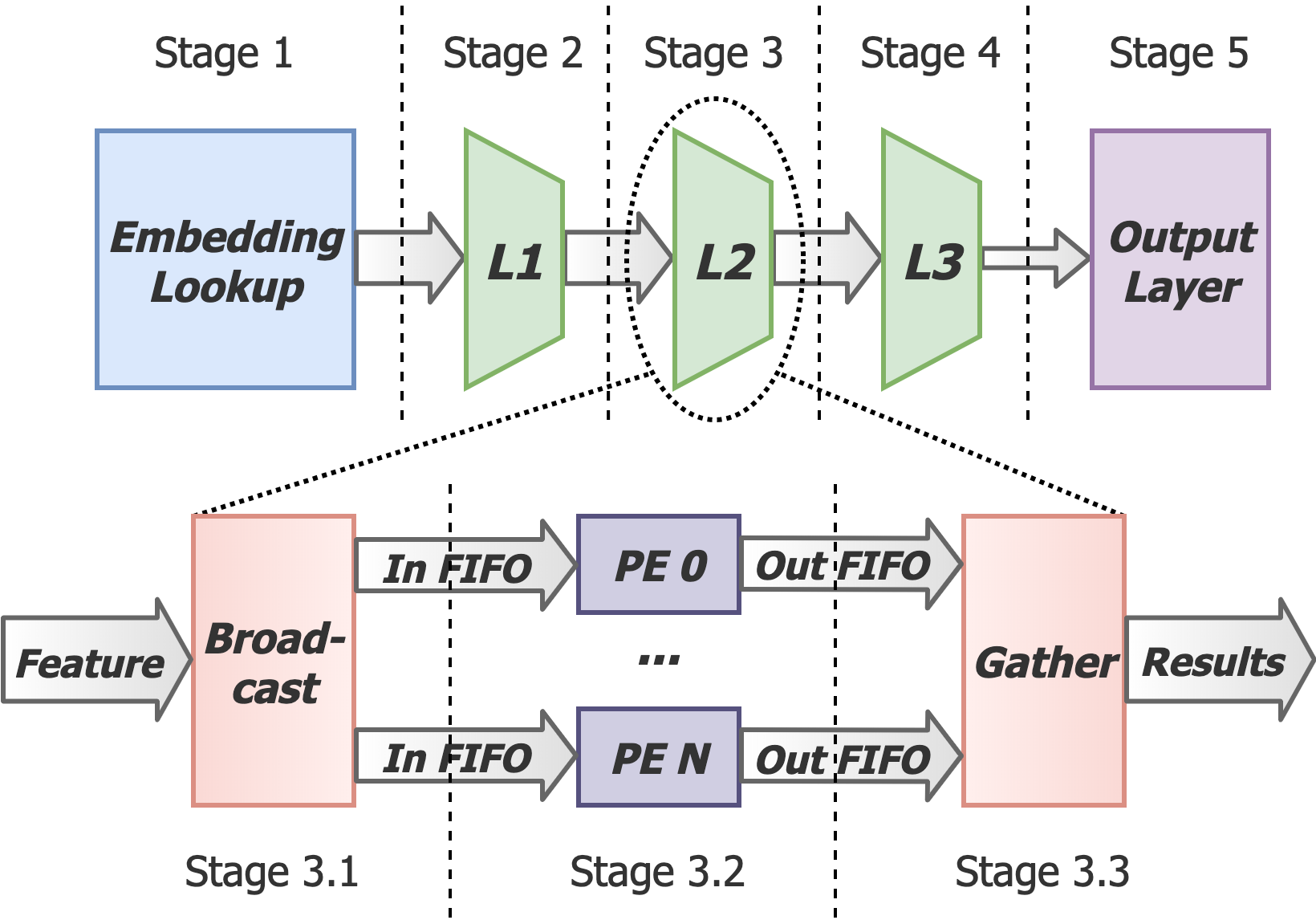}
  \caption{Highly pipelined and parallelized hardware design.}
\label{fig_microrec:FPGA_design}
\end{figure}

As shown in Figure~\ref{fig_microrec:FPGA_design}, I apply a highly pipelined accelerator architecture where multiple items are processed by the accelerator concurrently in different stages. In this design, the embedding lookup stage and three computation stages are pipelined. Each DNN computation module is further divided into three pipeline stages: feature broadcasting, computation, and result gathering. BRAMs or registers are applied to build pipes (FIFOs) as inter-module connections.

Latency concerns (SLA requirements) are eliminated by this highly pipelined design for two reasons. First, input items are processed item by item instead of batch by batch, thus the time to wait and aggregate a batch of recommendation queries is removed. Second, the end-to-end inference latency of a single item is much less than a large batch. 

\subsection{Embedding Lookup Module}
\label{sec_microrec:fpga_embedding_module}

The embedding lookup module gathers and prepares concatenated dense features for fully-connected layers. After receiving lookup indexes, the module concurrently retrieves embedding vectors from HBM, DDR, and on-chip memory banks. The concatenated embeddings are then fed to DNN computation modules through FIFOs.

\subsection{DNN Computation Module}
\label{sec_microrec:fpga_computation_module}

The lower half of Figure~\ref{fig_microrec:FPGA_design} presents the computation flow for a single FC layer, which consists of three pipeline stages: input feature broadcasting, general matrix-matrix multiplication (GEMM) computation, and result gathering. Partial GEMM is allocated to each processing unit (PE) for better routing design and potentially higher performance~\cite{johannes2020matmul_hls}. Each PE conducts partial GEMM through parallelized multiplications followed by an add tree~\cite{chen2014dadiannao}.

\section{Evaluation}

I evaluate the performance of MicroRec for both end-to-end recommendation inference and embedding lookups alone. Given real-world models from Alibaba and the recent recommendation inference benchmark~\cite{facebook_benchmark}, MicroRec outperforms the optimized CPU baseline significantly under all experiment settings.

\subsection{Experiment Environment}
\label{sec_microrec:exp_hardware_platform}

I employ Xilinx Alveo U280 FPGA~\cite{xilinx_u280}, a high-end card equipped with 8GB of HBM2 (32 channels) and 32 GB of DDR4 (2 channels).
I program the FPGA by Vivado HLS~\cite{vivado_hls}, which can translate C++ programs to hardware description language (HDL).
The code is then deployed on Vitis~\cite{vitis} to generate FPGA bitstream.

The software baseline performance is tested on an AWS server with Intel Xeon E5-2686 v4 CPU @2.30GHz (16 vCPU, SIMD operations, i.e., AVX2 FMA, supported) and 128 GB DRAM (8 channels). 
I apply an open-source solution on deep recommendation systems~\cite{git_wide_deep}, where \textit{TensorFlow Serving}~\cite{olston2017tensorflow-serving, tensorflow} supports highly optimized model inference. 

\subsection{Model Specification}

I experiment the performance of MicroRec on two classes of models from different sources. The first class contains production models deployed in Alibaba, while the second class comes from the recent recommendation inference benchmark~\cite{facebook_benchmark}. 

\paragraph{Production Models}

I experiment two deep recommendation models from Alibaba in the experiments. Both of them are memory-access intensive: they contain 47 and 98 embedding tables respectively, much more than current benchmark models~\cite{facebook_benchmark}, among which the largest model consists of only 12 tables. Table~\ref{tab_microrec:model_size} shows the parameters of the evaluated models. For example, the smaller recommendation model retrieves one vector from each of the 47 tables and gathers them into a 352-dimensional dense vector to be fed to fully-connected layers. The models I experiment do not contain bottom fully-connected layers, which are adopted in some systems to process dense input features~\cite{facebook_benchmark, ke2020recnmp}.

\begin{table}

  \caption{Specification of the production models. }
\begin{center}
\label{tab_microrec:model_size}
\begin{small}
  \begin{tabular}{@{}  l c c c c @{}  }   
\toprule
    Model & Table Num & Feat Len & Hidden-Layer & Size \\
\midrule
    Small & 47 & 352 & (1024,512,256) & 1.3 GB \\
    Large & 98 & 876 & (1024,512,256) & 15.1 GB \\
\bottomrule
\end{tabular}
\end{small}
\end{center}
\vskip -0.1in
\end{table}

\paragraph{Facebook Recommendation Benchmark}

I also experiment MicroRec on the recent recommendation inference benchmark by Facebook~\cite{facebook_benchmark}. The benchmark published three classes of recommendation models and their performance breakdown.  
Although I target to experiment these models for real-world-deployment, the benchmark only published a range of parameters for each type of model.
For example, the model class DLRM-RMC2 can contain from 8 to 12 tables, yet no numbers about table sizes and embedding vector lengths are provided.
Without such information, it is difficult to compare the inference performance, because some of the parameters are decisive to the inference workload. For instance, embedding vector lengths decide the number of operations to be performed in fully-connected layers.

Therefore, I compare the performance of the embedding layer: given the narrow range of table numbers \citet{facebook_benchmark} published, I can conduct multiple experiments and identify a speedup range of MicroRec.

\subsection{End-to-End Inference}
\label{sec_microrec:exp_entire}

\begin{table*}[t]  
  \caption{MicroRec performance on end-to-end recommendation inference. MicroRec achieves 2.5$\sim$5.4$\times$ speedup compared to the optimized CPU baseline (the speedup is compared to batch latency of FPGA, which consists of both the stable stages in the middle of the pipeline as well as the time overhead of starting and ending stages). Besides, the end-to-end latency to infer a single input item is as low as a couple of tens of microseconds: the latency concern of online model serving is eliminated.}
  \label{tab_microrec:exp_entire}
\begin{center}
\scalebox{0.75}{ 
\begin{tabular}{@{} L{10em} M{3.8em} M{3.8em} M{3.8em} M{3.8em} M{3.8em} M{3.8em} M{3.8em} M{3.8em} @{}} 
\toprule
    & CPU B=1	& CPU B=64& 	CPU B=256& 	CPU B=512& 	CPU B=1024& 	CPU B=2048& 	FPGA fp16 &	FPGA fp32 \\
\midrule
    \multicolumn{9}{c}{Smaller Recommendation Model} \\
    \midrule
    Latency (ms)&	3.34 &	5.41	&8.15&	11.15	&17.17 &28.18	&\textbf{1.63E-2} & 	\textbf{2.26E-2} \\
    Throughput (GOP/s)	&0.61&	24.04&	63.81&	93.32&	121.16&	147.65&	\textbf{619.50}&	\textbf{367.72}\\
    Throughput (items/s)&	299.71 &	1.18E+4&	3.14E+4	&4.59E+4&	5.96E+4&	7.27E+4	&3.05E+5&	1.81E+5\\
    \midrule
    Speedup: FPGA fp16  &204.72$\times$  &   	24.27$\times$& 9.56$\times$& 6.59$\times$&  5.09$\times$ &\textbf{4.19$\times$}&  -&  -\\
    Speedup: FPGA fp32& 147.54$\times$&
      14.58$\times$& 5.69$\times$ &3.91$\times$& 3.02$\times$&  \textbf{2.48$\times$}& -&  - \\
    \midrule
    \multicolumn{9}{c}{Larger Recommendation Model} \\
    \midrule
    Latency (ms)&	7.48 &	10.23	&15.62	&21.06&	31.72&	56.98&	\textbf{2.26E-2} & 	\textbf{3.10E-2}\\
    Throughput (GOP/s)&	0.42&
           	19.48&	51.03&	75.66&	100.49&	111.89&	\textbf{606.41}&	\textbf{379.45}\\
    Throughput (items/s)&	133.68&	6.26E+3& 1.64E+3&	2.43E+4	&3.23E+4&	3.59E+4	&1.95E+5&	1.22E+5\\
    \midrule
    Speedup: FPGA fp16  & 331.51$\times$ &   29.56$\times$& 11.73$\times$& 7.96$\times$&  6.02$\times$ &\textbf{5.41$\times$}&  -&  -\\
    Speedup: FPGA fp32& 241.54$\times$  &  18.67$\times$& 7.36$\times$&  4.99$\times$&  3.77$\times$&  \textbf{3.39$\times$}& -&  - \\
\bottomrule
  \end{tabular}
}
\end{center}
\end{table*}

Table \ref{tab_microrec:exp_entire} compares the performance of end-to-end recommendation inference on production models between the CPU baseline and MicroRec (both Cartesian and HBM are applied). 
On the CPU side, performance increases as batch size grows, so I select a large batch size of 2048 as the baseline (larger batch sizes can break inference latency constraints). 
On the FPGA side, MicroRec infers items without batching as discussed in Section \ref{sec_microrec:design_pipeline}.
Besides, I evaluate the FPGA performance of different precision levels, i.e., 16-bit and 32-bit fixed-point numbers.

\textit{MicroRec achieves significant speedup under all experimented settings.} In terms of throughput, it is 2.5$\sim$5.4$\times$ better than the baseline under two precision levels and two model scale. Moreover, the end-to-end latency to infer a single input item is 16.3$\sim$31.0 microseconds, 3$\sim$4 orders of magnitude lower than common latency requirements (tens of milliseconds). Note that the throughput of MicroRec is not the reciprocal of latency, since multiple items are processed by the deep pipeline at the same time.

\subsection{Embedding Lookup Performance}
\label{sec_microrec:exp_embedding}

I highlight the performance boost of embedding lookups brought by Cartesian products and HBM in this section on both the production models and the benchmark models.

\paragraph{Lookups on Production Models}
\begin{table*}[t]  
  \caption{Benefit and overhead of Cartesian products. It only costs marginal extra storage to achieve significant speedup.}
  \label{tab_microrec:cartesian_benifits}
\begin{center}
\scalebox{0.75}{ 
  \begin{tabular}{@{} L{9em} M{5em} M{7em} M{10em} M{5em} M{8em} @{}}
\toprule
    & Table Num & Tables in DRAM & DRAM Access Rounds & Storage & Lookup Latency \\
\midrule
    \multicolumn{6}{c}{Smaller Recommendation Model} \\
    \midrule
    Without Cartesian & 47 & 39 & 2 & 100\% & 100\% \\
    With Cartesian & 42& 34& 1 & \textbf{103.2\%} & \textbf{59.2\%} \\
    \midrule
    \multicolumn{6}{c}{Larger Recommendation Model} \\
    \midrule
    Without Cartesian & 98 & 82 & 3 & 100\% & 100\% \\
    With Cartesian & 84 & 68 & 2 & \textbf{101.9\%} & \textbf{72.1\%} \\
\bottomrule
  \end{tabular}
}
\end{center}
\end{table*}

\begin{table*}[t]
\label{tab_microrec:embedding}
  \caption{MicroRec performance on the embedding layer. Given the same element data width of 32-bits, it outperformed the optimized CPU baseline by over one order of magnitude. Besides, it only took no more than one microsecond to finish lookups and concatenations even in embedding-intensive models (47 and 98 tables).}
  \label{tab_microrec:exp_embedding}
\begin{center}
\scalebox{0.72}{ 
  \begin{tabular}{@{} L{12em} M{3.5em} M{3.5em} M{3.5em} M{3.5em} M{3.5em} M{3.5em} M{4em} M{6.5em} @{}} 
\toprule
    & CPU B=1	& CPU B=64& 		CPU B=256& 	CPU B=512& 	CPU B=1024& CPU B=2048& 	FPGA: HBM & 	FPGA: HBM + Cartesian \\
\midrule 
    \multicolumn{9}{c}{Smaller Recommendation Model} \\
    \midrule
    Latency (ms) &	2.59&	3.86&	4.71&	5.96 &8.39&	12.96&	7.74E-4&	\textbf{4.58E-4} \\

    Speedup: HBM  &3349.97$\times$ &77.91$\times$  &23.75$\times$&  15.04$\times$&  10.59$\times$&  \textbf{8.17$\times$}&  -&  - \\

    Speedup: HBM + Cartesian &  5665.07$\times$& 131.76$\times$ &40.16$\times$&  25.44$\times$& 17.91$\times$& \textbf{13.82$\times$}&  -&  - \\
    \midrule
    \multicolumn{9}{c}{Larger Recommendation Model} \\
    \midrule
    Latency (ms) &	6.25&	8.05	&10.92&	13.67&	18.11	&31.25&	1.38E-3& 	\textbf{1.03E-3} \\

    Speedup: HBM  &4531.23$\times$ &  91.29$\times$&  30.94$\times$& 19.36$\times$& 12.83$\times$& \textbf{11.07$\times$}& - &- \\

    Speedup: HBM + Cartesian &  6019.37$\times$ & 121.28$\times$&  41.10$\times$& 25.72$\times$& 17.04$\times$& \textbf{14.70$\times$} &  - &-\\
\bottomrule

  \end{tabular}
}
\end{center}
\end{table*}

\textit{MicroRec outperforms CPU baseline significantly on production models as shown in Table~\ref{tab_microrec:exp_embedding}}. Same as Section~\ref{sec_microrec:exp_entire}, a large batch size of 2048 is selected for the CPU baseline to achieve high throughput, while the accelerator always processes inputs item by item (no concept of batch sizes). 
This latency excludes streaming input features from CPU side memory as mentioned in footnote \ref{foot_CPU_input}. The result shows that MicroRec outperforms the baseline by 13.8$\sim$14.7$\times$  on the embedding layer (in addition to DRAM accesses, the many embedding-related operator calls in TensorFlow also leads to large consumption in the CPU baseline). 
Some detailed result interpretation includes:


\textit{Though HBM can achieve satisfying performance on its own, Cartesian products further speed up the process.} For the smaller model, as shown in Table~\ref{tab_microrec:cartesian_benifits}, except those tiny tables stored on-chip, there are still 39 tables left to be allocated to DRAM. Considering there are 34 DRAM channels in total (32 for HBM, 2 for DDR), it takes two DRAM access rounds to lookup 39 tables. Cartesian products can reduce the table number to 34, so that only one round of DRAM access is required. The experiment shows that, with Cartesian products, the latency of embedding lookup is only 59.17\% of the HBM-only solution (458 ns vs 774 ns). Similarly, for the larger model, Cartesian products reduce the memory access rounds from 3 to 2, consumed only 72.12\% of the time (1.63 us vs 2.26 us).

\textit{The storage overhead of Cartesian products is fairly low.} As shown in table \ref{tab_microrec:cartesian_benifits}, the products only lead to 3.2\% and 1.9\% storage overhead on the two models respectively. This is because only small tables are selected for Cartesian products as introduced in section \ref{sec_microrec:microrec_algorithm}, so that the products are still considerably small compared to a single large table.

\textit{By Cartesian products and HBM, the memory bottleneck caused by embedding lookup is eliminated.} Since the embedding lookups only cost less than 1 microsecond in MicroRec (as in Table~\ref{tab_microrec:exp_embedding}), the bottleneck shifts back to computation, in which the most expensive stage takes several microseconds. 

\begin{figure}
\centering
\includegraphics[width=0.75\linewidth]{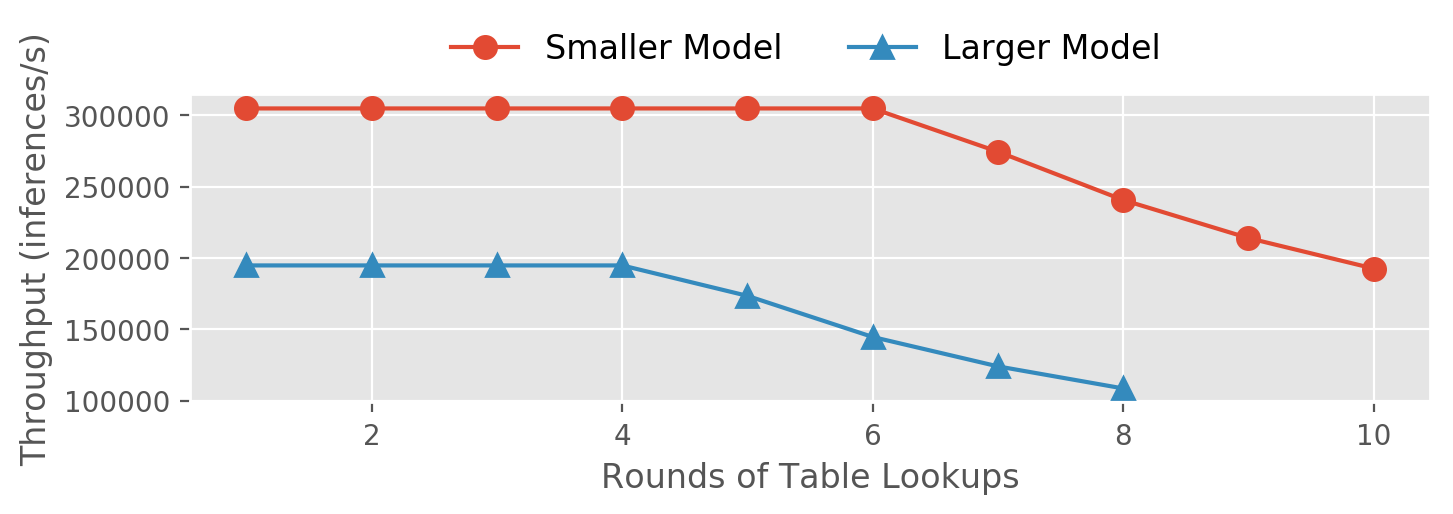}
\caption{End-to-end inference throughput of MicroRec. It allows multi-rounds lookup without sacrificing performance.}
\label{fig_microrec:multi-round-lookups}
\end{figure}

\textit{The accelerator performance is robust even as multiple rounds of lookups are required.} Although the production models only involves one lookup operations per table, alternative DNN architectures may require multiple rounds of lookups~\cite{facebook_benchmark}. Figure~\ref{fig_microrec:multi-round-lookups} proves the performance robustness of MicroRec in such scenarios by assuming more rounds of embedding retrievals on the two production models --- the smaller and larger models can tolerate 6 and 4 rounds of lookups without downgrading the end-to-end inference throughput at all using 16-bit fixed-points, because the DNN computation and embedding lookup stages are overlapped. Once more rounds of lookups are assumed, the performance starts to depend on the total memory access latency which is proportional to the rounds of DRAM accesses.

\paragraph{Performance on Benchmark Models}

\begin{table}
  \caption{MicroRec achieves 18.7$\sim$72.4$\times$ embedding lookup speedup compared to the Facebook's recommendation baseline.}
  \label{tab_microrec:facebook_benchmark}
\begin{center}
\begin{small}
  \begin{tabular}{@{} L{6em} c c c c c @{}} 
\toprule
\multirow{2}{*}{Performance} & \multicolumn{5}{c}{Embedding Vector Length} \\
\cmidrule{2-6}
     & 4	& 8 & 16 & 	32 & 64  \\
\midrule 
\multicolumn{6}{c}{8 Tables (Speedup Upper Bound)}\\
\midrule
\vspace{0.2em} Lookup (ns) &	334.5 & 353.7 & 411.6 &  486.3 & 648.4 \\
\vspace{0.2em} Speedup & 72.4$\times$ & 68.4$\times$ & 58.8$\times$ & 49.7$\times$ & 37.3$\times$ \\
\midrule
\multicolumn{6}{c}{12 Tables (Speedup Lower Bound)}\\
\midrule
\vspace{0.2em} Lookup (ns) & 648.5 & 707.4 & 817.4 & 972.7 & 1296.9 \\
\vspace{0.2em} Speedup & 37.3$\times$ & 34.2$\times$ & 29.6$\times$ & 24.8$\times$ & 18.7$\times$ \\
\bottomrule

  \end{tabular}
\end{small}
\end{center}
\end{table}

I compare the embedding lookup performance of MicroRec to the recent recommendation inference benchmark~\cite{facebook_benchmark}. Although the referred paper does not expose all model parameters, I can still identify the embedding lookup performance range on MicroRec by experimenting a range of table settings.
To be more specific, I experiment the embedding-dominated model class DLRM-RMC2, which contains 8$\sim$12 small tables and each table is looked up 4 times (thus $32\sim48$ lookups in total). 
Several assumptions are made for the missed information. First, by ``small tables'', I assume each table is within the capacity of an HBM bank (256MB). Second, I assume common embedding vector lengths from 4 to 64. Third, no Cartesian products are applied in the experiments, since the table sizes are assumed by us. 


Table \ref{tab_microrec:facebook_benchmark} shows the embedding lookup performance on MicroRec: it achieves 18.7$\sim$72.4$\times$ speedup compared to the published baseline performance (2 sockets of Broadwell CPU @2.4GHz; 14 cores per socket; AVX-2 supported; 256 GB 2400MHz DDR4 DRAM; batch size=256). This performance range is identified by experimenting table numbers from 8 to 12 and vector lengths from 4 to 64. The highest speedup occurred when there are only 8 embedding tables (32 lookups) with a short vector size of 4, for which only one round on HBM lookup is required.
The lowest speedup happens when there are 12 tables with a long vector size of 64, where 2 rounds of HBM accesses are necessary.  

\section{Related Work}
\label{sec_microrec:related_work}

This section introduces to hardware solutions for recommendation systems. 
According to Facebook, recommendation workloads can consume up to 79\% of total AI inference cycles in data centers~\cite{facebook_benchmark}.
However, little research has been focused on serving personalized recommendations efficiently. In order to provide enough background knowledge to the research community and tackle this important problem, \citet{facebook_benchmark} analyzed the recommendation workload comprehensively, open-sourced several models used by Facebook, and set up a performance benchmark.
\citet{tensordimm_kaist} is the first hardware solution for high performance recommendation inference. They reduced the memory bottleneck by introducing DIMM-level parallelism in DRAM and supporting tensor operations, e.g., gather and reduction, within the DRAM. \citet{ke2020recnmp} extended the idea of near-memory-processing and added memory-side-caching for frequently-accessed entries. \citet{gupta2020deeprecsys} took into account the characteristics of query sizes and arrival patterns, and developed an efficient scheduling algorithm to maximize throughput under latency constraints by using both CPUs and GPUs. \citet{hwang2020centaur} implemented an FPGA accelerator (without HBM) for deep recommendation inference, and the speedup was significant for models with few embedding tables. Compared to previous work, MicroRec is the first system that introduces data structure solution, i.e., Cartesian products, to reduce the number of DRAM accesses. It is also the first work resorting to HBM so as to parallelize embedding lookups.

\section{Conclusion}

I design and implement MicroRec, a high-performance deep recommendation inference engine. On the data structure side, MicroRec applies Cartesian products to reduce sparse memory accesses. On the hardware side, HBM is adopted to scale up embedding lookup concurrency, and the deeply pipelined architecture design on FPGA enables low inference latency. 
By the three strategies I propose, the memory bottleneck caused by embedding lookups is almost eliminated, and the latency requirements of recommendation inference are easily met.



\chapter{FleetRec: A Hybrid GPU-FPGA System for DLRM Serving}
\label{chap:fleetrec}

This chapter further improves recommender model inference efficiency by introducing a heterogeneous hardware system, building upon the foundations of efficient embedding table lookups established in Chapter~\ref{chap:microrec}.

\section{Introduction}

Due to the embedding table architecture and the need for real-time recommendations, three challenges are faced to build efficient inference systems for recommendations. 
\textit{First}, the embedding table architecture becomes a performance bottleneck. Due to the tiny size of each embedding vector (usually 4 to 64 dimensions) and the large number of embedding tables (tens to hundreds), the embedding table lookup operations are costly because they induce massive random DRAM accesses, leading to low memory bandwidth utilization and significantly downgraded performance. Even worse, these lookup operations result in extra overhead if one resorts to state-of-the-art machine learning frameworks such as TensorFlow and PyTorch. For example, even in an inference-oriented framework such as \textit{TensorFlow Serving}, there are tens of types of operators involved in the embedding layer and each operator is invoked multiple times. The many operator invocations significantly degrade performance, especially as small batches are often required for real-time inference.
\textit{Second}, the scale of recommendation models can reach over 100 GB since some embedding tables are huge, e.g., account information encodings. Such sizes exclude the option of using hardware accelerators, e.g., FPGAs and GPUs, as the inference engine because of the lack of memory on the device.
\textit{Third}, the latency requirement is stringent (usually tens of milliseconds).
Large batch sizes usually lead to better throughput for CPUs and GPUs because of the better utilization of the single instruction multiple data (SIMD) architecture and the amortization of function call overheads. 
In real-time recommendation systems, the \textit{Performance Metric} is throughput under service-level agreement (SLA) constraints, limiting the batch sizes usable in practice.
Although huge batch sizes are beneficial for CPUs and GPUs to improve throughput (inferences per second), recommendation systems require small batches due to the latency constraints.

Although much effort has been invested into accelerating deep recommendation models~\cite{tensordimm_kaist, gupta2020deeprecsys, hwang2020centaur, ke2020recnmp, jiang2020microrec}, they all fail to solve some of the challenges above, making them suitable only for a subset of use cases. 
For instance, \citet{gupta2020deeprecsys} suggests GPUs could be useful in recommendation for large batches compared to regular CPU-based engines, but the embedding performance bottleneck remains and GPUs cannot serve large models for the lack of memory capacity. Similarly, hybrid CPU-GPU and CPU-FPGA designs are evaluated but without solving the memory bottleneck~\citet{hwang2020centaur}.
 \citet{jiang2020microrec} resort to the high-bandwidth memory (HBM) available on FPGAs for high-performance embedding lookups, but its applicability is heavily limited to small models because of the 8 GB of HBM available on the board.  
\citet{tensordimm_kaist} and \citet{ke2020recnmp} propose to redesign DRAM at the micro-architectural level; however, it takes years to put such new DRAM chips in production even if they are eventually adopted, making the solution interesting from a research perspective but not from a practical stand point.

\textbf{Goal.} This work targets to build an end-to-end high-performance recommendation inference system that can (a) achieve high throughput (inferences per second) under SLA (latency) constraints of tens of milliseconds, and (b) adapt for various models with minimal usage of hardware devices (the model sizes can range from hundreds of MB to hundreds of GB, and the workload characteristic can be embedding-lookup-intensive or computation-intensive).

\textbf{Solution.} Based on the careful analysis of three production-scale models, I design and implement \textit{FleetRec}, a high-performance and configurable heterogeneous computing cluster for recommendation inference. 
On the embedding table lookup side, I resort to (a) FPGAs equipped with high-bandwidth memory (HBM) to enable highly concurrent lookup operations and (b) CPU servers with sufficient DRAM capacity for a few large tables (e.g., tens of GB). On the computation side, I use GPUs exclusively for DNN computation to avoid the irregular memory lookup operations that degrade the SIMD performance. 
These hardware resources (GPUs, FPGAs, and CPU servers) are regarded as end devices connected through a high-speed network (100 Gbps per link), so that one can configure the node type and quantity to support various size scales (up to hundreds of Gigabytes), number of embedding tables, and computation density.

\textbf{Key Results.} I evaluate FleetRec on three production models from Alibaba covering size scales from 1 GB to over 100 GB. FleetRec achieves 15.5$\sim$49.0$\times$ speedup in terms of throughput over the CPU-baseline and 7.4$\sim$16.1$\times$ speedup over FPGA accelerators. Besides, FleetRec lowers the inference latency by 21.0\%$\sim$92.5\% percent compared to CPUs. As a result, FleetRec is an ideal candidate for real-time inference --- it outperforms a CPU based system by 41.8$\sim$387.2$\times$ given a 10 ms latency bound. Besides the three industry models, one can also generalize FleetRec to any recommendation models: the performance interpretability of FleetRec enables to estimate the performance of any model without the need to implement the hardware. With this, the contributions of the chapter include:
\begin{itemize}
    \item I design and implement FleetRec, a high-performance and configurable recommendation engine supporting a wide range of model size scales and architectures. The design is based on the performance characteristics of three production models used by Alibaba.
    \item I implement an efficient dataflow architecture on an FPGA for high-throughput embedding table lookups. I also integrate a 100 Gbps TCP/IP stack into Xilinx's Vitis development platform, enabling FPGAs to serve as smart disaggregated memory for recommendations. I further develop an optimized software infrastructure on the GPU server, allowing a seamless and high-performance coordination between the memory and computation nodes.
    \item I test FleetRec on three production models. Compared to an optimized CPU baseline, FleetRec shows more than one order of magnitude speedup in terms of throughput while significantly lowering latency --- a significant advantage in real-time recommendations.
    
\end{itemize}

\section{Background \& Motivation}

This section points out the challenges to design a high-performance inference system for it, namely the embedding-vector-lookup bottleneck, the latency constraints, and the model size scale. Then existing solutions and their shortcomings are discussed. 

\subsection{Inference Challenges}
\label{sec_fleetrec:motivation_challenges}


\begin{figure}[t]
  \centering
  \includegraphics[width=0.85\linewidth]{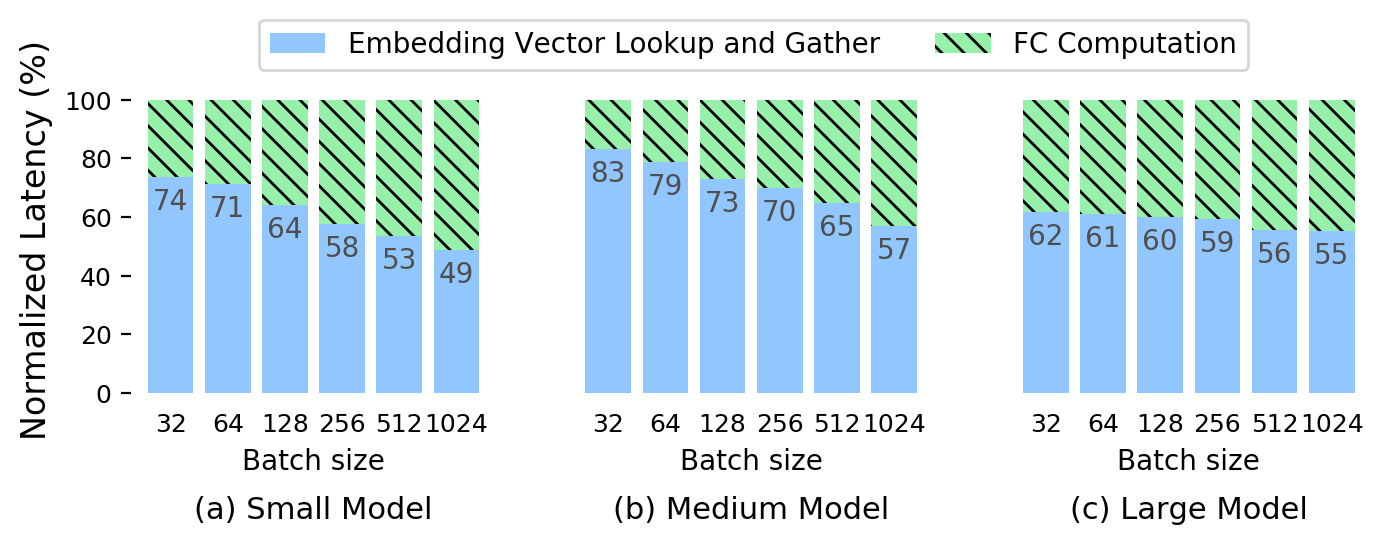}
  \caption{Latency breakdown of three production recommendation models ranging from 1 GB to over 100 GB.}
  \label{fig_fleetrec:CPU-embedding-breakdown}
\end{figure}

\textbf{Challenge 1: embedding table lookup.} The lookup of the embedding tables is a unique bottleneck in recommendation inference compared to regular DNN workloads. Figure~\ref{fig_fleetrec:CPU-embedding-breakdown} shows the cost of embedding layers during inference on three production models ranging from 1 GB to over 100 GB.

These lookup operations cause performance issues for two reasons. 
First, looking up embedding vectors on many tables causes random DRAM accesses, leading to significantly under-utilized memory bandwidth and low lookup performance. 
Second, embedding lookups result in operator-call overhead if one resorts to machine learning frameworks such as TensorFlow and PyTorch. According to the performance observations on \textit{TensorFlow Serving} which is optimized for inference, the embedding layer involves 37 types of operators (e.g., concatenation and slice) and these operators are invoked multiple times during inference, resulting in a large overhead especially for small batches. Unfortunately, small batch sizes are usually required in CPU-based recommendation engines to meet the latency requirements of tens of milliseconds. 

\textbf{Challenge 2: serving models over 100 GBs.} Embedding tables usually contribute to the majority the memory requirements in recommendation systems. At industrial scale, they can contain up to hundreds of millions of entries, consuming tens or even hundreds of gigabytes of memory. For example, the largest model in the experiments contains 377 embedding tables and requires 114 GB of memory. The single largest table contains 2 million entries of 64-dimensional encoded vectors: over 50 GB for a single table. Such sizes pose a challenge for specialized hardware. For example, the DRAM capacity of GPUs and FPGAs is around a few tens of GB, thus unable to serve large recommendation models.

\textbf{Challenge 3: optimizing throughput under SLA constraints.} Large batch sizes usually lead to better throughput for CPUs and GPUs because of the better utilization of the SIMD architecture and the amortization of function call overheads. 
In real-time recommendation systems, the \textit{Performance Metric} is throughput under SLA constraints, limiting the batch sizes usable in practice.


\subsection{Existing Approaches \& Limitations}

\begin{table*}
  \caption{FleetRec compared with existing solutions.}
  \label{tab_fleetrec:existing_solutions}
\begin{center}
\scalebox{0.61}{
\begin{tabular}{@{} L{9em} M{9em} M{9em} M{11em} M{11em} M{9em}@{}} 
    \toprule
    Solution & Embedding Lookups & DNN Computation & Supported Model Size  & Throughput under SLA & Inference Latency \\
    \midrule
    CPU~\cite{tensordimm_kaist, facebook_benchmark, hsia2020cross} & Slow & Slow &  Large & Low & Medium \\
    GPU~\cite{hsia2020cross, gupta2020deeprecsys, hwang2020centaur} & Slow &  Fast &  Medium$\sim$Large & Low$\sim$Medium & Medium$\sim$High \\
    CPU-FPGA~\cite{hwang2020centaur} & Slow & Medium  & Large & Low$\sim$Medium & Very Low \\
    FPGA with HBM~\cite{jiang2020microrec} &  Fast &  Medium & Medium & Medium & Very Low \\
    \midrule
    FleetRec (Ours) &  Fast  & Fast &  Large and Scalable &  High & Low \\
    \bottomrule
  \end{tabular}
}
\end{center}
\end{table*}

\begin{figure*}
  \centering
  \includegraphics[width=\linewidth]{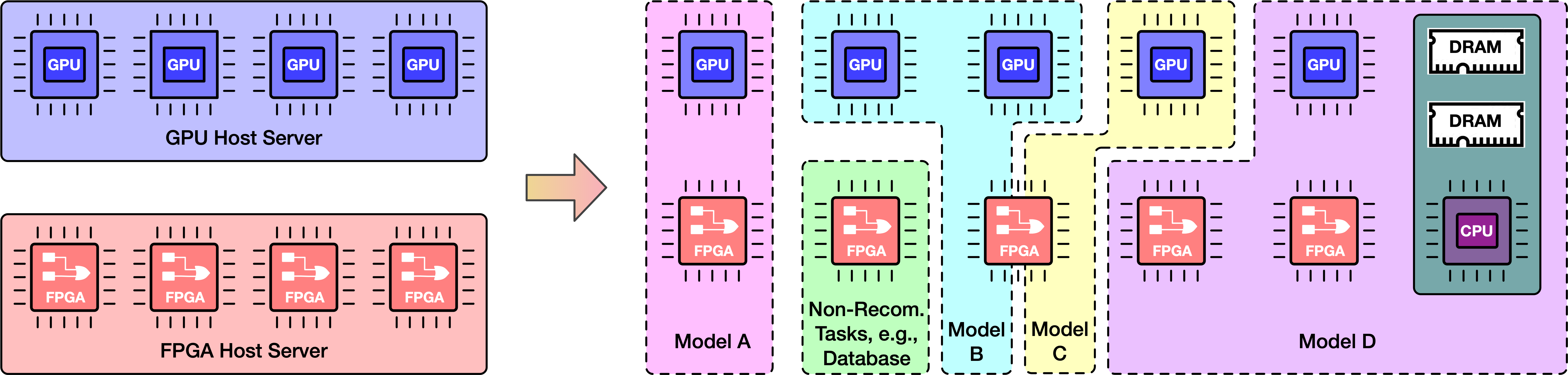}
  \caption{System overview of FleetRec. It is built upon heterogeneous computing clusters consists of CPUs, GPUs, and FPGAs without modifying the server setups. FPGAs and CPU servers are embedding table lookups engines while GPUs provide DNN computation firepower. These hardware resources can be bridged flexibly by a high-speed network to adapt various recommendation workloads (model A$\sim$D) while leaving the remaining hardware resources for non-recommendation tasks.}
  \label{fig_fleetrec:FleetRec_system_overview}
\end{figure*}

Although several solutions have been proposed to serve recommendation models, they all fail to meet some of the challenges just described. This section introduces existing solutions, summarises their pros and cons, and points to the need for a novel system capable of meeting all outstanding challenges.

\textit{CPU-based.} CPU-based recommendation inference is the go-to choice in industry~\cite{facebook_benchmark, gupta2020deeprecsys}, because (a) deployment on CPU servers requires no extra investment on novel hardware; (b) the DRAM capacity installed on CPU servers is typically enough to serve large recommendation models; and (c) inference latency on small batches is generally lower than in GPUs. Though widely deployed, CPUs are not known for their DNN inference performance compared to those of GPUs, and the memory bottleneck caused by the embedding lookups worsens the situation.

\textit{GPU-based.} Two deployments have been explored on GPU. Leaving both the embedding lookup and the DNN computation to the GPU is one option~\cite{hsia2020cross, gupta2020deeprecsys}, but it cannot serve large models because of the limited GPU DRAM capacity. Besides, although GPUs use high memory bandwidth (HBM), their SIMD architecture is not suitable for irregular operations such as individual table lookups, losing the bandwidth advantage of HBM. Another option is to do the embedding lookups on the CPU, and then transfer the concatenated vector to the GPU through the PCIe bus~\cite{hwang2020centaur}. 
In general, the speedup of GPUs improves the DNN computation but the memory bottleneck remains. In addition, GPUs can deliver high throughput for large batches but do less than optimally for the small batches needed to meet the latency constraints.

\textit{FPGA-based.} FPGAs can be viewed as application-specific integrated circuits once they are programmed, thus are suitable for latency-sensitive applications such as recommendation. \citet{hwang2020centaur} is an FPGA solution in which the FPGA accesses the CPU-side DRAM for embedding lookups and performs the DNN inference on the FPGA. This solution provides enough memory capacity but still suffers from the embedding bottleneck and the speedup is mainly obtained from the fast DNN computation. \citet{jiang2020microrec} provides an alternative solution taking advantage of the High-Bandwidth Memory (HBM) available on the latest FPGA models, thus providing the highest embedding lookup performance among existing solutions. However, it is restricted to small model sizes since the available memory capacity is only 8 GB of HBM plus 32 GB of DDR4 DRAM.

\textbf{Goal.} 
As summarized in Table~\ref{tab_fleetrec:existing_solutions}, all existing solutions have their limitations, thus are only suitable for a subset of inference scenarios. In this chapter, I aim to obtain a single solution that can (1) minimizes the memory bottleneck caused by embedding table lookups; (2) achieves high end-to-end inference throughput under SLA constraints; and (3) supports recommendation models larger than 100 GB.

\section{FleetRec}

\textbf{Key Advantages.} I introduce FleetRec, a high-performance and configurable heterogeneous computing cluster for recommendation inference. FleetRec provides several advantages over current solutions. First, it combines the strengths of heterogeneous hardware (FPGAs and GPUs) while avoiding the weaknesses of each platform. Second, it scales out to large models by simply plugging in more memory nodes (FPGAs and CPU servers). Third, by configuring the ratio of the two types of nodes (memory and computation), the cluster is adaptable to various workloads, regardles of whether the models are computation-intensive or table-lookup-intensive. 
In the following, I present the key insights of FleetRec's design abstracting away the low level details of the hardware implementation. 

\subsection{System Overview}

\textbf{System Components.} Figure~\ref{fig_fleetrec:FleetRec_system_overview} shows the architecture of FleetRec, built on a cluster of heterogeneous hardware. The embedding vector lookups are performed on FPGAs equipped with high-bandwidth memory and CPU servers with sufficient DRAM, while the DNN computation happens in the host servers equipping GPUs.
FleetRec regards each accelerator as an individual end device connected by 100 Gbps TCP/IP network. FleetRec can be adapted to a wide range of workloads by using different configurations that vary the number and interconnection topology between CPUs, GPUs, and FPGAs (Figure~\ref{fig_fleetrec:FleetRec_system_overview}).

\textbf{Recommendation Query Processing Flow.} The inference starts by retrieving the embedding vectors (the sparse features) through lookup indexes on the embedding tables residing in the memory nodes (FPGAs and, when needed, CPU servers). Each memory node completes the table lookup and concatenates the retrieved vectors before sending them to the GPU server. The GPU server concatenates all the received embedding vectors with dense features and runs them by the DNN in a batch. 

\textit{The key design philosophy of FleetRec is two-fold:}

\textbf{First, FleetRec takes advantage of the strengths of different types of hardware.} The latest FPGA equipped with HBM enables highly concurrent embedding table lookups (a few tens of lookups in parallel), yet it has limited memory capacity (8 GB HBM + 32 GB DDR) and insufficient DNN computation performance. According to the experiments implementing FPGA accelerators for recommendation inference, the DNN computation module is one order of magnitude slower than the HBM-fueled embedding lookup module \cite{jiang2020microrec}. GPUs are great for pure computation but suffer from irregular memory access patterns such as the embedding table lookups.
FleetRec combines the best of both worlds: FPGAs implement the embedding table lookups while GPUs run the DNN computation. FleetRec can also include CPU servers as memory nodes to provide sufficient DRAM capacity for a few large tables: the largest embedding table in the evaluated models is more than 50 GB. Keeping such tables in a CPU server is a more efficient choice than FPGAs.

\textbf{Second, FleetRec disaggregates computation and memory, leading to high scalability and flexibility.}
Instead of plugging a certain number of FPGAs and GPUs to the same host server, FleetRec treats these accelerators as individual end devices connected by a network, enabling flexible combinations between computation and memory resources. To scale out and support large recommendation models, more FPGAs or CPU nodes can be added to the cluster. To adapt to different models, the resources allocated to computation or embedding lookups can be independently adjusted to balance performance between the two components. For DNN-computation-intensive model architectures, installing more GPUs on the computation node matters more than having multiple memory nodes. For models with hundreds of embedding tables, pairing several FPGA nodes with one GPU can be a more reasonable choice. 

\subsection{The FPGA as Smart Disaggregated Memory}

\begin{figure}
  \centering
  \includegraphics[width=0.8\linewidth]{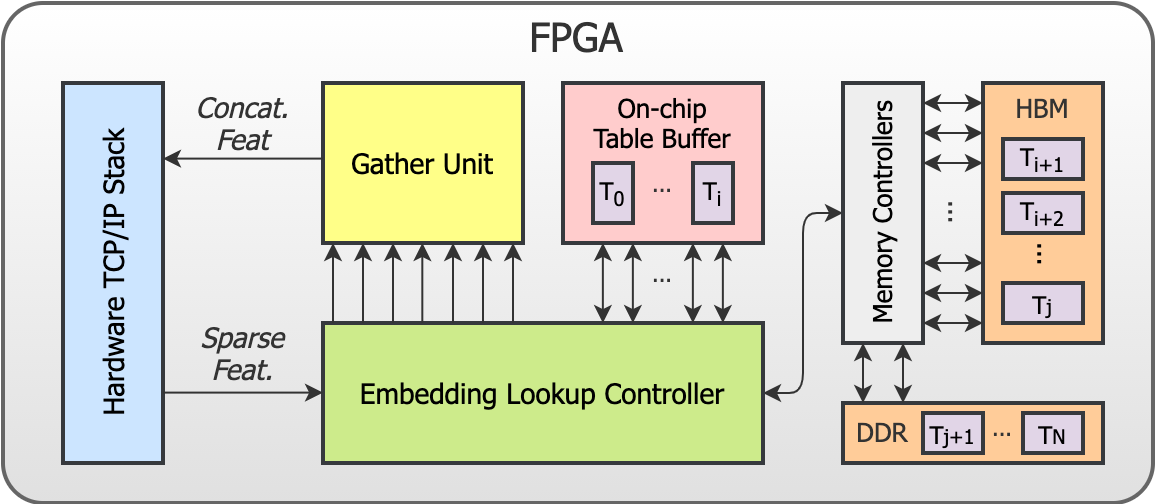}
  \caption{The hardware design of a FleetRec FPGA.}
  \label{fig_fleetrec:fpga_design}
\end{figure}

I use the latest FPGAs equipped with hybrid memory system as smart disaggregated memory to enable highly parallel embedding lookups. 
The Xilinx Alveo U280 FPGA cards used in the experiments contain three types of memory: high-bandwidth memory (HBM), DDR4 DRAM, and on-chip memory (BRAM and URAM). The HBM system on the U280 offers improved concurrency (32 independent memory banks) and bandwidth (up to 425 GB/s) compared to conventional DRAMs~\cite{jun2017hbm, o2014highlights, zeke2020benchmark_hbm}. Thus, embedding tables can be distributed across these banks so that each bank only contains one or a few tables, and up to 32 tables can be looked up in parallel\footnote{I use most (28 of 32) HBM banks to hold embedding table in the experiments. Another 2 HBM channels serve as network cache, while the rest 2 channels are not used because they overlap with the PCIe region of the card and using them can lead to routing issues and degraded performance~\cite{choi2020hls}.}. 
The 2 DDR4 DRAM channels on the board provide higher memory volume (32 GB DDR vs 8 GB HBM), although it only supports 2 parallel lookup operations at a time.
Besides DRAM (HBM and DDR) for which the memory access latency is around a couple of hundreds of nanoseconds, the U280 card also contains tens of MB of on-chip SRAM, with low-latency access comparable to that of a CPU cache.

To maximize the embedding lookup performance, a FleetRec FPGA allocates the tables to its hybrid memory system in the following manner. 
There are no embedding data exchange between the memory hierarchy because the model is known at the system development stage.
First, it stores as much small tables as possible in SRAM since that allows low-latency and high-concurrency vector retrieval. 
Second, it distributes the rest tables in HBM and DDR banks. The largest tables are stored in DDR banks because of its higher capacity (16 GB per DDR bank compared to 256 MB of an HBM bank).
Because the random access latency to HBM and DDR are close  (200$\sim$300 ns), FleetRec FPGA ties the number of embedding tables stored in each bank to balance the workload. 
During the embedding lookup process, vectors stored in different banks can be read in parallel and the performance is decided by the rounds of DRAM access. For example, a model contains 90 tables: 30 of them are stored on-chip while the rest 60 are evenly distributed to HBM and DDR banks (30 banks available in total). Then the FPGA will concurrently gather all on-chip vectors and issue 2 rounds of parallel DRAM accesses to retrieve all embedding vectors.

Figure~\ref{fig_fleetrec:fpga_design} illustrates the hardware design of an FPGA node in FleetRec: it takes sparse feature (lookup indexes) as input and outputs the concatenated embedding vectors through the network to the computation node. FleetRec uses an open-source 100 Gbps network stack~\cite{100gbps} integrated into the Vitis FPGA development platform~\cite{vitis}, so that the integrated network stack supports the Xilinx U280 FPGA cards as well as High-Level Synthesis, an FPGA development flow allowing programming hardware in C/C++. I then implement the components enabling the FPGA to serve as smart disaggregated memory for recommendation, including a table lookup controller to handle memory accesses and a gather unit to concatenate all the retrieved vectors.

\subsection{The GPU as DNN Engine}

Figure~\ref{fig_fleetrec:gpu_server} shows the working flow on the FleetRec GPU server. The inference starts by receiving the concatenated input features sent by the memory nodes. The batched input features are stored into page-locked memory in the CPU side DRAM (GPUs do not have direct network access). The GPU reads a batch of input features, runs the DNN computation, and returns the predicted CTR. I optimize performance to maximize throughput as follows:

\begin{figure}
  \centering
  \includegraphics[width=0.85\linewidth]{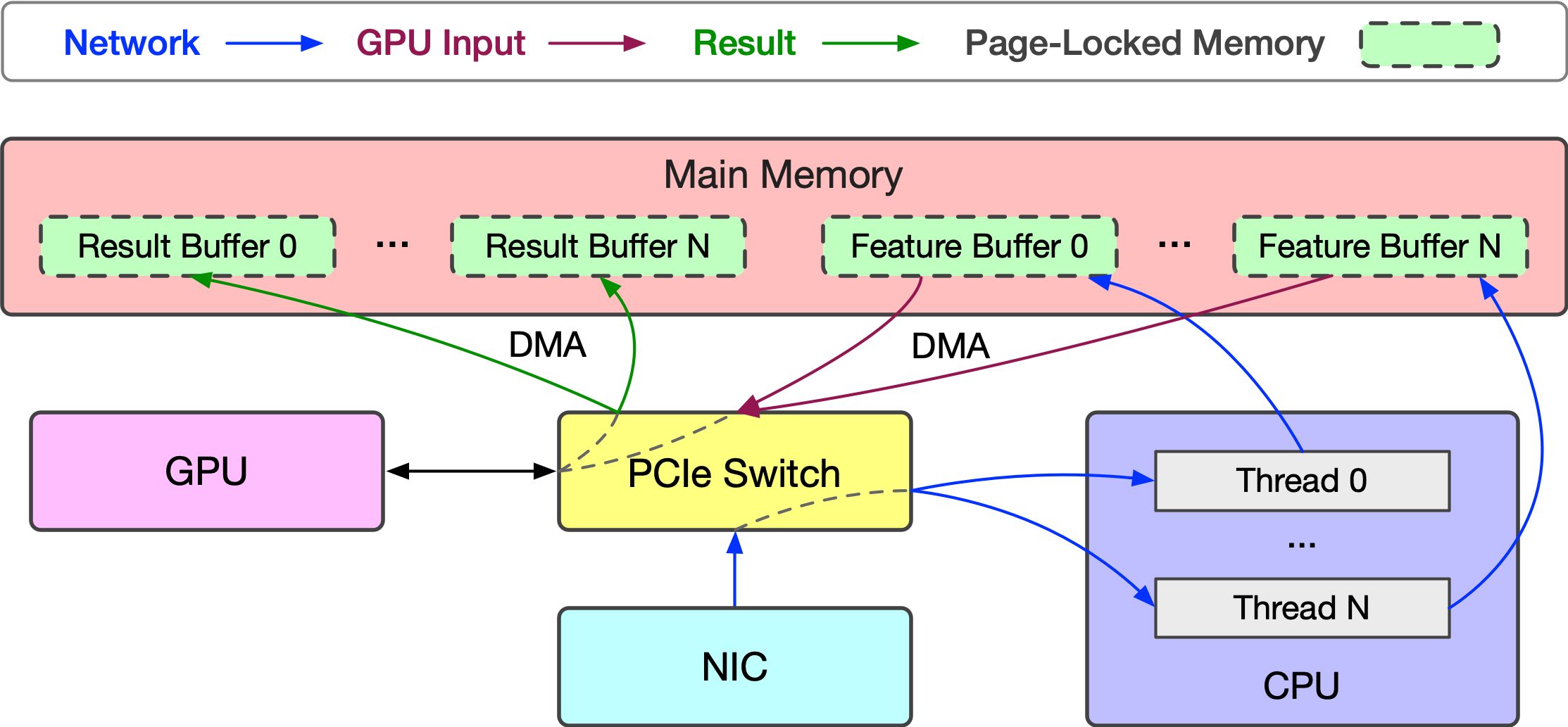}
  \caption{The GPU server receives concatenated feature from memory nodes over network and finishes inference.}
  \label{fig_fleetrec:gpu_server}
\end{figure}

\begin{figure}
  \centering
  \includegraphics[width=0.85\linewidth]{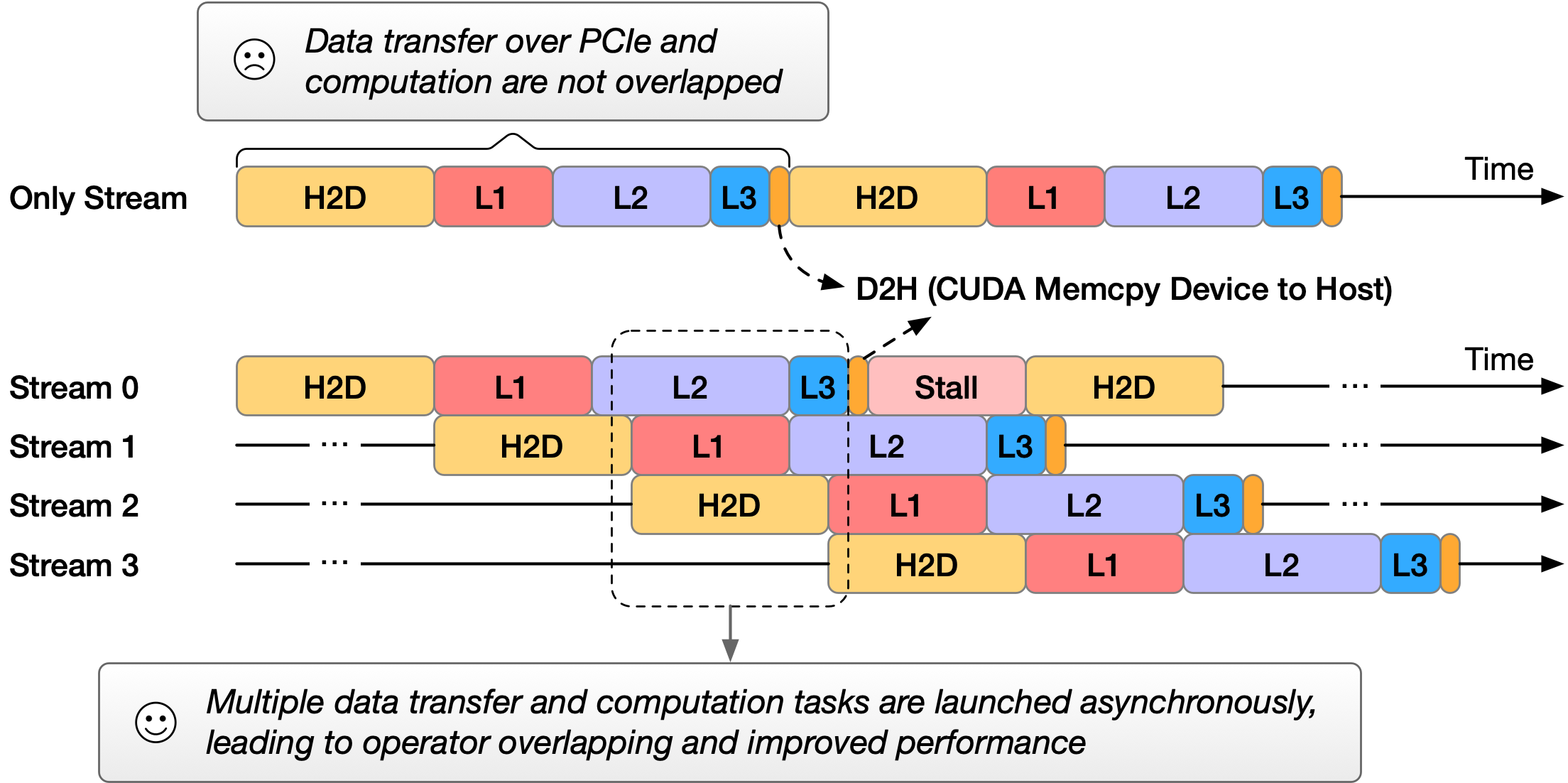}
  \caption{Maximize the GPU performance by overlapping data transfer and computation using multiple streams.}
  \label{fig_fleetrec:gpu_stream}
\end{figure}

\textit{GPU operator scheduling}. To maximize GPU utilization, 
I use multiple \textit{CUDA streams} for inference.
As shown in the lower half of Figure~\ref{fig_fleetrec:gpu_stream}, employing multiple streams (a) enables operator-level concurrency and (b) overlaps computation (3 layers of DNN in the form of general matrix-matrix multiplication) and communication between host and device (H2D and D2H).
In this case, the GPU throughput is maximized with a marginal latency overhead.

\textit{From network to main memory.} I use multiple threads on the CPU side for network packet processing, memory management, and issuing tasks to the GPU. As discussed above, multiple CUDA streams are launched to maximize inference throughput and I utilize individual threads on the host CPU to handle each task stream. 
The jobs of a thread include: (a) establishing TCP/IP connections to memory nodes, (b) receiving the network packets and storing the input feature into main memory, and (c) issuing the GPU commands including data transfer and computation. 
Page-locked memory serves as input feature buffer in main memory, so that direct memory access (DMA) between the CPU main memory and the GPU is possible (with DMA, a GPU can access the CPU side memory without involving CPU, thereby increasing throughput and reducing latency).
After each batch of DNN computations, the GPU writes the predicted CTRs to the host memory.

\section{Evaluation}

\begin{table}

  \caption{Specification of the three production models.}
\begin{center}
\label{tab_fleetrec:model_size}
\begin{small}
  \begin{tabular}{ @{} L{3em} M{5em} M{5em} M{7em} M{5em} @{} }   
\toprule
    Scale & \multicolumn{1}{c}{Table Num} & \multicolumn{1}{c}{Feature Len} & \multicolumn{1}{c}{Hidden-Layer} & \multicolumn{1}{c}{Size} \\
    
\midrule
    Small & 47 & 352 & (1024, 512, 256) & 1.3 GB \\
    Medium & 98 & 876 & (1024, 512, 256) & 15.1 GB \\
    Large & 377 & 3968 & (2048, 512, 256) & 114.4 GB \\
\bottomrule
\end{tabular}
\end{small}
\end{center}
\vskip -0.1in
\end{table}

\textbf{Results Overview.} I first evaluate FleetRec on three production models from Alibaba. FleetRec shows significant throughput improvement over both the CPU and FPGA baselines while also significantly reducing latency. Due to the improved throughput and reduced latency, FleetRec is especially good at real-time inference under strict SLA constraints  --- it achieves two orders of magnitude speedup over the CPU based system given a 10 ms SLA.
I then show how to generalize and configure FleetRec for other recommendation models by estimating the speedup of FleetRec and balancing computation and lookup performance to maximize performance while minimizing hardware usage.

\subsection{Model Specification}

I experiment with three deep recommendation models of different sizes from Alibaba. All three models involve heavy embedding vector lookups, thus showing different workload characteristics compared with non-recommendation DNN architectures. 
Table~\ref{tab_fleetrec:model_size} shows the parameters of the models.
These models contain 47, 98, and 377 embedding tables respectively, and each table is looked up once during inference.  For example, during each inference, the largest model gathers the embedding vectors retrieved from 377 tables into a 3968-dimensional dense vector, and feeds it to three fully-connected layers. Though not shown in the table, the single largest embedding table contains 200M entries consuming 51.2 GB memory footprint.

\begin{figure*}[t]
  \label{tab_fleetrec:throughput_under_SLA}
  \centering
  \includegraphics[width=1.0\linewidth]{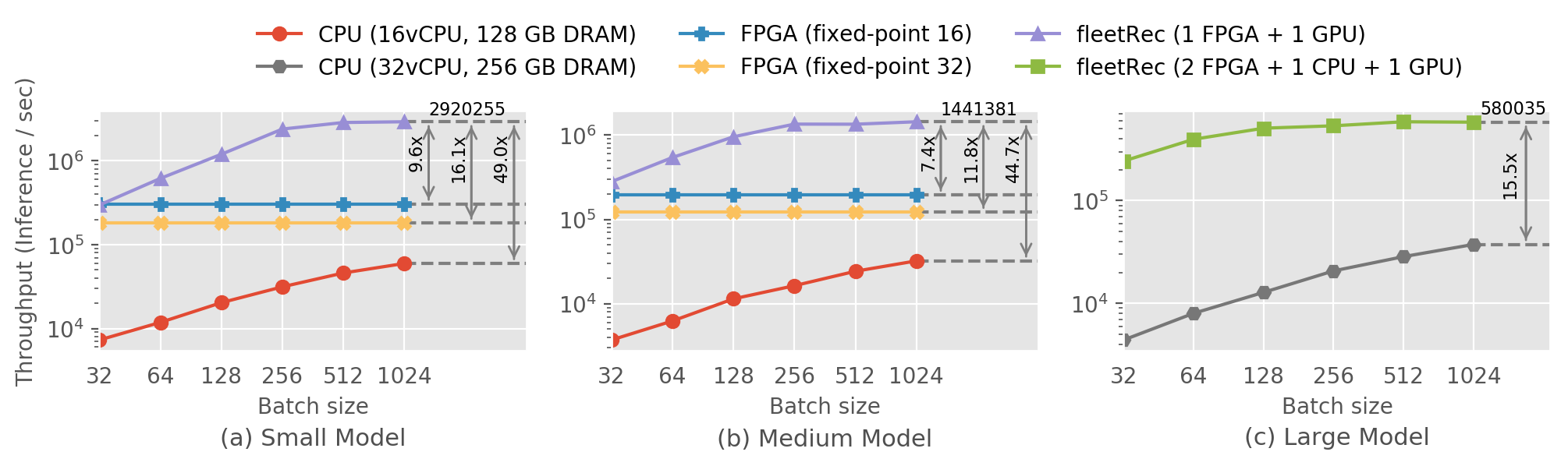}
  \caption{FleetRec significantly outperforms the CPU and FPGA baselines in terms of throughput.}
  \label{fig_fleetrec:throughput}
\end{figure*}

\begin{figure*}[t]
  \centering
  \includegraphics[width=1.0\linewidth]{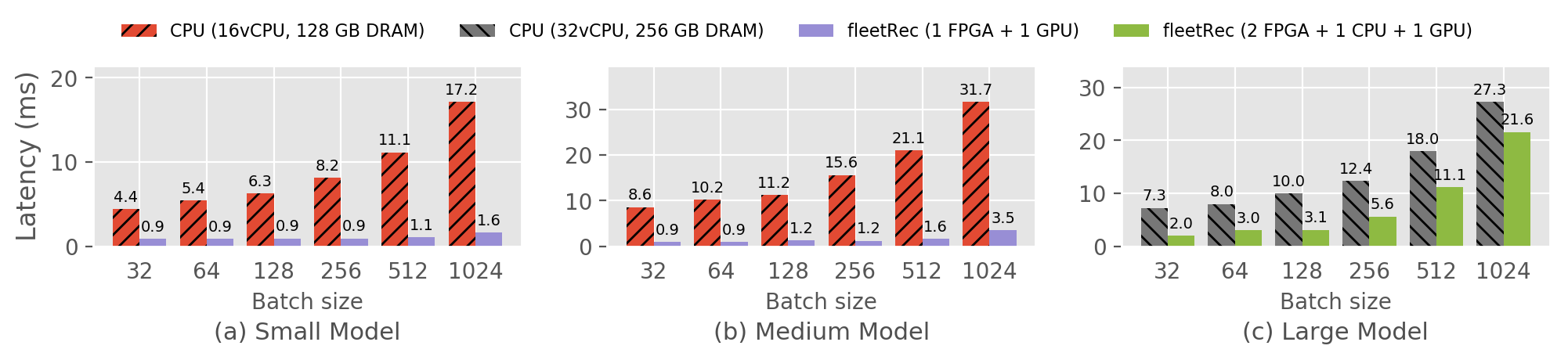}
  \caption{FleetRec achieves much lower latency compared to CPU engines given the same batch sizes.}
  \label{fig_fleetrec:latency}
\end{figure*}

\subsection{Experimental Setup}
\label{sec_fleetrec:exp_hardware_platform}

\textit{CPU baseline.} I run the models on two types of CPU servers on Amazon's AWS. The small and medium models are tested on a server with Intel Xeon E5-2686 v4 CPU @2.30GHz (16 vCPU, Broadwell, SIMD operations, i.e., AVX2 FMA, supported) and 128 GB DRAM (8 channels). A more powerful server with Intel Xeon Platinum 8259CL CPU @ 2.50GHz (32 vCPU, Cascade Lake, SIMD supported) and 256 GB DRAM is used for the larger model. For the ML framework, I use \textit{TensorFlow Serving} which is optimized for model inference. 

\textit{FPGA baseline.} Besides the CPU baseline, I also compare FleetRec with a set of FPGA recommendation accelerators (single FPGA implementation) optimized for each individual model. The FPGA accelerator is responsible for not only the embedding lookups but also the DNN computation. The computation module is implemented by constructing several general matrix-matrix multiplication (GEMM) blocks (each standing for one layer) connected by FIFOs. Each GEMM block is further composed by a set of processing elements (PEs) which is the basic unit to perform parallelized vector multiplications. Such modularized design maximizes hardware resource usage and avoids the performance degradation caused by placement and routing issues~\cite{johannes@matmul_hls}. Since FPGAs are naturally good at fixed-point computation rather than floating point, I perform quantization on the models and test the performance under two level of precision, i.e., 16-bit and 32-bit fixed-point numbers. Both the CPU baseline and FleetRec are tested with 32-bit floating point. Due to the memory capacity limitations in the  FPGA, the FPGA accelerator experiments only include the small and medium models.

\textit{FleetRec setups.} Similar to the CPU baseline, I use two FleetRec configurations. One FPGA plus a GPU are enough to serve the small and medium models. The larger models contains hundreds of tables consuming 114 GB memory, thus I upgrade the disaggregated memory to 2 FPGAs and 1 CPU server.
I use Xilinx Alveo U280 FPGAs equipped with 8GB of HBM2 DRAM (32 channels) and 32 GB of DDR4 DRAM (2 channels). I implement the FPGA hardware logic using Vitis HLS and set the clock frequency to 180 MHz. For the computation node, I use a Titan RTX GPU containing 4608 CUDA cores. The DNN computation flow is constructed by the cuBLAS library. I test 1$\sim$16 CUDA streams for all three models to maximize the GPU performance and presents the results with the highest throughput. The GPU server uses a Mellanox ConnectX-5 NIC with a 100 Gbps Ethernet connection. 
The FPGAs use an open-source 100 Gbps TCP/IP network kernel~\cite{100gbps}.

\subsection{End-to-End Inference Performance}

\begin{table*}

\caption{Throughput under strict SLA requirements. FleetRec shows huge advantage over CPU for real-time inference.} 
\scalebox{0.67}{
\begin{tabular}{@{}
L{9em} 
R{4em} R{4em} R{4em} 
M{0em}
R{4em} R{4em} R{4em} 
M{0em}
R{4em} R{4em} R{4em} 
@{}}\toprule
& \multicolumn{3}{c}{Small Model} & \phantom{}& \multicolumn{3}{c}{Medium Model} & \phantom{} & \multicolumn{3}{c}{Large Model}\\
\cmidrule{2-4} \cmidrule{6-8} \cmidrule{10-12}
\multicolumn{1}{l}{SLA (ms)} & \multicolumn{1}{c}{5} & \multicolumn{1}{c}{10} & \multicolumn{1}{c}{20} && \multicolumn{1}{c}{5} & \multicolumn{1}{c}{10} & \multicolumn{1}{c}{20} && \multicolumn{1}{c}{5} & \multicolumn{1}{c}{10} & \multicolumn{1}{c}{20} \\ 
\midrule
\multicolumn{3}{l}{Throughput (inferences / sec)} \\
\hspace{3em} CPU & 7.30E+3 & 3.14E+4  & 5.96E+4 && N/A &  3.72E+3 & 1.64E+4  && N/A  & 1.28E+4 & 2.85E+4 \\ 
\hspace{3em} FPGA & 3.05E+5 & 3.05E+5 & 3.05E+5  && 1.95E+5 & 1.95E+5 & 1.95E+5  && N/A & N/A & N/A \\ 
\hspace{3em} FleetRec & 2.92E+6 & 2.92E+6 & 2.92E+6  && 1.44E+6 & 1.44E+6 & 1.44E+6  && 5.07E+5 & 5.35E+5 & 5.80E+5 \\ 

\multicolumn{3}{l}{Speedup of FleetRec over} \\
\hspace{3em} FPGA & 9.57$\times$ & 9.57$\times$ & 9.57$\times$ && 7.39$\times$ & 7.39$\times$ & 7.39$\times$ && $+\infty\times$ & $+\infty\times$ & $+\infty\times$\\  
\hspace{3em} CPU & 400.07$\times$ & 92.97$\times$ & 48.96$\times$ && $+\infty\times$ & 387.24$\times$ & 87.92$\times$ && $+\infty\times$ & 41.76$\times$ & 20.34$\times$ \\  
\bottomrule \end{tabular} 
}
\label{tab_fleetrec:throughput_under_SLA} 
\end{table*}

\begin{figure*}[t]
  \centering
  \includegraphics[width=1.0\linewidth]{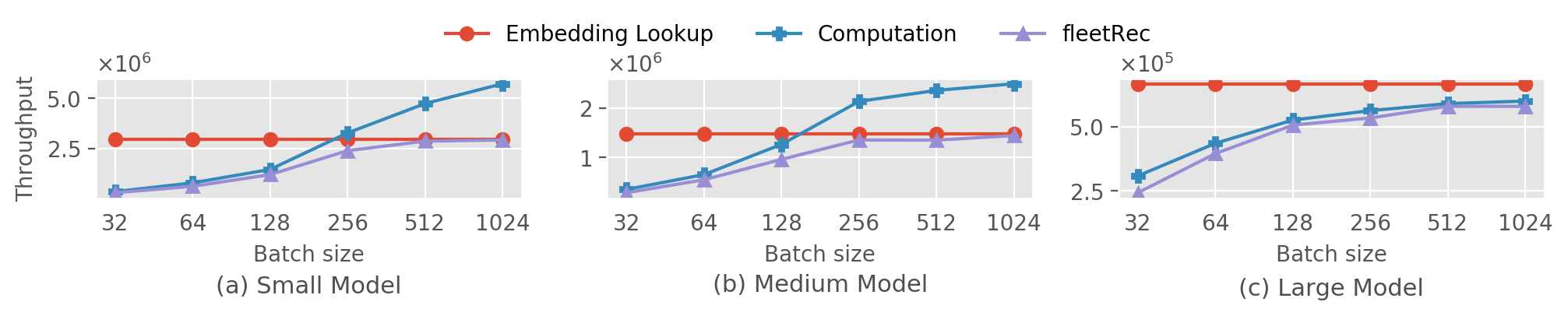}
  \caption{The performance of FleetRec can be estimated by taking the minimum of the lookup and computation module.}
  \label{fig_fleetrec:performance_estimation}
\end{figure*}

\begin{figure}[t]
  \centering
  \includegraphics[width=0.5\linewidth]{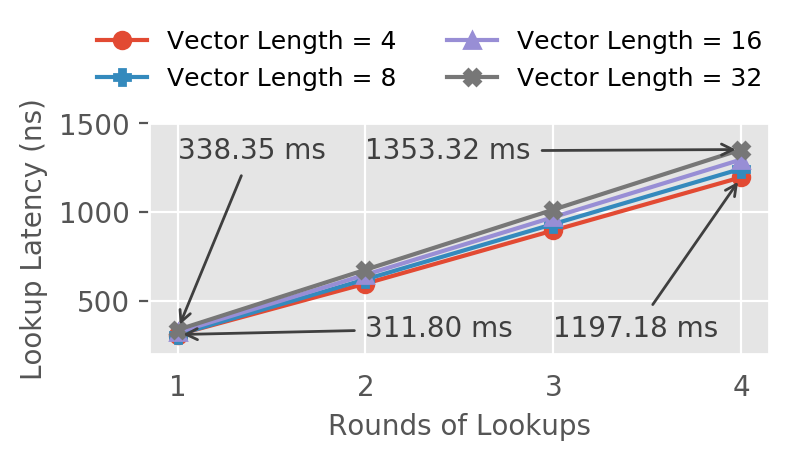}
  \caption{The embedding lookup performance can be estimated given the DRAM access rounds.}
  \label{fig_fleetrec:embedding_performance}
\end{figure}

\textbf{FleetRec shows more than one order of magnitude speedup in terms of inference throughput over the CPU baseline.} 
In Figure~\ref{fig_fleetrec:throughput}, I compare the throughput of CPUs, FPGAs, and FleetRec using batch sizes ranging from 32 to 1024 (larger batch sizes can violate the SLA of 10 ms). The throughput of the CPU-baseline and FleetRec increases with the batch size, while the throughput of the FPGA accelerators remains constant because the dataflow architecture used in the FPGA processes inference item by item instead of in batches. According to the peak throughput on the three models, FleetRec achieves 15.5$\sim$49.0$\times$ speedup over the CPU baseline, and 7.4$\sim$16.1$\times$ over the FPGAs. The significant speedup over FPGAs justifies introducing heterogeneous hardware in the form of a GPU in addition to the FPGAs. FleetRec exhibits even higher speedups over CPUs due to both the fast DNN computation enabled by the GPU and the highly concurrent embedding lookups provided by FPGAs equipped with HBM.
The speedups over CPUs on small and medium models are more significant than in the large one, because I use a more powerful CPU server (with twice as many cores and memory channels as well as the latest micro-architecture design) for large model inference, while only one GPU is deployed in FleetRec across the three experiments. 

\textbf{FleetRec exhibits much lower latency compared to CPUs.} 
As shown in Figure~\ref{fig_fleetrec:latency}, the latency reduction achieved by Fleet given the same batch sizes ranges from 21.0
\% to 92.5\% with an average of 76.6\%.
Note that FleetRec can perform single-millisecond-level inference on small and medium models using medium batch sizes (e.g., 256), while the CPU needs around 10 ms. Using a medium batch size of 256 already allows FleetRec to achieve close-to-peak throughput (around 85\% of the peak performance), while for the CPU only achieves around half of the throughput reached with batch sizes of 1024. Though still better than the CPU baseline, the latency of FleetRec on the large model is higher compared with the medium and small ones, because (a) the network traffic to transfer a batch of inputs of the large model is far heavier due to the feature length as shown in Table~\ref{tab_fleetrec:model_size}; and (b) each thread on the GPU server needs to maintain individual connections with several memory nodes and receive data from them in a round-robin manner.

\textbf{For real-time recommendations under strict SLA (latency) constraints, the speedup of FleetRec is even more significant.} Achieving high throughput and guaranteeing low latency can be contradictory on CPU-based inference engines, since throughput is increased by employing large batches. 
Table~\ref{tab_fleetrec:throughput_under_SLA} presents the throughput of several systems under different SLAs by translating the data presented in Figure~\ref{fig_fleetrec:throughput} and~\ref{fig_fleetrec:latency}. 
Given a latency constraint of 10 ms, FleetRec can achieve  41.8$\sim$387.2$\times$ speedup in throughput, even more significant than the 15.5$\sim$49.0$\times$ achieved when not considering the latency constraint.

\subsection{Generalizing and Configuring the System}

\textbf{FleetRec can be generalized and configured beyond the three industry models:} given any recommendation models, one can (a) decide how to configure the system (number of GPUs, FPGAs, and CPU memory nodes); and (b) estimate FleetRec's performance, without the need to implement the hardware at all.
To prove this, I first show that the performance of FleetRec can be estimated once the performance of the computation and embedding lookup modules is known. I then present how to estimate the performance of the two components without actually implementing them.

The performance of FleetRec can be estimated as shown in Figure~\ref{fig_fleetrec:performance_estimation}. Once the performance of the computation and embedding lookup modules is known, I can estimate the throughput of FleetRec under different batch settings by taking the minimum throughput of the two components. There is a tolerable performance gap between FleetRec and the lower performance bound of the two components, because FleetRec involves network while the performance of the computation and embedding lookup modules is tested without network.

The performance of both computation and embedding lookup modules can be estimated easily without the need to implement hardware. 
On the GPU side, one can resort to existing ML programming frameworks, e.g., TensorFlow, PyTorch, or MXNex, and remove the embedding layer to test only the DNN computation performance on a GPU.
On the FPGA side, the lookup performance is decided by DRAM (HBM and DDR) access rounds and influenced by embedding vector lengths. As shown in Figure~\ref{fig_fleetrec:embedding_performance}, given the same embedding vector length, the lookup latency is proportional to the rounds of DRAM access. The lookup latency of different vector lengths are very close (within 40 ns per round): the embedding vectors are short and cannot fully take advantage of the spatial locality within the DRAM, thus these accesses are almost random, and the FPGA only needs to pay a few more clock cycles to read a longer vector. 
Note that the embedding lookup is issued item by item no matter what the batch size is, thus the throughput is simply the reciprocal of the latency shown in Figure~\ref{fig_fleetrec:embedding_performance}. 
This predictability allows us to estimate the embedding lookup performance given a recommendation model without actually implementing it. For example, given a model with 90 tables and 30 of them small enough to be stored on-chip, I can allocate the rest 60 tables to DRAM (28 available HBM channels plus 2 DDR channels), and the FPGA can finish the lookup process with 2 rounds of DRAM access as long as the model size is within the capacity of the DRAM. This estimation also works for multi-FPGA lookup modules. I first estimate the performance of each individual nodes, and the lookup performance is bound by the one with the lowest throughput.

Once the performance of computation and embedding lookups is known, one can configure FleetRec in a way that maximizes performance while minimizing resource usage by balancing the performance of the two components. For example, one can couple an FPGA node with several GPUs for computation-intensive models. On the contrary, the design employing multiple FPGAs and a single GPU could fit models with many embedding tables and a set of light-weight DNN layers.



\section{Conclusion}


To popularize FleetRec in a wide range of deployments, I expect an end-to-end development flow from ML frameworks to hardware to be very useful. Though showing attractive speedups, FleetRec involves manually optimized hardware design for each individual recommendation models. This requires seamless collaboration between an ML team and a hardware team. Fortunately, the hardware logic of embedding table lookups follows a rather fixed pattern, thus it is possible to prepare a set of hardware code templates, and compile the look up logic to hardware using an automated code generator. 
Once this can be achieved, one can integrate FleetRec to existing ML frameworks such as TensorFlow and PyTorch, allowing an end-to-end development experience for ML engineers, who can then focus on the DNN architecture design and can deploy this high-performance system with a single button.


\chapter{Conclusions}

\section{Summary}


This thesis explores the efficiency of machine learning systems in response to two major trends.  
First, modern machine learning systems extend beyond the training and serving of a single model --- vector data systems are becoming increasingly essential in various AI workloads.  
Second, with the end of Moore's Law, modern hardware architectures are becoming increasingly heterogeneous, introducing new challenges and opportunities for system optimization.  
To address these developments, \textbf{this thesis focuses on improving the efficiency of vector-centric machine learning systems across the computing stack, investigating the intricate interplay between algorithms, systems, and hardware.} In summary:

Chapter~\ref{chap:rago} introduces \textit{RAGO}~\cite{jiang2025rago}, the first systematic study on performance optimization for RAG serving.  
To structure the complex RAG algorithm landscape, \textit{RAGO} introduces \textit{RAGSchema}, an abstraction that encapsulates performance-related attributes of RAG pipelines.  
Through case studies on various RAG paradigms, \textit{RAGO} highlights how workload characteristics influence system design.  
Finally, \textit{RAGO} implements a scheduling framework that optimizes task placement, resource allocation, and batching policies to enhance performance across different RAG configurations.

Chapter~\ref{chap:chameleon} introduces \textit{Chameleon}~\cite{jiang2023chameleon}, a heterogeneous accelerator system designed for efficient RAG serving.  
This work is driven by two key insights: (1) LLM inference and vector search exhibit fundamentally different computational patterns, and (2) large-scale retrieval often becomes a bottleneck in RAG pipelines, particularly with the rapid evolution of model accelerators.  
To address these challenges, Chameleon pairs specialized retrieval accelerators for vector search with GPUs for LLM inference in a disaggregated architecture, optimizing both components for their respective workloads, while enabling adaptable resource allocation for inference- and retrieval-heavy workloads.

Chapter~\ref{chap:piperag} investigates algorithm-level optimizations for RAG serving with \textit{PipeRAG}~\cite{jiang2024piperag}, the first approach to co-design iterative retrieval-augmented generation across algorithms and systems.  
Frequent retrievals from large databases can introduce stalls, leaving ML accelerators underutilized while awaiting data. To address this inefficiency, \textit{PipeRAG} introduces several key optimizations.  
It employs approximate data prefetching to enable retrieval and generation to run concurrently, reducing latency and improving throughput. Additionally, it adjusts retrieval intervals dynamically to maximize pipeline efficiency based on workload demands. Finally, a performance model balances retrieval quality and latency by adapting to the current generation state and hardware constraints.

Chapter~\ref{chap:fanns} presents \textit{FANNS}~\cite{jiang2023co}, an algorithm-hardware co-design framework for optimizing IVF-PQ, a widely used large-scale vector search algorithm. Due to the vast design space of IVF-PQ accelerators, \textit{FANNS} leverages FPGA reconfigurability to explore different hardware-algorithm configurations. Given a dataset, target recall requirement, and FPGA device, \textit{FANNS} automatically selects the optimal parameter-hardware combination and generates a deployable accelerator.  

Chapter~\ref{chap:falcon} presents \textit{Falcon}~\cite{jiang2024accelerating}, leveraging both algorithm- and hardware-level solutions to enable low-latency graph-based vector search. Due to the fine-grained nature of graph traversal, achieving high performance requires both algorithmic enhancements and specialized hardware. In addition to developing the \textit{Falcon} accelerator, this work introduces \textit{delayed-synchronization traversal} (DST), a hardware-efficient traversal algorithm that improves both search speed and recall by relaxing strict traversal order, thereby maximizing accelerator utilization.  

Chapter~\ref{chap:microrec} presents \textit{MicroRec}~\cite{jiang2021microrec}, a system that accelerates recommendation inference by optimizing embedding data structures and utilizing a heterogeneous memory hierarchy.  
\textit{MicroRec} restructures the data layout of the embedding tables to minimize lookup operations and intelligently distributes embedding tables across different memory tiers, including DDR memory, HBM, and SRAM. By aligning table placement with access frequency, \textit{MicroRec} ensures that frequently used embeddings reside in high-speed memory while less critical embeddings are stored more cost-effectively. The system is implemented on an FPGA platform, supporting both embedding lookups and end-to-end inference.

Chapter~\ref{chap:fleetrec} introduces \textit{FleetRec}~\cite{jiang2021fleetrec}, a high-performance, configurable heterogeneous computing cluster for recommendation inference. While \textit{MicroRec} accelerates embedding lookups, the FPGA-based deep neural network (DNN) inference stage becomes a new bottleneck. 
To address this, \textit{FleetRec} employs an accelerator heterogeneity strategy, combining FPGAs for efficient embedding table lookups with GPUs dedicated to DNN computation.  
The accelerators are connected via the network, enabling flexible resource allocation based on the specific recommendation workloads.

Overall, this thesis provides a comprehensive approach to optimizing vector-centric machine learning systems through cross-layer optimizations spanning algorithms, system design, and hardware architecture.
The principles and methodologies developed here lay a solid foundation for future machine learning system designs.

\section{Future Work}

We are at the early stages of an era where designing compound machine learning systems, optimizing vector data systems, and developing specialized hardware are becoming increasingly essential. The research presented in this thesis is only a step toward addressing these challenges, and numerous open problems remain to be explored. Some potential research directions include:

\textbf{Performance-Quality Pareto Frontier in RAG Systems.}  
While this thesis has extensively studied RAG performance optimization, RAG is still in its infancy. Currently, machine learning researchers and systems researchers largely work independently --- ML researchers focus on scaling generation quality through algorithmic innovations, while systems researchers are beginning to explore performance optimization. However, the key to practical deployment in large-scale data centers lies in identifying the optimal balance between performance and quality, rather than focusing solely on one aspect.

A major challenge in this direction is the vast number of tunable parameters at the algorithmic level. For instance, should we scale the size of knowledge databases, the number of retrieved documents, or the retrieval frequency? Alternatively, should we continue increasing model sizes? Each of these choices impacts both system performance and generation quality in different ways. Furthermore, these factors evolve independently over time --- models improve, retrieval algorithms advance, and hardware capabilities expand. As a result, navigating this performance-quality trade-off remains an evolving and complex research problem, with significant opportunities for future work.

\textbf{Scalability Challenges in Large-Scale Vector Data Systems.}  
Currently, academic research on vector search primarily focuses on datasets under a terabyte. As a result, most retrieval frameworks assume a single-node setup, where data can fit in main memory or even accelerators. Looking ahead, as machine learning models increasingly connect to vast external knowledge sources, the scale of retrieval systems will expand by several orders of magnitude beyond what is studied today. This shift necessitates a fundamental rethinking of how retrieval algorithms and systems should be designed at such unprecedented scales.  

Several key challenges emerge in this context: how should data be indexed efficiently across a large-scale infrastructure? How can retrieval performance be optimized while balancing the cost of maintaining large indexes? How should workloads be distributed across hundreds of nodes to ensure efficient query execution? Moreover, with the growing complexity of modern memory hierarchies—including CPU caches, main memory, CXL memory pools, and persistent storage --- how should database vectors be mapped to different levels of memory to optimize both performance and cost? These are open research problems that remain largely unexplored.

\textbf{Hardware Specialization and Algorithm Co-Design.}  
Hardware specialization is not simply a matter of implementing existing algorithms on new hardware. Instead, we should actively think about the interplay between algorithmic innovations and hardware advancements, ensuring that current hardware limitations do not restrict future algorithmic breakthroughs.  

A notable example is product quantization (PQ), a widely used retrieval algorithm designed around the constraints of conventional CPU architectures. PQ compresses database vectors by approximating multiple dimensions using a fixed one-byte representation. However, this approach imposes uniform compression across all dimensions, despite the fact that data distributions vary, and certain dimensions may hold more critical information than others. To improve retrieval quality within the same memory footprint, algorithms could adopt distribution-aware, variable-precision quantization, such as 5-bit or 9-bit encodings, rather than adhering to rigid, byte-aligned compression. While current hardware lacks support for such flexible encoding, this should not dictate the future of algorithm design. 
Instead, future research should explore how emerging hardware can enable more efficient algorithms that break away from existing architectural constraints.

\cleardoublepageempty

%
%
%
%
\pdfbookmark{Lists of Tables}{lot}
\listoftables
\cleardoublepageempty
%
\pdfbookmark{Lists of Figures}{lof}
\listoffigures
\cleardoublepageempty
%
%
\pdfbookmark{Bibliography}{book:bibliography}
\bibliographystyle{plainnat}  
\bibliography{ref_merged}
\cleardoublepageempty
%


\clearpage
\cleardoublepageempty
\end{document}